# Analyzing Non-Textual Content Elements to Detect Academic Plagiarism

Doctoral thesis for obtaining the academic degree of

Doctor of Engineering Sciences (Dr.-Ing.)

submitted by

Norman Meuschke

at the

Universität Konstanz

Faculty of Sciences
Department of Computer and Information Science

Date of the oral examination: March 5, 2021

1st Reviewer: Prof. Dr. Bela Gipp
2nd Reviewer: Prof. Dr. Harald Reiterer
3rd Reviewer: Prof. Dr. Michael L. Nelson

Konstanz, 2021

To my family

# Contents

# List of Figures

**Appendices**

# List of Tables

**Appendices**

# Abstract

Identifying academic plagiarism is a pressing problem, among others, for research institutions, publishers, and funding organizations. Detection approaches proposed so far analyze lexical, syntactical, and semantic text similarity. These approaches find copied, moderately reworded, and literally translated text. However, reliably detecting disguised plagiarism, such as strong paraphrases, sense-for-sense translations, and the reuse of non-textual content and ideas, is an open research problem.

The thesis addresses this problem by proposing plagiarism detection approaches that implement a different concept—analyzing non-textual content in academic documents, such as citations, images, and mathematical content.

The thesis makes the following research contributions.

It provides the most extensive literature review on plagiarism detection technology to date. The study presents the weaknesses of current detection approaches for identifying strongly disguised plagiarism. Moreover, the survey identifies a significant research gap regarding methods that analyze features other than text.

Subsequently, the thesis summarizes work that initiated the research on analyzing non-textual content elements to detect academic plagiarism by studying citation patterns in academic documents.

To enable plagiarism checks of figures in academic documents, the thesis introduces an image-based detection process that adapts itself to the forms of image similarity typically found in academic work. The process includes established image similarity assessments and newly proposed use-case-specific methods.

To improve the identification of plagiarism in disciplines like mathematics, physics, and engineering, the thesis presents the first plagiarism detection approach that analyzes the similarity of mathematical expressions.

To demonstrate the benefit of combining non-textual and text-based detection methods, the thesis describes the first plagiarism detection system that integrates the analysis of citation-based, image-based, math-based, and text-based document similarity. The system's user interface employs visualizations that significantly reduce the effort and time users must invest in examining content similarity.

To validate the effectiveness of the proposed detection approaches, the thesis presents five evaluations that use real cases of academic plagiarism and exploratory searches for unknown cases. Real plagiarism is committed by expert researchers with strong incentives to disguise their actions. Therefore, I consider the ability to identify such cases essential for assessing the benefit of any new plagiarism detection approach. The findings of these evaluations are as follows.

Citation-based plagiarism detection methods considerably outperformed text-based detection methods in identifying translated, paraphrased, and idea plagiarism instances. Moreover, citation-based detection methods found nine previously undiscovered cases of academic plagiarism.

The image-based plagiarism detection process proved effective for identifying frequently observed forms of image plagiarism for image types that authors typically use in academic documents.

Math-based plagiarism detection methods reliably retrieved confirmed cases of academic plagiarism involving mathematical content and identified a previously undiscovered case. Math-based detection methods offered advantages for identifying plagiarism cases that text-based methods could not detect, particularly in combination with citation-based detection methods.

These results show that non-textual content elements contain a high degree of semantic information, are language-independent, and largely immutable to the alterations that authors typically perform to conceal plagiarism. Analyzing non-textual content complements text-based detection approaches and increases the detection effectiveness, particularly for disguised forms of academic plagiarism.

# Kurzfassung

Die Erkennung wissenschaftlicher Plagiate ist, unter anderem, für Forschungseinrichtungen, Verlage und Forschungsförderer ein dringliches Problem. Bislang vorgestellte Erkennungsansätze analysieren lexikalische, syntaktische und semantische Textähnlichkeit. Diese Ansätze erkennen kopierten, mäßig umformulierten und wörtlich übersetzten Text. Die zuverlässige Erkennung von verschleierten Plagiaten wie starken Paraphrasen, sinngemäßen Übersetzungen und der Übernahme von nicht-textuellen Inhalten und Ideen ist jedoch ein ungelöstes Forschungsproblem.

Die vorliegende Dissertation adressiert dieses Problem durch die Vorstellung von Erkennungsansätzen, die ein neues Konzept umsetzen — sie analysieren nicht-textuelle Inhalte in wissenschaftlichen Dokumenten, wie zum Beispiel Zitate, Abbildungen und mathematische Ausdrücke.

Die Dissertation leistet die folgenden Forschungsbeiträge.

Sie präsentiert den bislang umfangreichsten Literaturüberblick zu Technologien für die Plagiatserkennung. Die Literaturstudie stellt die Schwächen derzeitiger Erkennungsverfahren bei der Identifizierung verschleierter Plagiate heraus. Darüber hinaus zeigt die Literaturanalyse eine signifikante Forschungslücke auf. Diese besteht im Hinblick auf Verfahren, die andere als textuelle Inhaltsmerkmale analysieren.

Im Anschluss fasst die Dissertation Arbeiten zusammen, welche die Forschung zur Analyse nicht-textueller Inhalte zur Erkennung wissenschaftlicher Plagiate initiierten, indem sie Zitatmuster in wissenschaftlichen Dokumenten untersuchen.

Um eine Plagiatsüberprüfung für Abbildungen in wissentlichen Dokumenten zu ermöglichen, präsentiert die Dissertation einen bildbasierten Erkennungsprozess, der sich an typische Formen der Bildähnlichkeit in wissenschaftlichen Arbeiten anpasst. Der Prozess integriert etablierte Verfahren zur Bildähnlichkeitserkennung sowie neu vorgeschlagene Methoden für den spezifischen Anwendungsfall.

Um die Erkennung von Plagiaten in Disziplinen wie Mathematik, Physik und Ingenieurwissenschaften zu verbessern, präsentiert die Arbeit den ersten Erkennungsansatz, der die Ähnlichkeit mathematischer Ausdrücke analysiert.

Um die Vorteilhaftigkeit einer Kombination von nicht-textuellen und textbasierten Erkennungsverfahren zu demonstrieren, beschreibt die Arbeit das erste Plagiatserkennungssystem, das die Analyse zitat-, bild-, mathematik- und textbasierter Dokumentähnlichkeit vereint. Die Benutzeroberfläche des Systems verwendet

Visualisierungen, die den Aufwand und die Zeit, die Benutzer investieren müssen, um ähnliche Inhalte zu untersuchen, erheblich reduzieren.

Um die Wirksamkeit der vorgestellten Erkennungsansätze zu validieren, präsentiert die Arbeit fünf Evaluationen basierend auf der Analyse bestätigter Plagiatsfälle und explorativer Suchen nach bislang unbekannten Fällen. Reale wissenschaftliche Plagiate werden von erfahrenen Forschern begangen, die starke Anreize haben ihre Handlung zu verschleiern. Für mich ist die Fähigkeit solche Fälle identifizieren zu können daher ein essenzielles Kriterium, um den Nutzen neuer Plagiatserkennungsansätze zu beurteilen. Die Ergebnisse der Evaluationen sind wie folgt.

Zitatbasierte Plagiatserkennungsverfahren erzielten deutlich bessere Ergebnisse als textbasierte Verfahren bei der Identifikation von Paraphrasen, Übersetzung- und Ideenplagiaten. Zudem fanden zitatbasierte Erkennungsverfahren neun zuvor unentdeckte Fälle wissenschaftlicher Plagiate.

Der bildbasierte Plagiatserkennungsprozess erwies sich als wirksam für die Identifikation häufiger Plagiatsformen für eine Vielzahl von Bildtypen, die typischerweise in wissenschaftlichen Dokumenten verwendet werden.

Mathematikbasierte Erkennungsverfahren konnten bestätigte Plagiatsfälle, in denen mathematische Inhalte übernommen wurden, zuverlässig finden und zusätzliche einen zuvor unbekannten Fall identifizieren. Mathematikbasierte Verfahren boten Vorteile für die Erkennung von Plagiaten, die textbasierte Methoden nicht erkennen konnten, besonders in Kombination mit zitatbasierten Verfahren.

Diese Ergebnisse zeigen, dass nicht-textuelle Inhalte ein hohes Maß an semantischen Informationen enthalten, sprachunabhängig und weitgehend robust sind gegenüber Änderungen, die Autoren typischerweise vornehmen, um Plagiate zu verbergen. Die Analyse von nicht-textuellen Inhalten ergänzt textbasierte Erkennungsansätze und erhöht die Erkennungsleistung, insbesondere für verschleierte Plagiatsformen.

# Acknowledgments

My doctoral research period has been the most memorable, rewarding, and, at times, challenging experience in my life so far. I want to thank the numerous individuals and organizations that supported me throughout this endeavor.

I am tremendously grateful to my doctoral advisor Bela Gipp who has inspired, guided, and supported my professional and personal development like no other person during that time. Our collaboration led us to four research institutions on three continents, resulted in a startup enterprise, and has become more fruitful and personally rewarding at every step of the way. I wholeheartedly thank Bela for his trustful, kind, and honest mentorship, motivating, unshakeable positivity in any situation, and steadfast support throughout the years.

Moreover, I emphatically thank Harald Reiterer, Michael Grossniklaus, and Michael L. Nelson for supporting my research and serving as the members of my doctoral committee. Harald, Michael, and the members of their research groups helped my colleagues and me immensely with settling in at the University of Konstanz. By co-advising multiple student projects, enabling collaborations in research and teaching, and providing valuable feedback on my dissertation proposal, Harald and Michael significantly contributed to the completion of this thesis. I am very grateful that Michael L. agreed to serve as the third reviewer of my thesis on short notice and highly appreciate his helpful feedback.

Likewise, I am indebted to Akiko Aizawa from the National Institute of Informatics Tokyo and Jim Pitman from the University of California, Berkeley. By inviting me to complete research stays in Tokyo and Berkeley, Akiko and Jim enabled two immensely valuable experiences from a professional and a personal perspective. I greatly enjoyed collaborating with Akiko and Jim and value the feedback they provided. Furthermore, I thank Akiko for hosting several students working on projects related to our shared research interests.

I extend my sincere appreciation to all the bright and passionate colleagues from all over the globe with whom I had the distinct pleasure of working over the years. I particularly thank Moritz Schubotz and Vincent Stange for their considerable contributions to Math-based Plagiarism Detection and the HyPlag prototype. Furthermore, I express my sincere gratitude to Corinna Breitinger and Tomáš Foltýnek, with whom I collaborated on multiple publications summarized in this thesis. I thank Joeran Beel, Howard Cohl, André Greiner-Petter, Cornelius Ihle, Felix Hamborg, Terry Lima Ruas, Malte Ostendorff, Philipp Scharpf, Timo Spinde,

Johannes Stegmüller, and Jan Philip Wahle for the opportunity to jointly work on exciting projects partially related to my doctoral research.

I am just as thankful for having had the opportunity to advise numerous students at the Universities of Berkeley, Konstanz, and Magdeburg, the University of Applied Sciences HTW Berlin, and the NII Tokyo. I gratefully acknowledge their contributions to the image-based approach to plagiarism detection, the HyPlag prototype, and other research projects not discussed in this thesis.

I am grateful to my many friends and relatives who kept in touch, rooted for me, and offered support regardless of the distance that often separated us. I cannot name everyone here, but I want them to know that I greatly appreciate all of them and everything they did for me.

My sincere gratitude also goes to several organizations that provided considerable financial support for the research presented in this thesis. I thank the German Research Foundation (DFG) for funding two projects on mathematical information retrieval and Math-based Plagiarism Detection that complement and extend the research presented in this thesis. My thanks in this regard also extend to the Young Scholar Fund (YSF) at the University of Konstanz. The YSF graciously funded the preliminary research conducted by Moritz Schubotz and me that enabled the successful application for the two DFG grants. Furthermore, I am deeply indebted to the German Academic Exchange Service (DAAD) for enabling my research stays at UC Berkeley and NII Tokyo, several conference participations, and research stays at NII Tokyo for several students with whom I worked. I thank the ACM Special Interest Group on Information Retrieval (SIGIR) for supporting multiple of my conference participation. My particularly heartfelt gratitude goes to SIGIR for choosing me as one of their student ambassadors to represent the community at the 50$^{th}$ Turing Award Ceremony in San Francisco, 2017. Meeting some of the greatest minds in computing was a once-in-lifetime experience.

I reserve my last and most important words of gratitude for the people closest to my heart. I am deeply grateful to my parents Christina & Jürgen, my grandparents Inge & Horst, and Waltraud & Hans, for their unconditional love, constant encouragement to pursue my dreams, and absolute support for making these dreams come true. I am immensely grateful to my sister Monique for being a continuous inspiration to me and someone I trust and rely on entirely. I thank my partner Anastasia for being the smart, loving, carrying, and incredibly positive person she is. I cannot count the times she went out of her way to help or simply do something nice to cheer me up during the stressful times of my doctoral research.

I am fortunate to have that many wonderful people in my life. I dedicate this thesis to my family and the people being a part of it without sharing the same blood.

1

Chapter 1

# Introduction

## Contents

This thesis improves the capabilities to detect disguised plagiarism in academic documents, which is an open research challenge in Information Retrieval and a pressing day-to-day problem, e.g., for publishers, research institutions, and funding organizations. Section 1.1 describes the problems arising from academic plagiarism. Section 1.2 summarizes the research gap that exists regarding the detection of disguised plagiarism forms. Section 1.3 presents the research objective and research tasks that guided my doctoral research. Section 1.4 outlines the presentation of my research in this thesis and presents the publications in which I shared my findings with the research community.

## 1.1 Problem

Academic plagiarism describes the use of ideas, words, or other work without appropriately acknowledging the source to benefit in a setting where originality is expected [138, p. 5], [173, p. 10]. Forms of academic plagiarism include copying content, reusing slightly modified content (e.g., mixing text from multiple sources), disguised content reuse (e.g., by paraphrasing or translating text), and reusing data or ideas without attribution [551, p. 6ff.]. The easily recognizable copy-and-paste type plagiarism is prevalent among students [331, p. 5ff.]. Disguised plagiarism is characteristic of researchers who have strong incentives to avoid detection.

Plagiarism is a severe form of research misconduct that has substantial negative impacts on academia and the public. Plagiarized research publications damage the scientific process by impeding the traceability of ideas, verification of assertions, replication of experiments, and correction of results [551, p. 22]. If researchers expand or revise earlier results, publications that plagiarized the original remain unaffected. Incorrect findings can spread and affect subsequent research or practical applications [173, p. 219f.]. For example, in medicine and pharmacology, systematic reviews of the literature are a vital tool to assess the efficacy and safety of medical drugs and treatments. Plagiarized publications can skew systematic reviews and thereby jeopardize patient safety [123, p. 974], [506, p. 638].

Furthermore, academic plagiarism wastes considerable resources. Even in the best case—if the plagiarism is discovered—reviewing and sanctioning plagiarized research publications, dissertations, and grant applications still cause a high effort for the examiners, affected institutions, and funding agencies.

The projects VroniPlag [403] in Germany and Dissernet [579] in Russia provide insights into the effort required for investigating plagiarism allegations and the prevalence of plagiarism among postgraduate researchers. Both projects are crowdsourced efforts of volunteers who analyze doctoral and postdoctoral theses for potential plagiarism. By July 2020, VroniPlag has published reports on 207[1] [538] and Dissernet on 652[2] [580] doctoral and habilitation theses, in which the volunteers found strong evidence of plagiarism. The investigations show that effective plagiarism checks of research-based theses often require dozens or even hundreds of work hours from the affected institutions. Both projects found plagiarism in theses from nearly all disciplines. Examiners had graded many of the theses that contained plagiarism as excellent or even exceptional.

Evidence also indicates that journals spent a significant amount of their resources on plagiarized manuscripts. The non-profit project Retraction Watch reports on retracted articles in scientific journals [377]. By July 2020, the Retraction Watch database contains 2,375 retractions issued due to plagiarism (11% of the 21,936 retractions in the database) [427]. Additionally, several studies, e.g., the References [47], [227], [333], [572], report on the findings of journals that routinely screened manuscripts using commercial text-matching software. Those studies found that 10%–20% of the submitted manuscripts contained unacceptable levels of text overlap. Even more alarming than the text overlap itself is that many of those manuscripts were published, nevertheless. Higgins et al. traced 57 manuscripts in which one journal had identified "unacceptable levels of plagiarized material" [227, p. 1] and rejected the submission. The authors found that 37 of those 57 manuscripts (65%) subsequently appeared in other journals. Of those 37 published articles, 34 still exhibited the text overlap identified previously [227, p. 4].

---

[1] In 80 of the 207 cases, the responsible universities rescinded the academic degree, or the author returned the degree after the unoriginal content was found. In 37 cases, the universities did not rescind the degree but often issued reprimands or other sanctions. In 90 cases, official investigations are either pending or the universities did not publish the results of their investigations [538].

[2] In 426 of the 652 cases, the responsible institutions rescinded the academic degree, in 168 cases the institutions did not rescind the degree, and in 58 cases investigations are pending [580].

If academic plagiarism remains undiscovered, the adverse effects are even more severe. Plagiarists can unduly receive research funds and career advancements as funding organizations may award grants for plagiarized ideas or accept plagiarized publications as the outcomes of research projects. The inflation of publication and citation counts through plagiarism can aggravate the problem. Studies showed that some plagiarized publications receive at least as many citations as the original [307, p. 1293]. This phenomenon is problematic as citation counts are widely-used indicators of research performance and often influence funding and hiring decisions.

From an educational perspective, academic plagiarism is detrimental to competence acquisition and assessment. Practicing is crucial to human learning. If students receive credit for work that others did, a critical extrinsic motivation for acquiring knowledge and competencies is reduced. Likewise, the assessment of competence is distorted, which can result in undue career benefits for plagiarists.

The rapid advancement of information technology, which offers convenient access to vast amounts of information, has made plagiarizing easier than ever. Given the enormous number of potential sources, systems that support plagiarism identification have become vital tools for safeguarding the scientific process and preventing the harmful effects of academic plagiarism.

## 1.2 Research Gap

Devising systems that support the detection of academic plagiarism has attracted extensive research. Most methods that researchers proposed for this task analyze the lexical, syntactic, and semantic similarity of text. These methods reliably identify copied and slightly edited text. Some methods can retrieve word-by-word translations. For strong paraphrases and sense-for-sense translations, the effectiveness of current methods is too low to be helpful in practice. Identifying the reuse of non-textual content and ideas is an unsolved research problem.

For production-grade plagiarism detection systems available to the public, the situation is even worse. All of those systems exclusively search for identical text [145].

Weber-Wulff, who performs regular performance evaluations of such systems, summarizes their capabilities as follows:

> [...] **PDS find copies, not plagiarism** [548, p. 6]
> [...] **for translations or heavily edited material,**
> **the systems are powerless** [...] [549]

I expect the limitations of current plagiarism detection systems result in a significant fraction of today's disguised academic plagiarism remaining undetected.

For economic reasons, commercial plagiarism detection systems focus on identifying the more easily recognizable copy-and-paste type plagiarism prevalent among students, despite the potentially severe consequences of disguised plagiarism in research publications. The higher number of students compared to researchers, the higher frequency of plagiarism among students than among researchers, and the availability of well-established, efficient methods to find verbatim text reuse currently make student plagiarism a more profitable market segment. Developing methods to identify disguised plagiarism committed by researchers requires research effort that the providers of commercial PDS currently avoid.

By openly providing methods to detect disguised academic plagiarism, I seek to advance the research on this applied problem and make tackling the problem more attractive for commercial providers.

## 1.3 Research Objective

The following **objective** guided my doctoral research:

**Devise, implement, and evaluate automated approaches capable of identifying previously non-machine-detectable forms of disguised academic plagiarism.**

To achieve my research objective, I derived the following four **research tasks**:

**Task 1** Identify the strengths and weaknesses of state-of-the-art methods and systems to detect academic plagiarism.

**Task 2** Devise detection approaches that address the identified weaknesses.

**Task 3** Evaluate the effectiveness of the proposed detection approaches.

**Task 4** Implement the proposed detection approaches in a plagiarism detection system capable of supporting realistic detection use cases.

## 1.4 Thesis Outline and Prior Publications

**Chapter 1** presents the problem of academic plagiarism, identifies the research gap that motivated this thesis, and describes how the thesis addresses the research objective and the four research tasks derived from the identified research need.

**Chapter 2** introduces the reader to relevant related work and derives the research idea of analyzing non-textual content elements in addition to textual features pursued in this thesis. The chapter addresses **Research Task 1**, i.e., identifies the strengths and weaknesses of current plagiarism detection approaches by presenting the most extensive review of research publications on the topic to date.

**Chapter 3** summarizes Citation-based Plagiarism Detection (CbPD)—the first detection approach that analyzed non-textual content elements. The chapter presents how the examinations of confirmed plagiarism cases guided the design of the citation-based detection methods. Large-scale retrieval experiments and a user study show CbPD's effectiveness and efficiency in identifying disguised forms of academic plagiarism and the positive effect CbPD has on user effort required for examining retrieved documents.

**Chapter 4** presents an image-based plagiarism detection process that exclusively analyzes figures in academic documents. The process combines novel contributions, such as new image analysis and scoring methods, with existing content-based image retrieval methods. An evaluation using confirmed cases of image plagiarism demonstrates the retrieval effectiveness of the process.

**Chapter 5** introduces a novel plagiarism detection approach that, for the first time, analyzes the similarity of mathematical expressions. Starting from an analysis of confirmed cases of plagiarism, the chapter describes the conceptualization of math-based detection methods and their integration with citation-based and text-based detection methods as part of a two-stage detection process. An evaluation using confirmed cases of plagiarism shows the approach's ability to retrieve many real-world examples. In an exploratory analysis of publications, the novel approach identified a previously unidentified case of reused mathematics considered an instance of plagiarism by the original author.

Chapters 3–5 jointly address **Research Task 2** and **Research Task 3**.

**Chapter 6** addresses **Research Task 4** by presenting the first plagiarism detection system that jointly analyses the similarity of academic citations, images, mathematical expressions, and text.

**Chapter 7** summarizes the research contributions of this thesis and gives an outlook on future research.

### 1.4.1 Publications

To subject my research to the scrutiny of peer review, I have published most of the content in this thesis in the publications listed in **Table 1.1**. The first column of the table shows in which chapters I reuse content from these publications. Additionally, **Table 1.2** lists publications partially related to the research presented in this thesis. For example, such publications address related information retrieval tasks, such as literature recommendation or news analysis. To indicate the rigor of the peer-review process, the second-to-last column in both tables shows the venue's rating using two widely accepted rankings. For conference publications, the table states the venue category in the Computing Research & Education (CORE) Ranking[3]. For journal articles, the table shows the Scimago Journal Rank (SJR)[4].

To acknowledge the fellow researchers with whom I published, collaborated, and discussed ideas, I will use "we" rather than "I" in the remainder of this thesis.

**Table 1.1.** Overview of core publications summarized in this thesis.

| Ch. | Venue | Year | Type | Length | Author Position | Venue Rating | Ref. |
|---|---|---|---|---|---|---|---|
| 1, 2 | CSUR | 2019 | Journal | Full | 2 of 3 | CORE A* | [140] |
| | IJEI | 2013 | Journal | Full | 1 of 2 | n/a | [338] |
| 3 | JASIST | 2014 | Journal | Full | 2 of 3 | SJR Q1 | [175] |
| | JCDL | 2014 | Conference | Short | 1 of 2 | CORE A* | [339] |
| | JCDL | 2011 | Conference | Short | 2 of 3 | CORE A* | [171] |
| | DocEng | 2011 | Conference | Full | 2 of 2 | CORE B | [170] |
| 4 | JCDL | 2018 | Conference | Full | 1 of 6 | CORE A* | [342] |
| 5 | JCDL | 2019 | Conference | Full | 1 of 5 | CORE A* | [344] |
| | CIKM | 2017 | Conference | Short | 1 of 5 | CORE A | [340] |
| 6 | SIGIR | 2018 | Conference | Short | 1 of 4 | CORE A* | [343] |
| | CICM | 2017 | Conference | Full | 2 of 5 | n/a | [454] |
| | SIGIR | 2013 | Conference | Demo | 2 of 5 | CORE A* | [172] |
| | IPC | 2012 | Conference | Full | 1 of 2 | n/a | [337] |

[3] http://portal.core.edu.au/conf-ranks, Ranks: A*—flagship conference, A—excellent conference, B—good conference, C—other ranked conferences

[4] https://www.scimagojr.com, Ranks: quartiles (Q1–Q4) of the SJR scores of journals in the field

**Table 1.2.** Overview of publications partially related to this thesis.

| Year | Venue | Type | Length | Author Position | Venue Rating | Ref. |
|---|---|---|---|---|---|---|
| 2020 | JCDL | Conference | Full | 5 of 6 | CORE A* | [445] |
| | JCDL | Conference | Short | 4 of 5 | CORE A* | [236] |
| | JCDL | Conference | Short | 4 of 5 | CORE A* | [63] |
| | JCDL | Conference | Poster | 3 of 5 | CORE A* | [141] |
| | JCDL | Conference | Poster | 3 of 4 | CORE A* | [457] |
| | iConf | Conference | Short | 4 of 7 | n/a | [142] |
| 2019 | CICM | Conference | Full | 4 of 5 | n/a | [456] |
| 2018 | JCDL | Conference | Full | 4 of 6 | CORE A* | [455] |
| | IJDL | Journal | Full | 2 of 3 | SJR Q1 | [218] |
| 2017 | JCDL | Conference | Full | 2 of 3 | CORE A* | [216] |
| | JCDL | Conference | Short | 3 of 4 | CORE A* | [179] |
| | JCDL | Workshop | Full | 1 of 4 | CORE A* | [341] |
| | WWW | Workshop | Full | 3 of 4 | CORE A* | [98] |
| | RecSys | Conference | Demo | 4 of 5 | CORE B | [459] |
| | CLEF | Conference | Full | 3 of 5 | n/a | [453] |
| | ISI | Conference | Full | 2 of 4 | n/a | [215] |
| | ISI | Conference | Short | 2 of 4 | n/a | [217] |
| 2016 | JCDL | Conference | Full | 3 of 6 | CORE A* | [458] |
| | JCDL | Workshop | Short | 2 of 4 | CORE A* | [335] |
| | JCDL | Workshop | Demo | 2 of 4 | CORE A* | [178] |
| | SIGIR | Conference | Full | 5 of 8 | CORE A* | [452] |
| | NTCIR | Conference | Full | 2 of 4 | n/a | [451] |
| 2015 | iConf | Conference | Full | 2 of 3 | n/a | [176] |
| | iConf | Conference | Short | 2 of 3 | n/a | [177] |
| 2014 | ICEIS | Conference | Demo | 2 of 5 | CORE C | [174] |

Preprints of all my publications are available at
**http://pub.meuschke.org**

My Google Scholar profile is available at
**http://scholar.meuschke.org**

2

Chapter 2

# Academic Plagiarism Detection

## Contents

This chapter provides background information on academic plagiarism and reviews technical approaches to detect it. Section 2.1 derives a definition and typology of academic plagiarism that is suitable for the technical research focus of this thesis. Section 2.2 provides a holistic overview of the research on academic plagiarism to contextualize the technically focused research areas that the subsequent sections present in detail. Sections 2.3 and 2.4 systematically analyze the research on plagiarism detection methods and describe production-grade systems that implement some of the presented methods. Section 2.5 presents datasets usable for evaluating plagiarism detection technology. Furthermore, the section discusses comprehensive performance evaluations of plagiarism detection methods and systems to highlight their weaknesses and demonstrate the research gap this thesis addresses. Section 2.6 summarizes the findings of the literature review and thereby fulfills Research Task 1. Section 2.7 derives the research idea pursued in this thesis.

## 2.1 Definition and Typology of Plagiarism

The term plagiarism originates from the Latin word *plagiarius*, meaning kidnapper or plunderer, and the verb *plagiare*, meaning to steal, [374], [489]. In 1601, the writer Ben Johnson introduced the expression *plagiary* into the English language to refer to a writer who steals words from another author [313]. The derived noun plagiarism entered the English language around 1620 [374].

Subsequently, the meaning of plagiarism broadened to its current definition of: "taking someone else's work or ideas and passing them off as one's own" [489]. This definition includes all types of intellectual property, including artistic and mechanical work products. This understanding of the term is too broad for this thesis, which focuses on identifying plagiarism in academic documents.

Therefore, we define **academic plagiarism** by building upon the definitions of Fishman [138, p. 5] and Gipp [173, p. 10] as:

> **The use of ideas, words, or other work**
> **without appropriately acknowledging the source**
> **to benefit in a setting where originality is expected.**

Other authors described plagiarism as theft, e.g., Bouarara et al. [59, p. 157], Ercegovac & Richardson Jr. [125, p. 304f.], Hussain & Suryani [233, p. 246], Kanjirangat & Gupta [260, p. 11], Park [390, p. 472], and Paul & Jamal [392, p. 223]. However, theft denotes the intentional and fraudulent taking of personal property of another without permission or consent. Theft also necessitates the intent to deprive others of using the item [138, p. 2], [297]. The definitions of Fishman and Gipp do not characterize academic plagiarism as theft for the following four reasons.

First, academic plagiarism can be unintentional. Authors may inadvertently fail to acknowledge a source, e.g., by forgetting to insert a citation or citing a wrong source [326, p. 1051]. Additionally, a memory bias called cryptomnesia can cause humans to attribute foreign ideas to themselves unconsciously [68].

Second, academic plagiarism does not deprive the creators of a work of the right to use the work [138, p. 3]. Even when the creators publish their work with the intent that others reuse it, the creators retain the right to use their work, and others are obliged to acknowledge the source when reusing the work in the academic context.

Third, academic plagiarists may act in consent with the creator of the original work but still commit plagiarism by not properly disclosing the source. The term collusion describes the behavior of authors, who write collaboratively, or copy from one another, although they are required to work independently [86, p. 2]. We include collusion in our definition of academic plagiarism.

Fourth, reused content must not necessarily originate from another person to constitute academic plagiarism. We include self-plagiarism, i.e., the insufficiently acknowledged reuse of own work, in the definition of academic plagiarism. Presenting updates and extensions, e.g., extending a conference paper to a journal article or making the work better accessible for researchers in a different field, can justify re-publishing own work, but still requires appropriate acknowledgment [64, p. 194ff.], [90, p. 94]. Unjustified reasons for reusing own content include the attempt to increase one's publication and citation counts.

### 2.1.1 Typologies of Academic Plagiarism

Aside from a definition, a typology helps structure the research [42, p. 943] and facilitates communication on a phenomenon [540, p. 42]. Researchers proposed a variety of typologies for academic plagiarism, which we briefly review to derive a typology that reflects the research objective of this thesis.

Walker proposed one of the first typologies, which he derived from a plagiarist's point of view [542, p. 102f.]. The typology distinguishes between:

1. **Sham paraphrasing** (presenting text reused verbatim as a paraphrase by leaving out quotations)
2. **Illicit paraphrasing**
3. **Other plagiarism** (plagiarizing with the consent of the original author)
4. **Verbatim copying** (without reference)
5. **Recycling** (self-plagiarism)
6. **Ghostwriting**
7. **Purloining** (copying another student's assignment without permission)

*The text in parentheses represents explanations we added for clarity.*

All typologies we encountered in our research categorize verbatim copying as a form of academic plagiarism. Aside from verbatim copying, the typology of Alfikri & Purwarianti distinguishes as separate forms of academic plagiarism the partial reproduction of smaller text segments, two forms of paraphrasing that differ regarding whether the sentence structure changes, and translations [14, p. 7886]. The typology of Velásquez et al. distinguishes verbatim copying and technical disguise but combines paraphrasing and translation into one type [515, p. 65].

Several typologies, e.g., in the References [84, p. 2f.], [515, p. 65], and [551, p. 10ff.], categorize referencing errors and the deliberate misuse of references as a form of academic plagiarism. Likewise, many typologies classify the unacknowledged reuse of ideas as a form of academic plagiarism, e.g., the typologies in the References [25, p. 134], [83, p. 18], [84, p. 3], [230, p. 1], [260, p. 11], [370, p. 3757].

Mozgovoy et al. presented a typology that consolidates other classifications into five types of academic plagiarism [357, p. 514ff.]:

1. **Verbatim copying**
2. **Hiding plagiarism** instances by paraphrasing
3. **Technical tricks** exploiting flaws of plagiarism detection systems
4. **Deliberately inaccurate use of references**
5. **Tough plagiarism**

In this typology, **tough plagiarism** subsumes the forms of academic plagiarism that are difficult to detect for both humans and computers, like idea plagiarism, structural plagiarism, and cross-language plagiarism [357, p. 515].

### 2.1.2 Our Typology of Academic Plagiarism

The research contributions of this thesis improve plagiarism detection technology. Therefore, we exclusively consider technical properties to derive a typology of academic plagiarism. We only distinguish forms of plagiarism as distinct types if their detection requires specialized methods. Some distinctions that are important from a policy perspective are irrelevant or less relevant from a technical perspective.

**Technically irrelevant properties** of academic plagiarism are:

» Whether the original author permitted the reuse of content;

» Whether the suspicious document and its source have the same author(s), i.e., whether similarities in the content may constitute self-plagiarism.

While we include collusion and self-plagiarism in our understanding of academic plagiarism, we do not devise specific methods to detect the two practices and thus do not distinguish them as distinct types of academic plagiarism.

**Properties of minor technical importance** are:

» The extent of content that represents potential plagiarism;

» Whether a plagiarist uses one or multiple sources.
(Researchers referred to plagiarizing from multiple sources as compilation plagiarism, shake and paste, patchwriting, remix, mosaic, or mash-up.)

Both properties are of little technical importance because they do not affect the conceptual design of the detection methods.

**Our typology of academic plagiarism** derives from the linguistic three-layered model of language consisting of lexis, syntax, and semantics. Ultimately, the goal of any language is communicating ideas [182, p. 2]. Thus, we extend the classic three-layered model to four layers and categorize forms of academic plagiarism by the layer they affect. We sort the resulting plagiarism forms in ascending order of their degree of obfuscation, as shown in **Figure 2.1**.

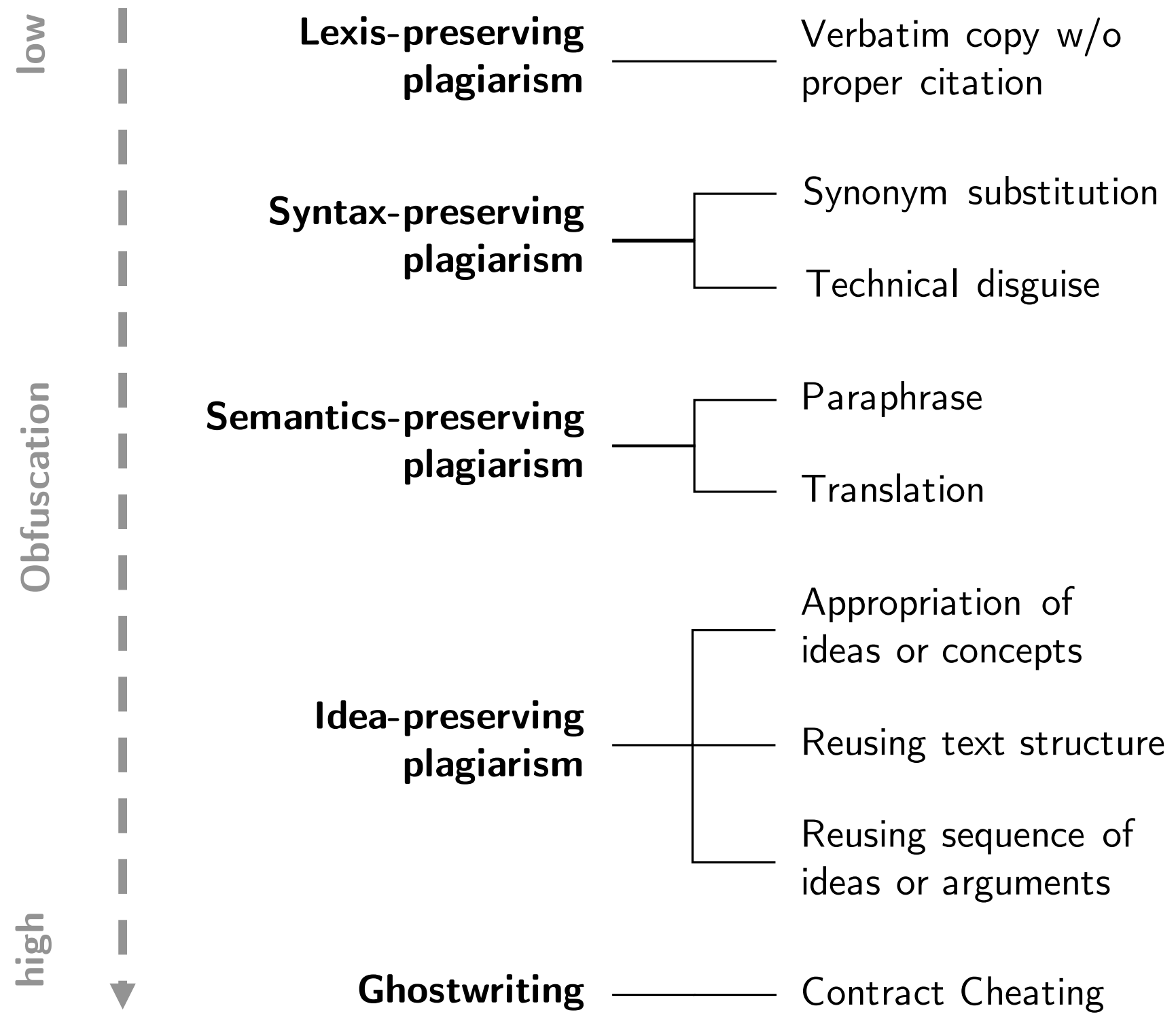

**Figure 2.1.** Typology of academic plagiarism used in this thesis.

**Lexis-preserving plagiarism** includes verbatim copying and plagiarism forms like shake and paste, pawn sacrifice, and cut and slide. Weber-Wulff described shake-and-paste plagiarism as combining sentences or paragraphs from multiple sources [551, p. 8f.]. Additionally, she proposed the term pawn sacrifice to describe citing a source but obfuscating the extent of content taken from the source [551, p. 10f.]. For example, the plagiarist may state the source in a footnote, only list the source in the bibliography without citing it in the text or cite only a portion of the content taken from the source. Weber-Wulff introduced the term cut and slide for putting copied content into footnotes or appendices [551, p. 12]. Likewise, plagiarists may integrate parts of the copied content in the main text and include the remainder in a footnote with or without naming the source.

**Syntax-preserving plagiarism** often results from employing simple substitution techniques, such as text string replacement using regular expressions or synonym substitution [408, p. 1001f.]. More sophisticated obfuscation techniques, which have become more widespread in recent years, employ, e.g., cyclic machine translation or automated text summarization [25, p. 234f.], [284, p. 355].

**Semantics-preserving plagiarism** refers to sophisticated forms of obfuscation that involve changing the words and sentence structure while preserving the meaning of passages. In agreement with Velásquez et al., we consider translation plagiarism a semantics-preserving form of academic plagiarism because one can consider a translation as the ultimate paraphrase [515, p. 65]. Section 2.4.5, p. 34, presents semantics-based detection methods and shows a significant overlap of paraphrase detection and cross-language plagiarism detection methods.

**Idea-preserving plagiarism** (also referred to as template plagiarism or boilerplate plagiarism) includes cases in which plagiarists reuse the concepts or structure of a source but describe them in their own words. This type of academic plagiarism is difficult to identify and even harder to prove.

**Ghostwriting** (also referred to as contract cheating) describes the hiring of a third party to write genuine text [97, p. 115f.], [542, p. 103]. Ghostwriting is the only form of academic plagiarism that is undetectable by comparing a suspicious document to a likely source [255, p. 188]. Currently, the only technical option for discovering ghostwriting is to compare the writing style features in a suspicious document with documents certainly written by the alleged author [255, p. 188f.].

Having established the definition of academic plagiarism and the typology of plagiarism forms we use in this thesis, the following section gives an overview of the research fields that address the phenomenon of academic plagiarism.

## 2.2 Research on Academic Plagiarism

For the two survey articles [140], [338], which we summarize in this chapter, we reviewed 376 publications from the 25-year period 1994–2019. The retrieved publications fall into three categories: plagiarism detection methods, plagiarism detection systems, and plagiarism policies. Ordering these categories by the level of abstraction at which they address the problem of academic plagiarism yields the three-layered model shown in **Figure 2.2**. We proposed this model to structure the extensive and heterogeneous literature on academic plagiarism [140, p. 6f.].

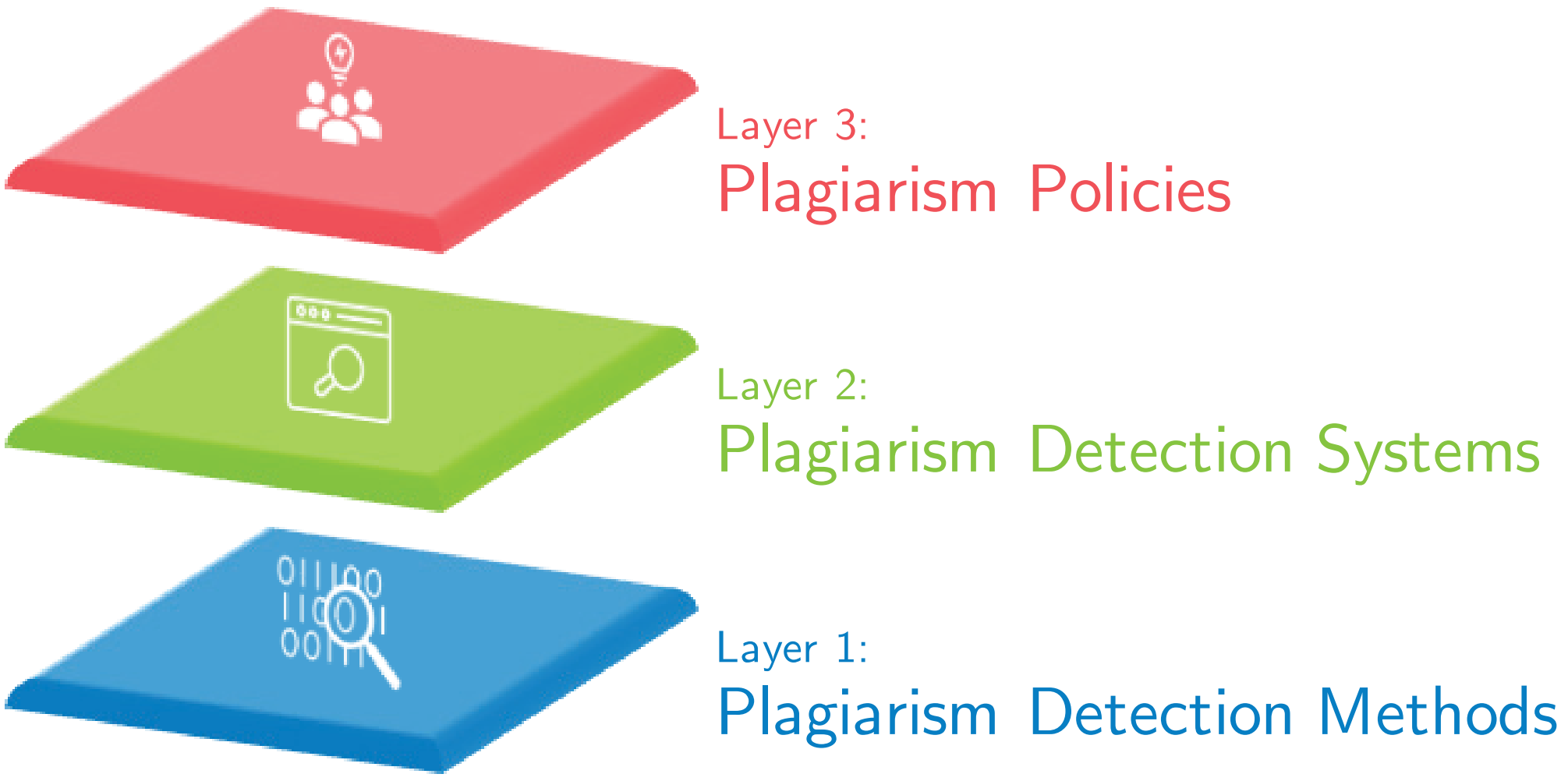

**Figure 2.2.** Three-layered model for addressing academic plagiarism.

#### Layer 1: Plagiarism Detection Methods

This layer subsumes research on information retrieval approaches that support the identification of plagiarism. Publications in this layer typically present methods that analyze lexical, syntactic, and semantic text similarity. The research on analyzing the similarity of non-textual content presented in this thesis also falls into this layer. To this layer, we also assign publications that address the evaluation of plagiarism detection methods, e.g., by contributing datasets. We refer to a class of conceptually related detection methods as a **plagiarism detection approach**.

#### Layer 2: Plagiarism Detection Systems

This layer encompasses applied research on production-grade software aiding in detecting academic plagiarism, as opposed to the research prototypes that publications assigned to Layer 1 typically present. Production-grade plagiarism detection systems implement the detection methods included in Layer 1 and visually present detection results to the users.

### Layer 3: Plagiarism Policies

This layer subsumes research addressing the prevention, detection, and prosecution of academic plagiarism. Typical publications in Layer 3 investigate students' and teachers' attitudes towards plagiarism (e.g., Foltýnek & Glendinning [139]), analyze the prevalence of academic plagiarism (e.g., Curtis & Clare [97]), or discuss the impact of institutional policies on academic integrity (e.g., Owens & White [383]).

The three layers of the model are interdependent and essential to analyze and address academic plagiarism. Plagiarism detection systems (Layer 2) depend on reliable plagiarism detection methods (Layer 1), which in turn would be of little practical value without production-grade detection systems that employ them. Using plagiarism detection systems in practice would be futile without the presence of a policy framework (Layer 3) that governs the investigation, documentation, prosecution, and punishment of plagiarism. The insights derived from analyzing the use of plagiarism detection systems in practice (Layer 3) also inform the research and development efforts for improving plagiarism detection methods (Layer 1) and plagiarism detection systems (Layer 2).

The research contributions of this thesis are novel plagiarism detection approaches, i.e., address Layer 1 of the model. Therefore, the remainder of this chapter reviews the research contributions in Layer 1. The following section specifies the plagiarism detection task and presents two paradigms for addressing it.

## 2.3 Plagiarism Detection Paradigms

From a technical perspective, the literature classifies plagiarism detection approaches into two conceptually different paradigms.

The **external plagiarism detection** paradigm encompasses detection approaches that compare suspicious input documents to documents assumed to be genuine (the **reference collection**). External detection approaches retrieve all documents that exhibit similarities above a threshold as potential sources [487].

The **intrinsic plagiarism detection** paradigm covers detection approaches that exclusively analyze the input document, i.e., do not perform comparisons to a reference collection. Intrinsic detection approaches examine the linguistic features of a text—a process known as **stylometry**. The goal is to identify parts of the text having a different writing style. Intrinsic detection approaches consider such differences as indicators of potential plagiarism [345]. Passages with linguistic differences can be analyzed using external plagiarism checks or be presented to examiners.

Hereafter, we describe the two plagiarism detection paradigms in more detail.

## 2.3.1 External Plagiarism Detection

External plagiarism detection methods typically follow a multi-stage process shown in **Figure 2.3**. The stages of the process are **candidate retrieval** (also called source retrieval [413, p. 2] or heuristic retrieval [486, p. 825]), **detailed analysis** (also known as text alignment [407, p. 2]), **postprocessing,** and **human inspection**. A multi-step process is necessary because the reference collection is typically extensive, e.g., the Internet. Therefore, pairwise comparisons of the input document to all documents in the collection are computationally infeasible.

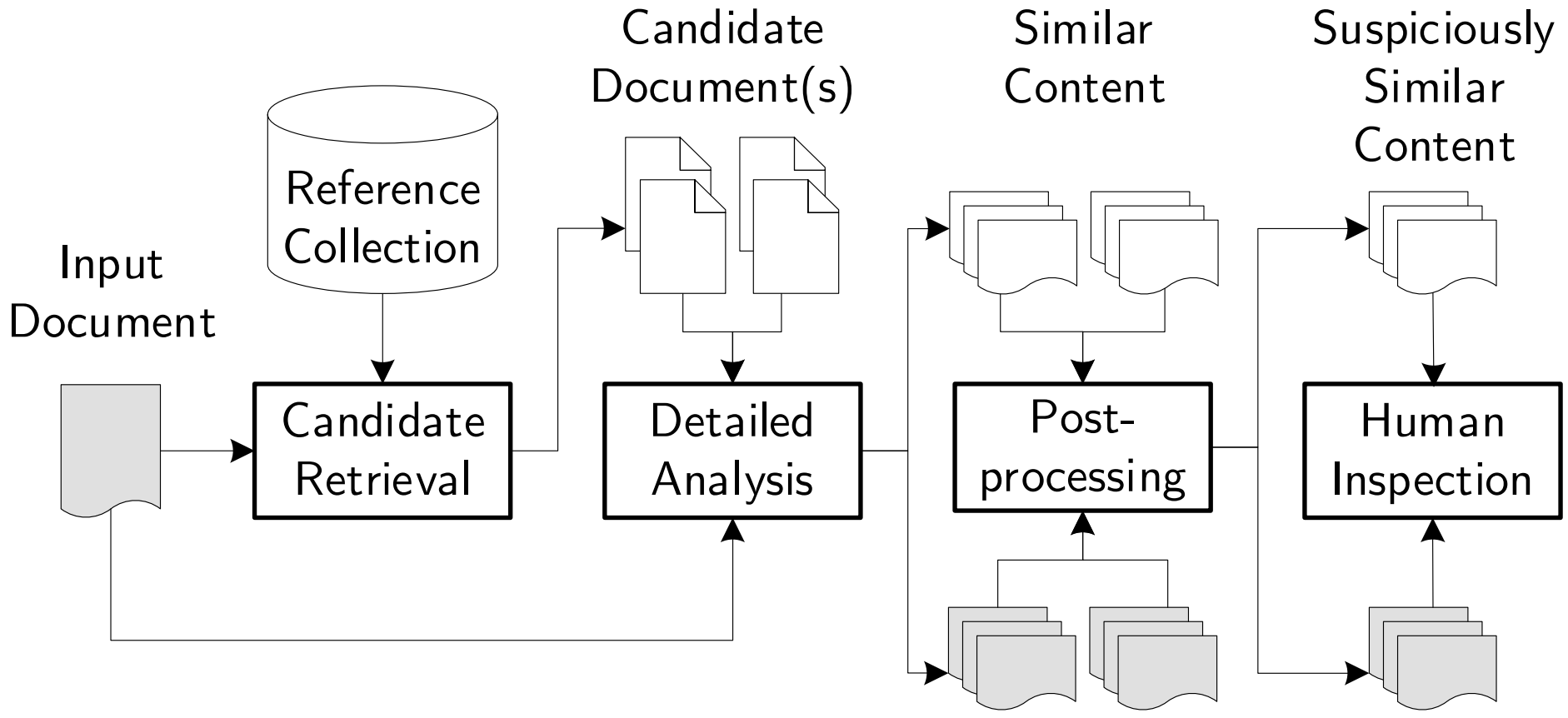

**Figure 2.3.** General external plagiarism detection process.

In the candidate retrieval stage, efficient algorithms limit the reference collection to a small subset of documents that could be the source for content in the input document. The algorithms in the detailed analysis stage then perform pairwise comparisons of the input document and each candidate document to identify similar content in both documents. In the postprocessing stage, content identified as similar undergoes a knowledge-based analysis to eliminate false positives, which the detection methods in the previous stages are prone to produce. Correctly cited content is a typical example of false positives. In the final stage, content identified as suspiciously similar is presented to the user for review.

The literature on academic plagiarism detection emphasizes the importance of the human inspection stage [293, p. 1], [327, p. 4452], [551, p. 71]. Without human review, no plagiarism detection system can prove the presence of plagiarism. Like all information retrieval systems, plagiarism detection systems can retrieve false positives that human reviewers need to recognize as such. Therefore, plagiarism detection systems cannot fully automate the identification of plagiarism; they are only the first step in a semi-automated detection and verification process that re-quires careful consideration on a case-by-case basis. Proving the absence of plagiarism is impossible using any system because the possibility that the system could not find or access the source cannot be eliminated [551, p. 113].

The following two subsections describe the candidate retrieval and detailed analysis stages in more detail because they are particularly relevant for the research presented in the remainder of this thesis.

### Candidate Retrieval

The task in this stage is retrieving from the reference collection all documents that share content with the input document [25, p. 137], [414, p. 3]. Presumably to reduce costs, many production-grade plagiarism detection systems use the Application Programming Interfaces (APIs) of web search engines to perform the candidate retrieval rather than maintaining their own collections and querying tools.

Recall is the critical performance measure for the candidate retrieval stage as the subsequent detailed analysis cannot identify source documents missed in the candidate retrieval stage. The number of queries issued is another typical metric to quantify the performance in the candidate retrieval stage [206, p. 6f.]. Keeping the number of queries low is particularly important if the candidate retrieval algorithm queries web search engines as such engines typically charge for issuing queries.

### Detailed Analysis

The documents retrieved in the candidate retrieval stage are the input to the detailed analysis stage. Formally, the detailed analysis task is defined as follows. Let $d_q$ be a suspicious input document. Let $D = \{d_s\} \mid s = 1 \dots n$ be a set of potential source documents. Determine whether a fragment $s_q \in d_q$ is similar to a fragment $s \in d_s$ ($d_s \in D$) and identify all such fragment pairs $(s_q, s)$ [407, p. 2]. An expert should determine if the identified pairs $(s_q, s)$ constitute plagiarism or false positives [407, p. 2]. The detailed analysis typically includes three steps [413, p. 16f.]:

1. **Seeding**: Finding parts of the content in the input document (the seed) within a document from the reference collection
2. **Extension**: Extending each seed as far as possible to find the complete passage that may have been reused
3. **Filtering**: Excluding fragments that do not meet predefined criteria (e.g., are too short) and handling overlapping passages

Rule-based merging is a common strategy for the extension step. The approach combines seeds adjacent in the suspicious and source document if the gaps between the passages are below a threshold [414, p. 18].

**Paraphrase identification** is often a separate step within the detailed analysis stages of external detection methods and a distinct research field. The task in paraphrase identification is determining semantically equivalent sentences in a set of sentences [131, p. 59]. SemEval is a well-known series of workshops that address paraphrase identification [7]. AL-Smadi et al. provided a thorough review of the research on paraphrase identification [19].

### 2.3.2 Intrinsic Plagiarism Detection

Intrinsic plagiarism detection is a sub-field of authorship analysis (see **Figure 2.4**).

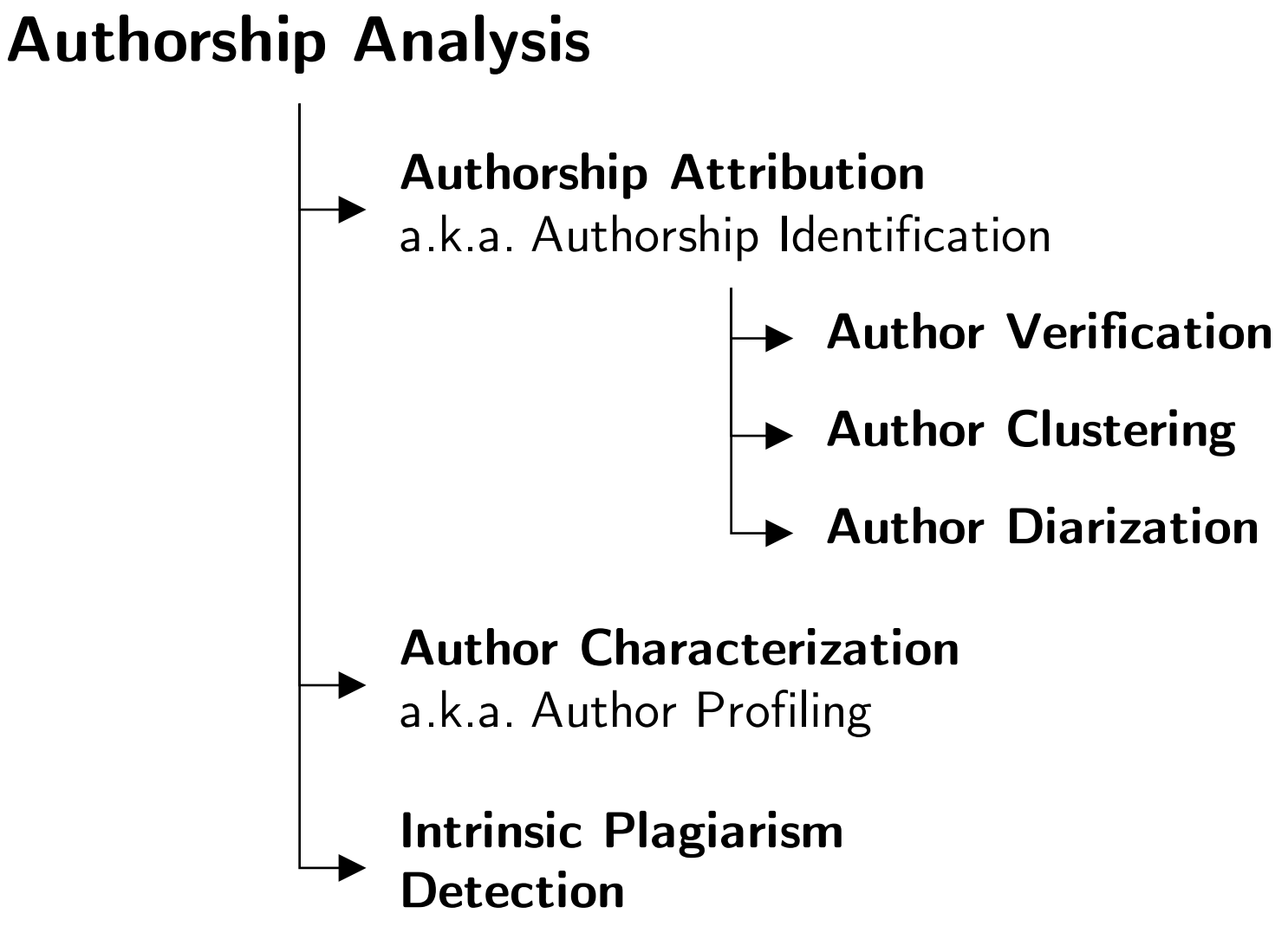

**Figure 2.4.** Taxonomy of research areas in authorship analysis.

**Authorship analysis** describes examining a document's linguistic features to deduce the document's author [117, p. 22].

**Authorship attribution** (also known as authorship identification) is a well-known authorship analysis problem. The task is finding the most probable author of a document with unknown authorship given a collection of documents with known authors from a small set of possible authors [61, p. 1], [253, p. 238].

**Author verification**, **author clustering**, and **author diarization** are variants of authorship attribution [209, p. 2], [431, p. 333]. The task in author verification is deciding if a single author, for whom writing samples are available, also authored a document with questionable authorship [209, p. 2], [484, p. 524]. The set of possible authors and writing samples is smaller in author verification scenarios than in authorship attribution scenarios. Author clustering and author diarization are open-set variants of the authorship attribution problem. The task in author clustering is grouping all documents so that each cluster corresponds to a different author. Author diarization represents the author clustering problem for a single document. The task is identifying different authors within the same document and grouping all passages written by the same author. The set of possible authors is initially unknown in author clustering and author diarization.

**Author characterization** seeks to derive sociolinguistic characteristics of the author, such as gender, age, educational background, and cultural background, from analyzing the text [117, p. 22], [480, p. 539].

**Intrinsic plagiarism detection** methods search for stylistic dissimilarities in a document and typically follow a three-step process [433, p. 1], [488, p. 68ff.]:

1. **Text decomposition**: Segmenting the text into equally-sized chunks (e.g., passages, character $n$-grams, or word $n$-grams), structural units (e.g., paragraphs or (overlapping) sentences), topical units, or stylistic units
2. **Style model construction**: Analyzing lexical, syntactic, and structural stylistic markers for each text segment, e.g., the frequencies of $n$-grams, punctuation marks, and word classes; Computing quantitative measures, e.g., reflecting the vocabulary richness, readability, and complexity of the text; Combining the measures into stylometric feature vectors
3. **Outlier identification**: Classifying the stylometric feature vectors of each text segment as members of the target class, i.e., likely being unsuspicious, or as outliers, i.e., likely being written by a different author; Typically, outlier identification approaches estimate the multi-dimensional probability density function of features of the target class, like Naïve Bayes.

Intrinsic plagiarism detection is closely related to authorship verification [209, p. 2], [488, p. 65]. Both tasks are one-class classification problems because a target class and examples of the target class exist [488, p. 65]. In authorship verification, the examples of the target class are explicit. Therefore, supervised approaches can be employed to train a model for recognizing target class members. In intrinsic plagiarism detection, the target class members, i.e., text passages that the alleged author wrote, are unknown [488, p. 65], [509, p. 1]. Thus, one requires unsupervised approaches for telling apart target class members from outliers.

The following section gives an overview of the broad spectrum of detection methods that follow the external or intrinsic plagiarism detection paradigms.

## 2.4 Plagiarism Detection Methods

We categorize **plagiarism detection methods** according to our typology of academic plagiarism presented in Section 2.1.2, p. 12. **Lexical detection methods** exclusively consider the characters in a document. **Syntax-based detection methods** analyze the sentence structure, i.e., the parts of speech and their relationships. **Semantics-based detection methods** compare the meaning of the text. **Idea-based detection methods** consider non-textual content like citations, images, and mathematical formulae. Before presenting each class of detection methods, we describe preprocessing steps and similarity measures relevant to all classes.

### 2.4.1 Preprocessing

The objective of preprocessing is to remove noisy input data while maintaining the information required for the analysis. The initial preprocessing steps of detection methods typically include format conversions and information extraction.

The decision on additional preprocessing operations heavily depends on the detection paradigm. For external text-based detection methods, typical preprocessing steps include lowercasing, punctuation removal, tokenization, segmentation, number removal or number replacement, named entity recognition, stop word removal, stemming or lemmatization, part of speech (PoS) tagging, and synset extension.

Detection methods employing synset extension typically use thesauri like WordNet [129] to assign the identifier of the class of synonymous words to which a word belongs. The methods then consider the synonymous words for similarity calculation. Lexical detection methods usually perform chunking as a preprocessing step. A chunking algorithm splits text elements into sets of given lengths, e.g., word $n$-grams, line chunks, or phrasal constituents in a sentence [83].

Intrinsic detection methods often limit preprocessing to a minimum for not losing potentially useful information [19], [128]. For example, intrinsic detection methods typically do not remove punctuation.

Most detection methods we review in this section employ well-established Natural Language Processing (NLP) software libraries for preprocessing, such as the Natural Language Toolkit Kit (Python) [56] or Stanford CoreNLP (Java) [321] libraries. Commonly applied syntax analysis tools include Penn Treebank [323], Citar [62], TreeTagger [447] and Stanford parsers [276]. Several researchers presented resources for Arabic [49], [50], [468] and Urdu [101] language processing.

## 2.4.2 Similarity Measures

All plagiarism detection methods require similarity measures. External detection methods compute the similarity of content in an input document and a potential source. The similarity scores often determine whether the potential sources proceed to the subsequent stages of the detection process. The scores also influence the ranking of the results. Intrinsic detection methods employ similarity measures to classify text passages as belonging to the target class or the outlier class (cf. Section 2.3.2, p. 19). The chosen similarity measure can strongly affect the effectiveness of a detection method [259]. Hereafter, we briefly describe the similarity measures frequently used as part of plagiarism detection methods.

### Set-based Similarity Measures

**Table 2.1.** Set-based similarity measures frequently applied for PD.

| Name | Mathematical Definition | Range |
|---|---|---|
| Jaccard Index | $s(A,B) = \frac{\lvert A \cap B \rvert}{\lvert A \cup B \rvert}$ | [0,1] |
| Dice Coefficient | $s(A,B) = \frac{2\lvert A \cap B \rvert}{\lvert A \rvert + \lvert B \rvert}$ | [0,1] |
| Containment Measure | $s(A,B) = \frac{\lvert A \cap B \rvert}{\lvert A \rvert}$ | [0,1] |
| Overlap Coefficient | $s(A,B) = \frac{\lvert A \cap B \rvert}{\min\ (\lvert A \rvert, \lvert B \rvert)}$ | [0,1] |
| Simple Matching Coefficient | $s(A,B) = \frac{\lvert A \cap B \rvert + \lvert \overline{(A \cup B)} \rvert}{\lvert A \cup B \rvert + \lvert \overline{(A \cup B)} \rvert}$ | [0,1] |

We consider two sets $A = \{a_1, a_{2,} \dots a_n\}$ and $B = \{b_1, b_{2,} \dots b_m\}$, e.g., representing the distinct words in two text passages or the sets of nodes or edges in two graphs. We assume that $A$ represents the suspicious document. Many set-based measures calculate the similarity of the sets $s(A,B)$ as a fraction. The numerator represents the size of the intersection of the two sets, i.e., the number of mutual elements in the sets. The denominator represents one of the following:

» The size of the union of both sets (**Jaccard Index** [242])

» The sum of the cardinalities of both sets (**Dice Coefficient** [106, p. 298])

» The cardinality of $A$ (**Containment Measure** [67, p. 23])

» The cardinality of the smaller set (**Overlap Measure** [500, p. 250])

The **Simple Matching Coefficient** [474, p. 1417] differs from the other measures in that it considers both the mutual presence and absence of elements. The measure resembles the Jaccard coefficient but is more suitable when the absence of features in equally sized sets is as informative as the presence of these features. Comparisons of stylometric markers are an example of such a situation.

**Table 2.1** summarizes the set-based similarity measures we described.

## Sequence-based Similarity Measures

**Table 2.2.** Sequence-based similarity measures frequently applied for PD.

| Name | Mathematical Definition | Range |
|---|---|---|
| Normalized Hamming Distance | $s(a,b) = 1 - \frac{d_H}{\lvert a \rvert}$ where $d_{\mathrm{H}} = \lvert \{ i \mid a_i \neq b_i \} \rvert$ | [0,1] |
| Normalized Levenshtein Distance | $s(a,b) = \frac{d_L(a,b)}{\lvert a \rvert + \lvert b \rvert}$ where $d_{\mathrm{L}}(a,b) = t_{a,b}(i,j) = \begin{cases} \max(i,j) & \text{if } \min(i,j) = 0 \\ \min \begin{cases} t_{a,b}(i-1,j)+1 \\ t_{a,b}(i,j-1)+1 \\ t_{a,b}(i-1,j-1)+1_{(a_i \neq b_j)} \end{cases} & \text{otherwise} \end{cases}$ | [0,1] |
| Normalized Longest Common Subsequence | $s(a,b) = \frac{\mathrm{c}_{a,b}}{\lvert a \rvert}$ where $\mathrm{c}_{a,b} = \max\{ l \mid \exists i : \forall j \in \{0,1,\ldots l-1\} : a_{i+j} = b_{i+j} \}$ | [0,1] |

We consider two sequences $a = (a_1, a_2, \ldots a_n)$ and $b = (b_1, b_2, \ldots b_m)$, of which $a$ represents the suspicious input document. To determine the similarity $s(a,b)$, sequence-based similarity measures consider the presence and absence of features as well as the features' position. Similarity measures for sequences frequently used in the context of plagiarism detection derive from the following measures:

» The **Hamming Distance** represents the number of positions at which elements differ in two sequences of the same length [220, p. 154f.].

» The **Levenshtein Distance** is a generalization of the Hamming distance for sequences of different lengths. The distance calculates the number of single-item edits, i.e., substitutions, insertions, and deletions, necessary to transform one sequence into the other [304].

» The **Longest Common Subsequence (LCS)** is the number of elements that occur in both sequences in the same order but can be interrupted by non-matching elements. To obtain similarity scores in the range [0,1], the LCS can be normalized. A common practice in plagiarism detection is normalizing the length of the longest common subsequence by the length of the sequence in the input document checked for plagiarism. In this case, the normalized LCS expresses the fraction of elements in the input document that are identical to elements in the comparison document.

**Table 2.2** summarizes the sequence-based similarity measures.

## Vector-based Similarity Measures

Many plagiarism detection approaches represent the content of documents as vectors $\mathbf{a} = (a_1, a_{2,} \dots a_n)$ and $\mathbf{b} = (b_1, b_{2,} \dots b_m)$, of which $\mathbf{a}$ represents the suspicious document. This representation allows employing well-established measures, which we summarize in **Table 2.3**, to quantify the similarity of the documents.

**Table 2.3.** Vector-based similarity measures frequently applied for PD.

| Name | Mathematical Definition | Range |
|---|---|---|
| Cosine Similarity | $s(\mathbf{a},\mathbf{b}) = \cos(\theta) = \frac{\mathbf{a} \bullet \mathbf{b}}{\lvert\mathbf{a}\rvert\lvert\mathbf{b}\rvert}$ | [0,1] |
| Euclidean Distance | $d_{\mathrm{E}}(\mathbf{a},\mathbf{b}) = \sqrt{\sum_{i=1}^{n}(a_i - b_i)^2}$ | $[0,\infty]$ |
| Minkowski Distance | $d_{\mathrm{M}}(\mathbf{a},\mathbf{b}) = \sqrt[p]{\sum_{i=1}^{n}\lvert a_i - b_i\rvert^p}$ | $[0,\infty]$ |
| Manhattan Distance (taxicab distance) | $d_{\mathrm{T}}(\mathbf{a},\mathbf{b}) = \sum_{i=1}^{n}\lvert a_i - b_i\rvert$ | $[0,\infty]$ |
| Canberra Distance | $d_{\mathrm{Ca}}(\mathbf{a},\mathbf{b}) = \sum_{i=1}^{n}\frac{\lvert a_i - b_i\rvert^2}{\lvert a_i\rvert + \lvert b_i\rvert}$ | $[0,\infty]$ |
| Chebyshev Distance | $d_{\mathrm{Ch}}(\mathbf{a},\mathbf{b}) = \max(\lvert a_i - b_i\rvert)$ | $[0,\infty]$ |

### 2.4.3 Lexical Detection Methods

This class of detection methods exclusively considers the lexical similarity of texts. The methods are best-suited for identifying copy-and-paste plagiarism that exhibits little to no obfuscation. Detecting disguised plagiarism requires combining lexical detection methods with more sophisticated NLP approaches.

Minor adaptions can enable lexical detection methods to identify homoglyph substitutions, which are a common form of technical disguise. Alvi et al. [23] is the only publication we retrieved that addressed this task. The authors used a list of confusable Unicode characters and performed approximate word $n$-gram matching using the Normalized Hamming Distance.

Lexical detection methods typically fall into one of the three categories that we describe in the following:

- $n$-gram comparisons
- Vector space models
- Querying web search engines

#### $n$-gram Comparisons

Comparing $n$-grams refers to determining the similarity of contiguous sequences of $n$ items, typically characters or words, and less frequently phrases or sentences. Brin et al. [65] and Pereira & Ziviani [395] are two of the few research papers that employed sentence-level $n$-gram comparisons.

External detection methods often employ $n$-gram comparisons for the candidate retrieval stage or the seeding phase of the detailed analysis stage. Also, intrinsic detection methods frequently use $n$-gram comparisons.

Detection methods using $n$-gram comparisons typically follow the **fingerprinting** approach illustrated in **Figure 2.5**. The methods first split a document into (possibly overlapping) $n$-grams, which they use to create a set-based representation of the document or passage—the "fingerprint." The elements of the fingerprint are called minutiae [228, p. 208]. Most detection methods store fingerprints in index data structures to enable efficient retrieval.

Some detection methods hash or compress the $n$-grams that form the fingerprints to speed up the comparison of fingerprints. Hashing or compression reduces the lengths of the strings to be compared and allows performing computationally more efficient numerical comparisons. However, hashing introduces the risk of false positives due to hash collisions. Therefore, hashed or compressed fingerprinting is more commonly applied for the candidate retrieval stage, in which achieving high recall is more important than achieving high precision.

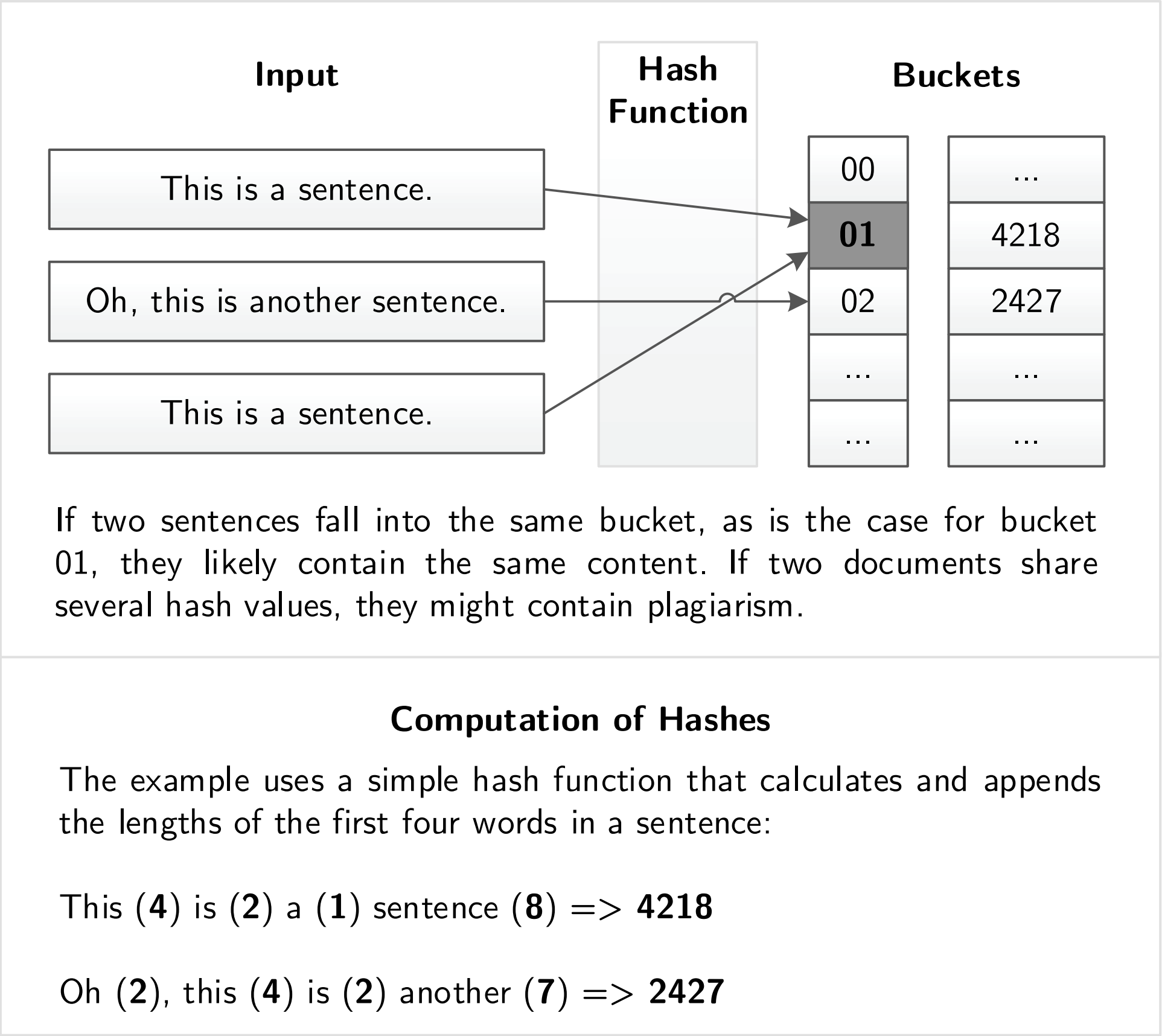

**Figure 2.5.** Concept of fingerprinting methods.

Fingerprinting is a widely-used approach for assessing lexical similarity [204, p. 71]. Recent research has focused on methods to detect disguised plagiarism. Such methods typically use $n$-gram fingerprinting in the preprocessing stage as in Alzahrani [24] or as a feature for machine learning, e.g., in Agirre et al. [8]. Character $n$-gram comparisons can be applied to perform cross-language plagiarism detection if the languages have a high lexical similarity, e.g., English and Spanish [411].

**Table 2.4** presents research publications employing word $n$-grams, **Table 2.5** lists publications using character $n$-grams, **Table 2.6** shows publications that employ hashing or compression for $n$-gram fingerprinting.

**Table 2.4.** Word $n$-grams detection methods.

| Parad. | Task | Method Variation | References |
|---|---|---|---|
| External | Document Level Detection | All word $n$-grams | [53], [136], [228], [314] |
| | | Sentence-bound WNG | [477] |
| | | Stop words removed | [24], [161], [191], [228], [250], [251], [367], [402], [469], [515] |
| | | Stop word $n$-grams | [192] |
| | Candidate Retrieval | All word $n$-grams | [265], [266], [492] |
| | | Stop words removed | [112] |
| | | Stop word $n$-grams | [481], [492] |
| | Detailed Analysis | All word $n$-grams | [23], [265], [266], [386], [465], [466] |
| | | $n$-grams variations | [3], [40], [387] |
| | | Stop words removed | [183], [186] |
| | | Stop word $n$-grams | [465], [481] |
| | | Context $n$-grams | [429], [430] |
| | | Named entity $n$-grams | [465] |
| | Paraphrase Identification | All word $n$-grams | [225], [503] |
| | | Combination with ESA | [539] |
| | Cross-language PD | Stop words removed | [111] |
| Intrinsic | Authorship Attribution | Overlap in LZW dict. | [78] |
| | Author Verification | Word $n$-grams | [155], [197], [272], [328], [355], [442] |
| | | Stop word $n$-grams | [272], [328], [355] |

**Legend:**
**ESA** Explicit Semantic Analysis, **dict.** dictionary, **LZW** Lempel-Ziv-Welch compression, **Parad.** Paradigm, **WNG** word $n$-grams

**Table 2.5.** Character $n$-grams detection methods.

| Parad. | Task | Method Variation | References |
|---|---|---|---|
| External | **Document Level Detection** | Pure CNG | [73], [226], [228], [320], [367], [554] |
| | | Overlap in LZW dict. | [476] |
| | | Machine learning | [59] |
| | | Bloom filters | [162] |
| | **Candidate Retrieval** | Pure CNG | [188], [368], [446], [577] |
| | | Sentence-bound CNG | [74] |
| | | Word-bound CNG | [462] |
| | **Detailed Analysis** | Pure CNG | [188] |
| | | Word-bound CNG | [462] |
| | | Hashed CNG | [23], [74] |
| | **Paraphrase Identification** | Machine learning | [503] |
| | **Cross-language PD** | Cross-language CNG | [41], [133], [134], [151] |
| Intrinsic | **Authorship Attribution** | Bit $n$-grams | [394] |
| | **Author Verification** | CNG as stylometric features | [29], [42], [77], [99], [155], [210]–[212], [243], [244], [247], [272], [286], [298], [317], [355], [442] |
| | **Author Clustering** | CNG as stylometric features | [10], [159], [201], [278], [443] |
| | **Style Breach Detection** | CNG as stylometric features | [51], [188], [266], [291] |

**Legend:**
**CNG** character $n$-grams, **dict.** dictionary, **LZW** Lempel-Ziv-Welch compression, **Parad.** Paradigm

**Table 2.6.** Detection methods employing compression or hashing.

| Parad. | Task | Method Variation | References |
|---|---|---|---|
| External | **Document Level Detection** | Hashing | [158], [162], [469], [554] |
| | | Compression | [476] |
| | **Cand. Retrieval** | Hashing | [24], [204], [358], [372] |
| | **Detailed Analysis** | Hashing | [21], [373], [386], [387] |
| Intr. | **Authorship Attribution** | Compression | [57], [78], [213], [224] |

**Legend:**
**Cand.** Candidate, **Intr.** Intrinsic, **Parad.** Paradigm

## Vector Space Models

Vector space models (VSMs) are a classic retrieval approach that represents texts as high-dimensional vectors. In plagiarism detection, words or word $n$-grams typically form the dimensions of the vector space, and the components of a vector undergo term frequency-inverse document frequency (tf-idf) weighting. Authors derived the idf values either from the suspicious document, e.g., Rafiei et al. [423], documents in the reference collection, e.g., Devi et al. [105], or an external corpus, e.g., Suchomel et al. [492]. The similarity of vector representations—typically quantified using the cosine similarity, i.e., the angle between the vectors—is used as a proxy for the similarity of the documents the vectors represent.

Most methods employ fixed similarity thresholds to retrieve documents or passages for subsequent processing. However, some methods, e.g., those that Kanjirangat and Gupta [258], Ravi et al. [426], and Zechner et al. [570] presented, divide the set of source documents into $K$ clusters by first selecting $K$ centroids and then assigning each document to the group whose centroid is most similar. The suspicious document serves as one of the centroids; the corresponding cluster becomes the input to the subsequent processing stages.

VSMs remain popular and well-performing approaches for detecting copy-and-paste plagiarism and identifying disguised plagiarism as part of semantic analysis methods. Intrinsic detection methods also employ VSMs frequently. A typical approach represents sentences as vectors of stylometric features to find outliers or group stylistically similar sentences.

**Table 2.7** presents publications that employ VSMs for external PD; **Table 2.8** lists publications using VSMs for intrinsic PD.

**Table 2.7.** External detection methods employing vector space models.

| Task | Scope | Unit | VSM Variation | Ref. |
|---|---|---|---|---|
| **Document Level Detection** | Doc. | WNG | Multiple WNG lengths | [251] |
| | | Word | Word lemmatization | [107] |
| | | | PD-specific term weighting | [346], [463] |
| | | | Includes synonyms | [367] |
| | Doc., Par. | Word | PD-specific term selection | [467] |
| | Sent. | Word | Avg. word2vec vector of sent. | [165] |
| | | | Hybrid similarity measures | [259] |
| | | | Standard tf-idf VSM | [223] |
| **Candidate Retrieval** | Doc. | WNG | Standard tf-idf VSM | [105] |
| | | | Length encoding of WNG | [43] |
| | | Word | Standard tf-idf VSM | [261] |
| | | | Z-order mapping of vectors | [13] |
| | Doc., Sent. | Word | Cluster pruning of vectors | [258], [426], [570] |
| | Text Chunk | Word | Bitmap vectors incl. synonyms | [448] |
| | | | Topic-based segmentation; PD-specific sim. measures | [112] |
| | Sent. | WNG | Multiple WNG lengths | [69] |
| | | Word | PD-specific term weighting and term selection | [165], [245] |
| **Detailed Analysis** | Doc. | Word | Transition point term weighting | [316], [401] |
| | Sent. | Word | Binary term weighting | [110], [260], [282], [387] |
| | | | Adaptive parameter tuning | [436], [437] |
| | | | Hybrid similarity measure | [543] |
| | | | Bitmap vectors incl. synonyms | [448] |
| **Paraphrase Identificat.** | Sent. | Word | Hybrid similarity measure | [131] |

**Legend:**
**Doc.** Document, **Par.** Paragraph, **Ref.** References, **Sent.** Sentence, **sim.** similarity, **VSM** vector space model **WNG** word $n$-grams

**Table 2.8.** Intrinsic detection methods employing vector space models.

| Task | Scope | Unit | VSM Variation | References |
|---|---|---|---|---|
| **Intrinsic PD (Style Breach Detection)** | Doc. | Feat. | Lexical and syntactic feat. | [264] |
| | Text Chunk | Word | Frequency term weighting | [370] |
| | SNG | Feat. | Lexical features | [428] |
| | Sent. | Feat. | Sent2Vec LSTM encoder | [433] |
| | | | Lexical features | [470] |
| **Author Verification** | Doc. | Feat. | Lexical features | [197], [243], [244], [328], [348] |
| | | | Lexical and syntactic feat. | [77], [211], [270], [399] |
| | | | Syntactic features | [405] |
| **Author Clustering** | Doc. | Feat. | Lexical features | [99], [247], [277], [278], [348], [514] |
| | | | Word embeddings | [443] |

**Legend:**
**Feat.** Features, **Doc.** Document, **LSTM** long short-term memory
**Sent.** Sentence, **Sent2Vec** Sentence to vector, **SNG** sentence $n$-grams

## Querying Web Search Engines

Many detection methods employ web search engines for candidate retrieval, i.e., finding potential source documents in the initial stage of the detection process. The strategy for selecting the query terms from the input document is crucial for the success of the approach. **Table 2.9** gives an overview of the strategies for query formulation, i.e., query term selection, employed by publications in our collection.

Intrinsic detection methods can employ web search engines to realize the **Impostors Method** first presented by Koppel et al. [286]. This method transforms the one-class verification problem regarding an author's writing style into a two-class classification problem. The method extracts keywords from the suspicious document to retrieve a set of topically related documents from external sources, the so-called impostors. The method then quantifies the "average" writing style observable in impostor documents, i.e., the distribution of stylistic features to be expected. Subsequently, the method compares the stylometric features of passages from the suspicious document to the features of the "average" writing style in impostor documents. This way, the method distinguishes the stylistic features that are author-specific from the features that are characteristic of the topic. Detection methods that used the Impostors Method achieved excellent results in the PAN[5] competitions, e.g., winning the competition in 2013 and 2014 [254], [482]. **Table 2.10** presents papers using the Impostors Method.

**Table 2.9.** External detection methods querying web search engines.

| **Query Formulation** | **References** |
|---|---|
| Words with highest tf-idf value | [166], [273], [423], [425], [494] |
| Words with highest tf-idf value & noun phrases | [121], [122], [578] |
| Terms & term $n$-grams with highest tf-idf value | [282] |
| Least frequent words | [207], [315] |
| Least frequent strings | [300] |
| Nouns and most frequent words | [417] |
| Nouns, facts, and most frequent words | [85] |
| Nouns and verbs | [285] |
| Nouns, verbs, and adjectives | [425], [556]–[558] |
| Variably sized word chunks | [73] |
| Increasing passage length | [517] |
| Keywords and longest sentence in a paragraph | [492]–[494] |
| Comparing different querying heuristics | [271] |
| Query expansion using UMLS thesaurus | [365] |

**Legend:**
**UMLS** Unified Medical Language System (https://uts.nlm.nih.gov/)

**Table 2.10.** Intrinsic detection methods employing the Impostors Method.

| **Task** | **References** |
|---|---|
| Author verification | [197], [272], [286], [356], [460] |

[5] PAN is a series of annual competitions for evaluating PD technology (see Section 2.5.1, p. 49).

### 2.4.4 Syntax-based Detection Methods

Syntax-based detection methods typically consider sentences and employ parts of speech tagging to determine the syntactic structure of the sentences. The syntactic information helps address morphological ambiguity during the lemmatization or stemming step of preprocessing, e.g., in Reference [234]. Other researchers employed syntactic analysis to reduce the workload of subsequent semantic analysis, typically by exclusively comparing word pairs that belong to the same parts of speech class, as Gupta et al. [192] exemplified.

Many intrinsic detection methods use the frequency of parts of speech tags as a stylometric feature. Other methods, e.g., Tschuggnall & Specht [507], rely solely on the syntactic structure of sentences.

**Table 2.11** presents publications presenting syntax-based detection methods.

**Table 2.11.** Syntax-based detection methods.

| Parad. | Method | Method Variation | References |
|---|---|---|---|
| External | PoS tagging | Reduce morphological ambiguity | [234], [235] |
| | | Compare words in same PoS class | [191], [350] |
| | | Analyze PoS and stop words | [192] |
| | | Compare PoS sequences | [566] |
| | | Combine PoS with compression | [224] |
| Intrinsic | PoS tags as style features | PoS frequency | [231], [292], [317], [384] |
| | | PoS $n$-gram frequency | [4], [42], [77], [164], [272], [355], [405], [518] |
| | | PoS freq., PoS $n$-gram freq., starting PoS class | [318] |
| | Syntactic trees | Direct comparison | [507], [508] |
| | | Syntactic graphs | [200] |

**Legend:**
**freq.** frequency, **Parad.** Paradigm, **PoS** Parts of speech

### 2.4.5 Semantics-based Detection Methods

Approaches to determine semantic similarity for natural language processing tasks, such as plagiarism detection, fall into two categories [184].

**Knowledge-based approaches** use information encoded in semantic networks, such as dictionaries and thesauri. The approach analyzes the connections between term nodes in the network to determine the relation of the terms. The major drawback of knowledge-based approaches is their domain specificity [185]. Most resources focus on lexical information about words but contain little information on the different word senses or encyclopedic information on terms. Creating and maintaining lexical resources requires expertise, time, effort, and money. The resources still only cover a small portion of the natural language lexicon [157].

**Corpus-based approaches** follow the idea of distributional semantics, i.e., terms co-occurring in similar contexts tend to convey a similar meaning. Reversely, distributional semantics assumes that similar distributions of terms indicate semantically similar texts. Word embeddings, Latent Semantic Analysis (LSA), and Semantic Concept Analysis (SCA) are established methods for language modeling and semantic text analysis derived from the idea of distributional semantics. The methods differ in the scope within which they consider co-occurring terms. Word embeddings consider the surrounding terms, LSA analyzes the entire document, and SCA uses an external corpus.

Recent detection methods often combine knowledge-based, corpus-based, and potentially lexical and syntactic analysis methods.

#### Vocabulary Expansion and Word Sense Disambiguation

Many detection methods use semantic networks for knowledge-based vocabulary expansion and word sense disambiguation to improve the detection effectiveness for non-lexis-preserving forms of plagiarism. WordNet is a semantic network that researchers frequently use for this purpose. The WordNet thesaurus groups terms by part of speech and assigns them to sets of synonyms (synsets). Additionally, WordNet contains many linguistic relations, making it especially suitable for the computation of semantic similarity.

Chen et al. [82], Palkovskii et al. [385], Álvarez-Carmona et al. [20], and Al-Shameri & Gheni [18] are representative examples of methods that consider WordNet synonyms in addition to verbatim term matches or use the graph-based distance of terms in WordNet to distinguish ambiguous words. We refrain from listing the large number of other detection methods that employ similar approaches.

## Semantic Role Labeling

Semantic Role Labeling (SRL) is a knowledge-based semantic analysis approach. SRL determines the roles of terms in a sentence, e.g., the actor, action, goal, event, and relations of these entities, using roles predefined in linguistic resources, such as PropBank [388] or VerbNet [275]. The goal is extracting the information on who did what to whom, where, and when [392]. The first steps in SRL are part-of-speech tagging and syntax analysis to obtain the dependency tree of a sentence. Subsequently, semantic annotation is performed [131].

**Table 2.12** summarizes detection methods that employ semantic role labeling as part of the detailed analysis task in external plagiarism detection.

**Table 2.12.** External detection methods using SRL for detailed analysis.

| Method Variation | References |
|---|---|
| SRL and WordNet graph distance for sentence ranking and similarity computation | [392] |
| SRL for syntactic and semantic similarity comparison | [379]–[381] |
| SRL as a feature for a machine learning approach | [131] |

## Word Embeddings

Word embeddings derive a vector representation of terms in a text by analyzing the words surrounding the term in question. The idea is that terms appearing in proximity to a term are more characteristic of the semantic concept represented by the term. Therefore, terms that frequently co-occur in proximity within texts also appear closer within the vector space [135, p. 418]. The scope of the embedding varies greatly from directly adjacent words, over sentences or passages, to entire documents. **Table 2.13** shows publications that employ word embeddings.

**Table 2.13.** Detection methods employing word embeddings.

| Paradigm | Task | References |
|---|---|---|
| **External** | Candidate Retrieval | [358] |
| | Cross-language PD | [181] |
| **Intrinsic** | Paraphrase Identification | [35], [133], [135], [225], [503] |
| | Author clustering | [443] |
| | Style-breach detection | [433] |

### Latent Semantic Analysis

Latent Semantic Analysis is a technique to reveal and compare the underlying semantic structure of texts [103]. To determine the similarity of term distributions in texts, LSA computes a matrix, in which rows represent terms, columns represent documents, and the matrix entries typically represent log-weighted tf-idf values [80, p. 112f.]. LSA then employs Singular Value Decomposition (SVD) or similar dimensionality reduction techniques to find a lower-rank approximation of the term-document matrix. SVD reduces the number of rows (i.e., prunes less relevant terms) while maintaining the similarity distribution between columns (i.e., the text representations). The terms remaining after the dimensionality reduction are assumed to be most representative of the semantic meaning of the text. Hence, comparing the rank-reduced matrix-representations of texts allows computing the semantic similarity of the texts [80, p. 112ff.]. LSA can reveal similar texts that traditional vector space models cannot express [233, p. 248]. While LSA handles synonymy well, its ability to reflect polysemy is limited [103, p. 400]. **Table 2.14** lists research papers employing LSA for plagiarism detection.

**Table 2.14.** Detection methods employing Latent Semantic Analysis.

| Parad. | Task | Method Variation | Ref. |
|---|---|---|---|
| External | Document Level Detection | LSA for phrases | [80] |
| | | LSA with tf-idf weighting for phrases | [234], [235] |
| | | LSA combined with other methods | [515] |
| | Cand. Retrieval | Pure LSA | [475] |
| | Paraphrase Identification | Pure LSA | [444] |
| | | LSA combined with machine learning | [14], [503], [519], [573] |
| | | Weighted matrix factorization | [12] |
| Intrinsic | Document Level Detection | LSA combined with stylometric feature analysis | [15] |
| | Author Attribution | LSA combined with machine learning | [16], [76] |
| | | LSA for character $n$-grams | [29] |

**Legend:**
**Cand.** Candidate, **LSA** Latent Semantic Analysis, **Parad.** Paradigm, **Ref.** References

Ceska was the first to apply LSA for plagiarism detection [80]. The ability of LSA to address synonymy is beneficial for paraphrase identification. Satyapanich et al. considered two sentences as paraphrases if their LSA similarity is above a threshold [444]. AlSallal et al. proposed a novel weighting scheme that assigns higher weights to the most common terms and used LSA as a stylometric feature for intrinsic plagiarism detection [16]. Aldarmaki & Diab used weighted matrix factorization—a method similar to LSA—for cross-language paraphrase identification [12].

### Semantic Concept Analysis

Semantic Concept Analysis summarizes detection methods that represent a text by mapping the terms in the text into a concept space (see **Table 2.15**). This model improves word sense disambiguation, expands the vocabulary, enables language independence, and reduces the dependence on lexical text similarity. SCA is resistant to synonym replacements and syntactic changes, which improves the detection of paraphrased plagiarism. Using multilingual resources to derive the semantic concepts allows applying SCA for cross-language plagiarism detection [150].

To derive the concept representations, several authors used monolingual or cross-language **thesauri** that systematically encode knowledge about linguistic entities and their relations. For example, Ceska et al. [79] and Gupta et al. [193] used the EuroVoc multilingual thesaurus [421]. Ceska et al. computed the position-aware Jaccard similarity of the concept sets that represent the texts. Gupta et al. represented texts as vectors of EuroVoc concepts and employed a use-case-specific adaption of the cosine measure to quantify the similarity of the vectors.

**Explicit Semantic Analysis (ESA)** is another strategy for deriving a semantic concept representation of texts. Instead of relying on systematically encoded resources, ESA uses the topics in a knowledge base (typically Wikipedia or other encyclopedias) to model texts as semantic concept vectors. Each article in the knowledge base is an explicit description of the semantic content of the concept, i.e., the topic of the article [156, p. 1606]. ESA derives concept vectors whose components reflect the relevance of the text for each of the semantic concepts, i.e., articles in the knowledge base [156, p. 1607]. Applying vector similarity measures to the concept vectors then allows determining the texts' semantic similarity. Using multilingual corpora, such as Wikipedia, enables applying ESA for cross-language plagiarism detection [411].

**Knowledge Graph Analysis (KGA)** is a graph-backed variant of Semantic Concept Analysis. KGA represents a text as a weighted directed graph in which the nodes represent the semantic concepts expressed by the words in the text, and the edges represent the relations between these concepts [150]. The relations are typically obtained from public databases, such as BabelNet [364] or WordNet. Determining the edge weights is the major challenge in KGA. Applying graph similarity measures yields a similarity score for documents or parts thereof (typically sentences). KGA achieves high detection effectiveness if the text is translated literally; for sense-for-sense translations, the results worsen [149].

Franco-Salvador et al. applied KGA using the BabelNet knowledge base for cross-language plagiarism detection [147]. In subsequent research, they presented improvements to their original method. For example, they improved the weighting procedure by using continuous skip-grams that additionally consider the context in which the concepts appear [150]. Dan & Bhattacharyya [100] constructed a semantic concept graph for sentences using the Universal Networking Language system [75] and enhanced the graph with lexical and syntactic similarity features.

**Table 2.15.** External detection methods employing Semantic Concept Analysis.

| Task | Method Variation | References |
|---|---|---|
| **Document Level Detection** | EuroVoc concept sets, Jaccard sim. | [79] |
| | EuroVoc concept vectors, cosine sim. | [193] |
| | ESA using Wikipedia combined with sequence similarity measures | [341] |
| **Paraphrase Identification** | KGA using UNL concepts enhanced with lexical and syntactic features | [100] |
| **Cross-language PD** | ESA using Wikipedia | [134], [406] |
| | KGA using BabelNet | [147]–[151] |

**Legend:**
**ESA** Explicit Semantic Analysis, **KGA** Knowledge Graph Analysis,
**sim.** similarity, **UNL** Universal Networking Language

## Alignment Methods

**Word Alignment** is a knowledge-based semantic analysis approach widely used for machine translation [152, p. 293] and paraphrase identification [319, p. 648]. Words are aligned if they are semantically similar according to their relations in the semantic network. The algorithms compute the semantic similarity of sentences as the proportion of aligned words. Word alignment approaches achieved the best performance for the paraphrase identification task at the SemEval workshop 2014 [495] and were among the top-performing approaches at SemEval-2015 [19], [496].

**Cross-language Alignment-based Similarity Analysis (CL-ASA)** is a corpus-based variation of the word alignment approach for cross-language semantic analysis. CL-ASA uses a parallel corpus to compute the probability that a term $x$ in the suspicious document is a valid translation of the term $y$ in a potential source for all terms in both documents. The sum of the term translation probabilities yields the probability that the suspicious document is a translation of the source [41, p. 213]. **Table 2.16** presents papers using Word alignment and CL-ASA.

**Table 2.16.** External detection methods employing alignment approaches.

| Task | Method Variation | References |
|---|---|---|
| **Document Level Detection** | Word alignment with modified Jaccard and Levenshtein similarity | [20] |
| **Paraphrase Identification** | Pure word alignment | [495], [496] |
| | Word alignment with machine learning | [222], [564] |
| **Cross-language PD** | CL-ASA | [41], [134] |
| | Translation and word alignment | [134] |

**Legend:**
**CL-ASA** Cross-language alignment-based similarity analysis

## Semantic Graph Analysis

Semantic graph analysis summarizes methods that use knowledge-based or corpus-based semantic features, potentially in combination with lexical and syntactic features, to represent the input text as a graph. Unlike semantic concept analysis methods, semantic graph analysis methods do not derive a language-independent representation of the text. Thus, the graphs address monolingual analysis tasks.

Researchers introduced a large variety of graph-based representations to model text for semantic analysis tasks. The graphs differ in the units chosen as nodes, e.g., words, phrases, or sentences, the number and type of edges, and the procedures for deriving edge weights. The edges can represent a plethora of lexical, syntactic, and semantic relations, as well as their combinations. For example, edges may indicate the co-occurrence of lexical units in the text or an external corpus. Likewise, edges can represent relations in a semantic network, e.g., synonymy or hypernymy. The possible schemes for deriving and assigning edge weights are virtually unlimited.

**Table 2.17** lists detection methods that employ semantic graph analysis. Kumar used semantic bi-word graphs for the seeding phase of the detailed analysis stage [290]. The edges expressed the semantic similarity of words based on the probability that the words co-occur in 100-word windows within a corpus of DBpedia [301] articles. Their method identified the similarity of text passages by computing the minimum weight bipartite clique cover. Momtaz et al. created word-based graphs for each sentence [352]. The edges expressed structural information derived from the sequence of words. Mohebbi et al. modeled sentences by constructing multiple graphs for different word classes [350]. For nouns and verbs, the edges represented several WordNet-based similarity metrics; for other word classes, the occurrence of the words in the analyzed sentences. To compute the similarity of sentences, the authors adapted the maximum matching approach for bipartite graphs.

**Table 2.17.** External detection methods employing graph-based analysis.

| Task | Method Variation | References |
|---|---|---|
| **Detailed Analysis** | Semantic bi-word graphs | [290] |
| | Word graphs for sentences | [352] |
| **Paraphrase Identification** | Multiple word-graphs for sentences | [350] |

## 2.4.6 Idea-based Detection Methods

Idea-based detection methods analyze non-textual content to complement methods that analyze lexical, syntactic, and semantic text similarity. Gipp & Beel introduced the idea for this class of detection methods [169]. They proposed analyzing the sequences of in-text citations in academic documents for similar patterns. Such patterns can indicate a high semantic similarity of the documents' content, regardless of whether the text has been paraphrased or translated.

In collaboration with Bela Gipp and others, I extended this initial work into the Citation-based Plagiarism Detection methods presented in Chapter 3, p. 79. This research laid the groundwork for the other idea-based detection methods analyzing images and mathematical content presented in this thesis.

**Table 2.18** lists publications on idea-based detection methods—asterisks (*) denote publications we summarize in subsequent chapters of this thesis. Therefore, the remainder of this section only discusses works by other authors.

**Table 2.18.** Idea-based methods for external plagiarism detection.

| Task | Method Variation | References |
|---|---|---|
| **Document Level Detection** | Analysis of in-text citation patterns | [169], [170]*, [172]*, [173], [175]*, [337]*, [339]*, [344]*, [397] |
| | Analysis of references | [196], [202], [329] |
| | Combination of text and citation analysis | [203], [204], [343]*, [398] |
| | Image analysis | [11], [32], [113], [115], [116], [232], [240], [342]*, [382], [422], [479] |
| | Analysis of mathematical sim. | [340]*, [344]*, [456]* |
| **Cross-language PD** | Analysis of in-text citation patterns | [171]*, [174]* |

**Legend:** * Publications we authored, **sim.** similarity

## Citation-based Detection Methods

HaCohen-Kerner et al. found that comparing bibliographic references detected potential plagiarism nearly as well as $n$-gram comparison methods yet required much less computational effort [202]. Combining the reference and $n$-gram comparisons improved the detection effectiveness. In a later work, they employed the reference comparison as a filtering heuristic for the candidate retrieval stage [203], [204].

Gureev & Mazov compared reference lists to identify translated and idea plagiarism in Russian research papers [196], [329]. Their objective was to adapt our work on Citation-based Plagiarism Detection, which requires full-text access for comparing in-text citations, to the limited availability of academic full-text databases in Russia. Pertile et al. extended our work on citation pattern analysis by introducing new similarity measures that consider the co-occurrence of citations in text segments [397]. In later research, they used machine learning to devise a hybrid approach that combines citation analysis and text-based detection methods [398].

## Image-based Detection Methods

Research on content-based image retrieval has yielded numerous approaches to identify similarities in images [516]. Several researchers adapted these methods to address the identification of plagiarized figures and images.

For natural images, Hurtik & Hodakova used higher degree F-transform to devise a highly efficient and reliable method to identify exact copies or cropped parts thereof [232]. The method does not consider image alterations aside from cropping. Iwanowski et al. evaluated the suitability of well-established feature point methods, such as SIFT [312], SURF [46], and BRISK [303], to additionally retrieve visually altered replications of natural images [240]. Srivastava et al. [479] addressed the same task using a combination of SIFT features and perceptual hashing [449]. Eisa et al. demonstrated that a text-based analysis of the caption and the sentences in the main text referring to a figure is often sufficient to reveal a potentially suspicious similarity between figures [115].

Other researchers addressed identifying specific types of plagiarized figures. Al-Dabbagh et al. identified copied or slightly modified bar charts by performing pairwise comparisons of the charts based on the numeric values of the bars and the text in the charts [11]. Rabiu & Salim detected plagiarized diagrams by computing the graph edit distance of matching components in two diagrams as part of pairwise comparisons of the diagrams [422]. They considered components as matching if their Jaccard similarity calculated for word unigrams was greater than 0.5. Arrish et al. detected similar flow charts, i.e., diagrams consisting of four shape types [32]. They represented the flow charts as vectors that reflected the occurrence frequencies of the four shape types and compared the vectors using the cosine similarity. Eisa et al. improved the similarity analysis for diagrams by considering the edges between components [113]. In later research, Eisa et al. extended their compound similarity computation for diagrams by including an analysis of the predicate-argument structure of the text in the components [116].

### 2.4.7 Hybrid Detection Methods

Each class of detection methods we described so far has characteristic strengths and weaknesses. Many authors showed that combining detection methods into hybrid methods achieved better results than applying the methods individually [8], [118], [151], [254], [271], [482], [484], [496], [569], [573].

There are three general approaches to combining plagiarism detection methods.

» **Adaptive algorithms** determine the obfuscation strategy, choose the detection methods, and set similarity thresholds accordingly.

» **Ensembles of detection methods** combine the results of individual detection methods using static weights.

» **Machine learning** determines the best combination of detection methods.

## Adaptive Algorithms and Ensembles

The winning detailed analysis method at PAN 2014 and 2015 [437] used an **adaptive algorithm**. After finding the seeds of overlapping passages, the authors extended the seeds using two different thresholds for the maximum gap. Based on the length of the passages, the algorithm recognized various plagiarism types and set the parameters for the VSM-based detection method accordingly.

The "linguistic knowledge method" of Abdi et al. [2] exemplifies an **ensemble of detection methods**. It combines the analysis of syntactic and semantic sentence similarity using a linear combination of two similarity metrics: i) the cosine similarity of semantic vectors and ii) the similarity of syntactic word order vectors [2]. The method outperformed other contesters on the PAN-10 and PAN-11 corpora (cf. Section 2.5.1, p. 49). **Table 2.19** lists other ensembles of detection methods.

**Table 2.19.** Ensembles of detection methods.

| Task | Method Variation | References |
|---|---|---|
| **Document Level Detection** | Combination of semantic and syntactic sentence vector representations | [2] |
| **Candidate Retrieval** | Combination of querying heuristics for web search engines | [271] |
| **Detailed Analysis** | Combination of sentence vectors using adaptive algorithms | [387], [436], [437] |

## Machine Learning

Machine learning-based detection methods typically train a classification model that combines several features. In intrinsic plagiarism detection, a combined analysis of stylometric features is the standard approach [484].

**Table 2.20** shows that many recent intrinsic detection methods employ machine learning to select the best performing feature combination [483]. A widely-used method for author verification is **unmasking** [482], which uses a classifier to distinguish the stylistic features of the suspicious document from the features in documents with known authorship. The idea is to train and run the classifier, remove the most significant features from the classification model and rerun the classifier. If the accuracy drops significantly, the documents likely have the same author; otherwise, different authors likely wrote the documents [482]. There is no consensus on the best features for authorship identification [318].

**Table 2.20.** Intrinsic detection methods using machine learning.

| Task | Features | Classif. | References |
|---|---|---|---|
| **Authorship Attribution** | Semantic (LSA) | SVM | [16] |
| **Author Verification** | Lexical | DT | [155] |
| | | EER | [212] |
| | | HBC | [197] |
| | | KNN | [210] |
| | | Misc. | [299], [442] |
| | | NB | [328] |
| | | RNN | [38] |
| | Lexical, semantic (LSA) | MLP | [17] |
| | Lexical, syntactic | GA | [355], [356] |
| | | KNN | [164] |
| | | Misc. | [354] |
| | | RF | [42], [318], [384] |
| | | SVM | [4], [130], [231], [286], [518] |
| **Author Clustering** | Lexical | RNN | [39] |
| | Lexical, syntactic | SVM | [576] |
| **Style Breach Detection** | Lexical, syntactic | GB, RT | [292] |

**Legend:**
**Classif.** Classifier, **DT** Decision Tree, **EER** Equal Error Rate, **GA** Genetic Algorithm, **GB** Gradient Boosting, **HBC** Homotopy-based Classification, **KNN** *k* Nearest Neighbors, **LSA** Latent Semantic Analysis, **MLP** Multilayer Perceptron, **NB** Naïve Bayes, **RF** Random Forrest, **RNN** Recurrent Neural Network, **RT** Regression Tree, **SVM** Support Vector Machines

**Table 2.21** shows that researchers also studied the application of machine learning for various components of the external plagiarism detection process. Kanjirangat & Gupta detected idea plagiarism using a genetic algorithm applied in a preprocessing step [260]. The method randomly chooses a set of sentences as chromosomes and combines the sentence sets that are most descriptive of the entire document. The combined sentence sets form the next generation. This way, the method gradually extracts semantic concepts, i.e., the sentences representing the document's idea. The semantic concepts then serve to retrieve semantically similar documents and passages. Gharavi et al. used machine learning to find the similarity thresholds for a vector space model [165].

Sánchez-Vega et al. proposed a method termed rewriting index to improve the detailed analysis stage [438]. Using five Turing machines, the method evaluates the degree of membership of each sentence in the suspicious document to a possible source document. Each Turing machine targets a specific plagiarism type, such as verbatim copying or basic word-level transformations (insertion, deletion, substitution). The output values of the Turing machines serve as the features for training a Naïve Bayes classifier, which identifies reused passages.

**Table 2.21.** External detection methods using machine learning.

| Task | Features | Classifier | Ref. |
|---|---|---|---|
| **Document Level Detection** | Citations | Several | [398] |
| | Lexical, semantic | SVM, NB | [14] |
| | | NB, SVM, DT | [262] |
| | Semantic | SVM | [127], [233] |
| | Syntactic | DT, KNN | [59] |
| **Candidate Retrieval** | Lexical | SVM | [283] |
| | Lexical, semantic, other | Several | [557] |
| | Lexical, syntactic | LDA | [558] |
| | | GA | [260] |
| **Detailed Analysis** | Lexical | NB | [438] |
| | | NB, DT, RF | [256] |
| | Lexical, semantic | SVM | [290] |
| | Lexical, syntactic, semantic | LoR | [284] |
| **Cross-language PD** | Semantic | ANN | [151] |

**Legend:**
**ANN** Artificial Neural Network, **DT** Decision Tree, **GA** Genetic Algorithm, **KNN** $k$ Nearest Neighbors, **LDA** Linear Discriminant Analysis, **LoR** Logistic Regression, **NB** Naïve Bayes, **Ref.** References, **RF** Random Forrest, **SVM** Support Vector Machines

**Table 2.22.** Paraphrase identification methods using machine learning.

| Task | Features | Classifier | References |
|---|---|---|---|
| Paraphrase Identification | Lexical | ANN | [118] |
| | | KNN, SVM, ANN | [490] |
| | | SVM | [128] |
| | Lexical, semantic | ANN | [225] |
| | | GPR | [400] |
| | | Miscellaneous | [131], [434] |
| | | RiR | [393] |
| | | SVM | [222], [519] |
| | Lexical, syntactic, semantic | DT | [133], [135] |
| | | LiR | [100] |
| | | LoR | [569] |
| | | Miscellaneous | [541], [573] |
| | | SVM | [69], [72], [263] |
| | | SVM, ME | [19] |
| | Lexical, syntactic, semantic, MT metrics | SVM, RF, GB | [503] |
| | MT metrics | SVM | [54] |
| | Semantic | DNN | [6] |
| | | DT | [35] |
| | | IsoR | [306] |
| | Semantic, MT metrics | RF | [420] |
| | Syntactic, semantic | DNN | [5] |
| | | SVM | [432] |

**Legend:**
**ANN** Artificial Neural Network, **DT** Decision Tree, **DNN** Deep Neural Network, **GB** Gradient Boosting, **GPR** Gaussian Process Regression, **IsoR** Isotonic Regression, **KNN** $k$ Nearest Neighbors, **LiR** Linear Regression, **LoR** Logistic Regression, **ME** Maximum Entropy, **MT** Machine Translation **RF** Random Forrest, **RiR** Ridge Regression, **SVM** Support Vector Machines

Particularly for paraphrase identification, the application of machine learning is a standard approach, as the publications listed in **Table 2.22** reflect. Zarrella et al. won the 2015 SemEval competition on identifying semantic similarity of Twitter tweets with an ensemble of seven algorithms [569]. Most of the algorithms in the ensemble used machine learning methods. In the experiments of Afzal et al., the linear combination of supervised and unsupervised machine learning methods outperformed each of the methods applied individually [5].

**Table 2.21** and **Table 2.22** indicate that machine learning is most beneficial when applied for the detailed analysis. The tables also show that SVM is the most popular model type for plagiarism detection tasks. SVM minimizes the distance of a hyperplane to the training data. Choosing the hyperplane is the main challenge [127]. In the experiments of Alfikri & Purwarianti, SVM classifiers outperformed Naïve Bayes classifiers [14]. In the experiments of Subroto & Selamat, the best performing model combined SVM and an artificial neural network [490]. El-Alfy et al. found that an abductive network outperformed SVM [118].

The following section presents plagiarism detection systems that implement (some of) the detection methods. Section 2.5 then presents insights into the effectiveness of state-of-the-art plagiarism detection methods and systems.

### 2.4.8 Plagiarism Detection Systems

The industry for plagiarism detection systems is vast, expanding, and fast-paced. Companies offer a growing number of software solutions [37, p. 3], [145, p. 4], [359, p. 50], [551, p. 71], but many services cease to exist after a short life cycle [37, p. 3], [145, p. 8], [359, p. 53]. Due to the quickly changing landscape of plagiarism detection systems, we explain characteristics common to many systems instead of describing specific systems. Foltýnek et al. [145], Weber-Wulff et al. [544]–[547], [549], [550], and Chowdhury & Bhattacharyya [84] described numerous plagiarism detection systems in detail. Foltýnek et al. and Weber-Wulff et al. also provided the results of comprehensive performance evaluations of the presented systems.

**Production-grade plagiarism detection systems**—as opposed to research prototypes—exclusively follow the external detection paradigm. Comparing input documents to a user-specified closed set of potential sources, e.g., all submission for an assignment, is a specialization of external plagiarism detection termed **collusion detection** [293, p. 10], [551, p. 75]. Most providers offer their plagiarism detection systems as web-based services. Some systems run on the user's computer. Typically, these locally running systems allow collusion detection only.

The size and coverage of the reference collection influence the system's effectiveness significantly. Intuitively, systems can only find sources included in the reference collection. Major providers of plagiarism detection services include subsets of the Internet, copyrighted material, such as journal articles or books, and documents previously submitted for plagiarism checks in their reference collections [551, p. 72ff.]. Other providers do not maintain reference collections but use the APIs of web search engines [551, p. 76]. This approach limits the detectable sources to content on the publicly accessible Internet.

Providers of plagiarism detection systems rarely publish information on the detection methods they employ. Therefore, estimating to what extent plagiarism detection research influences practical applications is challenging. Due to the extensive reference collections that the systems must analyze and the systems' characteristic detection capabilities, which we discuss in Section 2.5.3, p. 68, we conclude that all systems rely on lexical detection methods, such as fingerprinting.

The use of plagiarism detection services is subject to several **legal restrictions**. For example, European Union (EU) data protection law requires higher education institutions to share data only with companies who store and process the data on servers within the EU [145, p. 33]. Many plagiarism detection services do not meet this criterion. Furthermore, EU copyright law requires students' permission to share content the students produced, e.g., essays or theses, with external parties [145, p. 33], [551, p. 73]. By interpreting US case law, Brinkmann argued that plagiarism detection services that permanently store students' work without the students' consent violate privacy law [66]. Moreover, Brinkman derived that US law entitles students to be fully informed if and how educators employ plagiarism detection services to check the students' work. Bilateral contracts, such as non-disclosure agreements for graduation theses compiled in cooperation with companies, can also prohibit the use of plagiarism detection services.

The need to disclose potentially sensitive content often raises concerns regarding the **confidentiality of the data**. Researchers often hesitate to share unpublished grant proposals or publication drafts due to privacy concerns. Data breaches are a risk for any cloud-based service. However, researchers and practitioners criticize plagiarism detection services for opaque data management and data protection procedures [126], [551, p. 73f.]. Providers have strong incentives to add documents submitted for checks to their reference collections to increase their detection capabilities. Many providers reserve the right to store and use content disclosed to them after completing the check for which the content was submitted. While several providers offer options to prevent the permanent storage of submitted content, these settings are often well-hidden or even ignored [551, p. 73].

Despite these challenges, plagiarism detection systems have become crucial support tools for academic institutions[6] and scientific publishers [245], [305]. The following section indicates the effectiveness of current plagiarism detection methods and production-grade plagiarism detection systems.

## 2.5 Evaluation of PD Methods and Systems

Sections 2.5.1 through 2.5.3 review evaluation efforts for plagiarism detection methods and plagiarism detection systems to answer the following questions:

1. Which datasets exist for evaluating the effectiveness of plagiarism detection methods and systems? (Section 2.5.1)
2. How effective are the state-of-the-art detection methods we presented in Section 2.4? (Section 2.5.2, p. 56)
3. How effective are production-grade plagiarism detection systems available from professional providers? (Section 2.5.3, p. 68)

### 2.5.1 Evaluation Datasets for Plagiarism Detection

The availability of datasets is essential for performing any empirical research on Natural Language Processing and Information Retrieval. The covert nature of academic plagiarism complicates creating datasets and conducting conclusive performance evaluations of plagiarism detection technology.

Evaluation datasets for plagiarism detection can either include **simulated plagiarism** or cases of **real plagiarism**. The two options have inherent advantages and disadvantages. In agreement with Potthast et al. [408, p. 1000], we see the following advantages of using datasets that include simulated plagiarism:

» **The lack of ground truth data for real plagiarism**: Academic plagiarists are highly motivated to avoid detection and meet the strict quality standards of peer-reviewed venues. Therefore, plagiarism is often disguised and hard to identify. Consequently, one can only approximate the presence or absence of real plagiarism in document collections. The extent of simulated plagiarism introduced into a collection is known, which enables computing fine-grained performance measures, e.g., at the level of words.

---

[6] Turnitin, a major provider of plagiarism detection software, states that 15,000 institutions in 150 countries use its service [510]

» **The bias towards less-obfuscated forms of plagiarism**: Due to the effort necessary to detect disguised forms of academic plagiarism, identified cases of real plagiarism typically exhibit a low level of disguise. Therefore, creating datasets from real plagiarism carries the risk of overrepresenting less obfuscated forms of plagiarism in the collection.

» **The limited reproducibility of studies using real plagiarism**: Academic documents are typically subject to copyright, which often prevents public sharing of datasets that include real plagiarism cases. This restriction impedes comparing a new approach to state-of-the-art methods and reproducing the results of other researchers.

However, datasets that use simulated plagiarism exhibit a **critical disadvantage**. Simulated plagiarism is typically created using automated methods, e.g., random text replacements and synonym substitutions, or by tasking non-experts, e.g., students, to create plagiarism. It is questionable whether these plagiarism instances are representative of the sophisticatedly disguised plagiarism committed by experienced researchers with a strong incentive to hide their actions.

Hereafter, we present datasets compiled for evaluating external plagiarism detection methods and systems. We focus on datasets that support the external detection paradigm because only these datasets are potentially relevant for evaluating the research contributions of this thesis.

### Monolingual Evaluation Datasets

**Table 2.23** lists datasets that researchers made available for evaluating monolingual external plagiarism detection methods. Datasets with the prefix PAN originate from the PAN workshop series, which started in 2007 [485]. The acronym reflects the title of the first workshop—**P**lagiarism Analysis, **A**uthorship Identification, and **N**ear-Duplicate Detection. Since 2009, the PAN workshops included a competition for evaluating extrinsic and intrinsic plagiarism detection methods [407]. From 2009 to 2015, the PAN competitions offered tasks on external detection [552]. From 2009 to 2011, the competitions used one dataset (PAN-PC) to evaluate the complete plagiarism detection process. From 2012 to 2015, the candidate retrieval (CR) and text alignment (TA) tasks were evaluated separately using task-specific datasets.

The PAN datasets mostly contain simulated monolingual plagiarism in English and, to a lesser extent, German-English and Spanish-English document pairs containing simulated cross-language plagiarism [206], [407]–[410], [412]–[414].

**Table 2.23.** Datasets for evaluating external monolingual plagiarism detection.

| T. | Name | Susp./Src. | Lan | Src. | Obfuscation | Ref. |
|---|---|---|---|---|---|---|
| **CR+TA** | PAN-PC* 2009-2011 | '09: 7.2K/7.2K '10: 15.9K/11.1K '11: 11.1K/11.1K | DE, EN, ES | Books | Automated, manual, translation | [407], [409], [410] |
| **CR** | PAN-SR 2012-2015 | 303/1B | EN | Webpages | Manual | [206], [412]–[414] |
| **Text Alignment** | PAN-TA* 2012 | 3.0K/3.5K | DE, EN, ES | Books | Automated, manual, translation | [412] |
| | PAN-TA 2013-2014 | 3.6K/4.7K | EN | Webpages | Automated | [413], [414] |
| | Mohtaj15 | 952/3,309 | EN | Wikipedia, SemEval | Automated | [351] |
| | Cheema15 | 500/500 | EN | Essays | Manual | [81] |
| | Clough11 | 95/5 | EN | Short answers | Manual | [87] |
| | Alvi15 | 4x 50/50 | EN | Fairy tales | Retelling, automated | [22] |
| | Khoshnavataher15 | 1K/1K | FA | Wikipedia | Automated | [273] |
| | Mahak Samim | 2.5K/2.5K | FA | Journal articles | Automated | [461] |
| | Siddiqui14 | 1,156/509 | AR | Essays | Manual | [468] |

**Legend:**
**AR** Arabic, **Auto.** Automated, **CR** Candidate Retrieval, **DE** German, **EN** English, **ES** Spanish, **FA** Farsi, **Lan.** Language, **Ref.** References, **Src.** Source, **Susp.** Suspicious, **T.** Task, **TA** Text Alignment

* Numbers refer to monolingual cases (10-15% of the cases are cross-language).

To obfuscate most of the simulated plagiarism instances, the organizers employed the following techniques:

» **Automated obfuscation**: randomly removing, inserting, or replacing words or phrases; substituting words with their synonyms, antonyms, hyponyms, or hypernyms selected at random; randomly rearranging words while keeping the parts of speech sequence of the original [408]; machine translating text segments; using semantically equivalent text segments in English-German and English-Spanish retrieved from a parallel corpus [412]

» **Manual obfuscation**: hiring workers to manually paraphrase a given text segment [408] or writing essays about a given topic [412]

Several researchers compiled additional datasets by adopting the procedures for creating the PAN datasets. Mohtaj et al. created an English dataset by applying automated obfuscation techniques similar to those of the PAN datasets to Wikipedia articles [351]. The authors used the datasets of the SemEval [34] semantic text similarity task to retrieve semantically equivalent text segments.

Khoshnavataher et al. [273] and Sharifabadi & Eftekhari [461] applied automated obfuscation techniques to articles from Wikipedia and Persian research journals to create a Farsi language dataset. Cheema et al. [81], Clough & Stevenson [88] hired students to write partially plagiarized essays or short answers for assignments in English. Siddiqui et al. did the same for Arabic [468]. Alvi et al. followed an innovative approach for creating an English dataset [22]. The authors used versions of Grimm fairy tales that exhibit linguistic differences from having been retold over time. In addition to these natural differences in the text, Alvi et al. also employed automated obfuscation techniques to mask the texts' similarity further.

The creators of the PAN datasets varied the length of documents, the amount of plagiarism in documents, the length of plagiarism instances, and the topical domain and obfuscation of plagiarism instances [408], [412].

## Cross-language Evaluation Datasets

**Table 2.24** presents datasets for evaluating cross-language plagiarism detection methods. For compiling these datasets, the creators followed approaches similar to those for creating monolingual datasets. For creating the PAN-PC corpora 2010 and 2011, Potthast et al. predominantly employed online machine translation services to translate passages from books in the public domain [409, p. 3]. For about 1% of the test cases in the PAN-PC-11 corpus, Potthast et al. tasked workers hired via a crowdsourcing service with manually obfuscating the machine-translated texts [410, p. 2]. The PAN organizers embedded the translated passages into lexically similar texts. For the PAN Text Alignment Corpus 2012, Potthast et al. changed the process for creating cross-language test cases because the previous competitions had shown that the machine-translated cases were too easy to find. The participating detection methods simply employed the same online machine translation services the organizers had used to create the test cases to translate all non-English documents in the corpus [410, p. 2], [412, p. 17]. In 2012, Potthast et al. used manually translated passages from the Europarl parallel corpus [279].

**Table 2.24.** Datasets for evaluating external cross-language PD methods.

| T. | Name | Susp./Src. | Lan. | Src. | Obfuscation | Ref. |
|---|---|---|---|---|---|---|
| Text Alignment | PAN-PC* 2010-2011 | 2010: 557/601 (9,598 cases) 2011: 555/550 (5,575 cases) | DE, EN, ES | Books | Manual and machine transl. | [409], [410] |
| | PAN-TA* 2012 | n.a./n.a. 500 cases | | Europarl paral. corp. | Dissimilarity of aligned passages | [412] |
| | Ferrero16 | 39K/39K | EN, ES, FR | Multiple | Automated | [132] |
| | Asghari15 | 7.1K/19.9K | EN, FA | Wikipedia, paral. corp. | Dissimilarity of aligned sent. | [33] |
| | Hanif15 | 500/500 | EN, UR | Essays | Manual transl. | [221] |
| | Kong15 | 20/55 | EN, ZH | Essays | Manual transl. | [285] |

**Legend:**
**Corp.** Corpus, **EN** English, **ES** Spanish, **FA** Farsi, **Lan.** Language, **n.a.** not available, **paral.** parallel, **Ref.** References, **sent.** sentences, **Src.** Source, **Susp.** Suspicious, **T.** Task, **TA** Text Alignment, **transl.** translation, **UR** Urdu, **ZH** Chinese
* Numbers refer to cross-language cases only.

Asghari et al. embedded sentences from a Persian sentence-aligned paraphrase corpus into topically related Wikipedia articles [33]. Hanif et al. [221] and Kong et al. [285] tasked students with translating text passages in Urdu into English and, respectively, plagiarizing English webpages for writing Chinese essays. Ferrero et al.

combined and rearranged data from six multilingual parallel and comparable corpora[7]. The corpora include legal texts and conversational transcripts of the EU, Wikipedia articles, Amazon product reviews, and research papers, which the original authors translated [132]. The texts are in English, German, and Spanish, translated either by automated methods or humans.

The purpose of all datasets in **Table 2.23** and **Table 2.24** is to enable highly accurate performance evaluations of plagiarism detection methods. Therefore, the datasets include all source documents from which the creators took content to generate the simulated instances of plagiarism. Additionally, about 50% of the documents in most of the datasets do not contain plagiarism. The datasets also include detailed information on the exact locations of all simulated plagiarism instances.

## Evaluation Datasets for Production-grade PD Systems

**Table 2.25.** Datasets for evaluating production-grade PD systems.

| Name | Susp./Src. | Lan. | Type of Susp. Docs. | Ref. |
|---|---|---|---|---|
| Weber-Wulff13 | 35/89 | DE, EN, HE | Essays manually plagiarized from webpages and digitally available sources, such as scientific publications, newspapers, and magazines. | [550] |
| ENAI20 | 98/n.a. | CZ, DE, EN, ES, IT, LV, SK, TR | | [145] |

**Legend:**
**CZ** Czech, **DE** German, **Docs.** Documents, **EN** English, **ES** Spanish, **FA** Farsi, **IT** Italian, **Lan.** Language, **LV** Latvian, **Ref.** References, **SK** Slovakian, **Src.** Source, **Susp.** Suspicious, **TR** Turkish

The datasets in **Table 2.25** serve to evaluate production-grade plagiarism detection systems. Because a provider's reference collection coverage is a critical evaluation criterion, these datasets do not include the source documents from which plagiarized content originates. However, the creators of the datasets selected sources that the plagiarism detection services could include in their reference collections. The suspicious documents are essays that humans wrote and partially plagiarized. The datasets do not include information on the exact extent and location of plagiarized content in the suspicious documents.

[7] Parallel corpora consist of texts in language A and the translations of the texts in language B. Comparable corpora consist of texts of the same type, e.g., news articles, or on the same topic written in different languages. The text are not translations of one another [269, p. 487].

## Collections of Real Plagiarism Cases

**Table 2.26.** Collections of confirmed cases of plagiarism.

| Name | Susp./Src. | Lan. | Type of Susp. Docs. | Ref. |
|---|---|---|---|---|
| GuttenPlag | 1/135 | DE, EN | Doctoral theses | [199] |
| VroniPlag | 207/3,965* | DE, EN** | Doctoral and habilitation theses | [538] |
| Retraction Watch | 2,375*/n.a. | EN** | Academic publications | [427] |

**Legend:**
**DE** German, **Docs.** Documents, **EN** English, **Lan.** Language,
**n.a.** not available, **Ref.** References, **Src.** Source, **Susp.** Suspicious
* As of July 2020
** Statistics on the languages of sources are not available. Sources may also be in other languages than those listed in the table.

**Table 2.26** completes our review of potential evaluation datasets by summarizing public collections of confirmed and alleged cases of academic plagiarism. The **GuttenPlag** and **VroniPlag** projects are crowdsourced efforts of volunteers who investigate alleged plagiarism in doctoral and habilitation theses publicly. The projects use wikis to coordinate their work and present results. Both projects only publish so-called fragments, i.e., excerpts of content in the suspicious document that appears plagiarized and the corresponding excerpt from the source. Details on each source are available; however, many sources are not publicly accessible.

The GuttenPlag project initiated the collaborative wiki-centered process for investigating plagiarism allegations. Volunteers started the project after a law professor had found uncited copies of text in the doctoral thesis of the—at that time—German minister of defense K.T. zu Guttenberg [137], [419]. The volunteers sought to substantiate the plagiarism allegations, which Guttenberg denied. The volunteers found 1,218 plagiarized fragments originating from 135 sources on 371 of 393 pages in the thesis [199]. In a subsequent review of the thesis, the responsible university confirmed the allegations and rescinded the doctorate [104], [280, p. 13ff.].

Members of the GuttenPlag project initiated the VroniPlag wiki as a follow-up investigation into allegations of plagiarism in the doctoral thesis of Veronika Saß [403]. VroniPlag's investigations follow the meticulous process for annotating fragments established in the GuttenPlag project. Opposed to GuttenPlag, VroniPlag has not restricted its efforts to one thesis. As of July 2020, the VroniPlag project has published the results of analyzing 207 doctoral and habilitation theses in which the volunteers found substantial content that the authors did not acknowledge according to academic standards. Nearly all allegations published on VroniPlag triggered official investigations. Most of these investigations ended with a reprimand of the thesis or the withdrawal of the conferred degree [538].

**Retraction Watch** is a non-profit project reporting on retractions and corrections of scientific publications in a blog [377]. Additionally, Retraction Watch offers a database accessible via a web-based search interface to look up information about reported cases [427]. Retraction Watch reports on retractions for any reason, i.e., not only plagiarism but also errors, new insights, falsification, and others. As of July 2020, the Retraction Watch database includes 21,936 retraction notices, thereof 2,375 that list plagiarized content as the reason for the retraction. Retraction Watch only links to the documents pertaining to a retraction but does not provide the full-text documents.

### 2.5.2 Evaluation of Plagiarism Detection Methods

Comparing the effectiveness of plagiarism detection methods proposed in research publications is difficult because researchers often use different datasets, e.g., self-created datasets or subsets of publicly available datasets.

The **PAN competition series** is the most comprehensive, comparable evaluation of plagiarism detection methods to date. Therefore, we present the results of comparing the effectiveness of plagiarism detection methods from the shared tasks on candidate retrieval and detailed analysis in the PAN competitions 2012–2015.

#### Monolingual Candidate Retrieval

The PAN shared tasks on candidate retrieval [206], [412]–[414] required participants to retrieve potential source documents for input documents using a search engine the organizers provided. The search engine indexed a crawl of the Internet consisting of approximately 1.0 billion web pages, of which 50% are in English.

The PAN competitions measured the effectiveness of detection methods in terms of **precision** ($P$), **recall** ($R$), and $F_1$**-measure** calculated for documents the methods downloaded. Given that plagiarism detection is an information retrieval task, researchers typically employ these well-established set-based performance measures ($P, R,$ and $F_1$) or ranked-based measures like Mean Reciprocal Rank (MRR) and Mean Average Precision (MAP) for evaluations. We refer readers unfamiliar with Information Retrieval evaluations to the excellent and openly available introductory texts by Manning et al. [322, Ch. 8, p. 151ff.] and Clough & Sanderson [89].

The PAN competitions additionally evaluated the efficiency of detection methods in terms of **runtime and costs** using the following metrics [206], [412]–[414]:

1. The number of queries a method submitted to the search engine
2. The number of web pages a method downloaded after being presented the results list of the search engine
3. The number of queries a method submitted until retrieving the first true source document
4. The number of downloads a method performed until downloading the first true source document
5. The total runtime a method required for processing the test set of input documents (2012: 32 documents, 2013: 58, 2014: 99, 2015: 99)

The metrics one and two in the list above reflect the factors that typically influence the price for using commercial web search engines, i.e., are most relevant for plagiarism detection systems that do not maintain reference collections. Metrics three and four indicate how fast a system could point a user to any suspicious result. These factors are relevant for production-grade detection systems. The final retrieval effectiveness and total processing time are more relevant for evaluating research contributions than the time required for reporting the first partial results. Therefore, we do not report the results for the third and fourth metrics.

**Table 2.27** presents the results for the five detection methods that participated in the PAN candidate retrieval tasks 2012–2015 and achieved the best recall scores for the PAN-SR-15 dataset (cf. **Table 2.23**, p. 51). Recall is the most critical performance measure for the candidate retrieval stage as failing to retrieve a source prohibits detecting content that originates from that source in the subsequent detailed analysis stage. The values shown in the table are averaged over all 99 test documents. Boldface indicates the best score(s) in each column.

**Table 2.27.** Effectiveness of candidate retrieval methods in PAN-PC 2012-2015.

| # | Name | ***P*** | ***R*** | $F_1$ | ***Q*** | ***L*** | $\|\overline{X}\|$ | ***t*** | **Ref.** |
|---|---|---|---|---|---|---|---|---|---|
| 1 | Kong 2013 | .01 | **.59** | .01 | **47.90** | 5,185.30 | **0** | 106h 13' 46" | [282] |
| 2 | Prakash 2014 | .38 | .51 | .39 | 60.00 | 38.80 | 7 | **19h 47' 45"** | [417] |
| 3 | Kong 2014 | .08 | .48 | .12 | 83.50 | 207.10 | 6 | 24h 03' 31" | [283] |
| 4 | Williams 2014 | .57 | .48 | **.47** | 117.10 | 14.40 | 4 | 39h 44' 11" | [558] |
| 5 | Williams 2013 | **.60** | .47 | **.47** | 117.10 | **12.40** | 7 | 76h 58' 22" | [556] |

**Legend:**
***P***: Precision, ***R***: Recall, $\boldsymbol{F_1} = 2(P \cdot R)/(P + R)$, ***Q***: Queries, ***L***: Downloads, $|\overline{\boldsymbol{X}}|$: Number of undetected sources, ***t***: Runtime, **Ref.** References

None of the methods achieved the best results for all performance measures. By retrieving 59% of all source documents of the 99 test documents, the method of Kong et al. for the competition 2013 (cf. Kong 2013 in **Table 2.27**) achieved the best recall. It was the only method that identified at least one of the sources for each suspicious document. This good retrieval effectiveness came at the cost of performing the most downloads and requiring the longest processing time. However, the method issued the fewest queries per input document.

In 2014, Kong et al. submitted an updated method that required 96% fewer downloads and 78% less processing time than their method in 2013. However, this method used 74% more queries per document, achieved 11% less recall than in 2013, and missed to retrieve the sources for six input documents. The methods of Prakash et al. and Williams et al. (cf. Prakash 2014, Williams 2013, and Williams 2014 in **Table 2.27**) exhibited similar performance. The method Prakash 2013 used 99.7% fewer downloads and 72% less processing time than the method Kong 2013 to achieve a recall of 51%. However, Prakash 2013 failed to retrieve any source documents for seven plagiarized documents (7% of all test documents). The methods Williams 2013 and Williams 2014 used the fewest downloads but the most queries per document. Both methods achieved the best balance between precision and recall, resulting in the best $F_1$ scores of 0.47.

In summary, the PAN experiments indicate that exceeding a recall of 60% for documents with simulated plagiarism is an open research challenge even in laboratory settings. A tradeoff between workload and retrieval effectiveness is observable for methods that use external search engines. The queries issued to the search engine are a significant cost factor, which is why methods in the PAN competitions tend to use few queries. For systems that maintain reference collections, the number of queries is less relevant. Using more queries would likely increase the recall.

However, for production-grade plagiarism detection systems, the coverage of the reference collection and the need to achieve low processing times may counteract the gains in recall from issuing more queries. The PAN setup guarantees that the reference collection contains the source documents, which is not the case for production-grade systems. Moreover, the best-performing candidate retrieval method in the PAN competitions (cf. Kong 2013 in **Table 2.27**) required more than one hour of processing time for retrieving potential source documents for a moderately sized text (5,000 words). We expect this workload is economically infeasible for production-grade plagiarism detection systems. The candidate retrieval stage typically requires less computing effort than the subsequent detailed analysis stage. Whether a plagiarism detection system maintains a reference collection or uses an external search engine, we assume that spending these resources for retrieving sources would not be economically viable. We substantiate our assumption by presenting evaluations of the effectiveness of production-grade plagiarism detection systems in Section 2.5.3, p. 68. Before that, we present results on the performance of plagiarism detection methods for the detailed analysis stage.

## Monolingual Detailed Analysis

To evaluate detection methods participating in the PAN detailed analysis tasks, Potthast et al. proposed use-case-specific extensions of precision and recall at the level of characters [408, p. 998f.], plagiarism cases, and documents [414, p. 15f.].

To assess the effectiveness at the level of characters, Potthast et al. introduced the **PlagDet score** [408, p. 998f.]. The measure considers the set of plagiarism cases $C$, and the set of detections $X$ a method reports[8]. The authors defined a plagiarism case $c \in C$ as a four-tuple $c = \langle c_{\text{plg}}, d_{\text{plg}}, c_{\text{src}}, d_{\text{src}} \rangle$. Here, $s_{\text{plg}}$ represents a plagiarized text passage in a document that contains plagiarism $d_{\text{plg}}$ and $c_{\text{src}}$ the corresponding source passage in a source document $d_{\text{src}}$. A plagiarism detection $x \in X$ for document $d_{\text{plg}}$ is then a four-tuple $x = \langle x_{\text{plg}}, d_{\text{plg}}, x_{\text{src}}, d'_{\text{src}} \rangle$. The detection $x$ associates an allegedly plagiarized passage $x$ in $d_{\text{plg}}$ with a passage $x_{\text{src}}$ in an alleged source document $d'_{\text{src}}$. Potthast et al. further defined that $x$ detects $c$ iff $x_{\text{plg}} \cap c_{\text{plg}} \neq \emptyset$, $x_{\text{src}} \cap c_{\text{src}} \neq \emptyset$, and $d'_{\text{src}} = d_{\text{src}}$. Based on these definitions, the authors defined the precision $P$ and recall $R$ as follows:

$$P = \frac{1}{|X|} \sum_{x \in X} \frac{\left| \bigcup_{c \in C} (c \sqcap x) \right|}{|x|} \qquad R = \frac{1}{|C|} \sum_{c \in C} \frac{\left| \bigcup_{x \in X} (c \sqcap x) \right|}{|c|}$$

$$\text{where } c \sqcap x = \begin{Bmatrix} c \cap x & \text{if } x \text{ detects } c \\ \emptyset & \text{otherwise} \end{Bmatrix}.$$

[8] We use different identifiers than Potthast et al. to be consistent with other formulae in this thesis.

A method may detect only a fragment of a plagiarism instance, report a coherent instance as multiple detections, or report overlapping detections. Neither the precision nor the recall metric reflects these undesired effects. To account for these possibilities, Potthast et al. introduced the granularity score $g$ as:

$$G = \frac{1}{|C_X|} \sum_{c \in C_X} |X_C| \, ,$$

where $C_X \subseteq C$ are detected cases and $X_C \in X$ are detections of $c$. In other terms, $C_X = \{c \mid c \in C \text{ and } \exists x \in X : x \text{ detects } c\}$ and $X_C = \{x \mid x \in X : x \text{ detects } c\}$.

To enable a performance ranking of detection methods according to a single score, Potthast et al. integrated the precision, recall, and granularity measures into the PlagDet measure $M_{\text{PD}}$ defined as:

$$M_{\text{PD}} = \frac{F_1}{\log_2\big(1 + g(C, X)\big)} \, ,$$

where $F_1$ is the equally weighted harmonic mean of precision and recall, i.e.,

$$F_1 = 2 \frac{P \cdot R}{P + R}.$$

Potthast et al. introduced two thresholds $\tau_1(R)$ and $\tau_2(P)$ to calculate precision, recall, and $F_1$ scores at the level of plagiarism cases and documents [414, p. 15f.]. The thresholds allow choosing the minimal precision and recall that a method needs to achieve to consider a case or a document as a true positive detection. For calculating the case-level performance measures, a detection $x \in X$ counts as a true positive detection of the case $c \in C$ if $x$ contributes to detecting at least $\tau_1 \cdot |c|$ characters of $c$ (character-based recall) and if at least $\tau_2 \cdot |x|$ characters reported as belonging to the detection $x$ contributed to detecting the case $c$ (character-based precision). Analogously, at the level of documents, the two thresholds define the minimal precision and recall with which a method has to identify the plagiarism cases in the document to consider the overall document a true positive detection.

The next five tables (**Table 2.28** through **Table 2.32**) present the effectiveness of detection methods that participated in the PAN detailed analysis tasks between 2012 and 2014 at the level of characters. Each table shows the results of the five methods that achieved the highest PlagDet score for a specific group of test cases. For better comparability, the organizers applied all methods that participated in PAN 2012-2014 to the PAN-TA-13 dataset (cf. **Table 2.23**, p. 51) [414, p. 21].

Note that the scores in all five tables implicitly assume a perfect recall during the candidate retrieval stage. As we present on p. 56ff., the recall of all candidate retrieval methods in the PAN competition was significantly lower. The average recall of the five best-performing methods was 51% (cf. **Table 2.27**, p. 58).

**Table 2.28.** Detailed analysis results for the entire corpus.

| # | **Name** | **Year** | $\boldsymbol{P}^{\%}$ | $\boldsymbol{R}^{\%}$ | $\boldsymbol{G}$ | $\boldsymbol{M}_{\boldsymbol{PD}}^{\%}$ | **Ref.** |
|---|---|---|---|---|---|---|---|
| 1 | Sanchez-Perez | 2014 | 88.17 | **87.90** | **1.003** | **87.82** | [436] |
| 2 | Oberreuter | 2014 | 88.60 | 85.78 | 1.004 | 86.93 | [371] |
| 3 | Palkovskii | 2014 | 92.23 | 82.64 | 1.006 | 86.81 | [387] |
| 4 | Glinos | 2014 | **96.25** | 79.33 | 1.017 | 85.93 | [183] |
| 5 | Shrestha | 2014 | 85.91 | 83.78 | 1.007 | 84.40 | [466] |

**Legend:**
$\boldsymbol{P}^{\%}$: Precision in percent, $\boldsymbol{R}^{\%}$: Recall in percent, $\boldsymbol{G}$: Granularity,
$\boldsymbol{M}_{\mathbf{PD}}^{\%}$: PlagDet score in percent, **Ref.** References

**Source:** [414, p. 22ff.]

**Table 2.28** presents the five detection methods with the best PlagDet scores when considering the entire PAN-TA-13 dataset. All methods were submitted in 2014, which indicates the methods evolved continuously during the annual iterations of the PAN competition. All five methods achieved excellent granularity scores, which reflect the methods' success in finding the boundaries of plagiarism cases. Sanchez-Perez et al. achieved the best overall result with nearly equal precision and recall scores of 88% and a near-perfect granularity score. Glinos et al. achieved the best precision (96%) at the cost of a slightly lower recall than the other methods.

**Table 2.29** shows the scores of the five detailed analysis methods that identified unaltered cases of plagiarism (copy and paste of text segments) most effectively. The methods of Oberreuter et al. and a naïve baseline the organizers of the PAN competitions provided achieved near-perfect recall. Due to its virtually equal precision and recall scores, the method of Glinos et al. achieved the best PlagDet score. All five top-ranked methods achieved perfect granularity scores. The results confirm that state-of-the-art detection methods find literal plagiarism reliably.

**Table 2.29.** Detailed analysis results for copy-and-paste plagiarism.

| # | Name | Year | $P^{\%}$ | $R^{\%}$ | $G$ | $M_{PD}^{\%}$ | Ref. |
|---|---|---|---|---|---|---|---|
| 1 | Glinos | 2014 | **96.45** | 96.03 | **1.000** | **96.24** | [183] |
| 2 | Palkovskii | 2014 | 95.58 | 96.43 | **1.000** | 96.00 | [387] |
| 3 | Oberreuter | 2012 | 89.04 | 99.93 | **1.000** | 94.17 | [369] |
| 4 | Baseline | n/a | 88.74 | **99.96** | 1.009 | 93.40 | [414] |
| 5 | R. Torrejón | 2014 | 89.90 | 96.72 | **1.000** | 93.18 | [430] |

**Legend:**
$P^{\%}$: Precision in percent, $R^{\%}$: Recall in percent, $G$: Granularity,
$M_{PD}^{\%}$: PlagDet score in percent, **Ref.** References

**Source:** [414, p. 22ff.]

**Table 2.30** and **Table 2.31** list the detection methods that achieved the best results for machine-obfuscated plagiarism cases. The results in **Table 2.30** refer to plagiarism cases the organizers masked by performing random text operations, such as insertions, deletions, and substitutions of words or characters. The operations do not consider the plausibility and legibility of the resulting text.

**Table 2.30.** Detailed analysis results for randomly obfuscated plagiarism.

| # | Name | Year | $P^{\%}$ | $R^{\%}$ | $G$ | $M_{PD}^{\%}$ | Ref. |
|---|---|---|---|---|---|---|---|
| 1 | Sanchez-Perez | 2014 | 91.02 | **86.07** | 1.001 | **88.42** | [436] |
| 2 | Oberreuter | 2014 | 90.61 | 83.25 | **1.000** | 86.78 | [371] |
| 3 | Shrestha | 2014 | 91.10 | 83.16 | 1.006 | 86.56 | [466] |
| 4 | Palkovskii | 2014 | **91.45** | 82.24 | 1.002 | 86.50 | [387] |
| 5 | Kong | 2012 | 89.37 | 77.90 | **1.000** | 83.24 | [281] |

**Legend:**
$P^{\%}$: Precision in percent, $R^{\%}$: Recall in percent, $G$: Granularity,
$M_{PD}^{\%}$: PlagDet score in percent, **Ref.** References

**Source:** [414, p. 22ff.]

The results in **Table 2.31** refer to plagiarism cases obfuscated via cyclic machine translation. This obfuscation approach exploits the variance of machine translators. The idea is to translate a text written in the original language $L_o$ to one or more intermediate languages $L_{i,1} \dots L_{i,n}$ and from $L_{i,n}$ back to $L_o$. During this process, machine translators often replace words in the original text with words that the language model of the translation engine considers related. Partially, the translators also change the syntax of sentences. The PAN organizers used three translation engines sequentially, i.e., the output of each engine became the input of the following engine. Each translation engine used two intermediate languages drawn randomly from two sets of languages with low and high linguistic distances to the languages of the input texts, i.e., English, German, Spanish.

**Table 2.31.** Detailed analysis results for cyclically translated plagiarism.

| # | Name | Year | *P* | *R* | *G* | $M_{PD}$ | Ref. |
|---|---|---|---|---|---|---|---|
| 1 | Sanchez-Perez | 2014 | 88.47 | **88.96** | 1.001 | **88.66** | [436] |
| 2 | Oberreuter | 2014 | 89.98 | 86.34 | **1.000** | 88.12 | [371] |
| 3 | R. Torrejón | 2014 | **90.09** | 82.08 | **1.000** | 85.90 | [430] |
| 4 | Palkovskii | 2014 | 89.94 | 82.03 | 1.001 | 85.75 | [387] |
| 5 | Kong | 2012 | 85.42 | 85.00 | **1.000** | 85.21 | [281] |

**Legend:**
$P^{\%}$: Precision in percent, $R^{\%}$: Recall in percent, $G$: Granularity,
$M_{PD}^{\%}$: PlagDet score in percent, **Ref.** References

**Source:** [414, p. 22ff.]

**Table 2.30** and **Table 2.31** show that the best performing methods achieve similar detection effectiveness for both automated obfuscation procedures. Compared to unaltered plagiarism, the PlagDet scores were about 10% lower, while the granularity score of all top-ranked methods remained excellent. All top-ranked methods detected machine-obfuscated plagiarism cases reliably. The predominantly lexical and, to a smaller extent, syntactical changes the obfuscation introduced appear not to have posed a significant obstacle for the detection methods.

**Table 2.32** presents the results of the five detection methods that performed best for plagiarism cases that underwent summary obfuscation. This obfuscation type simulates idea plagiarism and is the only obfuscation that uses manually rewritten text to create plagiarism instances. The PAN organizers used a dataset of news articles that includes a human-made summary for each article. This summary was embedded into other news articles. To increase the topical similarity of the inserted summary and the surrounding text, the organizers replaced named entities in the surrounding text with named entities occurring in the summarized text.

The results in **Table 2.32** shows that all detection methods achieved significantly lower effectiveness in identifying summary obfuscation cases. Particularly the recall of all methods dropped drastically compared to cases that the organizers obfuscated using automated methods. Only the detection method of Suchomel et al. [492] identified more than 50% of the summary obfuscation cases.

**Table 2.32.** Detailed analysis results for summary obfuscation plagiarism.

| # | **Name** | **Year** | $P$ | $R$ | $G$ | $M_{PD}$ | **Ref.** |
|---|---|---|---|---|---|---|---|
| 1 | Glinos | 2014 | 96.45 | 48.61 | 1.051 | **62.36** | [183] |
| 2 | Suchomel | 2013 | 67.09 | **56.30** | 1.005 | 61.01 | [492] |
| 3 | Sanchez-Perez | 2014 | **99.91** | 41.27 | 1.059 | 56.07 | [436] |
| 4 | Suchomel | 2012 | 87.48 | 35.31 | 1.006 | 50.09 | [491] |
| 5 | R. Torrejón | 2012 | 92.67 | 29.01 | **1.000** | 44.18 | [430] |

**Legend:**
$P^{\%}$: Precision in percent, $R^{\%}$: Recall in percent, $G$: Granularity,
$M^{\%}_{\mathrm{PD}}$: PlagDet score in percent, **Ref.** References

**Source:** [414, p. 22ff.]

The PAN results indicate that state-of-the-art detection methods struggle to find manually rewritten texts even under optimal conditions, i.e.:

1. The reference collection contained all source documents.
2. The organizers did not limit the runtime of the methods.
3. The writers who summarized the news articles had no reason to purposefully reduce or mask the similarity of the summary to the source text. The opposite is true for academic plagiarists.

Due to these circumstances, we see the summary obfuscation cases in the PAN datasets as an optimistic approximation of the obfuscation strength that one can expect for actual cases of idea plagiarism in academic documents. Consequently, we expect that a recall of approximately 60% represents an upper bound on the effectiveness of current plagiarism detection methods in identifying manually rewritten texts with high semantic but low lexical similarity.

Notably, hybrid detection methods achieved the best results for each of the three obfuscation types. The method of Sanchez-Perez [436] obtained the best results for copied and cyclically translated cases as well as the best recall and overall PlagDet score when considering the entire dataset. Their detection method used an adaptive algorithm, which we describe in Section 2.4.7, p. 42. Glinos et al. [183] achieved the best PlagDet score for summary obfuscation cases and the best precision when considering the entire dataset. Their hybrid detection method combined an alignment method to identify weakly obfuscated cases with a clustering method to find more strongly disguised cases.

## Cross-Language Plagiarism Detection

To indicate the effectiveness of cross-language plagiarism detection methods, we refer to the results of a study by Franco-Salvador et al. [151]. The study evaluated methods that represent all major detection approaches, specifically:

**Lexical Cross-language Detection Methods**

- The Cross-Language Character 3-Gram model **(CL-C3G)** is a basic tf-idf-weighted vector space model that uses character 3-grams as the term unit and the cosine measure for similarity calculation [334], [411]. The only adaption of the model for the cross-language setting is the removal of diacritics. The model relies on the lexical and syntactic similarities of languages. Hence, the model is best-suited for languages from the same or linguistically close language families, such as Germanic languages, e.g., English and German, and Romanic languages, e.g., Spanish and Italian.
- The Cross-Language Vector Space Model (**CL-VSM**) represents texts by concatenating the tf-idf-weighted term vector representations of a text in two languages [148]. The vector entries for the language that differs from the language of the text are found using statistical machine translation.

**Semantic Concept Analysis Methods** (cf. Section 2.4.5, p. 37)

- The Cross-Language Knowledge Graph Analysis (**CL-KGA**) method uses BabelNet concepts and relations weighted according to the scheme proposed by Franco-Salvador et al. [147], [150].
- The Cross-Language Explicit Semantic Analysis (**CL-ESA**) method derives its concept vector representation from 10,000 comparable Wikipedia articles in English, German, and Spanish using a tf-idf-weighted vector space model with cosine similarity as proposed by Potthast et al. [406], [411].

**Cross-lingual Word Embeddings** (cf. Section 2.4.5, p. 35)

For deriving the cross-lingual word embeddings, Franco-Salvador et al. employed the machine learning and deep learning approaches we briefly characterize hereafter. The training dataset for all approaches comprised 250,000 parallel English-Spanish and English-German sentences.

» The Siamese neural network architecture (**S2Net**) proposed by Yih et al. [565] trains two neural networks concurrently on the aligned input data.

» The bilingual autoencoder (**BAE**) proposed by Gupta et al. [194] learns a dimensionality-reduced representation of the input data.

» The External-data Composition Neural Network (**XCNN**) proposed by Gupta et al. [195] is a deep neural network architecture that first trains a monolingual latent semantic model from external relevance data. In a second step, the architecture uses a comparably smaller amount of parallel training data to derive the latent cross-lingual representation.

**Hybrid Detection Methods**

» The Continuous Word Alignment-based Similarity Analysis (**CWASA**) method proposed by Franco-Salvador et al. [151] combines the idea of word embeddings and word alignment approaches (cf. Section 2.4.5, p. 38). The method aligns each continuous representation of a word in a document $d$, i.e., typically a real-valued vector, with the most similar word representation of another document $d'$ in a different language. The similarity of $d$ and $d'$ is computed as the average cosine similarity of the aligned representations. Franco-Salvador et al. evaluated this alignment procedure for each of the cross-lingual word embeddings S2Net, BAE, and XCNN.

» The Knowledge-Based document Similarity (**KBSim**) method proposed by Franco-Salvador et al. is an ensemble (cf. Section 2.4.7, p. 42) of the CL-KGA and CL-VSM methods [148].

Franco-Salvador et al. evaluated these detection methods for the cross-language subset of the PAN-PC-11 corpus (cf. Section 2.5.1, p. 52). The subset comprises aligned German-English and Spanish-English test cases. Most of the cases were obtained via statistical machine translation. For approximately 1% of the test cases, human workers additionally obfuscated the machine-translated texts.

A limitation of the study by Franco-Salvador et al. is that the authors simplified the cross-language candidate retrieval task in their experiments. In practice, a plagiarism detection system must compare a suspicious input document to the entire reference collection. Franco-Salvador et al. compared suspicious documents only to the known source documents for cross-language plagiarism instances, i.e., a relatively small subset of the PAN-PC-11 corpus. Therefore, the reported results represent an optimistic upper bound on the detection effectiveness of the evaluated detection methods. We report the results of this study nevertheless as it is the most comprehensive, recent comparison of cross-language detection methods.

**Table 2.33.** Detection results for cross-language plagiarism.

| Obf. | # | | $P$ | $R$ | $G$ | $M_{PD}$ |
|---|---|---|---|---|---|---|
| Machine translation | 1 | KBSim (XCNN) | **81.60** | 59.40 | **1.00** | **68.80** |
| | 2 | CL-KGA | 74.20 | **59.50** | 1.00 | 66.00 |
| | 3 | CWASA (XCNN) | 73.20 | 58.50 | 1.00 | 65.00 |
| | 4 | CL-VSM | 67.30 | 55.30 | 1.01 | 60.30 |
| | 5 | CL-ASA | 73.60 | 47.90 | 1.08 | 55.20 |
| | 6 | S2Net | 78.40 | 47.10 | 1.11 | 55.00 |
| | 7 | CL-ESA | 57.10 | 47.90 | 1.05 | 50.30 |
| | 8 | CL-C3G | 60.20 | 34.70 | 1.16 | 39.80 |
| MT + manual | 1 | KBSim (S2Net) | **22.40** | **17.40** | **1.00** | **19.60** |
| | 2 | CL-KGA | 20.70 | 14.30 | 1.00 | 16.90 |
| | 3 | CWASA (XCNN) | 21.20 | 14.00 | 1.00 | 16.80 |
| | 4 | S2Net | 17.30 | 8.60 | 1.00 | 11.50 |
| | 5 | CL-ASA | 14.60 | 7.60 | 1.00 | 10.00 |
| | 6 | CL-ESA | 10.70 | 8.10 | 1.00 | 9.20 |
| | 7 | CL-C3G | 10.40 | 5.40 | 1.00 | 7.20 |
| | 8 | CL-VSM | 14.70 | 8.60 | 1.00 | 10.90 |

**Legend:**

**Obf.** Obfuscation, $P^{\%}$: Precision in percent, $R^{\%}$: Recall in percent,

$G$: Granularity, $M^{\%}_{PD}$: PlagDet score in percent, **MT** Machine Translation

**Source:** [151, p. 97].

**Table 2.33** shows the results divided according to whether test cases underwent manual obfuscation after the machine translation. Franco-Salvador et al. evaluated the German-English and Spanish-English test cases separately. However, we are only interested in estimating the expectable maximum performance of cross-language detection methods. Therefore, the table only shows the results for the set of test cases for which a method performed better. Moreover, we only show the cross-lingual word embedding model and the combination of CWASA with a word embedding model that achieved the highest PlagDet scores at the character level.

For purely machine-translated test cases, a recall of approximately 60% seems to be an optimistic upper bound. We observed the same bound for summary obfuscation cases, i.e., paraphrases, in the PAN evaluations (cf. **Table 2.32**, p. 64). The average precision of the five best-performing detection methods for identifying purely machine-translated cases (73.98%) is lower than the respective result for paraphrases, which was 88.72%. We hypothesize that finding the boundaries of plagiarized fragments is more difficult in the cross-language than in the monolingual setting. This difficulty could explain the lower character-based precision score.

For test cases that humans obfuscated after the machine translation step, the detection effectiveness of all methods dropped drastically. Franco-Salvador et al. noted that one reason for performance drop is the significantly lower number and shorter length of manually obfuscated test cases [151, p. 96].

As for monolingual test cases, a hybrid detection method (KBSim) achieved the best results for both subsets of cross-language test cases. Except for CL-ESA, detection methods that (in part) employ a language-independent model, i.e., KBSim, CL-KGA, and CWASA, performed better than methods that exhibit a stronger dependence on lexical features.

In summary, we note that identifying translated plagiarism is a significant challenge for state-of-the-art detection methods, particularly if humans obfuscated the cases. Given that Franco-Salvador et al. simplified the detection task in their experiments, one can assume that detection effectiveness in realistic use cases will be lower.

### 2.5.3 Evaluation of Plagiarism Detection Systems

To assess the capabilities of production-grade plagiarism detection systems, we refer to the most recent evaluation of such systems, which the European Network on Academic Integrity (ENAI) published in February 2020 [145]. ENAI is an association of 30 universities and research institutions from Europe and Asia.

The evaluation included 15, primarily commercial, plagiarism detection systems. The objective was to assess the detection effectiveness and usability of the systems in a higher education setting. For this purpose, the organizers compiled test documents in eight languages; all documents include the same amount of simulated plagiarism of a specific form. The test cases are publicly available [144].

The **types of simulated plagiarism** the organizers created manually are:

- Copying content verbatim
- Replacing words with synonyms
- Paraphrasing passages
- Translating (50% manually, 50% using machine translation)

The sources for all plagiarism instances had to be available on the public Internet, e.g., Wikipedia, open access research publications, and graduation theses. This requirement eliminated a potential benefit of system providers whose reference collections include content with restricted access.

To quantify the **detection effectiveness**, the organizers used a 5-point ordinal scale that considers the amount of similar text a system identifies [145, p. 11]:

- **5 points**: all or almost all of the similar text
- **4 points**: a major portion of the similar text
- **3 points**: more than 50% of the similar text
- **2 points**: 50% or less of the similar text
- **1 point**: a minor portion of the similar text
- **0 points**: one sentence or less

For false positives, i.e., original text flagged as plagiarized, the organizers assigned the negative point value that reflects the amount of original text flagged incorrectly.

To quantify the **usability of systems**, the organizers derived 23 criteria from the literature and their experience as educators [145, p. 21]. The criteria address:

- The **analysis workflow**, e.g., whether systems support uploading multiple input documents or limit the amount of analyzable text
- The **presentation of results**, e.g., whether systems offer side-by-side views of input documents and sources or downloadable report files
- **Other factors**, e.g., whether systems support learning management systems or offer phone support

## Detection Effectiveness

**Table 2.34.** Detection effectiveness by plagiarism type and language group.

| | Akademia | Copyscape | Docol©c | Dupli Checker | DPV | intihal.net | PlagAware | PlagiarismCheck | Plagiarism Software | PlagScan | StrikePlagiarism | Turnitin | Unicheck | Urkund | Viper |
|---|---|---|---|---|---|---|---|---|---|---|---|---|---|---|---|
| **Cop.** | 2.1 | 3.1 | 3.8 | 1.2 | 0.4 | 1.1 | **4.3** | 2.7 | 3.0 | 3.4 | 3.8 | 3.5 | 3.7 | 4.0 | 3.3 |
| L1 | 2.7 | 3.4 | 4.3 | 1.0 | 0.7 | 0.7 | 4.6 | 2.3 | 2.5 | 4.3 | 4.1 | 4.5 | 3.8 | 4.3 | 3.3 |
| L2 | 1.7 | 3.0 | 3.4 | 1.4 | 0.2 | 1.5 | 4.1 | 3.0 | 3.4 | 2.8 | 3.5 | 2.6 | 3.5 | 3.8 | 3.4 |
| **Syn.** | 1.9 | 2.2 | 2.0 | 0.6 | 0.3 | 0.8 | 3.6 | 2.5 | 1.7 | 3.5 | 3.0 | 3.0 | 2.0 | **3.8** | 1.4 |
| L1 | 2.7 | 2.3 | 2.8 | 0.7 | 0.6 | 0.5 | 3.6 | 2.2 | 2.0 | 4.3 | 3.7 | 3.8 | 2.8 | 4.0 | 1.8 |
| L2 | 1.3 | 2.1 | 1.1 | 0.5 | 0.0 | 1.0 | 3.5 | 2.8 | 1.5 | 2.6 | 2.3 | 2.3 | 1.2 | 3.5 | 1.1 |
| **Para.** | 1.1 | 0.6 | 0.6 | 0.1 | 0.1 | 0.4 | 1.4 | **1.6** | 0.5 | **1.6** | 1.1 | 1.3 | 0.7 | **1.6** | 0.4 |
| L1 | 1.6 | 0.7 | 0.7 | 0.1 | 0.2 | 0.3 | 1.5 | 1.7 | 0.4 | 2.4 | 1.6 | 1.8 | 0.9 | 1.6 | 0.4 |
| L2 | 0.8 | 0.5 | 0.5 | 0.1 | 0.0 | 0.5 | 1.4 | 1.5 | 0.5 | 0.8 | 0.6 | 0.8 | 0.5 | 1.6 | 0.3 |
| **Tran.** | **1.5** | 0.0 | 0.1 | 0.0 | 0.0 | 0.3 | 0.1 | 0.0 | 0.1 | 0.3 | 0.3 | 0.1 | 0.2 | 0.1 | 0.1 |
| L1 | 2.0 | 0.0 | 0.3 | 0.0 | 0.0 | 0.3 | 0.3 | 0.0 | 0.3 | 0.7 | 0.7 | 0.3 | 0.5 | 0.3 | 0.3 |
| L2 | 1.3 | 0.0 | 0.0 | 0.0 | 0.0 | 0.3 | 0.0 | 0.0 | 0.0 | 0.0 | 0.0 | 0.0 | 0.0 | 0.0 | 0.0 |
| **Total** | 6.6 | 5.9 | 6.5 | 1.9 | 0.8 | 2.6 | **9.5** | 6.8 | 5.3 | 8.8 | 8.1 | 7.9 | 6.5 | **9.5** | 5.3 |

**Legend:**

**Cop.** Copies, **Para.** Paraphrases, **Syn.** Synonyms, **Tran.** Translations

**L1:** English, German, Italian, Spanish; **L2:** Czech, Latvian, Slovak, Turkish

**Source:** [143]

**Table 2.34** shows the effectiveness scores $M$ of the 15 tested systems for the four types of simulated plagiarism. Additionally, we present the scores for two groups of languages. The first language group comprises English, German, Italian, and Spanish, whereas the second language group comprises Czech, Latvian, Slovak, Turkish. Boldface indicates the best result for each type of plagiarism, whereas underlining highlights the best result per language group. We list the average score a system achieved for all plagiarism instances of the specified type. In their report, Foltýnek et al. used the number of suspicious documents to compute the average scores [145]. We decided on normalizing by plagiarism instance to make the results better comparable to the results of the PAN competitions. For this purpose, we accessed the raw evaluation data that ENAI published [143].

The results are consistent with those of the PAN competitions. Intuitively, all systems achieved their best results for identifying copy-and-paste plagiarism. Ten of the 15 systems achieved an average score $M \geq 3$, i.e., typically found more than 50% of the copied text. Three systems found only a minor portion of the copied text (scores of $0.4 \leq M \leq 1.2$), which is surprising because all sources are accessible online. Overall, the top-ranked systems found verbatim text copies reliably.

All systems achieved better results for language group one, i.e., the Germanic (English, German) and Romanic languages (Italian, Spanish), than for the Balto-Slavic (Czech, Slovak, Latvian) and Turkic (Turkish) languages in group two. This effect is consistent for all types of simulated plagiarism.

Synonym replacements affected the systems' detection effectiveness differently. While PlagScan performed slightly better for synonym replacement test cases than for literally copied test cases, several systems exhibited a sharp drop in their overall detection effectiveness, e.g., Viper ($\Delta M = -1.9$), Docol©c ($\Delta M = -1.8$), and Unicheck ($\Delta M = -1.7$). Other systems exhibited only a slight decrease in detection effectiveness ($0 \leq M \leq 0.5$) for synonym replacement test cases compared to copy-and-paste test cases, e.g., Urkund, Turnitin, and PlagiarismCheck.

For paraphrased test cases, the overall scores of all systems were lower than two, i.e., the systems typically detected less than 50% of the plagiarized text. Six systems achieved scores below one, i.e., identified hardly any of the plagiarized content. These results suggest that all systems exclusively employ lexical detection methods, which reach their limits for human-made paraphrases. Consequently, Foltýnek et al. summarize: "For paraphrased texts, none of the systems was able to provide satisfactory results." [145, p. 26].

The detection results for translated test cases additionally support the assumption that almost all systems exclusively search for lexical similarity. Except for Akademia, none of the systems identified translated plagiarism instances aside from flagging spurious literal matches by chance. According to Foltýnek et al., Akademia was the only system in the ENAI test that provides users with an option to check for translated plagiarism [145, p. 26]. Interestingly, Foltýnek et al. describe that Akademia's correct detections of translated plagiarism instances resulted from **identifying similarities in the references**, not the text itself [145, p. 26].

All systems achieved less than 50% of the possible score for effectiveness. These results indicate that production-grade detection systems do not (yet) employ the numerous semantic, syntactic, and ensembles of detection methods we presented in Section 2.4. Foltýnek et al. referred to the systems as **text-matching tools**. In conclusion, the ENAI testers classify none of the systems as "useful tools," for which the testers required an average overall effectiveness score $M \geq 3.75$. Only five systems were classified as "partially useful tools" ($2 \leq M < 3.75$) [145, p. 27].

## Usability

**Table 2.35.** Usability scores of tested plagiarism detection systems.

| | Akademia | Copyscape | Docol©c | Dupli Checker | DPV | intihal.net | PlagAware | PlagiarismCheck | Plagiarism Software | PlagScan | StrikePlagiarism | Turnitin | Unicheck | Urkund | Viper |
|---|---|---|---|---|---|---|---|---|---|---|---|---|---|---|---|
| **Wfl.** (6) | 2 | 3 | 6 | 2 | **6** | 3 | 4.5 | 5 | 2 | 6 | 4 | 5 | **6** | **6** | 5 |
| **Pres.** (9) | 5 | 1 | 6 | 3 | 6 | 3 | 5.5 | 5 | 3 | **8** | 6 | 7 | 6 | **8** | 6 |
| **Oth.** (8) | 2 | 5 | 6 | 3 | 2 | 3 | 6.5 | **8** | 5.5 | 7 | 6.5 | 5.5 | **8** | 6 | 3 |
| **Tot.** (23) | 9 | 9 | 18 | 8 | 14 | 9 | 16.5 | 18 | 10.5 | **21** | 16.5 | 17.5 | 20 | 20 | 14 |

**Legend:**
**Oth.** Other factors, **Pres.** Presentation of results, **Tot.** Total, **Wfl.** Workflow

**Source:** [145, p. 21ff.]

**Table 2.35** summarizes the scores the tested systems achieved for the 23 usability criteria, which the test organizers grouped into the three categories **analysis workflow**, **presentation of results**, and **other factors**. The testers assigned a score of one if a system fulfilled the criterion, a score of 0.5 if a system offered the desired functionality, but three testers could not find it without detailed guidance, and a score of zero otherwise [145, p. 12]. We state the maximum score for each category in brackets below the category name. Readers can find the scores for each of the 23 criteria in the ENAI test report [145, p. 21ff.].

Most systems achieved high scores for the criteria that strongly influence the effectiveness of the user interface, i.e., the analysis workflow and the presentation of results. Three systems achieved perfect scores for their **analysis workflow**, and only three systems received less than half of the points possible in that category. The three workflow-related criteria that the fewest systems fulfilled were:

1. Allowing the upload of multiple input documents: Six systems did not offer this functionality.
2. Not requiring the user to fill in metadata for input documents: Five systems required such input from the user.
3. Using the original filename for input documents: Five systems changed the file name during the document upload.

Two systems fulfilled all but one criterion related to the **presentation of results**. The three presentation-related criteria the fewest systems fulfilled were:

1. Offering side-by-side views of the input document and a potential source as part of the downloadable result report: Only one system (Urkund) offered this functionality.
2. Presenting detection results in a side-by-side view of the input document and a potential source as part of the user interface: Only four of the 15 systems offered a side-by-side comparison.
3. Using the format of the input document to present the detection results: 10 systems changed the format of the input document.

That only four systems visualize detection results in a **side-by-side view** of the input document, and a potential source is surprising. Most systems solely highlight the parts of the input document that likely originate from other sources and display excerpts of the potential source [145, p. 23]. Some systems only provide a list of links to potential source documents. Those systems leave it to the user to find and examine the similarity of the potential source and the input document. We agree with the ENAI testers that the side-by-side layout is most intuitive and enables users to inspect the severity of identified content similarities quickly and effectively.

Regarding other factors that influence the functionality of the systems and the quality of the service, two systems received the maximum number of points. However, five systems received less than half of the points possible, indicating that the providers' service quality differs significantly. The three criteria in the **other factors** category that the fewest systems fulfilled were:

1. Offering integration with the learning management platform Moodle: Eight systems do not offer this functionality.
2. Providing user support in English: Eight systems did not offer this service.
3. Offering a free trial: Seven systems did not offer this functionality.

In summary, most of the tested systems received significantly **higher scores for their usability than for their detection effectiveness**. The top-ranked systems in the usability evaluation are mostly well-designed software products marketed by professional vendors that typically offer high-quality support services.

## 2.6 Findings of the Literature Review

By reviewing 376 publications from the 25-year period 1994–2019, we presented the most comprehensive survey of plagiarism detection technology to date. Our survey shows that plagiarism is a longstanding problem at all academic levels and negatively affects many stakeholders in academia and society. The rapid advancement of information technology offers convenient access to vast amounts of information, which has made plagiarizing easier than ever. However, information technology also facilitates the detection of plagiarism and thus initiated a cat-and-mouse game [568] between plagiarists and those tasked with safeguarding academic integrity. Plagiarism detection technology has become an essential component of an integrated socio-technical approach to counteracting academic plagiarism.

Sections 2.4.3 through 2.4.7 that extensive research on plagiarism detection has yielded a broad spectrum of external and intrinsic detection methods. Most external methods draw on NLP research to analyze the lexical, syntactic, and semantic similarity of text for the monolingual or cross-language use cases.

Intrinsic detection methods mostly use lexical and syntax-based text analysis. However, the intrinsic detection paradigm exhibits shortcomings from a practitioner's perspective. First, intrinsic detection methods are inherently error-prone for documents written by multiple authors, as these documents often exhibit differing writing styles [428, p. 4f.]. This shortcoming is critical since most scientific publications have multiple authors [180, p. 262]. Second, intrinsic methods are not well-suited for detecting paraphrased plagiarism. Third, the methods are not reliable enough for practical applications yet. Author identification methods achieve a precision of approximately 60%, author profiling methods of approximately 80% [416]. These values suffice for raising suspicion and triggering further examination but not for proving plagiarism. Methods for automated author obfuscation aggravate the problem. The most effective obfuscation methods can mislead the detection systems in almost half of the cases [415]. Fourth, intrinsic plagiarism detection methods cannot point an examiner to the source document of potential plagiarism. If a stylistic analysis raised suspicion, extrinsic detection methods or other search and retrieval approaches are necessary to discover the potential source document(s).

Therefore, most plagiarism detection research addresses the external paradigm. Until circa 2010, researchers and practitioners focused on detecting literal and slightly obfuscated plagiarism in web-scale collections. Among many other evaluations, the results of the PAN competitions (cf. Section 2.5.2, p. 56) and the ENAI comparison of production-grade plagiarism detection systems (cf. Section 2.5.3, p. 68) show that these efforts were successful. State-of-the-art plagiarism detection methods and systems achieve $F_1$ scores of 88%–96% for forms of academic plagiarism with no or little disguise (cf. Table 2.29, p. 62 and Table 2.30, p. 62).

Improving the detection of disguised forms of academic plagiarism has been the focus of plagiarism detection research since 2010 at the latest. To accomplish this objective, researchers increasingly investigated semantic analysis (cf. Section 2.4.5, p. 34) and hybrid detection methods (cf. 2.4.7, p. 42). For many detection tasks, hybrid methods have outperformed individual methods [8], [118], [151], [254], [271], [484], [496], [569], [573]. The evaluation results we presented in Section 2.5.2, p. 56, are in line with this observation. Machine learning approaches represent the logical evolution of the idea to combine various detection methods. Machine learning and deep learning methods have found increasingly widespread adoption in plagiarism detection research and significantly increased detection effectiveness.

Despite the advances in plagiarism detection research, significant challenges remain regarding the detection of disguised forms of academic plagiarism, such as strong paraphrases, translations, and idea plagiarism. The best detection methods in the PAN competitions achieved a recall of 60% for the candidate retrieval stage, which presents an upper bound for the detection rate in the subsequent detailed analysis stage (cf. Table 2.27, p. 58). For the detailed analysis, the best methods achieved approximately 60% recall for identifying strong paraphrases (cf. Table 2.32, p. 64) and manually disguised translations in a simplified scenario (cf. Table 2.33, p. 67). Notably, these results were obtained under ideal conditions, i.e., all source documents are accessible, and no limits on computing time exist. The ENAI comparison of production-grade detection systems shows that under more realistic conditions, the detection effectiveness for these strongly disguised forms of plagiarism is much lower (cf. Table 2.34, p. 70). These results suggest that the detection capabilities of external detection methods that analyze textual features have reached a plateau.

We and others have proposed that **analyzing non-textual content in academic documents**, such as citations, figures, tables, and mathematical content, has a large potential to improve the effectiveness of plagiarism detection methods. In 2010, Mozgovoy et al. concluded their survey of plagiarism detection technology by proposing a roadmap for the future development of plagiarism detection systems [357]. They suggested the inclusion of syntactic parsing, querying synonym thesauri, employing LSA to discover "tough plagiarism," intrinsic plagiarism detection, and tracking citations and references [357, p. 527]. As our review shows, all these suggestions have been realized. In 2015, Eisa et al. summarized their review of plagiarism detection methods by praising the effort invested into improving text-based methods. Still, they noted a critical lack of: "techniques capable of identifying plagiarized figures, tables, equations and scanned documents or images." [114, p. 396].

## 2.7 Research Approach

The idea this thesis investigates for improving the detection of disguised academic plagiarism, i.e., semantics-preserving and idea-preserving plagiarism forms, is to analyze non-textual content elements, specifically, academic citations, images, and mathematical content. The idea reflects the findings of our literature review, which showed that the detection effectiveness of methods that analyze the text alone has reached a plateau. We expect that considering additional content features can help to overcome this relative standstill in detection effectiveness. Moreover, non-textual content analysis enables plagiarism checks for documents that contain content that current detection methods ignore, such as mathematical expressions and images.

The rationale of our research is that non-textual content elements are semantically rich, language-independent, and not easily substitutable. For example, academic citations and mathematical notation allow representing comprehensive semantic concepts, such as the Mass-Energy Equivalence, in a condensed form, i.e., as an equation $E = mc^2$ or a citation to a publication by Einstein.

Moreover, images, citations, and mathematical content are independent of the language and script of the text. Customarily, authors of academic documents include citations in the Roman script using established citation styles, such as numeric or author-year styles, even if the text is in a language with a non-Roman script, e.g., Chinese, Arabic, or Russian. Likewise, mathematical notation uses a specific, reasonably standardized script. These notational conventions, which offer fewer degrees of freedom than natural language, facilitate parsing, disambiguating, and mapping the respective content features.

Substituting citations or mathematical expressions causes more effort than paraphrasing text because it requires expert knowledge. Replacing or leaving out highly relevant citations or formulae—if possible at all—can make a document immediately suspicious in the eye of a domain expert. Therefore, including similarity assessments of non-textual content elements as part of plagiarism checks increases the effort plagiarists must invest and the chance of being detected.

Building upon this research idea, the following chapters of this thesis:

» Describe the analysis methods for citations, images, and mathematical content in academic documents we devised;

» Present the integration of these non-textual content analysis methods with well-performing lexical, syntactic, and semantic text analysis methods;

» Show that these research contributions increase the detection capabilities for strongly disguised forms of academic plagiarism.

3

Chapter 3

# Citation-based Plagiarism Detection

## Contents

This chapter presents Citation-based Plagiarism Detection (CbPD)—the first plagiarism detection approach that analyzed non-textual content elements. The author started researching CbPD as part of his final year thesis [336], which Bela Gipp supervised after proposing the approach [169]. Bela Gipp's doctoral thesis [173] presents the approach in detail.

The structure of this chapter is as follows. Section 3.1 defines relevant terminology and presents related work that used academic citations for semantic analysis tasks. Section 3.2 describes how the observations we made for confirmed cases of plagiarism influenced the design of the CbPD methods. Section 3.3 describes the citation-based detection methods we devised. Sections 3.4 and 3.5 present the methodology and results of a large-scale evaluation using the PubMed Central Open Access Subset. Section 3.6 summarizes our research contributions in this chapter.

## 3.1 Citation-based Document Similarity

### 3.1.1 Terminology

The use of the terms citation and reference is ambiguous in the literature [295, p. 42]. Many researchers employ the terms interchangeably to denote both the sources a document cites and the citations a document receives.

We refer to entries in the bibliography of academic documents as **references**. We use **citations** to refer to short strings in the text of the documents that point to references or to denote the number of times other documents reference a document. We clarify the intended meaning by providing context information. In the absence of errors, a multiple-to-one relationship exists between citations and references. We use the verbs **cite** and **reference** synonymously to express that a document refers to a source. If a distinction between a citation and a reference is unnecessary, we use the more common expression, citation. For instance, we refer to methods that use citations, references, or combinations thereof as citation analysis or citation-based methods [173, p. 44f.], [336, p. 35].

### 3.1.2 Related Work

Researchers and practitioners have long recognized that citations and references convey valuable information on the semantic content and relations of academic documents. In 1955, Garfield proposed tracking references in a central index to support researchers in exploring relevant literature, increase mutual awareness, and foster research collaboration [160, p. 108]. He also suggested counting the citations a publication received to measure research impact [160, p. 109]. In 1958, Kessler proposed **Bibliographic Coupling (BC)**, i.e., the number of identical references in two documents, as a measure of the documents' relatedness [553].

In 1973, Small [471] and Marshakova-Shaikevich [325] criticized that Bibliographic Coupling cannot reflect changes in the semantic relation of documents as the bibliographies of the documents are static after publication. They suggested the **Co-Citation (CoCit)** measure as an alternative. CoCit assumes a more substantial semantic relation between documents that later documents frequently cite together.

In 2006, Gipp proposed **Co-Citation Proximity Analysis (CPA)** as an improvement to CoCit made possible by the increasing availability of full texts [167]. CPA weights the co-citation relationship according to the smallest distance between the citations within a citing document. The idea is that the semantic relation of co-cited documents is stronger if the citing document refers to the cited documents in close proximity, e.g., in the same sentences, rather than, e.g., in different sections. In the first study on CPA, Gipp & Beel used static weights to compute CPA scores, e.g., 1 if the citations occur in the same sentence and 1/2 if they occur in different paragraphs. However, the authors suggested that future research should analyze dynamic weighting, e.g., dependent on the research field [168].

**Figure 3.1** illustrates how Bibliographic Coupling, Co-Citation, and Co-Citation Proximity Analysis quantify the relatedness of documents A and B or A, B and C in the case of CPA. Documents 1 and 2 appeared earlier, documents 3 and 4 later than documents A, B, and C. Blue color indicates that the scores of CoCit and CPA can change over time. In contrast, the red shading used for BC signifies that the score is static after the citing document has been published.

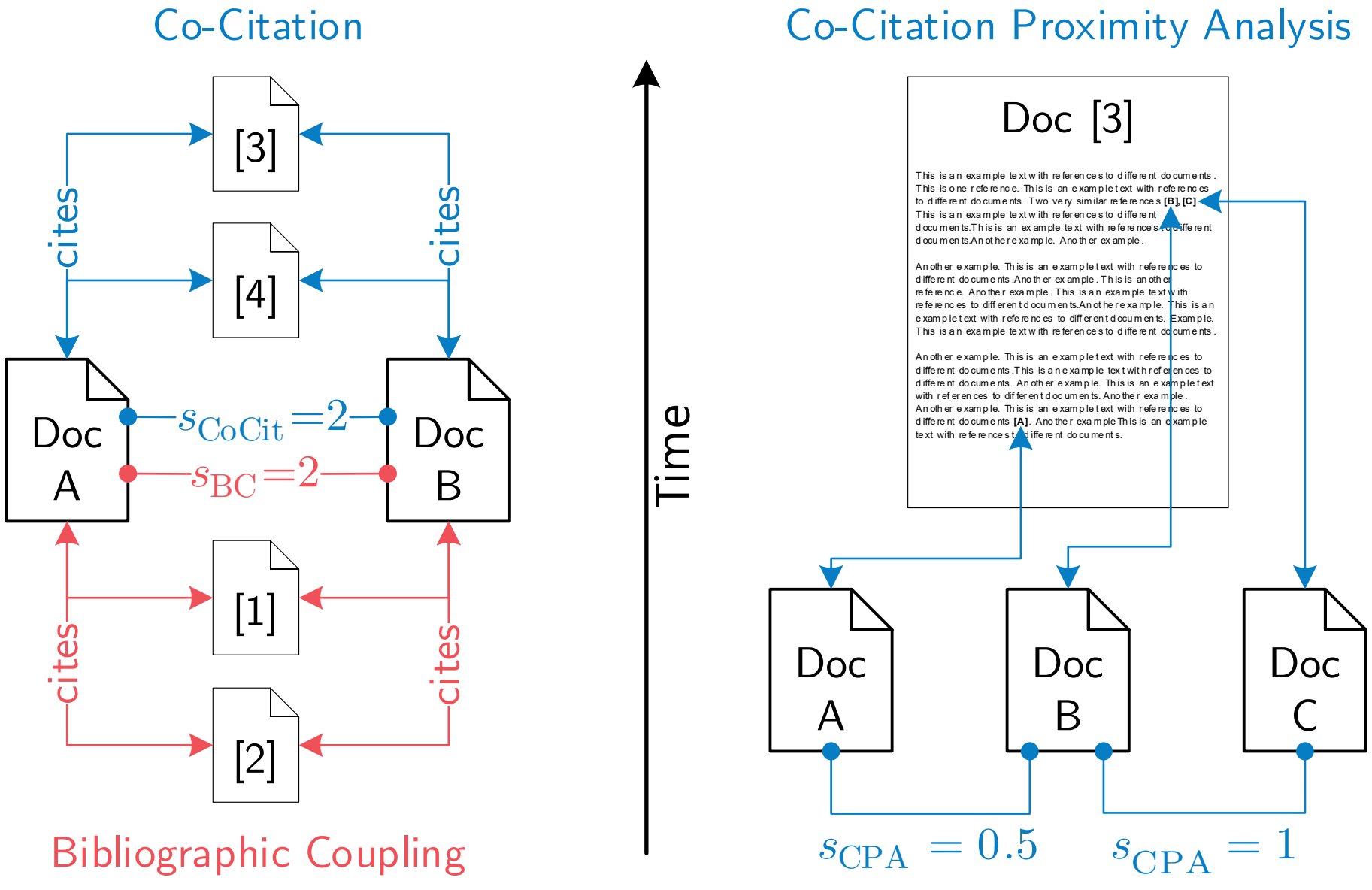

**Figure 3.1.** Citation-based similarity measures.

## 3.2 Citation-based PD Concept

The idea of Citation-based Plagiarism Detection is to identify and use distinctive citation patterns for document retrieval. Citation patterns are sequences of citations in two documents A and B, partially or entirely linked to shared references of A and B. The **distinctiveness** of a citation pattern indicates how frequently the citation pattern occurs within the collection. The distinctiveness of a citation pattern depends on the **overlap**, **order**, and **proximity** of the citations that form the pattern. The distinctiveness reflects the number of citations two documents share, the distance of the shared citations within the documents' texts, and to which degree the order of the shared citations is similar in the documents.

**Figure 3.2** illustrates the CbPD concept; it shows the documents A and B that cite C, D, and E. Given the three shared references, documents A and B likely discuss semantically related content. More interestingly, however, documents A and B cite the sources C, D, and E in a similar order (see the matching colors in the pattern comparison), resulting in a pattern agreement of length three. Thus, the CbPD approach allows computing document similarity and retrieving similar documents, even in the absence of matching text.

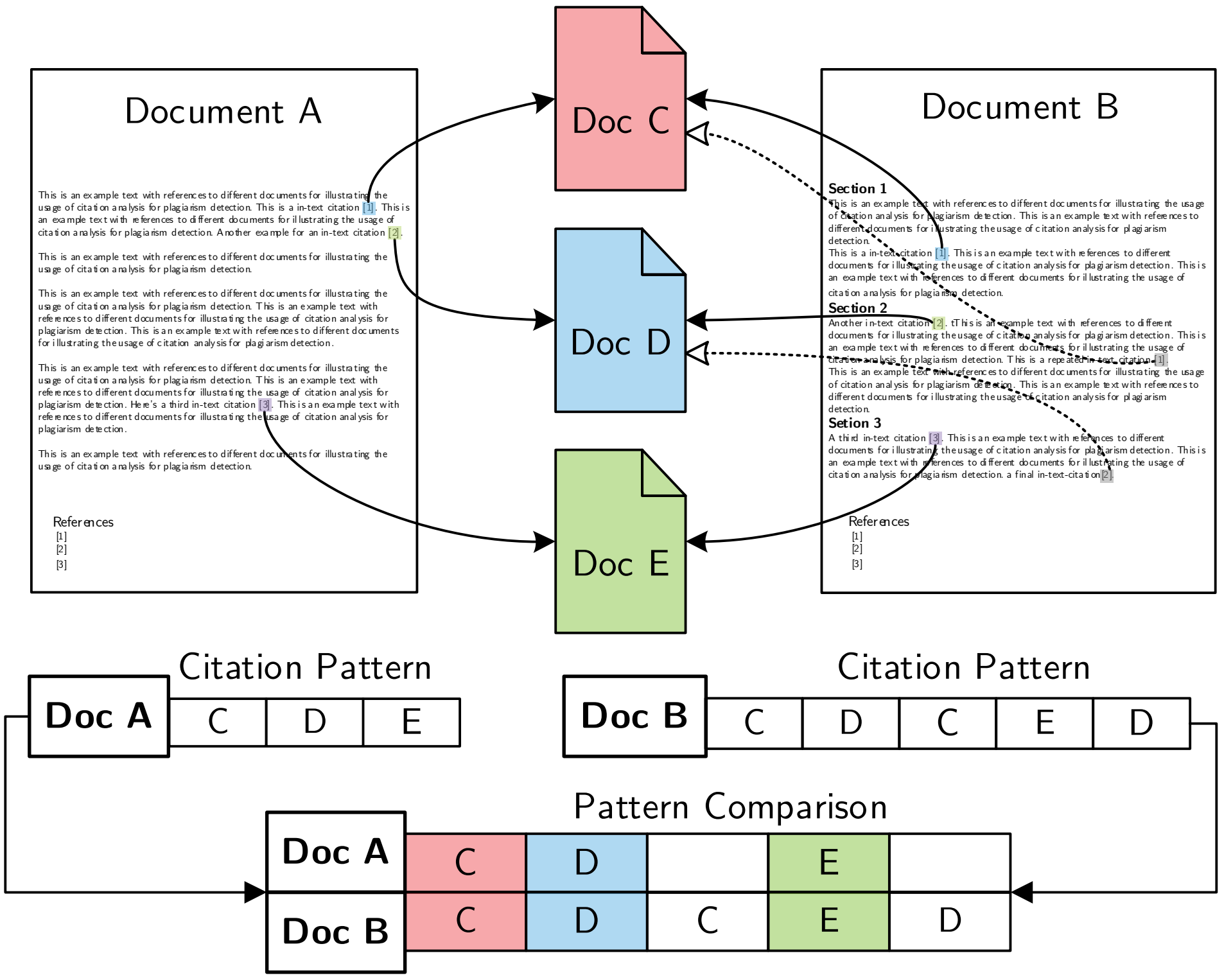

**Figure 3.2.** Concept of Citation-based Plagiarism Detection.

### 3.2.1 Conceptualization of the Detection Approach

To test the hypothesis that analyzing citation patterns can reveal academic plagiarism, we manually examined confirmed cases of plagiarism from the GuttenPlag, VroniPlag, and Retraction Watch collections (see Section 2.5.1, p. 55, for details on the collections). For cases from all three collections, we observed that plagiarists often paraphrase copied text but typically do not change the citations copied from the source document, which supported our hypothesis.

The GuttenPlag project was particularly interesting because it brought together hundreds of examiners who thoroughly analyzed a single thesis. The analysis revealed reused content on 371 of the 393 pages in the thesis affecting 63.8% of all lines [199]. As for all cases of non-artificial plagiarism, certainty about the extent of unoriginal content is virtually impossible. However, due to the unprecedently meticulous investigation, we expect the project's findings to be the most accurate approximation of the true amount of unoriginal content in any thesis to date.

Our study of the GuttenPlag findings focused on investigating whether CbPD is better suitable than existing detection methods for identifying translated plagiarism. As discussed in Sections 2.5.2, p. 65, and 2.5.3, p. 70, detecting translated plagiarism has been particularly challenging for plagiarism detection methods. The GuttenPlag project had identified 16 text passages on thirty-one pages of Guttenberg's thesis that had been translated into German from one or multiple English sources. We studied the citation patterns for those passages and analyzed the thesis using widely-used text-based plagiarism detection systems.

**Figure 3.3** shows the citations contained in the 16 translated passages. The third column illustrates the citation patterns we observed in Guttenberg's thesis (denoted in the figure as Guttenberg06) and the respective source document of the passage. Identically colored cells represent citations to the same source, whereas intermediate blank cells indicate one or more citations to non-shared sources.

Thirteen of the 16 passages contain citation patterns that are distinctive enough to retrieve the passage as suspicious. Cleaning the identified citation patterns of citations that both documents do not share at the respective positions of their citation sequences makes the high similarity of the patterns particularly apparent. The lower part of **Figure 3.3** exemplifies this "cleaning" of citation patterns for pages 242–244 of Guttenberg's thesis.

None of the three plagiarism detection systems we tested identified any of the 16 passages. This result supported our expectation that CbPD holds much promise for improving the detection of disguised forms of academic plagiarism.

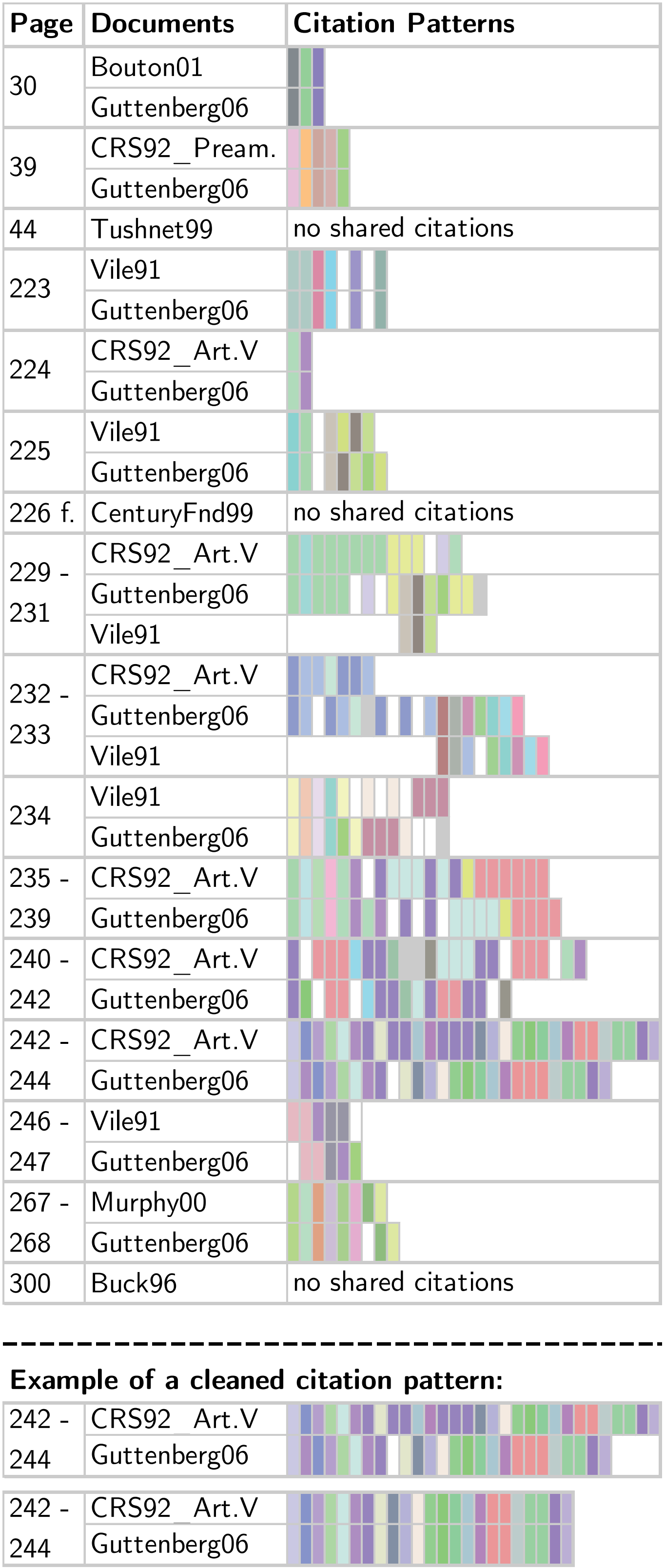

**Figure 3.3.** Citation patterns for translated passages in Guttenberg's thesis.

### 3.2.2 Challenges to Citation Pattern Identification

Detecting citation patterns is a non-trivial task because the diverse forms of academic plagiarism result in different citation patterns. The following challenges exist for devising methods to identify distinctive citation patterns.

**Unknown pattern constituents**: Unlike in text string matching, the subsequences of citations to extract from the suspicious document and search for within the reference collection are initially unknown. Citations that two documents share are identifiable easily. However, that all shared citations represent unoriginal text segments is unlikely. Semantically related documents often share citations legitimately. Two documents may share some citations due to semantic relatedness, while other shared citations may result from undue text reuse.

**Transpositions**: The order of citations in a reused text segment can differ from the source segment. A trivial cause can be that both documents use numeric citation styles, one of which sorts the bibliography alphabetically and the other by publication date. An author may also have reordered a reused text segment.

**Insertions, substitutions, and scaling of citations**: While paraphrasing reused text, authors may include citations from other documents, insert additional non-shared citations, substitute the citations of the source with semantically similar non-shared citations, or use shared citations more than once (referred to as scaling).

**Figure 3.4** illustrates the challenges to citation pattern identification. The figure displays the citation sequences in a source document (Src. Doc.) and a suspicious document (Susp. Doc.). The documents share eight citations (1–8), of which only three citations (1, 2, 3) occur within a text segment that the author of the suspicious text reused from the source. The other citations occur together with non-shared citations (X) in original text distributed over the length of the documents. The author of the suspicious document changed the order of shared citations in the reused text segment, used the citation to the first reference twice, and inserted two non-shared citations. In this case, to distinguish the suspicious citation pattern (1, 2, 3) from the legitimately shared citations, a detection method would also have to examine the distance of the citations in the text.

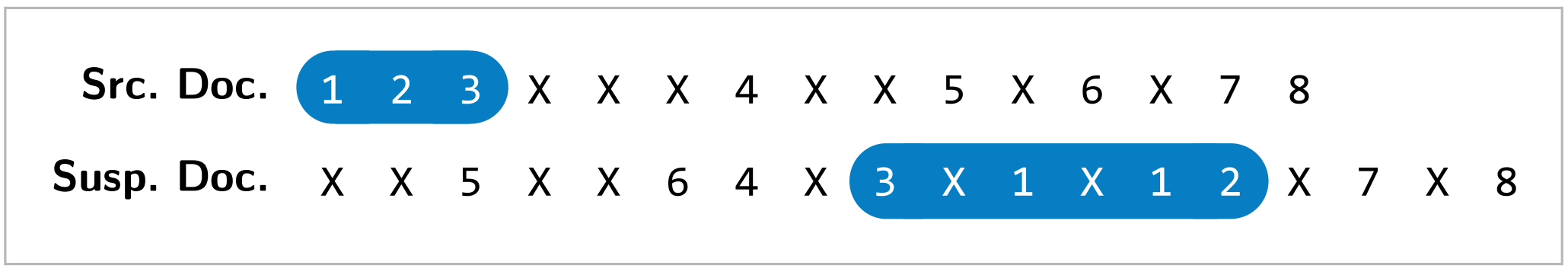

**Figure 3.4.** Hard-to-identify citation pattern in a reused text segment.

## 3.3 Citation-based Detection Methods

Motivated by the results of our preliminary, manual examination of confirmed plagiarism cases, we devised multiple algorithms to identify citation patterns. We tailored each algorithm to the properties we had observed for specific forms of academic plagiarism. Our goal was to evaluate a balanced mix of detection methods that analyze documents' global and local similarities. To address the challenges of identifying citation patterns, we included methods that rely on set-based and sequence-based similarity measures. Set-based measures can handle transpositions and scaling of citations. Sequence-based measures reflect the similarity in the order of citations, which can be a strong indicator of potentially suspicious similarity.

**Table 3.1** shows the categories of similarity assessments, global vs. local and set-based (order-agnostic) vs. sequence-based (order-observing), according to which we devised the detection methods, which we describe hereafter.

**Table 3.1.** Categorization of citation-based detection methods.

| | **Global Similarity Assessment** | **Local Similarity Assessment** |
|---|---|---|
| **Set-based** | Bibliographic Coupling | Citation Chunking |
| **Sequence-based** | Longest Common Citation Sequence | Greedy Citation Tiling |

### 3.3.1 Bibliographic Coupling

We evaluated Bibliographic Coupling Strength, i.e., the number of references two documents share, as a coarse-grained measure of global document similarity. BC ignores the order and positions of citations within the text. Thus, BC Strength alone is typically an insufficient indicator of potential plagiarism and does not allow pinpointing potentially plagiarized text segments.

### 3.3.2 Longest Common Citation Sequence

We included the **Longest Common Citation Sequence (LCCS)**, i.e., the Longest Common Subsequence (cf. Section 2.4.2, p. 22) of the citation sequences in two documents as a global, order-focused detection method. **Figure 3.5** presents an example of two documents that share a LCCS of length three. Arabic numerals represent shared citations, and the letter X represents non-shared citations. We also devised a more restrictive measure—LCCS distinct—which only considers the first occurrence of a shared citation within the LCCS.

Intuitively, considering the LCCS yields high similarity scores if an author reused longer parts of a document without altering the contained citations. The method allows for arbitrarily sized gaps of non-matching citations and can compute a similarity score despite potential transpositions of citations.

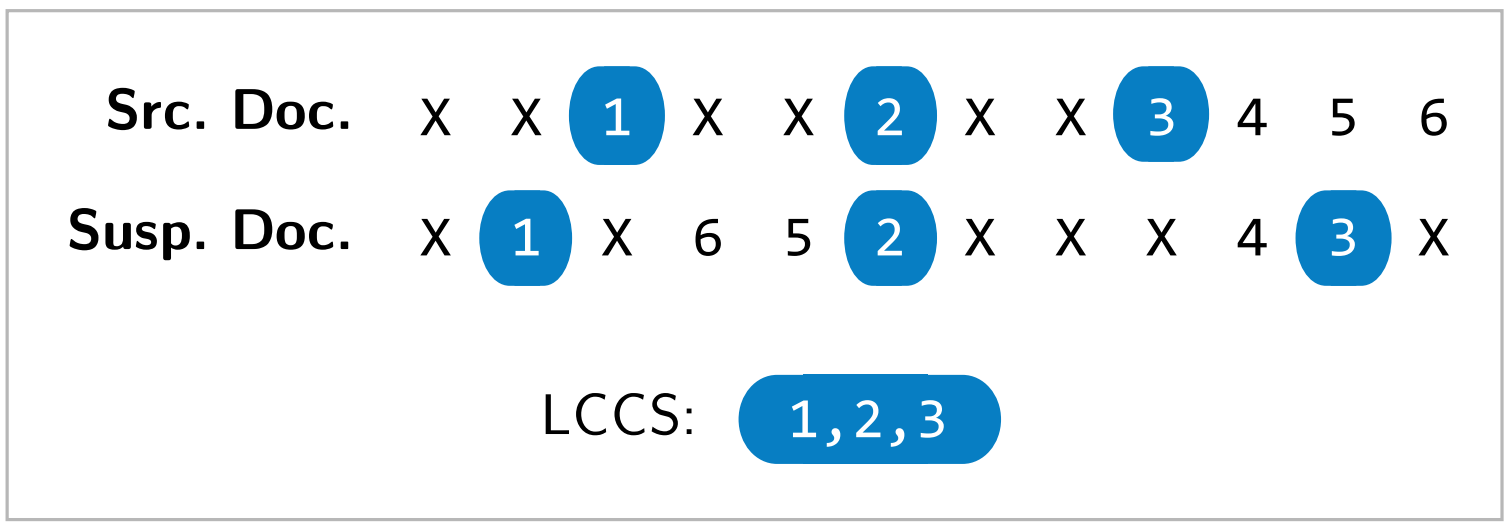

**Figure 3.5.** Longest Common Citation Sequence in two documents.

### 3.3.3 Greedy Citation Tiling

Greedy Citation Tiling (GCT) adapts an algorithm that Wise proposed for computing the similarity of text strings, primarily for the plagiarism detection use case [559]. Researchers successfully applied the original Greedy String Tiling (GST) algorithm for plagiarism detection in text documents [248] and source code [418].

Greedy Citation Tiling identifies all individually longest blocks of consecutive shared citations in identical order, so-called **citation tiles**. We represent citation tiles as triples $t = (s_1, s_2, l)$, in which $s_1$ and $s_2$ denote the starting positions of the match in the first and second document, and $l$ indicates the length of the match.

**Figure 3.6** exemplifies citation tiles in two documents. Arabic numerals denote shared citations, the letter X non-shared citations, and colored highlights with Roman numerals citation tiles using the notation of triples representing start, end, and length of the tile. As the figure illustrates, the tiling approach can cope with arbitrarily sized gaps between citation tiles and transpositions in the order of individual citation tiles. Users can choose a minimum size of matching citation tiles. Using this option causes the GCT algorithm to ignore all citation tiles shorter than the specified minimum length. Finding long citation tiles, i.e., patterns of shared citations in the same order, provides a strong indication of content similarity.

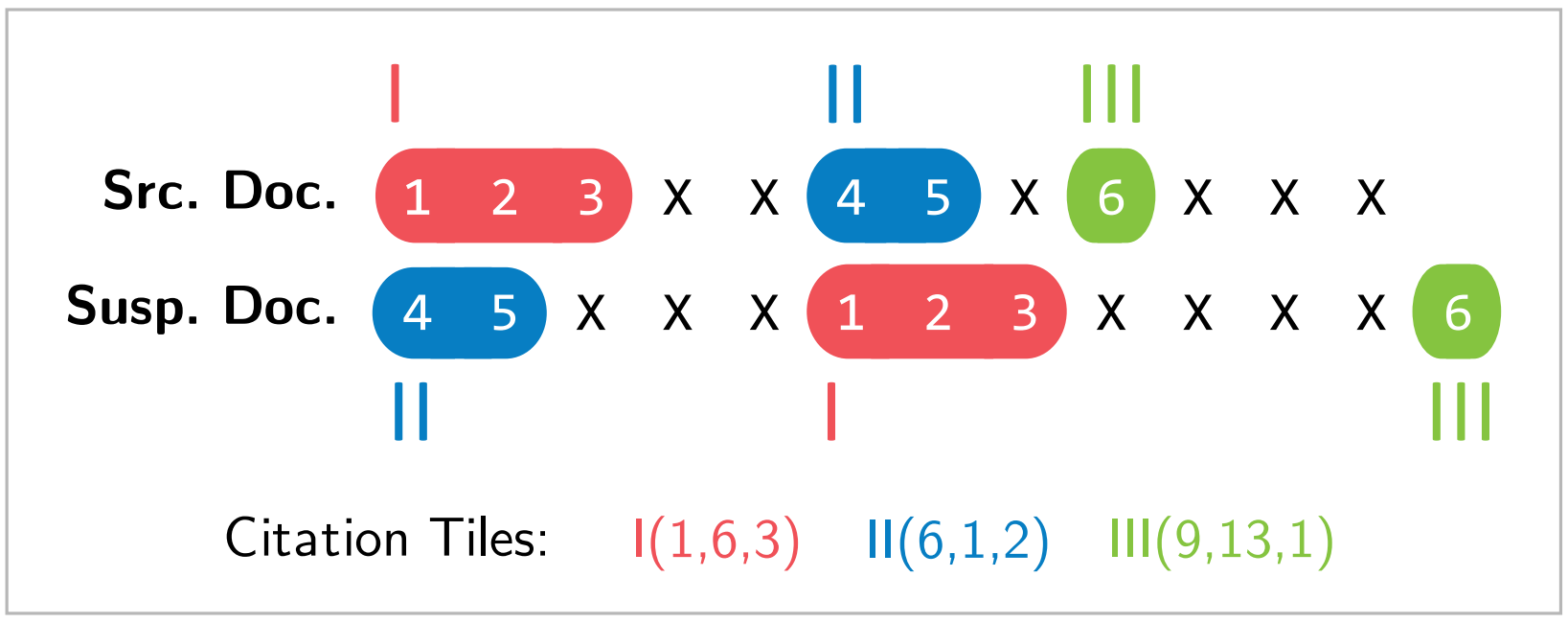

**Figure 3.6.** Greedy Citation Tiles in two documents.

Wise designed the Greedy String Tiling algorithm primarily to find copy-and-paste and shake-and-paste plagiarism. Greedy Citation Tiling could serve the same purpose but opposed to the text-based algorithm may also identify paraphrased cases of shake-and-paste plagiarism. GCT can handle transpositions of citations that result from rearranging text segments, which is typical for shake-and-paste plagiarism. However, GCT cannot identify citation scaling. To find such cases, we devised another class of detection methods, which we explain in the following section.

### 3.3.4 Citation Chunking

Citation Chunking (CC) describes a collection of algorithms that aim at identifying citation patterns regardless of whether the order of matching citations differs in both documents, i.e., regardless of whether matching citations have been transposed or scaled. We named the method Citation Chunking because it resembles the feature selection strategies of text-based fingerprinting methods. Citation Chunking algorithms select a variably sized substring of a document's citation sequence and consider the selected citations as a single unit of comparison—a chunk.

The idea of citation chunking is to consider shared citations as anchors at which citation patterns can exist. Starting from an anchor, the chunking algorithm constructs citation chunks by increasing the considered citations depending on the current characteristics of the chunk and the succeeding citations.

#### Chunking Strategies

Determining the best size of a citation chunk is a non-trivial task. Larger chunks are more suitable for detecting the global similarity of documents and can better compensate for the transposition and scaling of citations. On the other hand, smaller chunks are more suitable for pinpointing specific areas in documents with high similarity. To capture the citation patterns we observed for different forms of academic plagiarism as part of our manual investigation, we devised different algorithms for forming citation chunks.

**Algorithm 1: Include consecutive shared citations only**

The first chunking algorithm forms chunks consisting entirely of shared citations that occur consecutively in both documents. Other than citation tiles, the consecutive shared citations do not have to be in the same order. **Figure 3.7** illustrates two pairs of citation chunks that chunking algorithm one forms for two documents.

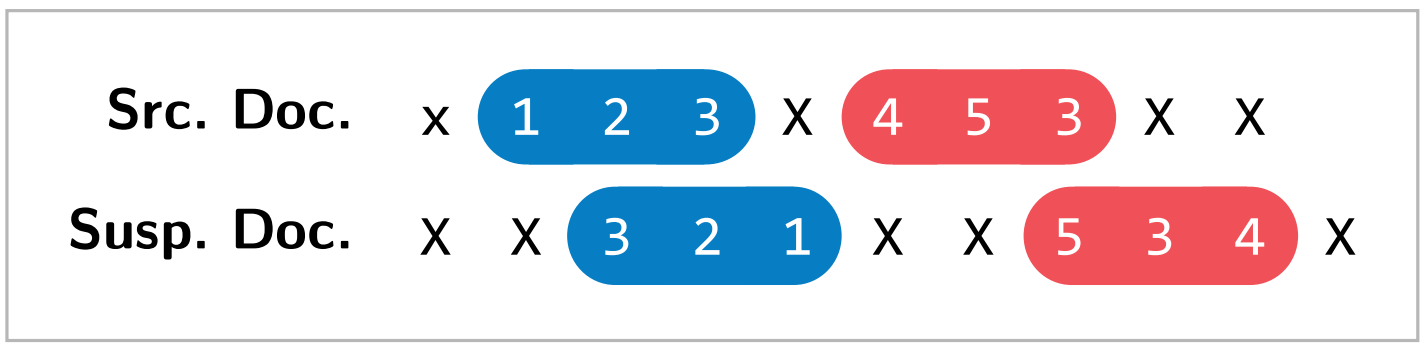

**Figure 3.7.** Chunking algorithm 1—consecutive shared citations only.

Algorithm one is the most restrictive chunking procedure we devised, intending to identify confined text segments with high citation-based similarity. The algorithm is best suited for detecting cases of copy-and-paste plagiarism that may have been concealed by rewording or translation.

**Algorithm 2: Inclusion depends on the previous chunk**

Chunking algorithm two includes a citation in a chunk if the number of non-shared citations that separate the citation under consideration for inclusion from the last shared citation is smaller than the number of citations (shared and non-shared citations) the chunk contains already. **Figure 3.8** exemplifies two pairs of citation chunks that chunking algorithm two yields for the documents shown in the figure.

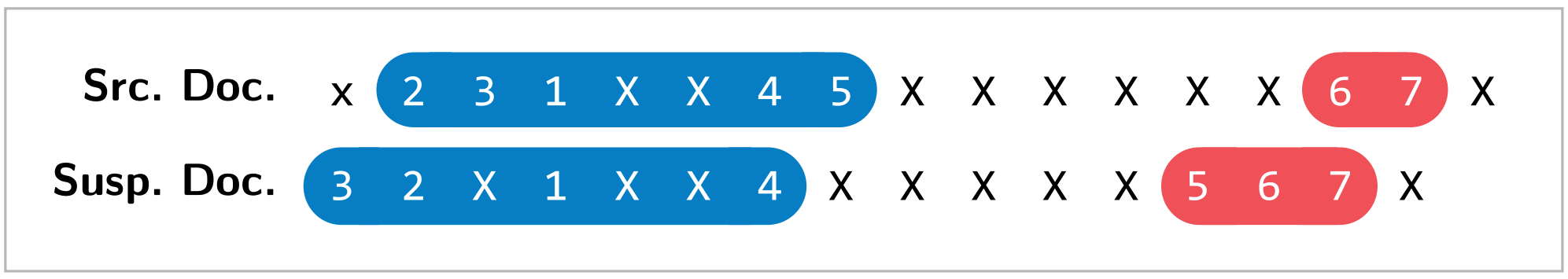

**Figure 3.8.** Chunking algorithm 2—depending on the prior chunk.

Chunking algorithm two seeks to uncover potential cases where a plagiarist took over text segments or logical structures from a source document. The algorithm allows for sporadic non-shared citations that plagiarists may have inserted to make their text appear more genuine. The algorithm can also reveal potential cases of concealed shake-and-paste plagiarism as it allows an increasing number of non-shared citations in a chunk, given that the chunk includes several shared citations already. The rationale is to capture a plagiarist's behavior of interweaving text segments (including the contained citations) from different source documents.

**Algorithm 3: Inclusion depends on the textual distance**

Chunking algorithm three defines a textual range in which possible plagiarism is deemed likely. Studies have shown that plagiarism more frequently affects confined text segments, e.g., a few paragraphs, rather than extended text passages or the entire document [274], [330]–[332], [424]. Given these findings, chunking algorithm three only includes citations in chunks if the citations occur within a specified textual range. The algorithm uses a sliding window to check which citations occur within the defined range. **Figure 3.9** illustrates chunking algorithm three.

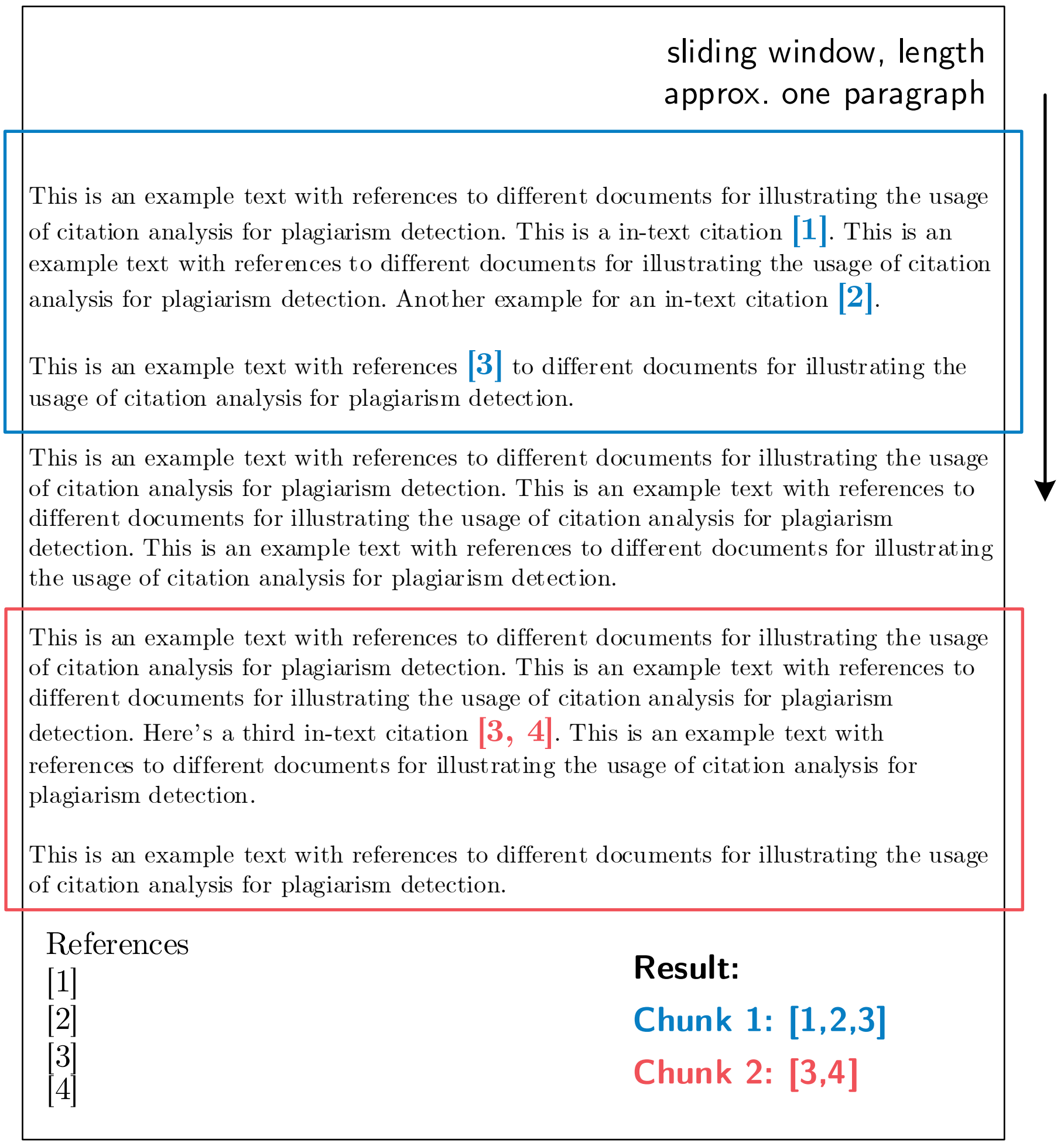

**Figure 3.9.** Chunking algorithm 3—citations in specified textual range.

Because plagiarists may change the segmentation of plagiarized text, chunking algorithm three analyzes the textual proximity of citations in terms of multiple text units, including characters, words, sentences, and paragraphs. Defining a suitable maximum distance for the proximity of citations in the text is highly dependent on the individual corpus analyzed. If document length is short and individual documents contain fewer sections and paragraphs, altering the text structure is more difficult for a plagiarist. Therefore, a relatively small maximum distance is most suitable to detect plagiarism in short documents with few sections. In contrast, reordering text usually becomes easier the longer the document.

To determine a suitable proximity threshold, we analyzed the average number of hierarchically subordinate text constituents (e.g., characters and words) contained within hierarchically superordinate text constituents (e.g., paragraphs). For example, in one document, a paragraph may, on average, contain 120 words and 720 characters. If less than 120 words separate one shared citation from another shared citation, chunking algorithm three would include the second shared citation in the chunk. Using this approach, a Citation Chunking method employing algorithm three can deal with artificially created paragraph split-ups.

## Merging of Chunks

We devised an optional merging step for chunks to experiment with larger chunk sizes, which may reveal longer passages of similar text. The merging step combines chunks that either of the three chunking algorithms formed if the number of non-shared citations that separate two chunks is less or equal to the number of shared citations in the first chunk. If the merging procedure merges citation chunks in one iteration, it performs another iteration to check whether the merged chunks still fulfill the criteria for merging with other chunks. If that is the case, the procedure merges the previously merged chunks once more to form even larger chunks.

**Figure 3.10** illustrates the repeated merging of chunks. In iterations one and two, the procedure merges the first three chunks. In iteration three, the procedure no longer merges any chunks because the distance of the last chunk (citations 6 and 7) to the merged chunk is larger than the number of shared citations in the merged chunk. The merging procedure terminates at that point.

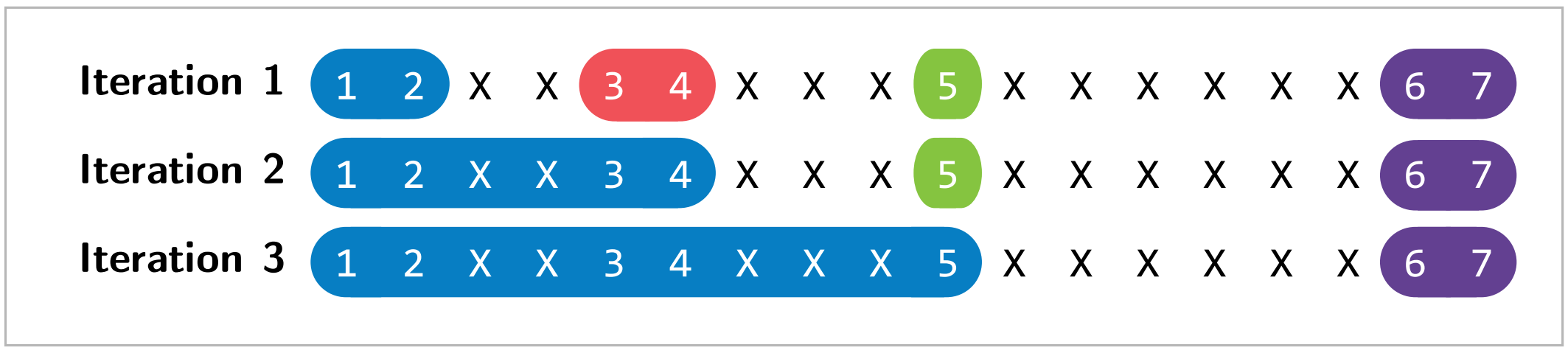

**Figure 3.10.** Merging of citation chunks.

**Figure 3.11** summarizes the chunk formation process as a flowchart.

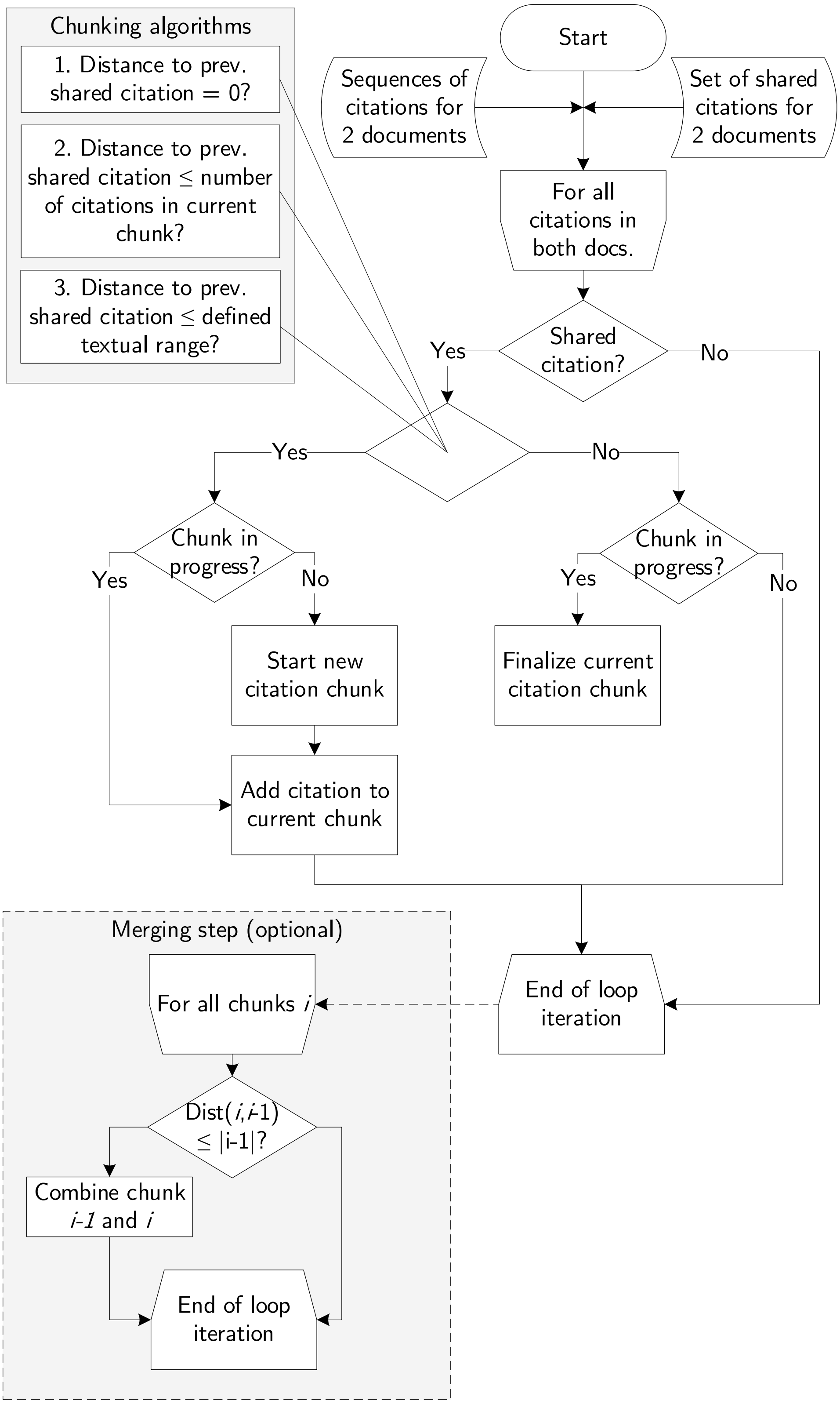

**Figure 3.11.** Process for forming citation chunks.

### Chunk Comparison

After forming and optionally merging citation chunks, the detection method compares chunks to each other regardless of the order of citations. The number of shared citations in two chunks determines the similarity of the chunks. Using this set-based comparison, the Citation Chunking method accounts for potential transpositions and scaling of citations.

We tested two strategies for comparing citation chunks in two documents.

1. The first strategy forms citation chunks for both documents and then performs pairwise comparisons of each chunk in the first document with each chunk in the second document. The comparison procedure identifies the chunk pairs with the highest overlap in citations and stores them as a match. If multiple chunks have an equal overlap in citations, the procedure stores all combinations with maximal overlap.
2. The second strategy only forms chunks for one document, which it compares to the unaltered citation sequence of the second document. The procedure "slides" each chunk of the first document over the entire citation sequence of the second document. The procedure identifies the subsequence of citations in the second document having the highest overlap with the chunk and stores both the chunk and the citation subsequence as a match.

## 3.4 Evaluation Methodology

To conclusively show that Citation-based Plagiarism Detection can improve the detection of disguised forms of academic plagiarism, we quantitatively evaluated the new detection approach using a large-scale dataset. The novelty of CbPD presented a significant challenge for a large-scale quantitative evaluation.

### 3.4.1 Dataset

CbPD demands properties of the dataset that the datasets for prior evaluations of plagiarism detection methods and systems do not offer (see Section 2.5.1, p. 49, for such datasets). An evaluation dataset suitable for CbPD ideally offers the following:

1. **Real plagiarism**: Test cases should not be fabricated, neither manually nor automatically, when the goal is to evaluate the effectiveness of CbPD on realistically disguised plagiarism containing potential citation copying.
2. **Citations**: The full text of documents must contain readily available citations to allow the parsing of citation positions.

3. **Size and diversity**: The dataset should contain a large number of documents from various authors to reflect different writing and citing styles.
4. **Ground truth**: A quantitative evaluation requires knowledge about the truth concerning a question, in our case, knowing whether retrieved documents and parts thereof represent potential academic plagiarism.

Given that the datasets of prior evaluations focus on text-based plagiarism detection systems, the creators of the datasets did not purposefully include academic citations. Furthermore, with plagiarized sections artificially fabricated, available datasets miss the full range of realistically disguised plagiarism we expect to find in real collections. For the 2010–2013 PAN competitions datasets, Potthast et al. made a significant effort to create realistic plagiarism by contracting writers to produce plagiarized articles using crowdsourcing platforms. This approach produced the most realistic test cases available so far, especially for the 2012 and 2013 collections, containing around 300 articles featuring disguised plagiarism [412], [413]. However, none of the test cases contains citations.

We argue that it remains doubtful whether articles written by contractors lacking expert knowledge are comparable in their degree of plagiarism disguise to the forms of plagiarism found in scientific publications. The motivation for disguise likely differs in a setting where authors work months or even years on a publication. Because the strength of the CbPD approach lies in detecting heavily disguised plagiarism, we used a real scientific document collection in our evaluation to reflect the full range of disguised plagiarism forms and potential citation copying.

We chose the PubMed Central Open Access Subset (PMC OAS) [511], an openly accessible collection of biomedical research publications for a large-scale evaluation of CbPD. The PMC OAS contains peer-reviewed publications, which leads us to assume a low level of plagiarism containment. However, if present, we assume that several plagiarism cases have been disguised, which allowed them to remain undetected, thus fulfilling the real plagiarism requirement number one. Given that the PMC OAS contains scientific publications, citations are readily available, fulfilling requirement number two. At the time of our investigation (April 2011), the PMC OAS contained 234,591 articles by approximately 975,000 authors from 1,972 peer-reviewed journals. Therefore, the collection fulfills the third requirement of a large-scale and diverse dataset. A desirable bonus of the PMC OAS is its XML document format, which offers machine-readable markup for metadata and citations.

We conducted a user study to establish a ground truth approximation on perceived plagiarism and its severity for a finite pool of documents. In summary, when combined with a user study, the PMC OAS collection is ideally suited for evaluating the detection effectiveness and efficiency of CbPD methods.

### 3.4.2 Evaluated Detection Methods

**Table 3.2** lists the detection methods for which we report results hereafter. We analyzed the PMC OAS collection using 19 variants of the citation-based detection methods described in Section 3.3, p. 86. The seven variants listed in the table achieved the best detection effectiveness, which is why we present their results in detail. We compared the citation-based detection methods to two state-of-the-art text-based detection methods—Encoplot (Enco) and Sherlock.

**Table 3.2.** Evaluated detection methods.

| | |
|---|---|
| **Citation-based detection methods** | |
| **BC abs.** | Absolute Bibliographic Coupling Strength |
| **BC rel.** | Relative Bibliographic Coupling Strength |
| **LCCS** | Longest Common Citation Sequence |
| **LCCS dist.** | Longest Common Sequence of distinct citations |
| **Max. GCT** | Longest Greedy Citation Tile |
| **CC-bcn** | Longest Citation Chunk – **b**oth documents chunked, considering **c**onsecutive shared citations only, **n**o merging step |
| **CC-bpn** | Longest Citation Chunk – **b**oth documents chunked, considering shared citations depending on the **p**rior chunk, **n**o merging step |
| **Text-based detection methods** | |
| **Enco** | Encoplot—exact character 16-gram string matching |
| **Sherlock** | Sherlock—probabilistic word-based fingerprinting |

**Encoplot** performs pairwise document comparisons using character 16-gram matching [187]. During each comparison of a document pair, the method extracts all character 16-grams from the two documents into two separate lists, sorts the lists, and uses a modified merge sort algorithm to identify matching 16-grams. To speed up the comparison, Encoplot exclusively matches the first occurrence of a 16-gram in the first document to the first occurrence of that 16-gram in the second document, the second occurrence to the second, and so on. If the number of 16-gram occurrences in the documents is different, Encoplot does not identify all possible matches. This more restrictive $n$-gram matching procedure achieves a worst-case time complexity of $O(n)$. Encoplot achieved the highest overall PlagDet score in the PAN competition 2009 [407, p. 7].

**Sherlock**[9] is an open-source plagiarism detection program that uses hashed word $n$-gram fingerprinting with semi-random fingerprint selection. Sherlock is representative of many plagiarism detection systems. Sherlock allows customizing the length of the word $n$-grams and the probability of retaining word $n$-grams.

The program consists of two separate processes for creating and comparing the hashed word $n$-grams, called signatures. During the signature creation, the program chunks the input text, hashes the word $n$-grams into unsigned long integer values (signatures), and semi-randomly selects signatures to retain. The selection criterion is whether the signature contains a customizable number of zero bits. Due to the properties of the hash function, the criterion ensures that the chunk retention rate is probabilistic. At the same time, the selection is deterministic for identical input, i.e., the decision to retain or discard a signature will be the same for identical word $n$-grams. Sherlock sorts retained signature for faster comparisons. By default, the program partitions the input texts into word 3-grams and selects, on average, one of 16 signatures. We increased the probability of retaining signatures to one out of eight on average to perform a finer-grained comparison for our experiment. In the second phase, Sherlock performs pairwise signature comparisons and reports the identified similarity as a percentage, computed as

$$s = \frac{100\ |m|}{|d_1| + |d_2| - |m|}$$

where $|m|$ denotes the number of matching signatures and $|d_1|$ and $|d_2|$ denote the number of signatures in the two input documents.

### 3.4.3 Corpus Preprocessing

The PMC OAS collection comprised 234,591 documents before preprocessing. We excluded 13,371 documents for being unprocessable. Such documents included scanned articles in image file formats, duplicates, and documents with multiple text bodies (e.g., summaries of all articles in conference proceedings). **Table 3.3** gives an overview of the excluded documents.

Of the 221,220 processable documents, we removed an additional 36,118 documents with no references or citations and 68 documents with inconsistent citations. Documents with no references or citations were typically short comments, letters, reviews, or editorial notes that cited no other documents.

---

9 The tool's website went offline recently. The source code and documentation are still available via: https://web.archive.org/web/20180219024142/http://web.it.usyd.edu.au/~scilect/sherlock/

**Table 3.3.** Number of PMC OAS documents excluded from evaluation.

| Criterion | Number |
|---|---|
| PMC OAS | 234,591 |
| Excluded documents | 13,371 |
| No text body | 12,783 |
| Duplicate files | 471 |
| Multiple text bodies | 117 |
| **Processable documents** | **221,220** |

The final test collection comprised of 185,170 documents. We could not acquire citation placement information for 16,866 documents because citations were not marked up in the XML source file or because the original text stated citations within figures or captions. An additional 10,746 documents listed the same reference multiple times in their bibliography, and 59 documents listed references that did not occur in the main text. We did not exclude these documents because the likelihood of false negatives, i.e., unidentified cases of plagiarism, is higher when removing the documents entirely than if we retain the incomplete citation information. **Table 3.4** summarizes the results of the preprocessing steps.

**Table 3.4.** Preprocessing results for PMC OAS collection.

| Criterion | Documents | Citations | References |
|---|---|---|---|
| Processable documents | 221,220 | 10,976,338 | 6,921,249 |
| No references and/or citations | 36,118 | 0 | 6,447 |
| Inconsistent citations | 68 | 11,405 | 4,722 |
| **Test collection** | **185,170** | **10,964,933** | **6,910,080** |
| References without citations | 16,866 | - | 65,588 |
| Citations without references | 59 | 474 | - |
| Non-unique references | 10,746 | - | 32,122 |

### 3.4.4 Applying Detection Methods and Pooling

The typical plagiarism detection task requires a one-to-many analysis, i.e., comparing a single input document to a reference collection. Because the set of potentially suspicious documents is unknown in our evaluation, the PMC OAS collection calls for a many-to-many analysis. Analyzing the collection in this fashion would require

$$\binom{\mathrm{n}}{2} = \binom{185{,}170}{2} = 17{,}143{,}871{,}865 \text{ comparisons.}$$

This number is practically infeasible to perform by any plagiarism detection system in a sensible timeframe without an initial limitation of the test collection.

Text-based detection methods typically reduce the retrieval space by comparing heuristically selected text fragments and imposing a minimum threshold for shared text. Such heuristics, however, have the inherent disadvantage of decreasing detection accuracy. On the other hand, the citation-based detection approach allows limiting the document collection without compromising detection accuracy. Because documents must be bibliographically coupled, i.e., share at least one reference, to qualify for a citation-based analysis, we reduced the collection size by filtering for Bibliographic Coupling Strength, $s_{\mathrm{BC}} \geq 1$.

**Figure 3.12** shows the distribution of Bibliographic Coupling Strength $s_{\mathrm{BC}}$ for document pairs $\partial_k = (d_1, d_2)$ in terms of the inverse cumulative frequency $\bar{f}_c = \sum_{i=n}^{1} \forall \partial_k \mid s_{\mathrm{BC}}(\partial_k) \geq i$ plotted on an absolute scale (black solid line) and a $\log_2$ scale (grey dashed line). Additionally, the figure shows the mean $\mu$, standard deviation $\sigma$, and quartiles $Q_{1,2,3}$ of the distribution below the abscissa.

Restricting the analysis to bibliographically coupled documents reduced the collection size to 39,463,660 document pairs. Due to the practical infeasibility of a collection-wide, text-based many-to-many analysis, we applied Encoplot and Sherlock only to the 6,219,504 document pairs with a Bibliographic Coupling Strength $s_{\mathrm{BC}} \geq 1$. This reduction may have excluded some true positives.

However, we argue that limiting the collection size using Bibliographic Coupling Strength is unlikely to affect text-based detection performance significantly adversely. We performed an ex-post many-to-many analysis of the top-20 most suspicious documents identified in our user study (cf. Section 3.4.6, p. 101) to substantiate this hypothesis. As we did not filter for BC Strength, computing the Encoplot scores for these 20 documents with all other documents in the PMC OAS collection took several weeks on a quad-core system.

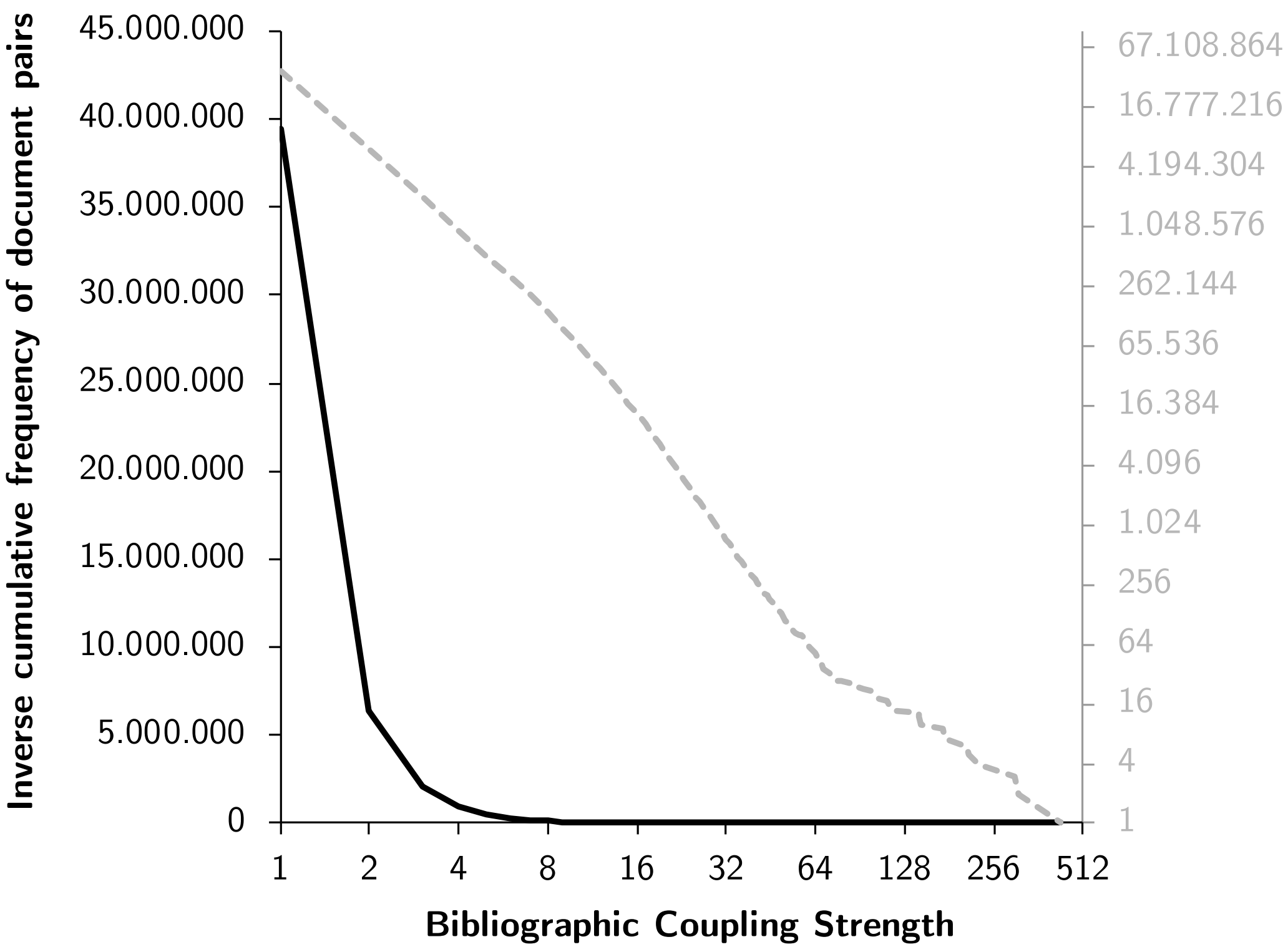

$\mu = 1.21$ $\sigma = 0.95$ $Q_1 = 1$ $Q_2 = 1$ $Q_3 = 1$

**Figure 3.12.** Bibliographic Coupling Strength for documents in PMC OAS.

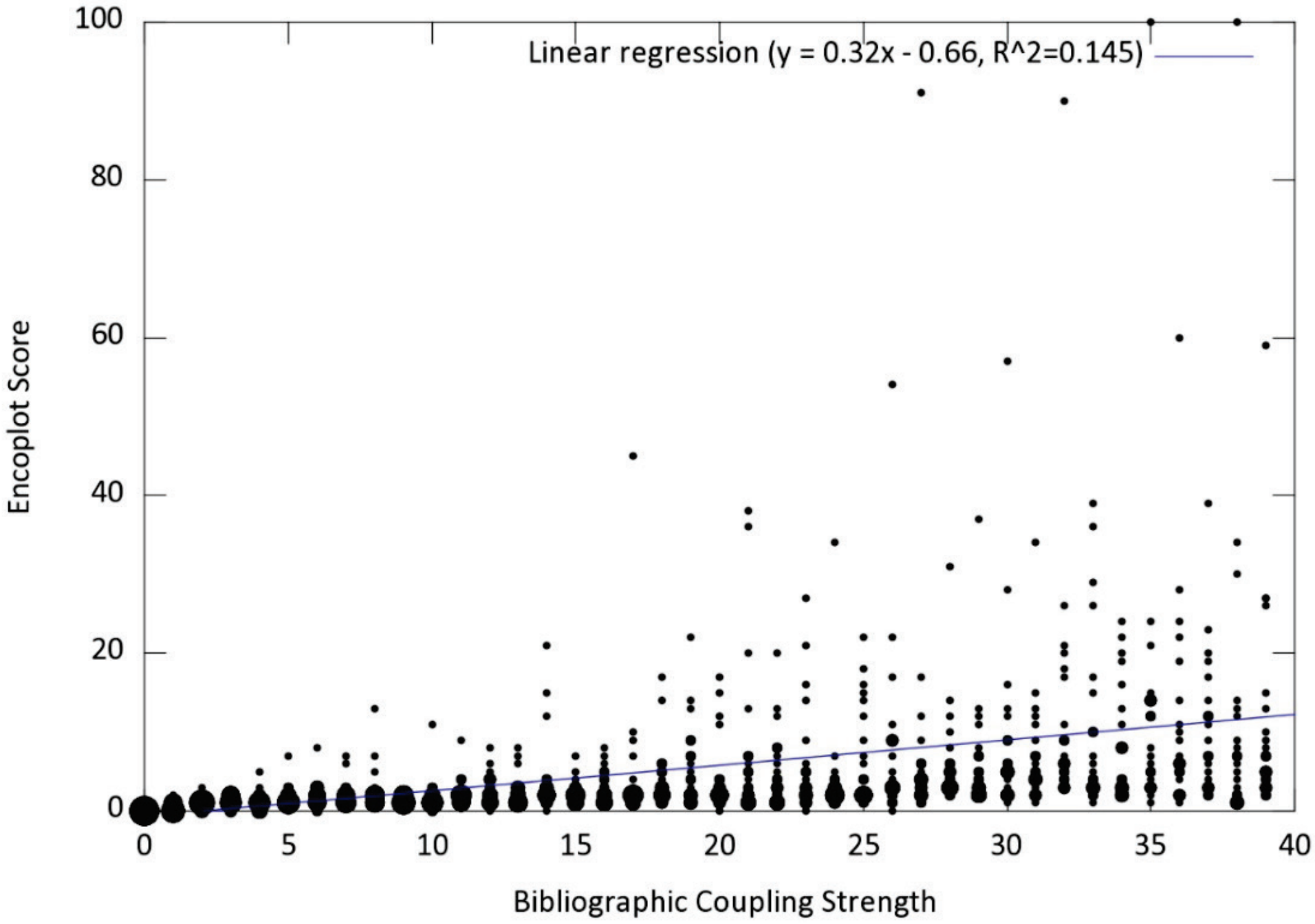

**Figure 3.13.** Correlation between BC Strength and Encoplot score.

**Figure 3.13** plots Bibliographic Coupling Strength and the text-based Encoplot score. The smallest dots represent single occurrences, while the largest dots represent up to 20 occurrences. The analyzed sample contained no document pair with an Encoplot score above 3% that was not bibliographically coupled. Given the correlation between bibliographic similarity and text-based similarity, we hypothesize that the loss of detection performance is minimal. Therefore, we consider requiring a minimum Bibliographic Coupling Strength as an acceptable compromise to limit the collection size and enable a many-to-many analysis of documents.

Because judging all results is infeasible, we pooled the top-30 ranked document pairs for the nine detection methods, as is common practice in Information Retrieval evaluations, such as TREC, NTCIR, or CLEF [70]. Pooling describes combining the top-$n$ results of $k$ retrieval systems. Only the pooled results are judged for relevance. Typical pooling approaches include using the top-$n$ results of the best state-of-the-art systems or from all systems participating in the experiment, which is the pooling approach we followed.

### 3.4.5 Addressing False Positives

The retrieval of false positives is a universal problem for plagiarism detection methods. For the PMC OAS collection, false positives presented a more significant challenge to text-based methods than citation-based methods as specific document types reused standardized expressions or boilerplate text. Several instances of high textual similarity were thus justified. For this reason, we applied a false positive reduction strategy to the pooled documents before collecting relevance judgments.

We excluded the collection-specific document types editorials and updates. Editorials were typically non-scientific texts written by journal editors or publishers, which provide publishing guidelines or descriptions of the journal's purpose and policies. Such text is often "recycled" as boilerplate text among journals without citing the source. Updates included revisions to published material and slight changes to annually published medical standards, best practices, or procedures, commonly published by medical associations, e.g., the American Diabetes Association. The References [26] and [27] exemplify such updates.

This exclusion of documents was necessary for a meaningful performance evaluation. Without it, the text-based detection methods, particularly Encoplot, would have retrieved among its top ranks almost exclusively legitimately similar documents. These documents would have caused an unwarranted high rate of false positives for the text-based methods. We also excluded PMC OAS publications that cited each other or shared authors to reduce false positives that referenced the source or were likely examples of legitimate collaboration.

Two additional factors contributed to a higher false positive rate in the case of the PMC OAS. First, we carried out the pooling process as a many-to-many document comparison, while the typical plagiarism detection use case is performing a one-to-many comparison. A many-to-many comparison of an extensive collection naturally results in retrieving high numbers of legitimately similar documents. Second, the relatively sparse amount of plagiarism in the PMC OAS makes the retrieval of legitimately similar documents more likely.

Despite our strategy to reduce false positives, some cases of legitimate text similarity remained, which we identified during the pooling step and removed before the user study. Collecting the top-30 similar documents for the text-based method Encoplot required examining 235 documents because Encoplot retrieved 205 collection-specific false positives, such as editorials and updates. Collecting the top-30 similar documents for the LCCS method required examining 31 documents because only one collection-specific false positive was retrieved.

The citation-based methods retrieved fewer false positives because many documents featured unique citation patterns despite the high textual overlap, e.g., in medical case studies and editorials. Other documents simply had insufficient citations due to their non-scientific nature. In conclusion, false positive rates are highly collection-dependent. Every collection contains different document formats and text from different disciplines, meaning the reuse of text or citations may be seen as legitimate in some cases but not in others.

### 3.4.6 Collecting Relevance Judgements

We performed a user study to collect judgments on the dominant form of user-perceived plagiarism and the document's level of suspiciousness, i.e., the document's relevance to a plagiarism detection scenario. The top-30 pooling method yielded 270 document pairs, of which 181 were unique. We presented the unique pairs to 26 participants using the PDS prototype CitePlag [172] to obtain relevance judgments. We grouped participants according to their level of biomedical expertise into the three groups:

1. **Medical experts**: 5 participants
2. **Graduate students** from the medical and life sciences: 10 participants
3. **Undergraduate students** from a variety of majors: 11 participants

Because no standard guidelines or thresholds exist for classifying a document as "plagiarism," we asked participants to assess documents keeping in mind the **information need** in a real plagiarism detection scenario:

> Consider viewing a retrieved document pair as relevant if similarities exist that an examiner in a real check for plagiarism would likely find valuable to be made aware of.

We instructed participants to rate presented document pairs on a **scale from zero to five**. A score of zero indicated a false positive, while scores of one through five described various levels of document suspiciousness. An online submission form provided uniform guidelines, including definitions of the four examined plagiarism forms and verbal descriptions of the suspiciousness scores, i.e., the relevance to the plagiarism detection scenario. For example, a score of five indicated extremely suspicious similarities with obvious plagiarism intent. In contrast, a score of one described noticeable similarities in some sections where an author may have found inspiration from the source but most likely did not plagiarize.

A participant from each of the three knowledge groups examined each document pair. If examiners found the presented document pair to fulfill the given information need, i.e., suspiciousness score, $s > 0$, we asked them to:

1. Indicate the most prevalent form of suspected plagiarism;
2. Assign a suspiciousness score between one and five;
3. Indicate if a text-based, citation-based, or hybrid document similarity visualization was most suitable to assess the suspiciousness of the document.

Our evaluation procedure was as follows. We retained and grouped by the user-perceived form of plagiarism all document pairs to which at least one examiner assigned a suspiciousness score $\mathrm{s} > 0$. If examiners disagreed on the dominant form of perceived plagiarism, we used the expert response. For each document pair to arrive at a single score, we calculated the weighted average of the suspiciousness scores assigned by the examiner groups as

$$\bar{s} = \frac{(\mathrm{s_u} + 1.25\ \mathrm{s_g} + 1.5\ \mathrm{s_e})}{3.75},$$

where $\mathrm{s_u}$ denotes the score assigned by undergraduates, $\mathrm{s_g}$ the score assigned by graduate students, and $\mathrm{s_e}$ the score assigned by medical experts. Finally, to derive a ground truth for the four examined forms of user-perceived plagiarism, we sorted the document pairs in each of the plagiarism categories by decreasing $\bar{\mathrm{s}}$. Then, we selected the top-10 documents with the highest user-assigned suspiciousness scores.

To confirm agreement among participants above the agreement rate to be expected by chance, we calculated inter-rater agreement using Fleiss' Kappa, $\kappa$, as follows

$$\kappa = \frac{\bar{P} - \bar{P}_e}{1 - \bar{P}_e} .$$

In this notation, $\bar{\mathrm{P}}$ represents observed agreement and $\bar{\mathrm{P}}_\mathrm{e}$ denotes the probability of chance agreement. Thus, $\bar{\mathrm{P}} - \bar{\mathrm{P}}_\mathrm{e}$ is the degree of agreement achieved above chance and $1 - \bar{\mathrm{P}}_\mathrm{e}$ the degree of agreement that is attainable above chance. Fleiss' Kappa for all assigned document scores was 0.65, indicating substantial inter-rater agreement on suspiciousness. The agreement was highest for user-perceived copy-and-paste plagiarism, $\kappa = 0.73$, and lowest for perceived structural and idea similarity, $\kappa = 0.59$. This observation matched our expectation of higher discrepancies in judgment for disguised plagiarism forms, which are often more controversial.

## 3.5 Results

We performed three evaluations to assess the utility of the citation-based plagiarism detection approach. The three evaluations addressed:

1. **Retrieval effectiveness** in terms of ranking quality
2. **Computational efficiency** in terms of time complexity
3. **User utility** in terms of subjective user ratings on the effort and objective measurements of the time required to examine retrieved documents

### 3.5.1 Retrieval Effectiveness

Typically, users of a plagiarism detection system can reasonably verify suspicious documents only if they are retrieved at the highest ranks. We, therefore, consider the rank at which a detection method retrieves the top-$n$ relevant results as a crucial measure of the method's effectiveness. We evaluated the effectiveness of the nine detection methods by comparing the methods' ranked results to the ground truth approximation on document suspiciousness derived in the user study for the four forms of user-perceived plagiarism we examined.

For each of the top-10 user-rated document pairs, we selected the more recent publication, i.e., the potentially suspicious document, and checked at which rank, if at all, a detection method identified the recent publication as similar to the earlier publication. If detection methods assigned the same score, and thus the same rank, $i$, to multiple documents, we calculated the mid-rank $\bar{\mathrm{r}}_i$ as

$$\bar{r}_i = \frac{r_{i-1} + (|d_i| - 1)}{2}$$

and assigned $\bar{r}_i$ to all documents $d_i$ with initial rank $i$. We found that the best-performing approach strongly depends on the form of user-perceived plagiarism. The following subsections describe the retrieval effectiveness for the four forms of user-perceived plagiarism we asked the participants to distinguish:

» Lexis-preserving plagiarism, i.e., **copy and paste** or **shake and paste**

» Semantics-preserving plagiarism, i.e., **paraphrases**

» Idea-preserving plagiarism like reusing **structures and ideas**

**Figure 3.14** shows the distribution of ranks for all four forms of user-perceived plagiarism. We omit to show whiskers for the box plots as some maximum values are much larger than the third quartile values. Showing whiskers would make the boxes hard to see. Instead, we show the raw data below each plot.

### Ranking Quality for Instances of Copy and Paste

The rank distribution for user-perceived copy-and-paste plagiarism shows that the text-based detection method Enco performed best at ranking this form of content similarity highly. The citation-based LCCS method performed second best, and the text-based method Sherlock ranked third. The upper quartile of the three best-performing methods equals one. That is, for at least 75% of the examined top-10 document pairs, the methods retrieved the source document at rank one.

Of the top-10 copy-and-paste document pairs, Enco identified all at rank one. LCCS and Sherlock retrieved nine at rank one. The results confirm that current detection methods reliably find documents containing verbatim text overlap. The citation-based methods, especially LCCS, performed better than expected for these document pairs. The reason may have been collection-specific, in that many document pairs with extensive text overlaps also featured long shared citation patterns.

### Ranking Quality for Instances of Shake and Paste

The distribution of ranks for user-perceived shake-and-paste plagiarism shows that Enco identified the suspicious document at rank one for all 10 document pairs. Sherlock and the two LCCS methods each identified nine pairs at rank one. The ranking quality of the other CbPD methods was slightly lower, yet the methods identified the source document for each of the user-classified top-10 document pairs. No third quartile of any citation-based method exceeded rank two.

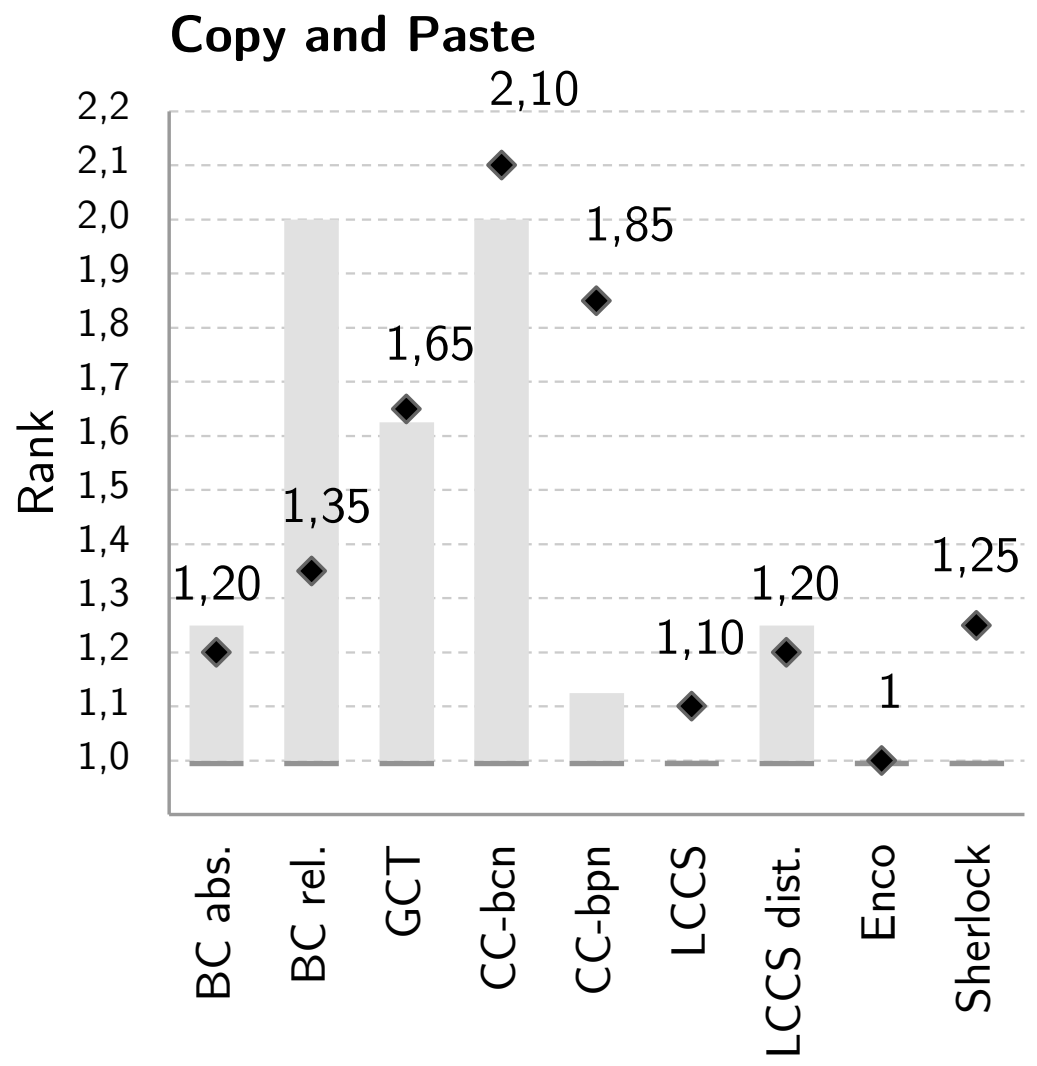

| | BC abs. | BC rel. | GCT | CC-bcn | CC-bpn | LCCS | LCCS dist. | Enco | Sher-lock |
|---|---|---|---|---|---|---|---|---|---|
| **Min** | 1.00 | 1.00 | 1.00 | 1.00 | 1.00 | 1.00 | 1.00 | 1.00 | 1.00 |
| **Q1** | 1.00 | 1.00 | 1.00 | 1.00 | 1.00 | 1.00 | 1.00 | 1.00 | 1.00 |
| **Median** | 1.00 | 1.00 | 1.00 | 1.00 | 1.00 | 1.00 | 1.00 | 1.00 | 1.00 |
| **Q3** | 1.25 | 2.00 | 1.63 | 2.00 | 1.13 | 1.00 | 1.25 | 1.00 | 1.00 |
| **Max** | 2.00 | 2.00 | 2.80 | 9.00 | 9.00 | 2.00 | 2.00 | 1.00 | 3.50 |
| **Mean** | 1.20 | 1.35 | 1.65 | 2.10 | 1.85 | 1.10 | 1.20 | 1.00 | 1.25 |

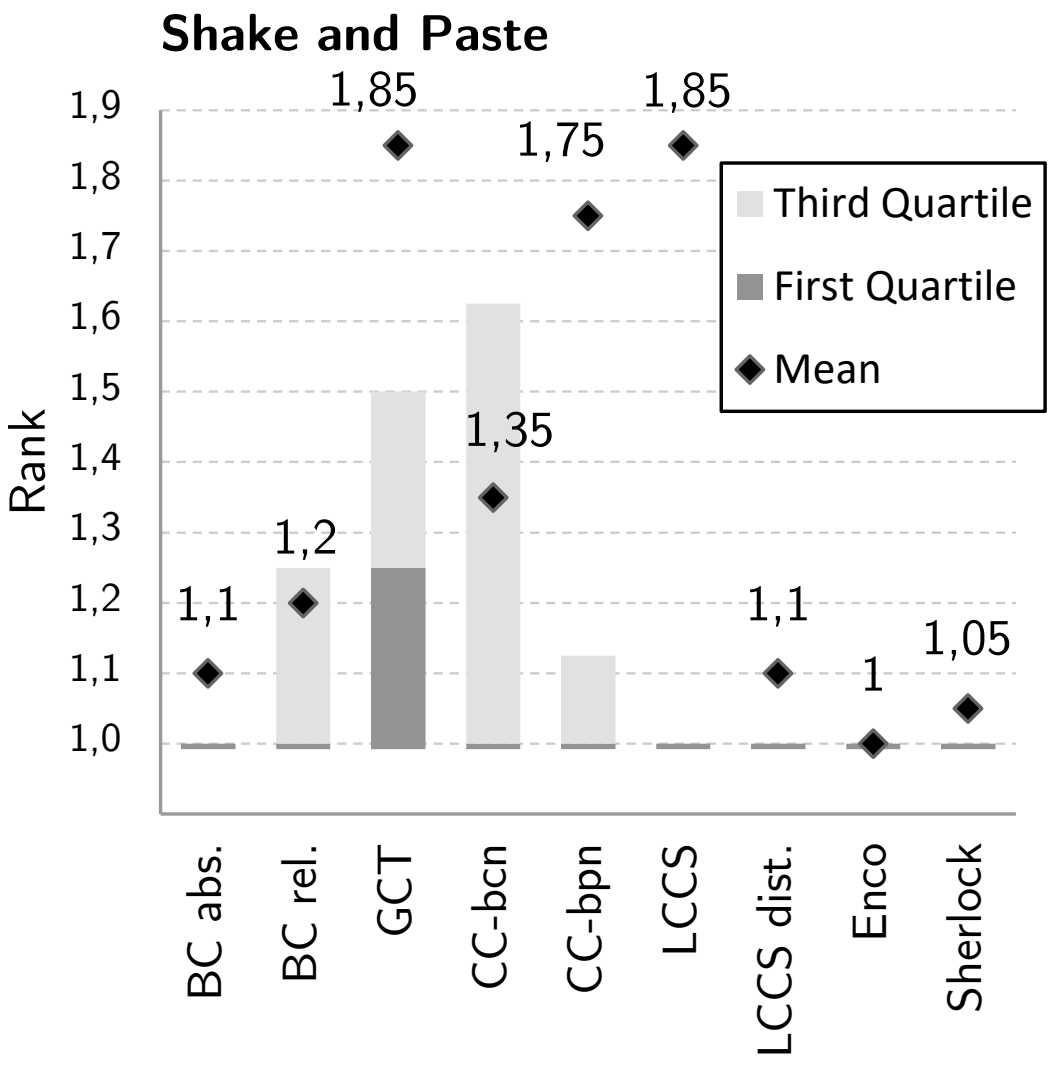

| | BC abs. | BC rel. | GCT | CC-bcn | CC-bpn | LCCS | LCCS dist. | Enco | Sher-lock |
|---|---|---|---|---|---|---|---|---|---|
| **Min** | 1.00 | 1.00 | 1.00 | 1.00 | 1.00 | 1.00 | 1.00 | 1.00 | 1.00 |
| **Q1** | 1.00 | 1.00 | 1.00 | 1.00 | 1.00 | 1.00 | 1.00 | 1.00 | 1.00 |
| **Median** | 1.00 | 1.00 | 1.25 | 1.00 | 1.00 | 1.00 | 1.00 | 1.00 | 1.00 |
| **Q3** | 1.00 | 1.25 | 1.50 | 1.63 | 1.13 | 1.00 | 1.00 | 1.00 | 1.00 |
| **Max** | 2.00 | 2.00 | 7.50 | 2.50 | 8.00 | 9.50 | 2.00 | 1.00 | 1.50 |
| **Mean** | 1.10 | 1.20 | 1.85 | 1.35 | 1.75 | 1.85 | 1.10 | 1.00 | 1.05 |

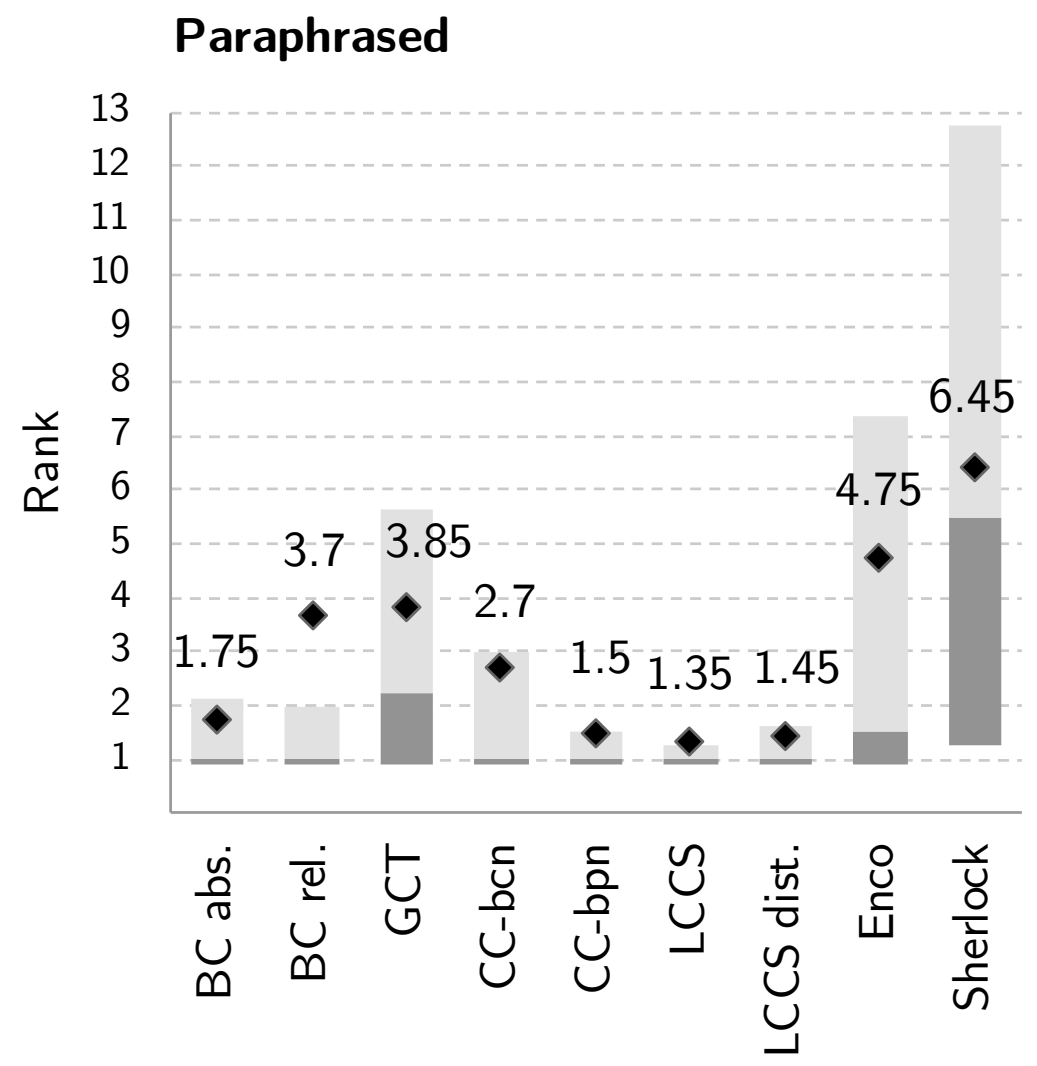

| | BC abs. | BC rel. | GCT | CC-bcn | CC-bpn | LCCS | LCCS dist. | Enco | Sher-lock |
|---|---|---|---|---|---|---|---|---|---|
| **Min** | 1.00 | 1.00 | 1.00 | 1.00 | 1.00 | 1.00 | 1.00 | 1.00 | 1.00 |
| **Q1** | 1.00 | 1.00 | 1.00 | 1.00 | 1.00 | 1.00 | 1.00 | 1.00 | 1.38 |
| **Median** | 1.00 | 1.00 | 2.25 | 1.00 | 1.00 | 1.00 | 1.00 | 1.50 | 5.50 |
| **Q3** | 2.13 | 2.00 | 5.63 | 3.00 | 1.50 | 1.25 | 1.63 | 7.38 | 12.75 |
| **Max** | 6.00 | 26.00 | 13.00 | 14.00 | 4.00 | 3.50 | 4.00 | 18.00 | 14.50 |
| **Mean** | 1.75 | 3.70 | 3.85 | 2.70 | 1.50 | 1.35 | 1.45 | 4.75 | 6.45 |

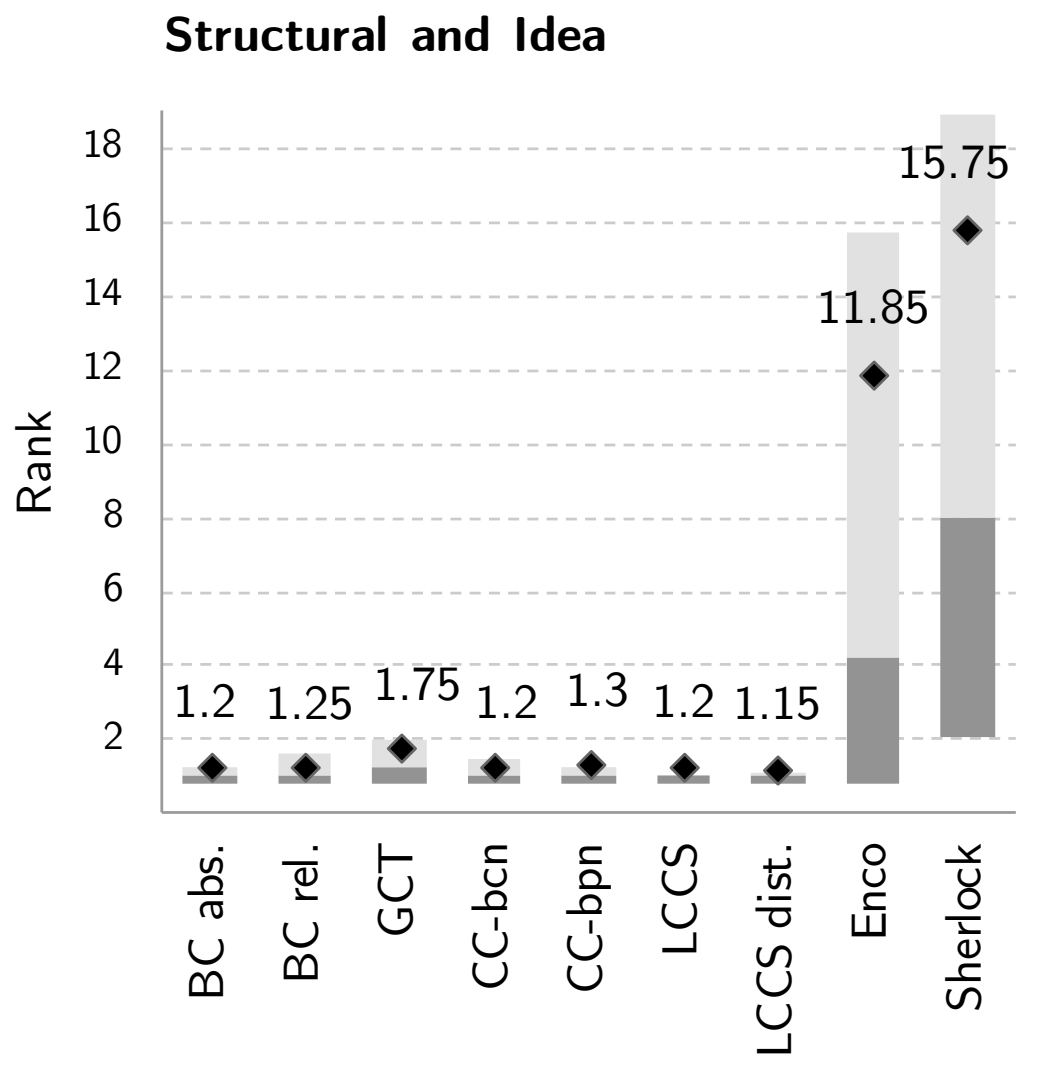

| | BC abs. | BC rel. | GCT | CC-bcn | CC-bpn | LCCS | LCCS dist. | Enco | Sher-lock |
|---|---|---|---|---|---|---|---|---|---|
| **Min** | 1.00 | 1.00 | 1.00 | 1.00 | 1.00 | 1.00 | 1.00 | 1.00 | 1.00 |
| **Q1** | 1.00 | 1.00 | 1.00 | 1.00 | 1.00 | 1.00 | 1.00 | 1.00 | 2.25 |
| **Median** | 1.00 | 1.00 | 1.25 | 1.00 | 1.00 | 1.00 | 1.00 | 4.25 | 8.00 |
| **Q3** | 1.25 | 1.63 | 2.00 | 1.50 | 1.25 | 1.00 | 1.13 | 15.75 | 18.88 |
| **Max** | 2.00 | 2.00 | 5.00 | 2.00 | 3.00 | 3.00 | 2.00 | 57.50 | 79.50 |
| **Mean** | 1.20 | 1.25 | 1.75 | 1.20 | 1.30 | 1.20 | 1.15 | 11.85 | 15.75 |

**Figure 3.14.** Ranking quality for user-perceived plagiarism forms.

The good performance of Enco's exact character 16-gram string matching and Sherlock's probabilistic word-based fingerprinting approach in identifying instances of shake-and-paste similarity was no surprise, given that many of the identified instances have high verbatim text overlap. The citation-based methods performed better than expected, mainly due to most instances of shake-and-paste similarity being concentrated in the introduction and background sections of publications, which also included a high number of shared citations.

### Ranking Quality for Paraphrases

The box plots for both paraphrases and structural and idea similarity show that the CbPD approach outperformed text-based methods in identifying these forms of user-perceived plagiarism. The two best-performing methods for paraphrases, LCCS and LCCS dist., identified eight and seven of the top-10 document pairs at rank one and ranked no document pair below rank four. Enco identified six and Sherlock eight of the document pairs below the top rank of one. The lowest ranks at which the two text-based methods retrieved one of the top-10 document pairs were at rank 18 for Encoplot and at rank 14.5 for Sherlock.

### Ranking Quality for Structural and Idea Similarity

For structural and idea similarity, the advantage of CbPD is even more apparent than for paraphrases. The citation-based methods, especially the variations of LCCS (LCCS and LCCS dist.), significantly outperformed the text-based methods in prominently ranking structural and idea similarity. LCCS identified nine and LCCS dist. eight document pairs at rank one, and the remaining document pairs no lower than rank three. On the other hand, Enco ranked six, and Sherlock nine document pairs at rank four or below.

One can see that the lowest ranks at which Enco and Sherlock retrieved the document pairs were at rank 57.5 for Enco and rank 79.5 for Sherlock. As opposed to the text-based methods, the median for six of the seven citation-based methods is equal to 1 and 1.25 for the seventh method (GCT).

### Detailed Comparison of Ranking Quality

**Figure 3.15** visualizes the ranking distribution for the citation-based and text-based detection methods in detail. The 16 scatter plots compare the two best-performing citation-based methods for each of the four forms of user-perceived plagiarism with the two text-based methods, Encoplot and Sherlock, without aggregating ranks. Non-aggregated ranks distinguish these scatter plots from the box plots in **Figure 3.14**, p. 105. The plots show the rank at which the text-based methods retrieved each of the top-10 document pairs on the horizontal axis. The vertical axis shows the rank for the citation-based methods. Larger dots represent multiple documents retrieved at the same combination of ranks.

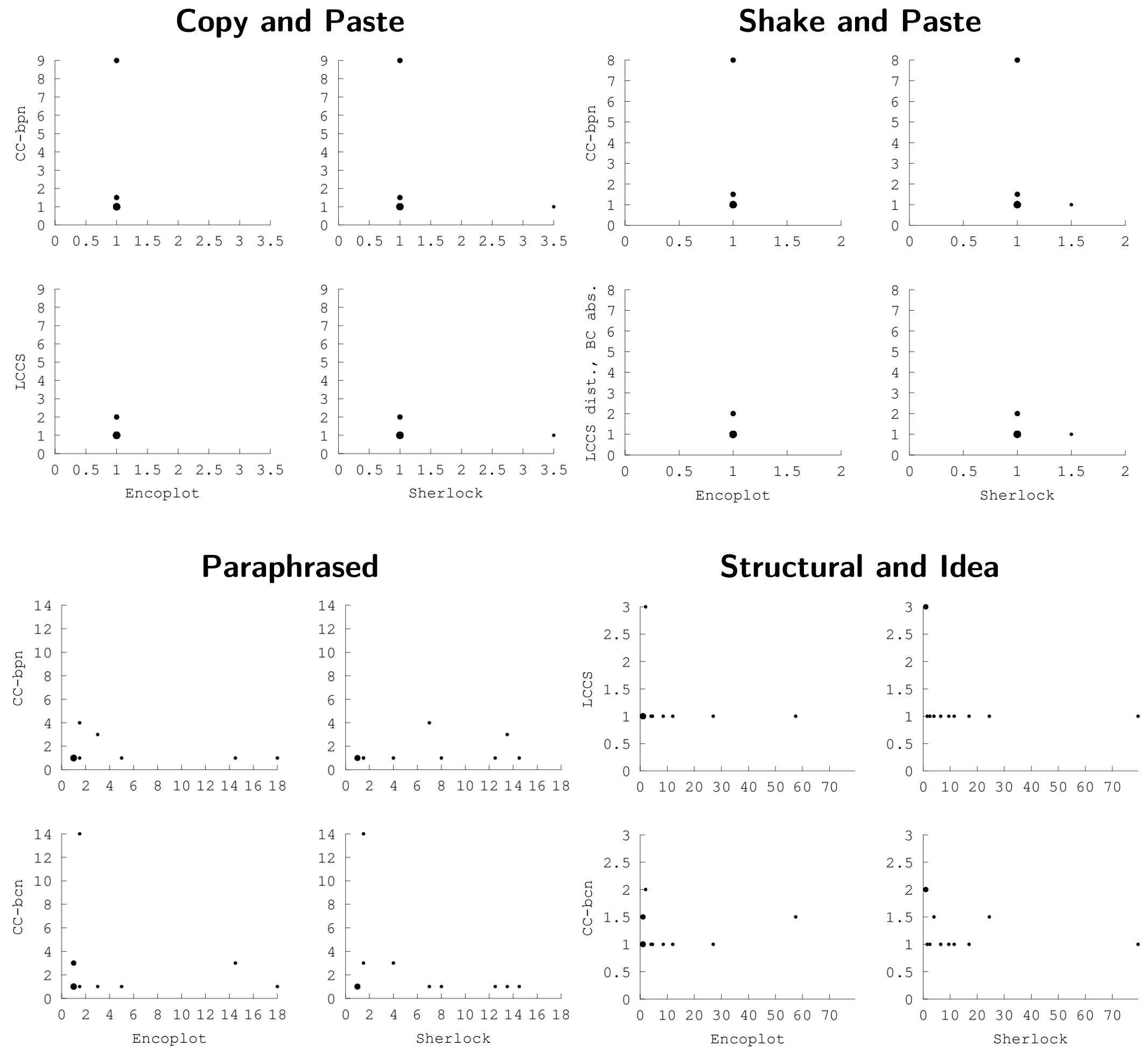

**Figure 3.15.** Ranks for top-10 results by user-perceived plagiarism form.

The scatter plots for instances of user-perceived copy-and-paste and shake-and-paste plagiarism show that the text-based and even the best performing citation-based methods prominently ranked these forms of similar content. Of the 10 document pairs in the copy-and-paste category, Encoplot identified all at rank one, LCCS and Sherlock retrieved nine at rank one. Similarly, Encoplot identified all 10 document pairs in the shake and paste category at rank one, Sherlock and the two LCCS measures identified nine pairs at rank one. The results confirm that current text-based detection methods have no difficulty identifying verbatim text overlap in realistic document collections.

The scatter plots for paraphrases and user-perceived structural and idea plagiarism show that the citation-based methods outperform text-based methods in identifying these forms of similarity, which typically have very little or no notable text overlap. For paraphrases, the citation-based methods CC-bcn and CC-bpn identified seven and eight of the 10 document pairs at rank one and ranked none of the document pairs lower than rank four. Encoplot and Sherlock identified six and eight of the document pairs below the top rank of one. The lowest ranks at which the two text-based methods retrieved one of the top-10 document pairs were at rank 18 for Encoplot and at rank 14.5 for Sherlock.

For user-perceived structural and idea plagiarism, the advantage of the citation-based methods in ranking quality is even more substantial. The citation-based methods CC-bpn and CC-bcn identified eight and seven document pairs at rank one, and the remaining document pairs no lower than rank three. Encoplot and Sherlock ranked six and nine document pairs at rank four or lower ranks. The lowest ranks at which Encoplot and Sherlock retrieved the document pairs were at rank 57.5 for Encoplot and 79.5 for Sherlock.

The scatter plots reflect that text-based and citation-based methods have complementary strengths. While the plots show dots mostly on vertical lines for copy and paste and shake and paste, they show dots mostly on horizontal lines for paraphrases and structural and idea similarity. These results show that text-based methods excel in identifying lexis-preserving forms of user-perceived plagiarism, while the citation-based detection methods more effectively detect semantics-preserving and idea-preserving forms of user-perceived plagiarism.

### 3.5.2 Computational Efficiency

Computational efficiency is crucial to PDS performance because many-to-many comparisons quickly become unfeasible for extensive collections. Text-based methods require a candidate retrieval heuristic if the collection is too large to perform pairwise document comparisons. However, the candidate retrieval step likely causes a loss in detection effectiveness. The CbPD approach retains only those documents in the reference collection, which share at least one reference with the examined document. Thus, we retained only the document pairs with Bibliographic Coupling Strength $s_{\mathrm{BC}} \geq 1$ in the PMC OAS collection for further analysis.

In general, the processing time for automated plagiarism detection consists of two elements—first, the time required for preprocessing, and second, the time required for document comparison. Preprocessing encompasses file system and database operations as well as document type conversions. For example, Enco and Sherlock required converting the XML format of PMC OAS documents to plain text. Preprocessing for citation-based methods includes text parsing to acquire references, determining the position of citations in the text, and extracting document metadata. Extracted data must be cleaned and disambiguated before being stored in the database. Because the restriction $s_{\mathrm{BC}} \geq 1$ limits collection size, we included the time required for computing Bibliographic Coupling Strength to the preprocessing time of citation-based detection methods.

Text-based detection methods require $O(n)$ time for preprocessing, as $n$ documents must be converted from XML to plain text. The citation-based methods also require $O(n)$ time for converting and parsing documents and for cleaning and disambiguating the parsed data. The additional calculation of $s_{\mathrm{BC}}$ requires $O(n \cdot log(n))$ time when using an index that allows comparing the references in documents in $O(log(n))$ time. All citation-based methods had similar overall runtimes. We, therefore, summarized all seven citation-based detection methods under the label "CbPD" and examined their mean processing time.

**Figure 3.16** plots both measured and extrapolated average case processing times for Enco, Sherlock, CbPD1, and CbPD5, where CbPDn stands for any citation-based method using an $s_{\mathrm{BC}}$ threshold $\geq n$. Processing time in hours is plotted on a $\log_{10}$ scale and assumes a 3.40 GHz quad-core processor with 16GB RAM. Shaded columns depict the size ranges of well-known, large-scale collections, namely PMC OAS, PubMed, and Google Scholar. The figure shows that for processing the PMC OAS, the CbPD5 algorithm required 14.7 hours, while Sherlock (without candidate retrieval) would require an estimated 140 years. For Enco and Sherlock, we measured processing times for sample sizes 10, 100, and 1,000 and extrapolated the processing times for the larger collection sizes with unfeasible runtime requirements. For the CbPD methods, we calculated processing times up to the size of the PMC OAS and extrapolated the times for more extensive collections.

The efficiency of the detection methods depends heavily on collection size. For analyzing a single document pair in a one-to-one comparison, the text-based methods are comparatively less expensive than the citation-based methods. The reason is that for smaller collections, citation parsing is computationally more intensive than a text-based many-to-many comparison. However, the break-even point is commonly reached at around five documents, depending on document length and number of citations. For larger collections, text-based methods are typically more expensive, given that they require $\binom{n}{2}$ comparisons or heuristic candidate retrieval.

In summary, the superior computational efficiency of the citation-based approach is advantageous, especially for extensive document collections, which remain analyzable in a many-to-many fashion without a loss in detection accuracy.

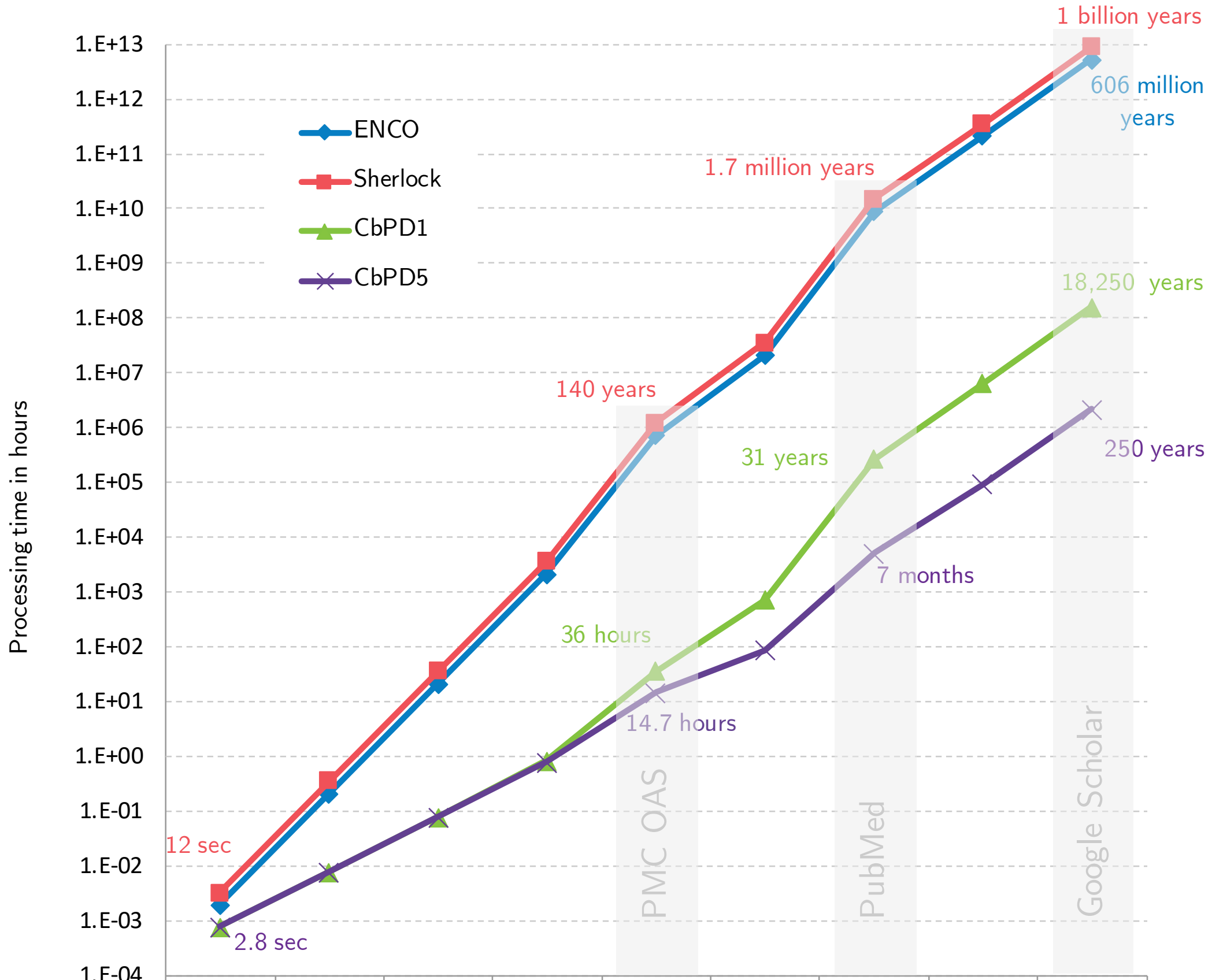

| | 10 | 100 | 1000 | 10000 | 185170 | 1.00E+06 | 2.05E+07 | 1.00E+08 | 5.00E+08 |
|---|---|---|---|---|---|---|---|---|---|
| ENCO | 1.9E-03 | 2.1E-01 | 2.1E+01 | 2.1E+03 | 7.3E+05 | 2.1E+07 | 8.9E+09 | 2.1E+11 | 5.3E+12 |
| Sherlock | 3.3E-03 | 3.6E-01 | 3.6E+01 | 3.6E+03 | 1.2E+06 | 3.6E+07 | 1.5E+10 | 3.6E+11 | 9.0E+12 |
| CbPD1 | 7.8E-04 | 7.8E-03 | 7.8E-02 | 8.4E-01 | 3.6E+01 | 7.2E+02 | 2.7E+05 | 6.4E+06 | 1.6E+08 |
| CbPD5 | 7.8E-04 | 7.8E-03 | 7.8E-02 | 7.8E-01 | 1.5E+01 | 8.6E+01 | 5.2E+03 | 9.3E+04 | 2.2E+06 |

The first row lists the collection size. The rows underneath show processing times in hours (partially extrapolated).

**Figure 3.16.** Processing times of detection methods by collection size.

## 3.5.3 User Utility

An effective plagiarism detection approach maximizes user utility by addressing the user's information need and minimizing user effort. We assessed utility by questioning the 26 participants on the similarity visualization method—text-based, citation-based, or hybrid—they found most suitable for examining the various forms of user-perceived plagiarism. We additionally examined if a reduction in user effort is attainable if the citation-based approach is combined with the strictly text-based similarity visualization of current plagiarism detection systems.

**Table 3.5** shows the visualization that participants indicated as most suitable, depending on the dominating form of user-perceived plagiarism. We collected these responses for the 461 document pairs that all three examiner groups judged.

**Table 3.5.** User-perceived suitability of document similarity visualizations.

| | Copy and Paste | Shake and Paste | Para-phrase | Structural and Idea | Transla-tion* |
|---|---|---|---|---|---|
| **Text-based** | 51% | 27% | 6% | 1% | - |
| **Citation-based** | 1% | 5% | 32% | 86% | 54% |
| **Hybrid** | 47% | 68% | 62% | 13% | 46% |

* examination of Guttenberg thesis only

Most participants indicated traditional text highlights as the single most suitable similarity visualization method to assist in document verification for user-perceived copy-and-paste plagiarism. For the heavily disguised structural and idea similarity, most participants rated the citation-based approach as the most effective visualization method. The participants perceived a hybrid approach combining text and citation pattern visualization as most suitable for detecting paraphrases and user-perceived shake-and-paste plagiarism.

Because the PMC OAS contains English publications only, we additionally asked 13 volunteers (out of 26 participants) to indicate the suitability of document similarity visualizations for an excerpt from the Guttenberg thesis [198]. We selected a passage that we had analyzed in our preliminary investigation of CbPD (cf. Section 3.2.1, p. 83) and for which the responsible university had confirmed the presence of translated plagiarism. All participants preferred the availability of a citation-based similarity visualization over exclusively visualizing textual similarity by highlighting literal text matches. Approximately half of the participants responded that visualizing similar citation patterns in addition to matching text was most helpful for them. The other half found that the citation-based visualization alone was most beneficial. However, with opinions on translated plagiarism collected only for a single plagiarism case, these results cannot be generalized.

In the subsequent evaluation, we examined if a reduction in user effort, measured as a time saving, is observable upon citation pattern visualization. We recruited eight participants to judge document suspiciousness—once with text similarity visualized and once with both text and citation pattern similarity visualized. We recorded the time examiners required to verify the first two instances they deemed plagiarism for each visualization option.

Each participant rated 25 document pairs, six pairs in each of the four user-perceived plagiarism categories, and a single document to represent translated plagiarism, i.e., an excerpt from the Guttenberg thesis. The six documents for each of the four user-perceived plagiarism forms were a random sample of the top-30 documents yielded by the pooling approach.

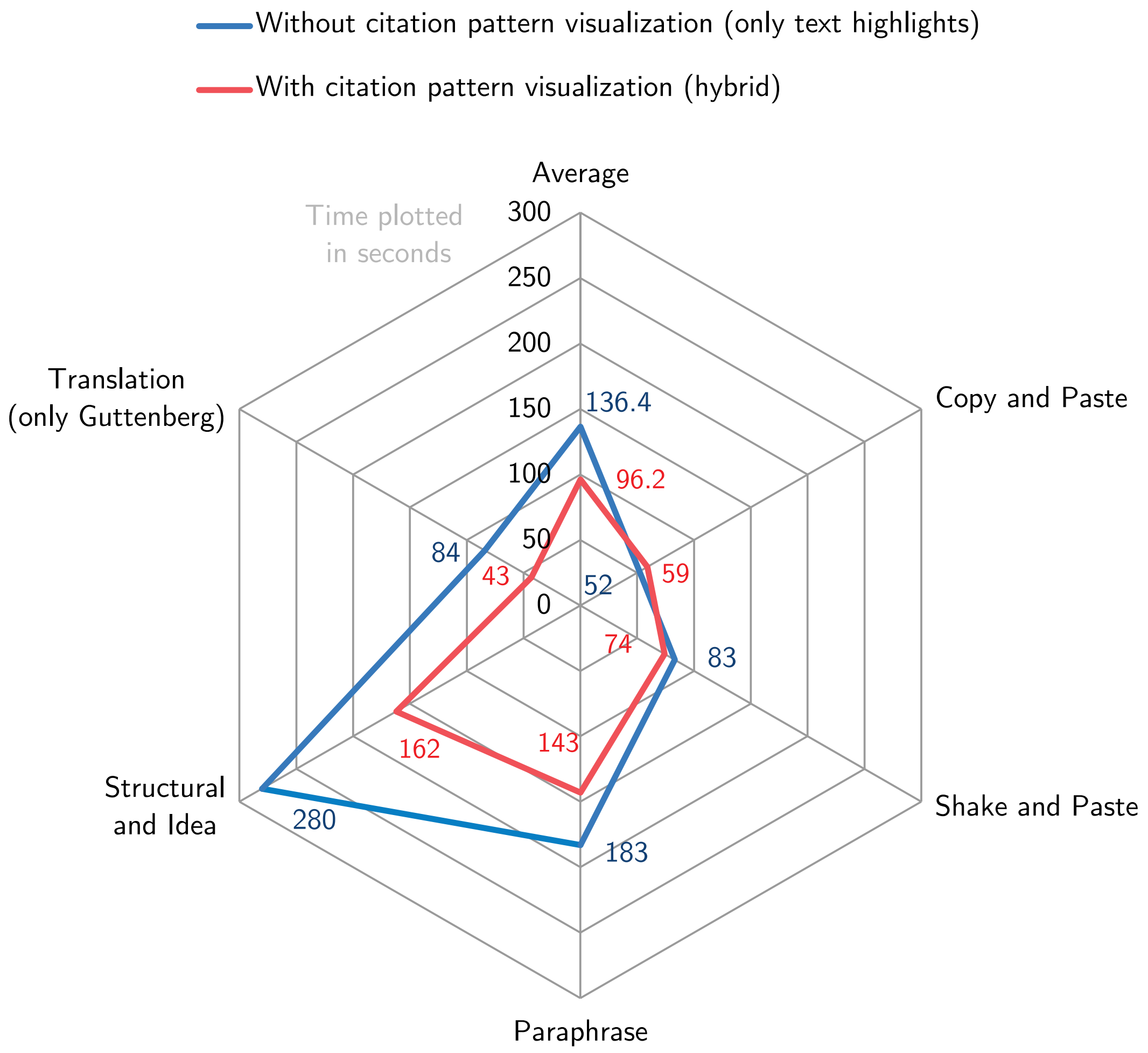

**Figure 3.17.** Mean times in seconds required for document verification.

**Figure 3.17** shows the mean times in seconds recorded for document verification with and without citation pattern visualization. We observed a notable reduction in the mean times required to identify suspicious similarity upon citation pattern visualization for the heavily disguised structural and idea similarity, with a 42.1% time reduction, followed by paraphrases, with a 21.8% reduction, and instances of user-perceived shake-and-paste plagiarism, with a 10.8% reduction. We also observed a lower user effort to verify translated plagiarism; however, this data should not be generalized, given that it represents only a single examined case.

The recorded time savings were in line with the user-classified suitability of the similarity visualizations. The citation pattern visualization was most helpful for verifying structural and idea similarity. For user-perceived plagiarism forms with high textual similarity, such as copy and paste, citation pattern visualization interestingly increased the time required for an examination compared to a text-only visualization. We suspect that some examiners clicked through sections with high citation pattern similarity more thoroughly and thus took longer to submit the first two instances of user-perceived plagiarism.

### Limitations of the Evaluation

General challenges faced when evaluating the performance of plagiarism detection methods using non-artificially created test collections are the lack of ground truth data and the subjectivity of human judgment. We addressed the first challenge by establishing a ground truth approximation for a pooled set of documents. We addressed the second challenge as best as possible by providing uniform definitions and guidelines to participants.

Although the citation-based detection methods considered citations and references to sources cited outside of the PMC OAS, the restricted access to full texts allowed searching for plagiarism only if the similar document was included in the PMC OAS. We assumed the PMC OAS collection to exhibit a relatively low frequency of plagiarism for two reasons. First, the journals' peer review process, which typically employs text-based plagiarism detection systems, likely prevented a substantial share of plagiarized manuscripts from being published. Second, prior studies performed text-based examinations of the published articles in the PMC OAS collection, e.g., the study of the Harold Garner Lab [497]. Because non-disguised plagiarism is more likely to be detected and removed, the results obtained from the PMC OAS may not be representative of other collections.

## 3.6 Conclusion Citation-based PD

With Citation-based Plagiarism Detection, we proposed the first approach that analyzes non-textual content in academic documents. Prior research has shown that academic citations contain much semantic information on the subject matter of academic documents, signify the documents' relations to other documents, and indicate the impact of the research a document presents. Moreover, citation styles impose a language-independent representation of citations in academic documents.

The idea of the CbPD approach is that these properties make academic citations valuable for identifying disguised plagiarism. First, because substituting citations requires significant domain expertise and thus causes additional effort for plagiarists who attempt to change citations copied from a source. Second, because deleting citations is often difficult, particularly for citations to seminal works, as not citing relevant related work would make a document immediately suspicious for a domain expert. Third, because the language-independent representation of citations qualifies them as features for cross-language plagiarism detection.

Manually investigating confirmed cases of academic plagiarism documented by the GuttenPlag, VroniPlag, and Retraction Watch projects supported our hypothesis. We found that academic plagiarists often alter text copied from another source but rarely alter citations included in the copied text.

Given the observations in our investigation of confirmed plagiarism cases, we devised four groups of citation-based detection methods to identify suspicious citation patterns. Three groups, i.e., Bibliographic Coupling, Longest Common Citation Sequence, and Greedy Citation Tiling, adapt successful citation and sequence analysis algorithms to the citation-based plagiarism detection use case. Moreover, we devised Citation Chunking as a new, use-case-specific class of algorithms.

The four groups include 19 variants of citation-based detection methods and complement each other for two reasons. First, the methods analyze either the local or global similarity of documents. Second, the methods focus on different citation pattern types that are characteristic of specific forms of academic plagiarism. By combining the four groups of detection methods, we cover a wide range of academic plagiarism forms, including semantics-preserving and idea-preserving plagiarism.

We demonstrated the effectiveness and practicability of Citation-based Plagiarism Detection by applying the approach to confirmed cases of cross-language plagiarism in the thesis of K. T. zu Guttenberg. The citation-based detection methods could identify 13 of 16 instances of cross-language plagiarism in the thesis, while text-based detection methods identified none of the instances.

To quantify the effectiveness and utility of Citation-based Plagiarism Detection, we evaluated the approach on a realistic, large-scale collection of scientific documents containing various degrees of plagiarism disguise. Our evaluation dataset, which we derived from the PubMed Central Open Access Subset, contained 185,170 publications. We evaluated the seven CbPD methods that performed best for this dataset and two popular text-based methods using human judgment and a top-$n$ results pooling procedure. We compared the ranks at which each detection method identified the top-30 suspicious document pairs for each user-perceived plagiarism form. Our findings showed that the citation-based detection methods significantly outperformed the text-based methods in retrieving among the top ranks documents containing paraphrases and structural and idea similarity. The text-based methods ranked the top results as judged by humans highest for user-perceived copy-and-paste and shake-and-paste plagiarism.

For user-perceived cases of plagiarism, we contacted the authors of the earlier published article. So far, three plagiarized medical studies have been retracted by the issuing journal, and six further publications were confirmed to contain plagiarism by the earlier authors. Without CbPD, several of these previously unidentified cases in the PMC OAS collection would have remained undetected as current text-based detection methods could not retrieve them.

For the PMC OAS collection, we approximated the advantage of Citation-based Plagiarism Detection over text-based methods in terms of processing time to be on the order of $3.6 \times 10^4$. In other terms, the citation-based methods required 14.7 hours for processing the PMC OAS collection, while the text-based methods would have required ~140 years.

The visualization of similar citation patterns significantly reduced user effort measured as the time examiners save. The positive effect was especially noticeable for semantics-preserving and idea-preserving forms of academic plagiarism.

In summary, we found that citation-based and text-based plagiarism detection methods possess complementary strengths. Text-based methods achieve high effectiveness for identifying even short instances of lexis-preserving plagiarism. On the other hand, citation-based methods were significantly more effective in detecting semantics-preserving and idea-preserving forms of user-perceived plagiarism that typically lack significant textual overlap. A hybrid approach that analyzed and visualized both text-based and citation-based document similarity achieved the highest detection effectiveness and utility rating by participants of our user study.

The data and code of our experiments are available at http://thesis.meuschke.org.

4

Chapter 4

# Image-based Plagiarism Detection

## Contents

This chapter presents research on analyzing similar images to support the detection of academic plagiarism. Our use-case-specific definition of images includes:

» Natural images, i.e., photographs and photo-realistic renderings;

» Data visualizations, e.g., bar charts, scatter plots, or graphs;

» Conceptual visualizations, e.g., flow charts, organigrams, and diagrams.

Like citations, images convey much information in a compact form independently from the text. Therefore images are promising features to analyze for identifying semantics-preserving and idea-preserving plagiarism. Even the plagiarism of data can be detectable if one can reconstruct the data values, e.g., from graphs.

This chapter is structured as follows. Section 4.1 presents related work on Image-based Plagiarism Detection to point out the research gap we address. Section 4.2 describes typical forms of image similarity we observed in practice. Building upon our observations, Section 4.3 derives the functional requirements on image-based plagiarism detection methods. Section 4.4 explains the image-based detection process we devised to meet these requirements. Section 4.5 presents the evaluation of our detection process using real cases of image reuse from the VroniPlag collection. Section 4.6 concludes the chapter by summarizing our research contributions.

## 4.1 Related Work and Research Gap

As we present in Section 2.4.6, p.41, researchers addressing Image-based Plagiarism Detection focused on specific image types, primarily natural images (e.g., Hurtik & Hodakova [232], Iwanowski et al. [240], Srivastava et al. [479]) or different chart types (e.g., Al-Dabbagh et al. [11], Rabiu & Salim [422], Arrish et al. [32]).

The detection methods proposed for natural images reliably retrieve exact and cropped image copies as well as images that underwent affine transformations, such as scaling or rotation. The methods focused on achieving good results even if the photo quality is reduced or modified, e.g., by blurring. For this purpose, the methods predominantly employ feature point methods and perceptual hashing.

Feature point methods identify and match visually interesting areas of a scene. The methods are insensitive to affine image transformations and relatively robust to changes in illumination and the introduction of noise [516, Ch. 1–3].

Perceptual hashing describes a set of methods that map the perceived content of images, videos, or audio files to a fixed-size value (pHash) [205], [449]. Images perceived as similar by humans also result in similar pHash values, in contrast to cryptographic hashing, where a minor change in the input results in a drastically different hash value. Thus, one can quantify the similarity of images as the similarity of their pHash values. If image components, such as shapes, are re-arranged, both feature point methods and perceptual hashing often fail.

Iwanowski et al. [240] found that the effectiveness of the feature point methods they tested decreases if the test images consist of multiple sub-images. We also observed this limitation. For example, the two images shown in **Figure 4.1** consist of six and four sub-images, respectively. The image in the later document omits two of the sub-images present in the image in the earlier document. Applying the combination of the SIFT feature extractor [312] and MSAC feature estimator [505] to compare these two compound images correctly identifies a high similarity between the two sub-images at the top in both compound images. However, the method cannot establish a similarity for the other sub-image pairs. Decomposing the compound image into sub-images and applying near-duplicate detection methods, such as perceptual hashing, could solve the problem.

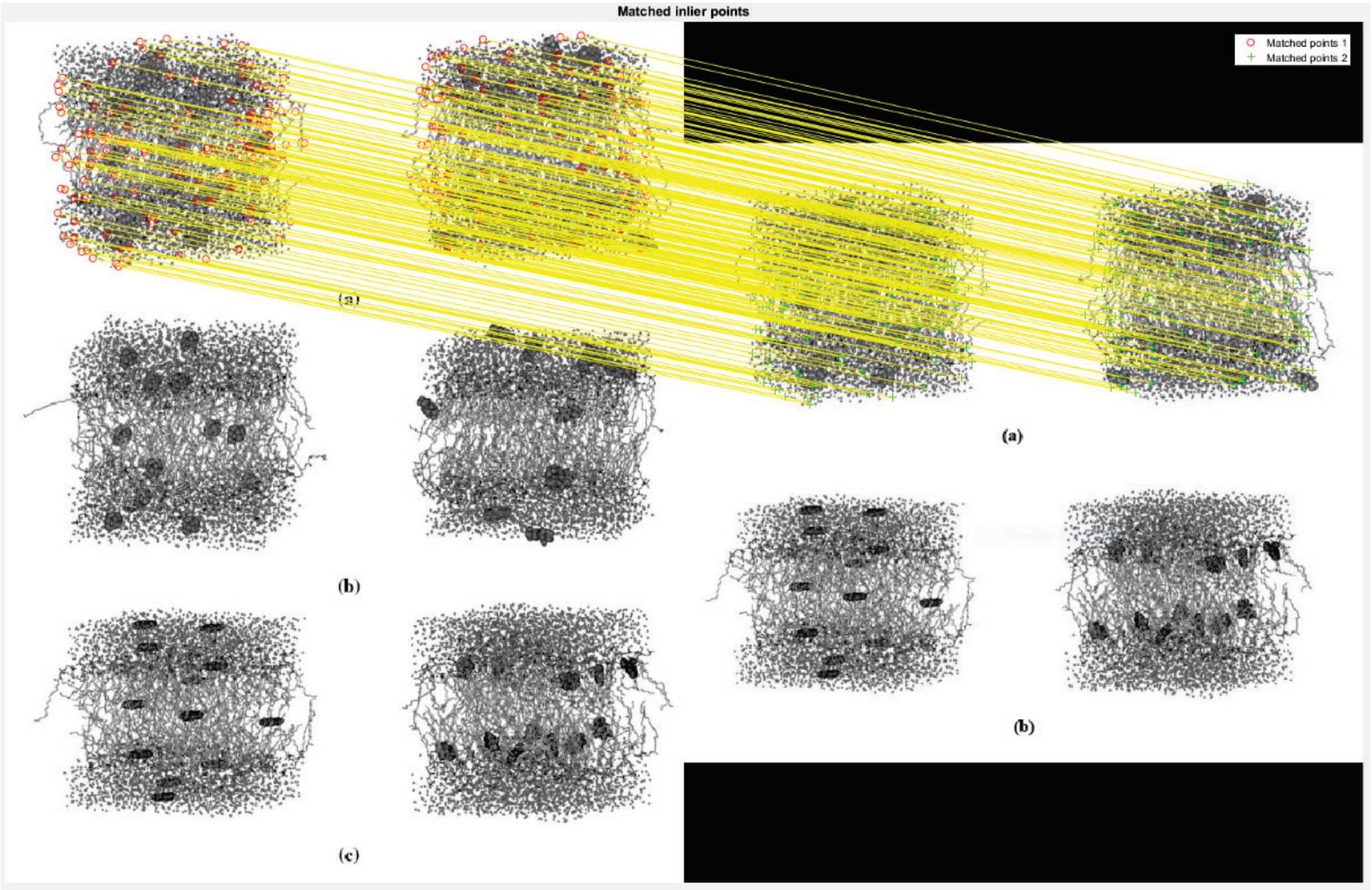

**Figure 4.1.** Comparison of compound images using SIFT+MSAC.

Furthermore, feature point methods and perceptual hashing typically fail to establish meaningful similarities for images primarily containing text, e.g., tables inserted as images. The methods frequently match the feature points for individual letters to multiple letters occurring at different positions in the comparison document, which prevents identifying meaningful clusters of matching features.

Detection methods specializing in specific chart types addressed some of those problems. For example, Rabiu & Salim analyzed the textual and structural similarity of diagrams [422]. To compute the textual similarity of diagrams, the authors applied word unigram text matching to diagram components. They classified all components having a Jaccard similarity of word unigrams above 0.5 as matching. To quantify the structural similarity of the diagrams, they computed the graph edit distance for all components considered a match.

Arrish et al. proposed a method to compute the similarity of flowcharts despite potentially missing or re-arranged shapes. It limits the shape types considered for analysis to four basic geometric forms. The method constructs a four-dimensional vector space model using the occurrence frequencies of the shape types and uses the cosine similarity to compute the similarity of flow charts.

In summary, prior research on Image-based Plagiarism Detection proposed methods that reliably retrieve exact and cropped image copies and images that underwent affine transformations. These methods focus on photographs and specific chart types, for which they achieve good results even if the image quality is low or the image underwent moderate modifications, e.g., by re-arranging shapes.

Our research aimed to offer an efficient, practice-oriented detection process for identifying potentially suspicious images in academic documents. Therefore, we could not limit our analysis to one image type, as most related works did, but sought to enable the analysis of several image types typically found in academic documents. Moreover, the detection process should be able to handle the challenges arising in a realistic detection scenario. To define the requirements for such a detection process, we examined alleged and confirmed cases of plagiarism.

## 4.2 Types of Image Similarity

We used the VroniPlag collection (cf. Section 2.5.1, p. 55) as a source for real cases of similar images in academic documents. Using a targeted web crawler, we retrieved all pages of the VroniPlag wiki that document fragments involving the reuse of similar images. Hereafter, we describe and classify the types of image similarities observed during our review of these fragments.

### 4.2.1 Exact Copies

We define images as exact copies if they have identical dimensions and pixel values. This type of similarity is scarce because authors who reuse images usually cannot access the original image file. We found no exact copies of images in our investigation. Changes due to cropping and alterations that authors inadvertently introduce when reusing images from Portable Document Format (PDF) or print versions of the source document are the main reasons why exact copies are scarce. Copying digital images from a PDF document will typically re-compress the images, resulting in re-arranged pixels and the loss of information.

### 4.2.2 Near-Duplicate Images

We classify images as near-duplicates if they share most of their visual content yet exhibit minor differences introduced by:

1. Removing non-essential content (e.g., numeric labels or watermarks)
2. Cropping or padding
3. Performing affine transformations (e.g., scaling or rotation)
4. Changing the resolution, contrast, or color space.

Especially changes of the categories three and four can be introduced inadvertently by extracting and reusing images from a PDF or a printed document. We frequently found near-duplicate images in our investigation. **Figure 4.2** shows a representative example. The author reused an illustration of a kidney [246] from the Wikipedia article on the kidney [555] without attribution. Some lines that connect labels to points in the illustration are missing in the reused image.

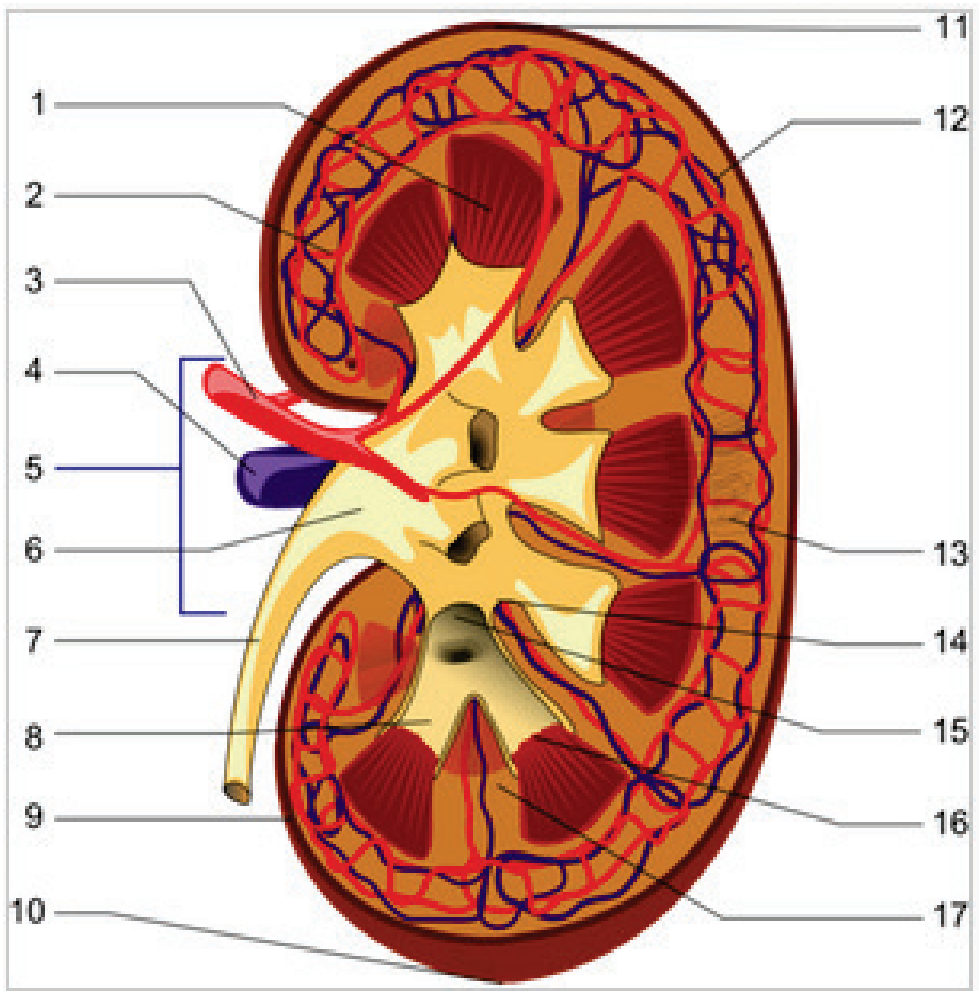

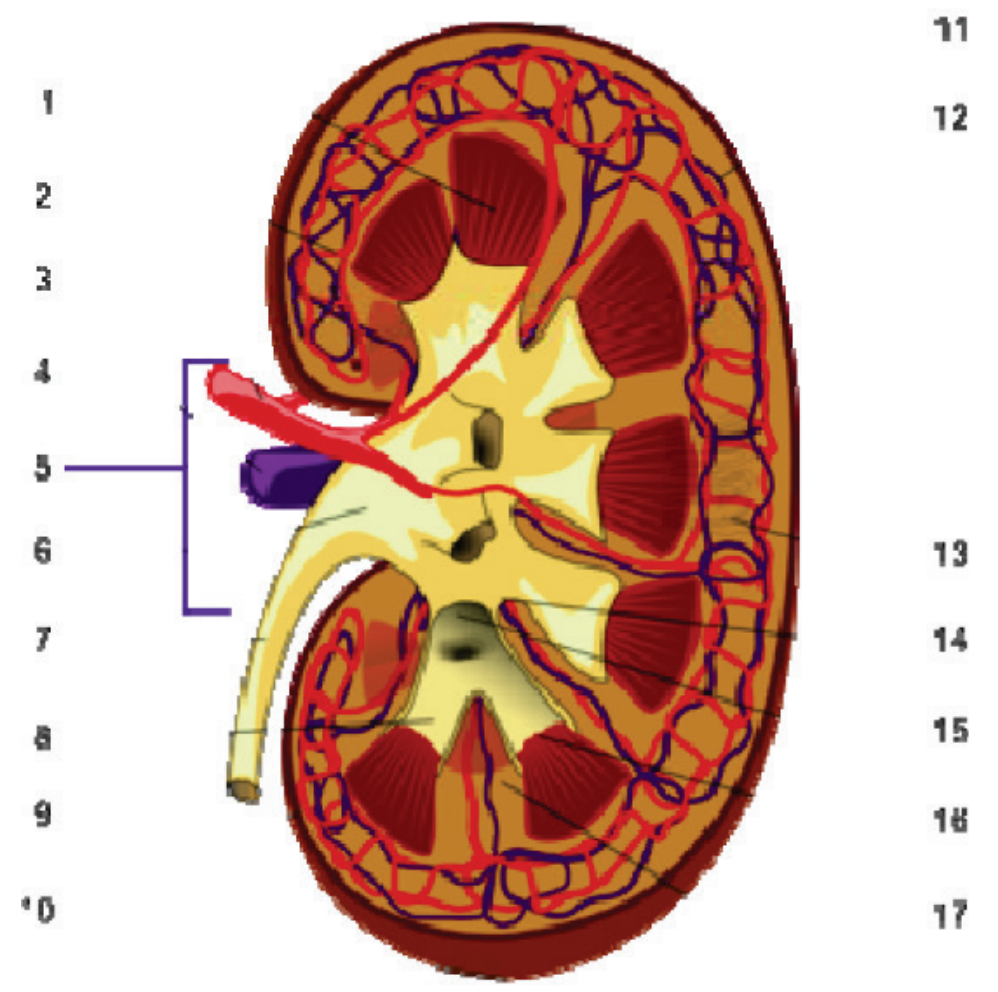

| Original image: [246] | Reused image: [440, p. 14] |
|---|---|
| VroniPlag report on the finding: [531] | |

**Figure 4.2.** Example of a near-duplicate image.

After investigating the case [571], the responsible university allowed the author to publish an erratum to the original thesis. The erratum cites the original image as the source of the reused image shown in **Figure 4.2** and lists 55 additional citations for content used without attribution in the original thesis [441].

### 4.2.3 Altered Images

We refer to reused images exhibiting differences that required purposeful actions to change the image as altered images. The reuse of altered images is hard to classify conclusively, given the virtually infinite possibilities for modifying an image. Given our observations, we distinguish three broad categories of altered images.

1. **Weakly altered images** typically reuse parts of an original image as near copies. **Figure 4.3** shows an example in which an author reused sub-images of a compound image. The responsible university confirmed substantial plagiarism in the doctoral dissertation that contains the reused image [512, p. 3 (TOP 6)] and rescinded the respective doctorate [513, p. 2 (TOP 3)].
2. **Moderately altered images** typically reuse most or all visual components of the original image yet re-arrange or re-draw the components. **Figure 4.4** shows a typical example. The author reused the shapes and composition of the original image. The responsible university allowed the author to publish a corrected version of the thesis that cites the sources of the reused image and other previously unattributed content [532].
3. **Strongly altered images** are typically completely re-drawn versions of the original image with significant changes made to image components' arrangement or visual appearance. **Figure 4.5** shows an example of this type of alteration. The two technical drawings show construction plans with identical dimensions, yet the arrangement of the sub-images and the placement of the labels and measurements differ. Because the author resigned his doctorate, the responsible university closed its investigation into the plagiarism allegations without issuing an official decision [71].

<table>
<tr><td>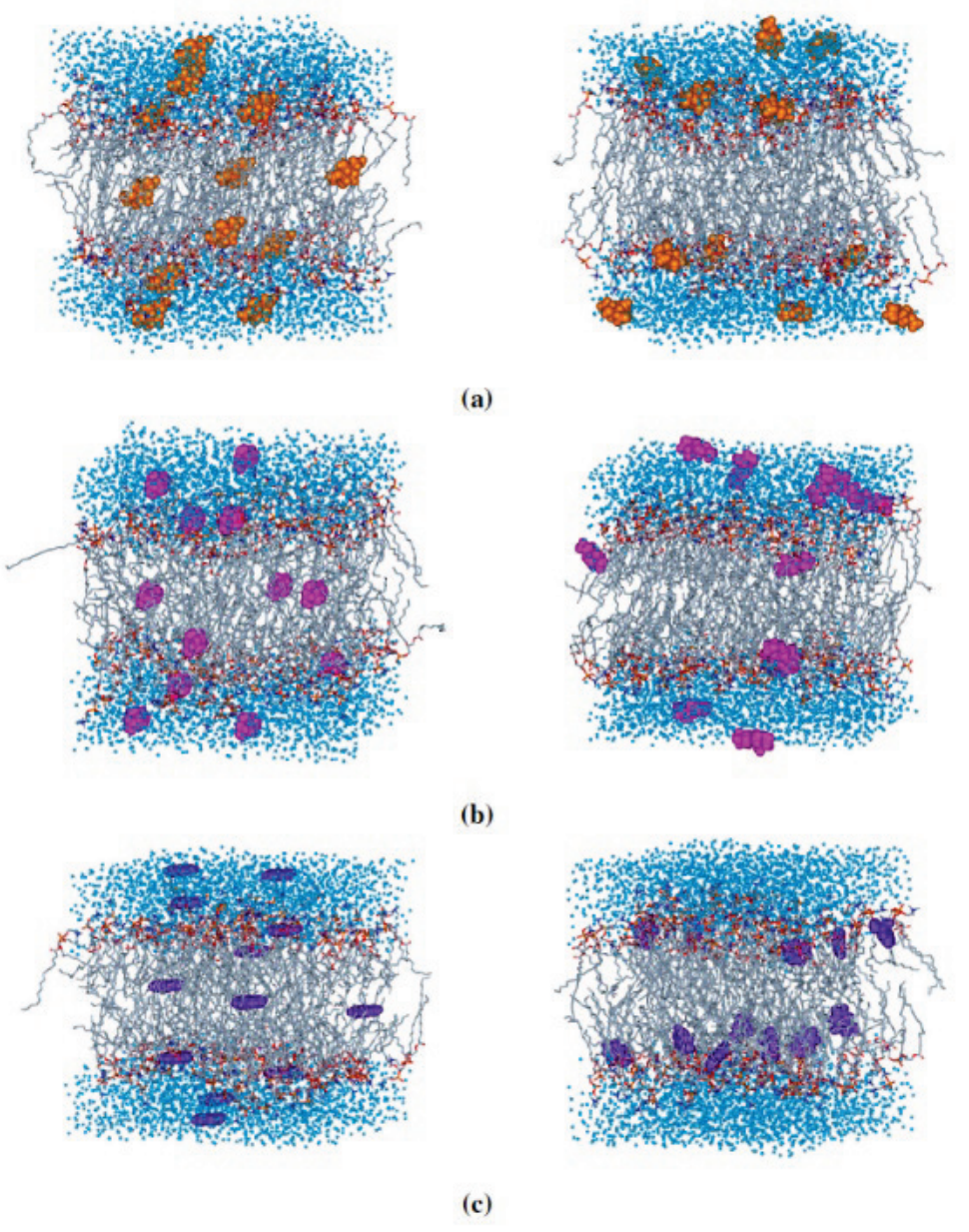
</td><td>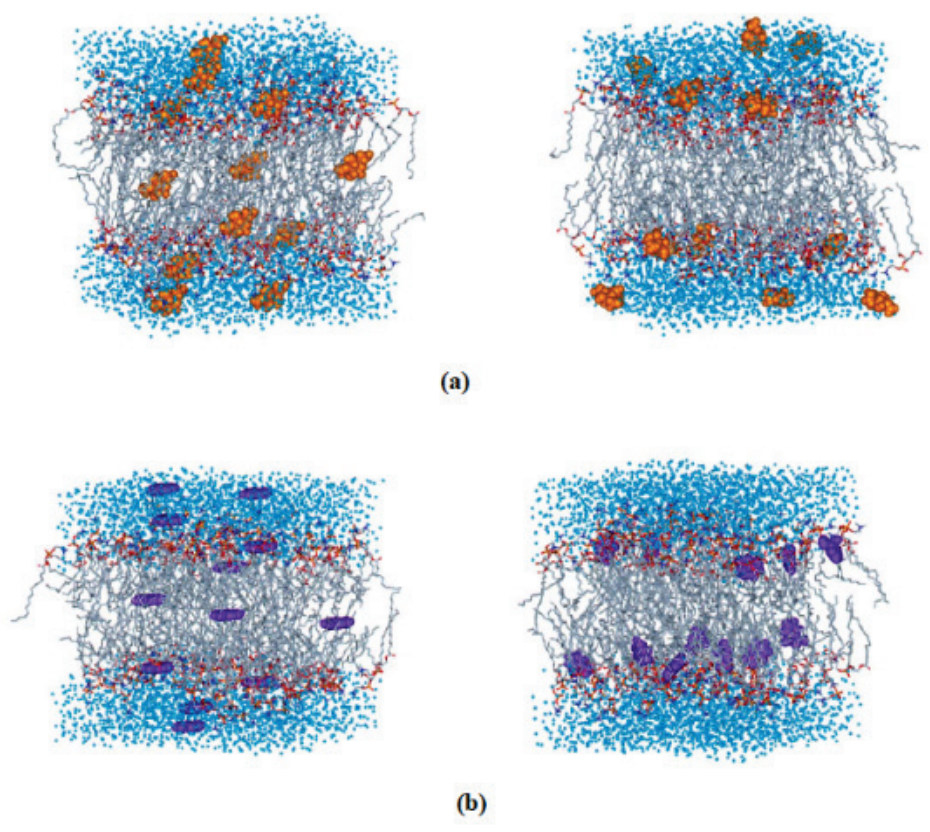
</td></tr>
<tr><td>Original image: [30, p. 160]</td><td>Reused image: [567, p. 73]</td></tr>
<tr><td colspan="2">VroniPlag report on the finding: [537]</td></tr>
</table>

**Figure 4.3.** Example of a weakly altered image.

<table>
<tr><td>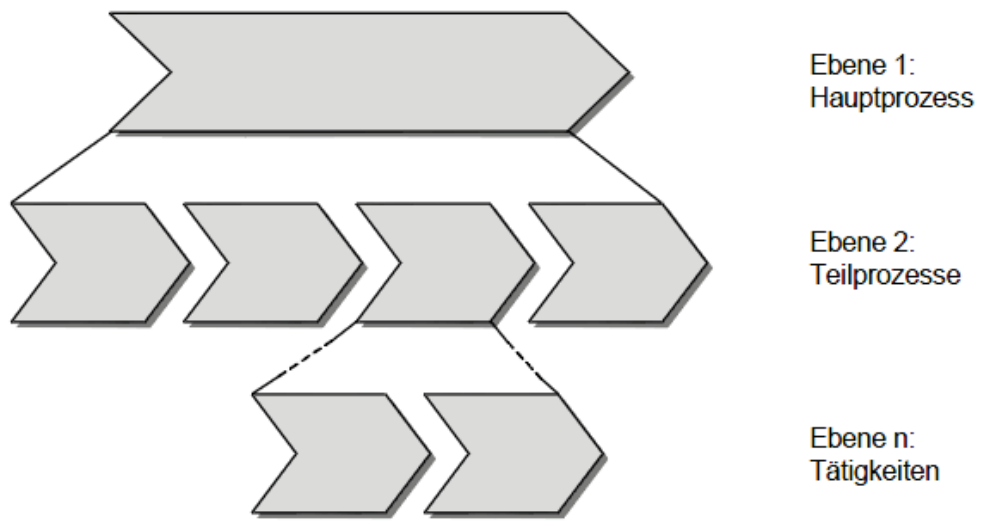
</td><td>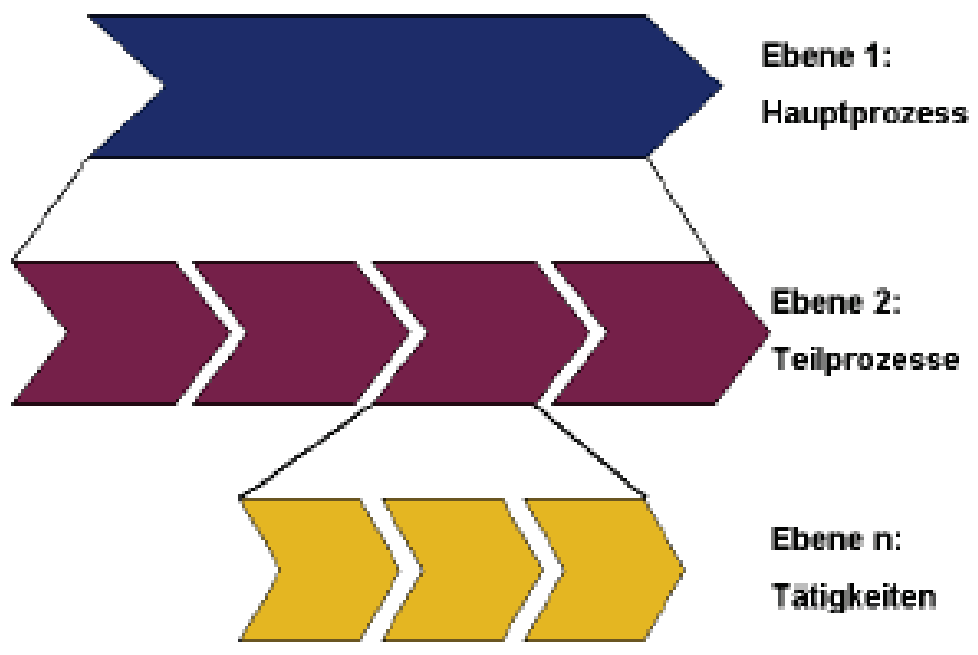
</td></tr>
<tr><td>Original image: [163, p. 23]</td><td>Reused image: [45, p. 14]</td></tr>
<tr><td colspan="2">VroniPlag report on the finding: [520]</td></tr>
</table>

**Figure 4.4.** Example of a moderately altered image.

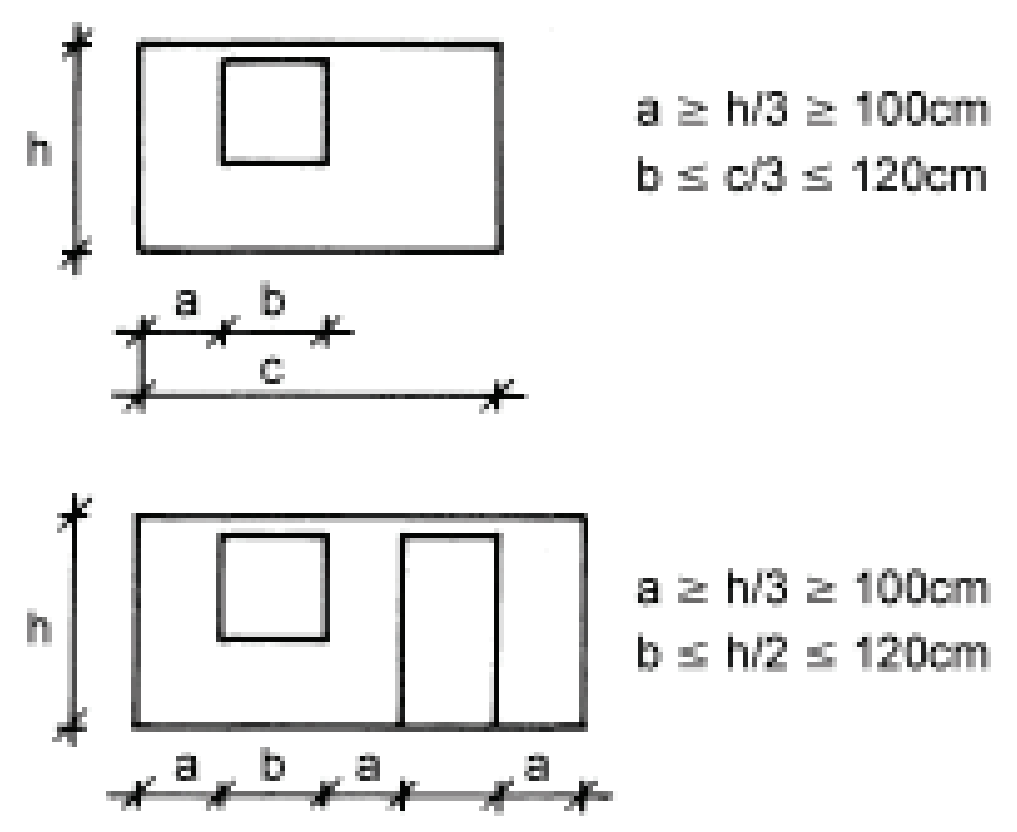

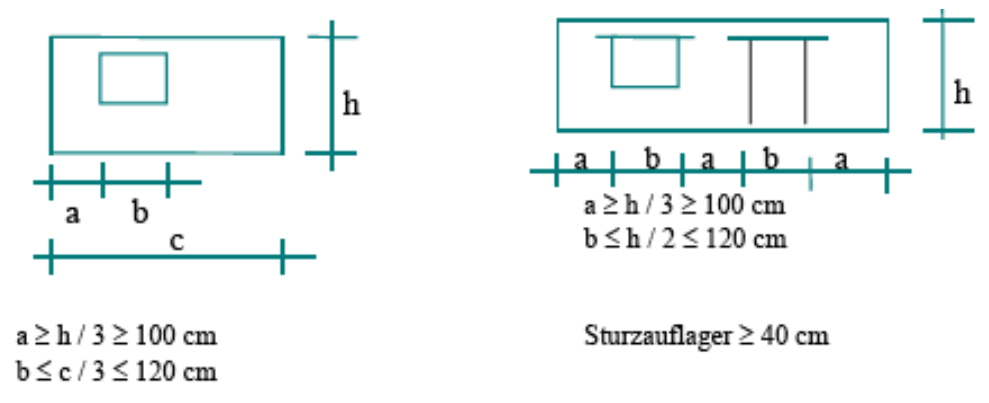

| Original image: [347, p. 3] | Reused image: [435, p. 193] |
|---|---|
| VroniPlag report on the finding: [522] | |

**Figure 4.5.** Example of a strongly altered image.

## 4.2.4 Visualizing Reused Data

Reusing data or data visualizations without attribution may constitute plagiarism or data fabrication if the data presented was not collected.

**Figure 4.6** and **Figure 4.7** show near-identical bar charts and line charts that the VroniPlag project found in a doctoral thesis. The scale of the vertical axis in the reused line chart differs slightly from the original. VroniPlag found content reused without proper attribution on 46 of 69 pages in the thesis [533]. However, by July 2020, the responsible university has not announced sanctions in this case.

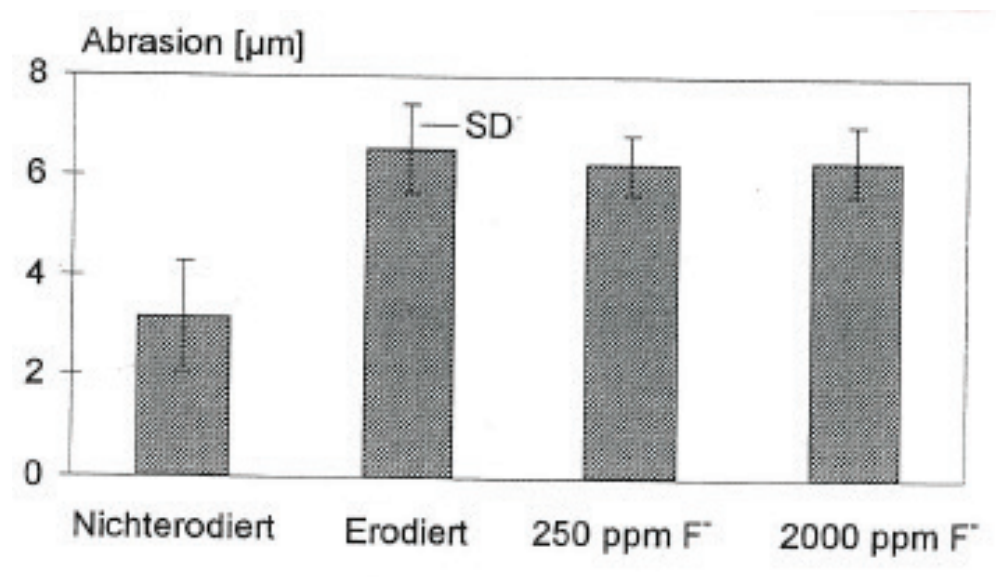

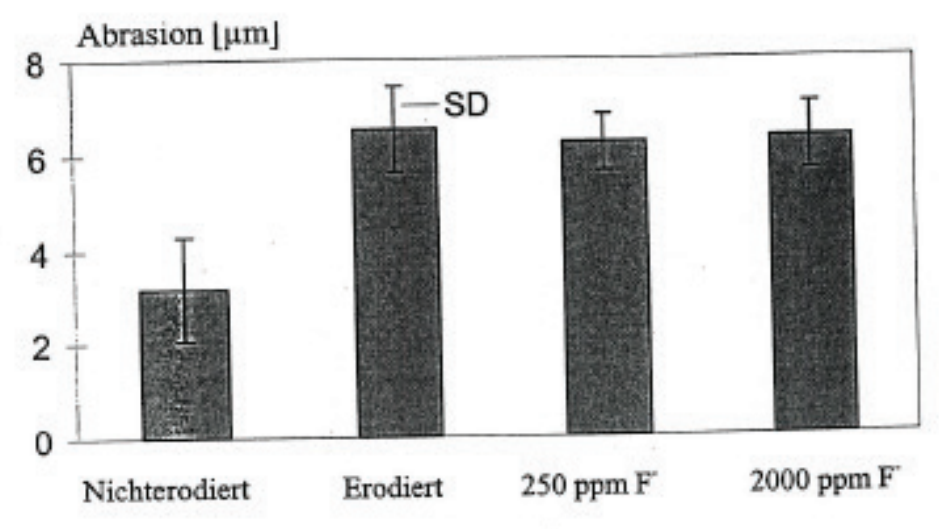

| Original image: [36, p. 117] | Reused image: [575, p. 47] |
|---|---|
| VroniPlag report on the finding: [527] | |

**Figure 4.6.** Example of a reused bar chart.

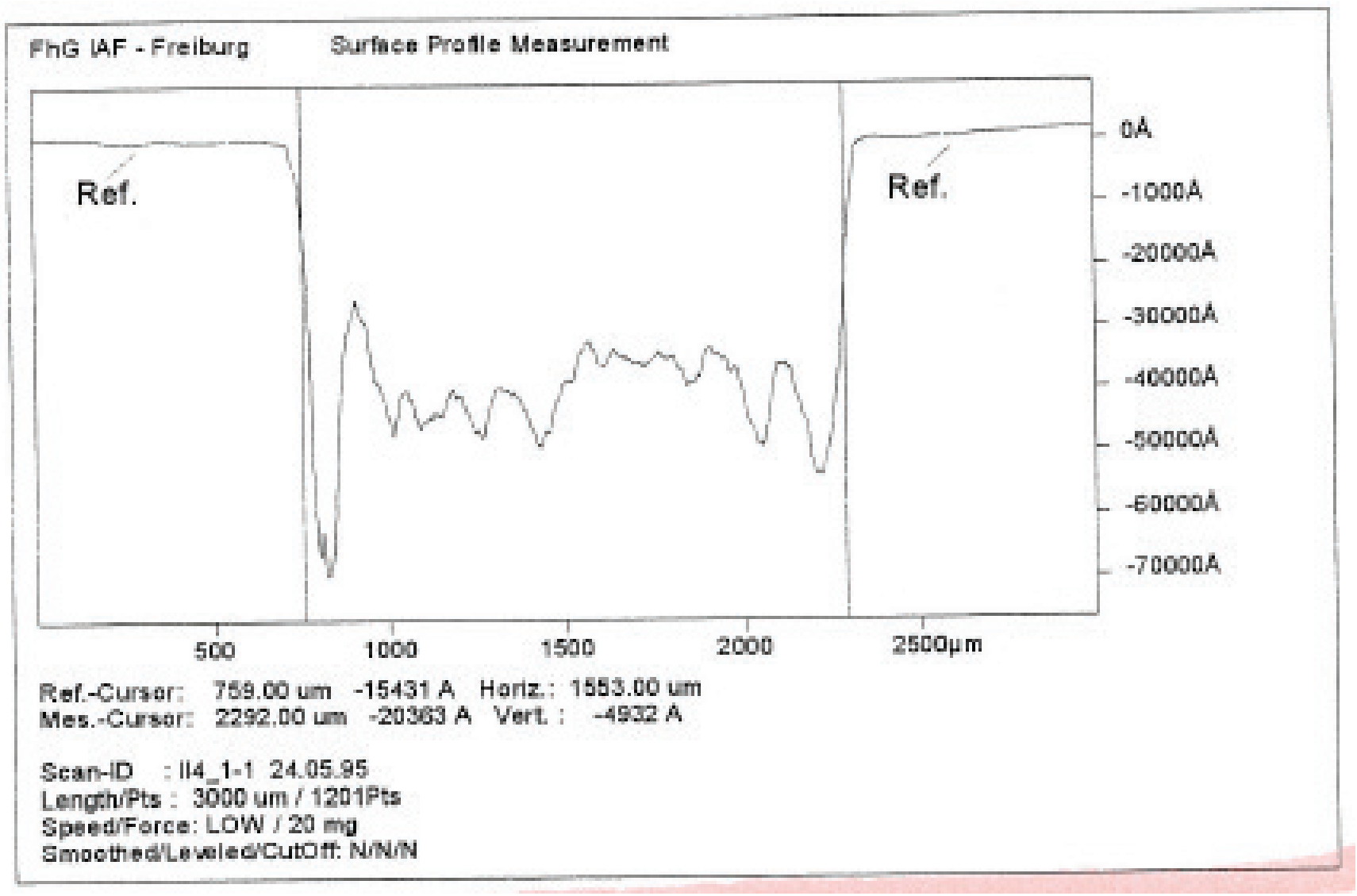

Original image: [36, p. 116]

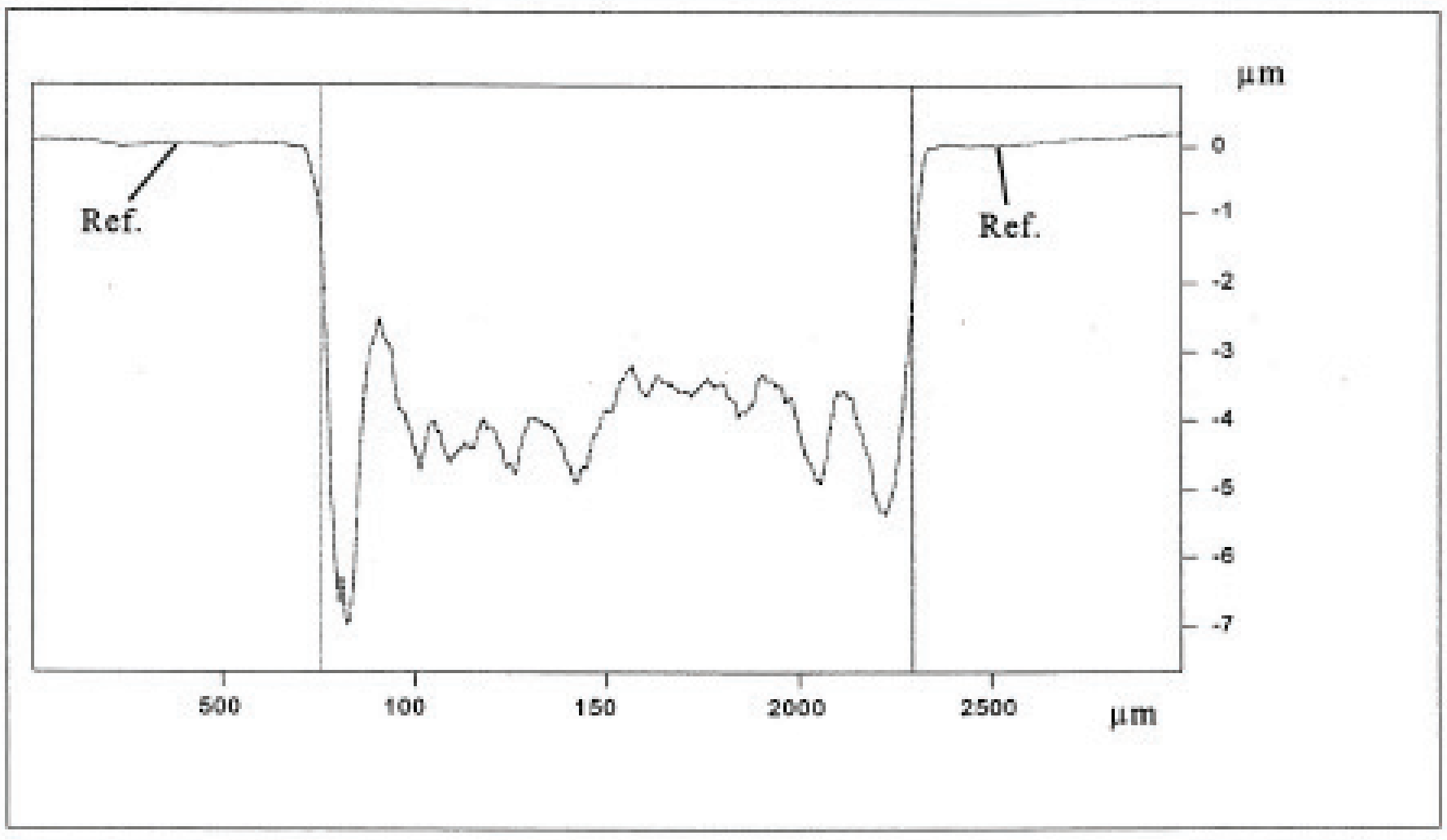

Reused image: [575, p. 44]

VroniPlag report on the finding: [527]

**Figure 4.7.** Example of a reused line chart.

In our investigation of the VroniPlag collection, we found no cases that visualize data differently. However, given that misuse of data is a well-known problem in academia [55], [563], we hypothesize that such cases do exist. We believe that our image-based detection process will make them identifiable.

## 4.3 Requirements Analysis

Given the types of image similarities we observed in the VroniPlag collection, we derived the following requirements for methods to detect such similarities.

Most images we reviewed fell into the category of near-duplicates. We expected this result because investigations of plagiarism allegations exhibit a known bias towards identifying less obfuscated, hence easier to spot, forms of content reuse.

Retrieving near-duplicate images requires methods to reliably identify and efficiently match visually apparent features. Obtaining a semantic understanding of the image composition and underlying data, e.g., by incorporating knowledge about the image type, is typically unnecessary. Robustness against minor, potentially unintentional variations in image quality and dimensionality are essential. Another requirement is computational efficiency because the plagiarism detection task typically requires comparing documents to extensive collections.

One can often reduce the task of retrieving weakly altered images to identifying near copies of image sections. For such cases, methods for identifying sub-images are vital to achieving high retrieval effectiveness. Detecting more strongly altered images often requires obtaining a deeper semantic understanding of the visualized data. Since the visual appearance of the data differs, additional features, such as labels and information from the text surrounding the images, should be considered.

Employing image analysis to identify data reuse is a challenging retrieval task because it requires bridging the semantic gap between the visual representation and the underlying data. We expect that identifying visually different representations of (near-)identical data requires methods tailored to analyzing specific types of visualizations, such as bar charts, box plots, line graphs, or scatter plots. Similar to how humans interpret data visualizations, the methods should consider all available information, e.g., size, shape, color, and position of data points, axes scales, and the content and position of labels and legend entries.

Given the variety of possible image similarities, we regard a combination of multiple analysis methods as most promising for covering the spectrum of similarities. The following section describes the detection approach we developed to address these requirements and the insights presented in related work.

## 4.4 Image-based Detection Process

**Figure 4.8** illustrates the adaptive image-based detection process, whose components we describe in the following sections. The input to the process is a PDF, from which we extract the images and check whether they contain meaningful sub-images (see Section 4.4.1, p. 128, for details on the sub-image extraction step). To reduce the computational load for the system, we use the convolutional neural network (CNN) described in Section 4.4.2, p. 129. The CNN classifies images according to their suitability for being analyzed using the different analysis methods.

Currently, the process includes four detection methods to identify image similarity. As we present in Section 4.4.3, p. 129, we employ perceptual hashing as a well-established, fast, and reliable method to find highly similar images.

To improve the identification of disguised image similarity, we employ two methods that perform text matching for the text extracted from images using Optical Character Recognition (see Section 4.4.4, p. 130, for our OCR approach). The first method, $n$-gram text matching, which we describe in Section 4.4.5, p. 130, is a widely used text-based detection method (see Section 2.4.3, p. 25) for details on $n$-gram matching methods). The second method, Positional Text Matching, is one of our research contributions, which we present in Section 4.4.6, p. 131.

Another contribution is the Ratio Hashing method (presented in Section 4.4.7, p. 133), which we proposed to identify highly similar bar charts. It is a specialized method to identify data reuse.

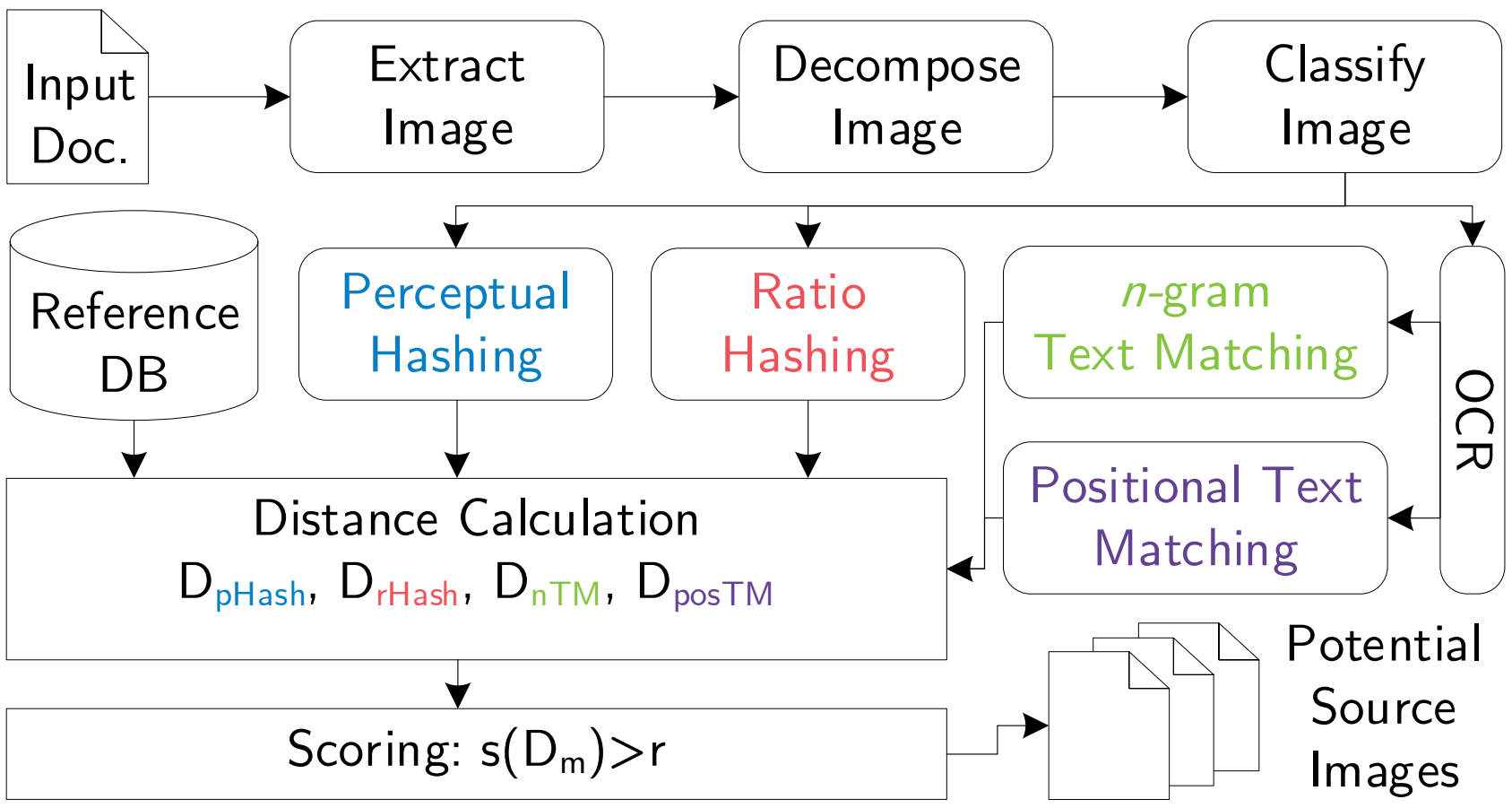

**Figure 4.8.** Overview of the adaptive image-based detection process.

Our process applies all detection methods independently of each other. The methods compute method-specific feature descriptors $K$ and compare them to the feature descriptors for all documents in the reference collection. We use a relational database to store feature descriptors. The comparisons of the descriptors return separate lists of distance scores $D_m$ for each analysis method $m$. The system sorts the lists in ascending order of the distance scores and provides them to a scoring method, described in Section 4.4.8, p. 135.

The scoring method computes method-specific suspiciousness scores $s(D_m)$ that indicate whether clear outliers exist within the lists of method-specific distance scores. The process then returns as potential sources for an image in the input document all images for which at least one method-specific suspiciousness score $s(D_m)$ is above a reporting threshold $r$.

### 4.4.1 Image Extraction and Decomposition

To extract the images contained in the input documents, we use poppler [153], an open-source library for PDF processing. To reduce the storage requirements, we convert all images to JPEG. We discard JPEG images whose file size is below 7,500 bytes to reduce the computational effort and avoid false positives. This threshold reflects our observation that images with fewer than 7,500 bytes typically contain single characters, logos, or decorative elements that are of little value for identifying potential instances of plagiarism.

To decompose compound images, such as in **Figure 4.3**, p. 123, we devised a heuristic process based on two assumptions. First, we assume that white pixels separate subimages. Second, we assume that sub-images are rectangular and aligned horizontally or vertically within the compound image. Although these assumptions exclude some images, we consider the approach a reasonable tradeoff between accuracy and computational effort. If successful, image decomposition can increase the detection effectiveness for sub-images. However, compound images, for which image decomposition fails, are still analyzable.

The decomposition process includes the following steps:

1. **Converting the image to grayscale** to reduce the runtime of the detection process. For most images in academic documents, the loss of information from ignoring colors is negligible.
2. **Padding the image with white pixels** on all sides. This step is necessary to remove a potential border in step five of the decomposition algorithm.

3. **Binarizing the image** using adaptive thresholding to obtain a black and white image. The thresholding method cross-correlates the neighborhood of the analyzed pixels with a Gaussian kernel [375].
4. **Dilating the image** to ensure black pixels are connected. Noise, e.g., due to scanning images, often disconnects pixels connected initially.
5. **Removing a potential border**. The decomposition algorithm flood fills white areas of the image with black pixels, subtracts the original image, and inverts the resulting image. If the image has a border, this step will reveal the area enclosed by the border as a black rectangle. If the algorithm detects such a rectangle, it crops the original image to a size slightly smaller than the rectangle's dimensions, thereby removing the image border.
6. **Estimating the bounding box(es)**. As in step five, the algorithm flood fills the image with black pixels, subtracts the original image, and inverts the resulting image. This step fills all closed shapes. The algorithm then uses the border-following method of Suzuki and Abe [499] to find contours in the image. Lastly, the algorithm estimates the bounding boxes of sub-images by looking for large contours aligned along the image axes.
7. **Extracting the sub-images** by cropping the image to the identified bounding box(es) and storing the results in the reference database.

### 4.4.2 Image Classification

To apply detection methods only to images for which the methods are suitable, we use a deep convolutional neural network that distinguishes photographs and bar charts from other image types. The detection process then applies Ratio Hashing exclusively to bar charts and perceptual hashing exclusively to photographs because they typically contain too little text to apply OCR text matching. All other image types are analyzed using perceptual hashing and OCR text matching.

The CNN implements the AlexNet architecture [289]. We used the Caffe framework [249] to train the CNN. Manually checking 100 classified images showed that the CNN achieves an accuracy of 92% for photographs and 100% for bar charts.

### 4.4.3 Perceptual Hashing

We included perceptual hashing in the detection process since prior research showed the method's suitability to reliably retrieve near-duplicate images [205], [479]. In the experiments of Srivastava, Mukherjee, and Lall, perceptual hashing achieved an accuracy of 0.84, which was the second-best result following SIFT, which achieved an accuracy of 0.95 [479]. Given that SIFT required approximately four times more runtime than perceptual hashing and perceptual hashing outperformed other feature point methods, such as SURF, FREAK, and KAZE [479], we consider the method to be a reasonable tradeoff between accuracy and computational effort.

We tested different variants of perceptual hashing and found that using a Discrete Cosine Transform and comparing pHash values using their Hamming distance achieved the best accuracy. The Hamming distance of two pHash values is the number of bits that differ in the hashes. Our detection process precomputes the pHash values for all images in the reference collection and stores the pHash values in the reference database. In its current state, our detection process employs pairwise comparisons of the pHash for an input image to all pHash values for images in the reference database. The pairwise comparisons can be replaced with a locality-sensitive hashing approach to speed up the process and enable comparing an image to vast collections, as demonstrated by Srivastava, Mukherjee, and Lall [479].

### 4.4.4 OCR Preprocessing

Including textual features in the similarity analysis requires a preprocessing step to extract the text from images. Research on OCR has provided a wide range of methods for this task. We chose the open-source OCR engine Tesseract [472] because it allows extracting characters and words, including their positions. Tesseract is widely-used, actively maintained, and repeatedly outperformed proprietary OCR engines for the task of recognizing English texts [391].

Before applying Tesseract, the system normalizes each image to a height of 800 pixels while maintaining the aspect ratio, which significantly improves the recognition. OCR is computationally expensive, and processing times vary greatly depending on the input image. Our process extracts the text for all images in the collection once and stores the information in the reference database.

### 4.4.5 $n$-gram Text Matching

Determining textual similarity by analyzing matching word or character $n$-grams is a well-established information retrieval method. As we describe in Section 2.4.3, p. 25, numerous text-based plagiarism detection methods employ variable-size or fixed-size $n$-grams. For regular texts, $n$-grams lengths equaling three to five words, i.e., about 15-30 characters, are used most frequently [53], [189], [266].

We considered two use-case-specific factors to choose an $n$-gram size for analyzing text in figures extracted using OCR. First, images typically contain smaller text fragments, such as labels or bullet points. Second, our detection process extracts the text content of images using OCR, which is likely to introduce noise, i.e., incorrectly recognized characters. Such recognition errors can significantly reduce detection effectiveness, especially for word $n$-gram detection methods.

To account for the likelihood that incorrectly recognized characters occur, we chose a comparably fine-grained $n$-gram resolution of three characters. Given the typically sparse presence of text in figures, our process retains all $n$-grams identified for an image as an unordered set that forms the $n$-gram descriptor of that image. Typically, $n$-gram-based detection methods that analyze entire documents employ some form of $n$-gram selection. We form the $n$-gram descriptors for all images in the collection during preprocessing and store the descriptors in the reference database.

Currently, our detection process performs pairwise comparisons of the $n$-gram descriptor of an input image to all $n$-gram descriptors of documents in the reference collection. To scale the image-based detection process to extensive collections, one can introduce an additional filtering step easily. One possibility is to index individual $n$-grams and requiring a minimum $n$-gram overlap to perform the complete comparison of the $n$-gram descriptors. To quantify the distance $d$ of two $n$-gram descriptors $K_1$ and $K_2$, our detection process uses the set-based distance function

$$d_{\text{nTM}} = \frac{K_1 \ominus K_2}{K_1 \cap K_2},$$

where $\ominus$ represents the symmetric difference.

### 4.4.6 Positional Text Matching

OCR errors are a severe threat to the retrieval effectiveness of text-matching methods. The typically sparse amount of text present in academic images further aggravates the problem. We proposed considering positional information as an additional analysis feature to improve the robustness of similarity assessments examining textual content in images. Specifically, we suggested considering text matches for computing the similarity of two images only if the matching text occurs in broadly similar regions in both images.

**Figure 4.9** illustrates the approach. We assume an input image (left) and a comparison image (right) that have been scaled to the same height or width while maintaining the aspect ratios of the images. The markers A, B, C, and D in the figure symbolize text identified by the OCR engine, e.g., characters, words, or $n$-grams. Each text fragment identified in the input image is considered the center point around which we define a proximity region. In the figure, we use a fixed-size circle as the proximity region. However, other shapes and dynamic sizing of the shape, e.g., dependent on the length of the text fragment, are also possible. The method projects the proximity regions of the input image into the comparison image. The similarity score computation only considers the text matches in corresponding proximity regions (A and D). Text matches outside of a proximity region (B) and non-matching text (C and X) do not influence the similarity score.

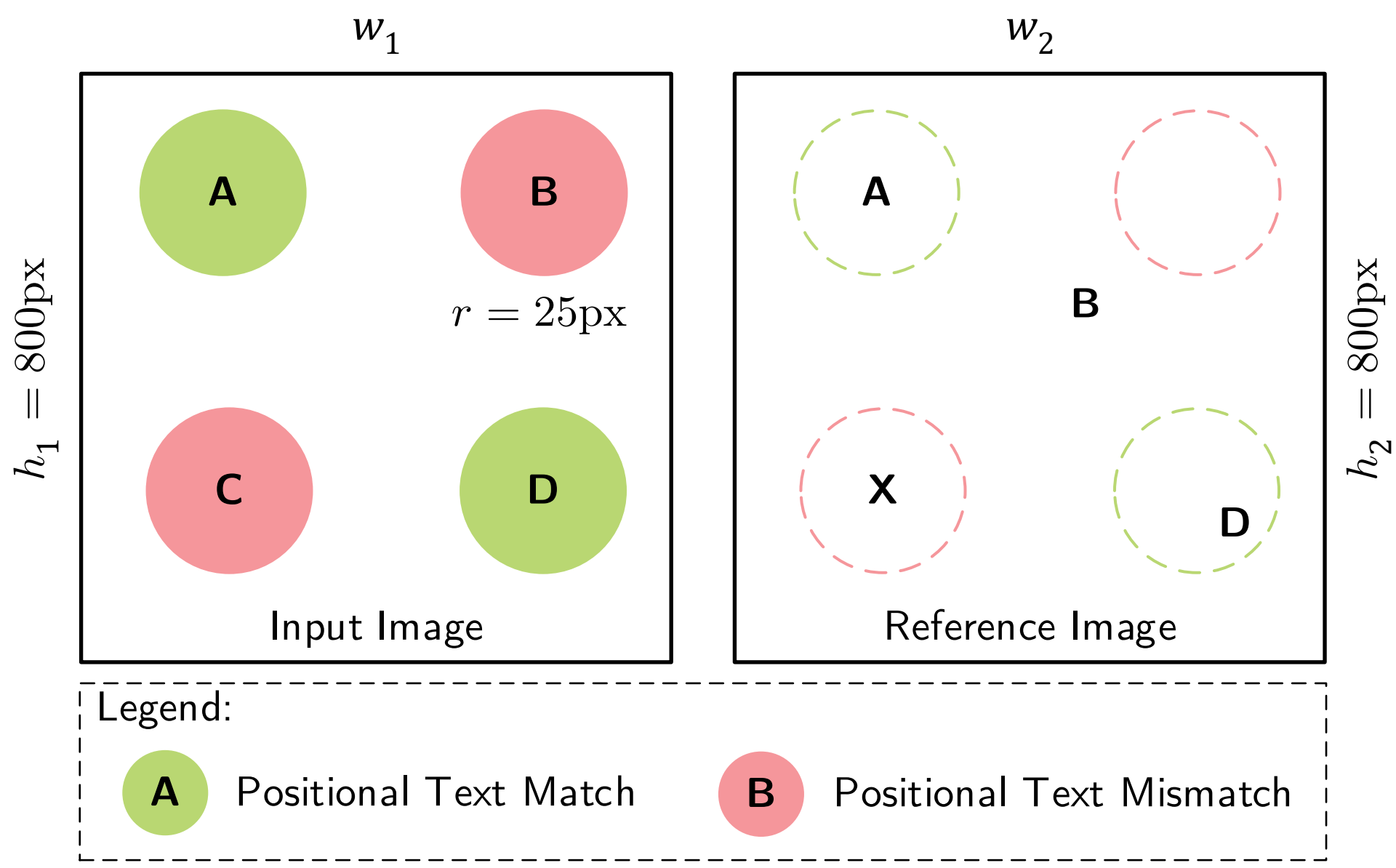

**Figure 4.9.** Illustration of Positional Text Matching.

The current implementation of Positional Text Matching uses the following parameters. The images are rescaled to a height of 800 pixels while maintaining their aspect ratio. We use single characters as the center points around which we define a fixed-size circular proximity region with a radius of 25 pixels. The distance function considers the number of positional text matches divided by the number of characters in the longer text. This normalization reflects the assumption that two images are less likely to be similar if their amount of textual content differs sharply.

The pairwise comparison of positional text matches is computationally more expensive than the set-based $n$-gram comparison. For our initial study, we employed no filtering of candidate images except for the classification by image type described in Section 4.4.2, p. 129. To scale the approach, one could add filtering heuristics, such as requiring a minimum Jaccard similarity of the $n$-gram sets.

### 4.4.7 Ratio Hashing

We proposed Ratio Hashing to identify semantically similar yet visually differing bar charts to demonstrate a detection method that targets the plagiarism of data and results. Due to the diversity of chart types, using specialized detection methods to analyze each type is advisable. Because authors frequently use bar charts in academic publications, we geared Ratio Hashing towards them.

The idea of Ratio Hashing is to compute a feature descriptor ("hash value") from the relative heights of bars compared to the height of the longest bar. To determine the distance of two ratio hashes, we compare the components of the hash, i.e., the relative bar heights, in decreasing order, and calculate the sum of the differences of the bar heights. **Figure 4.10** illustrates the Ratio Hashing algorithm.

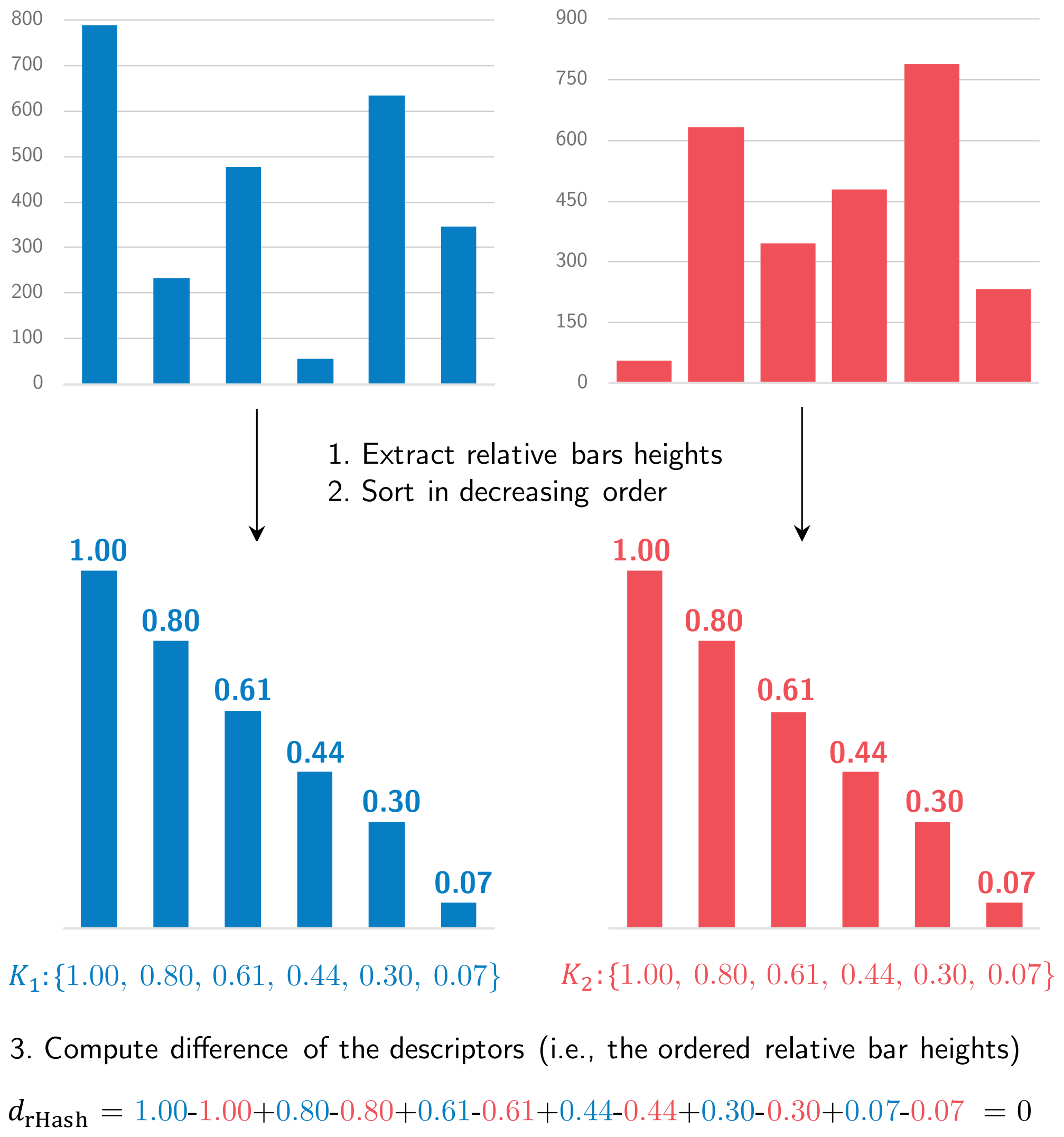

**Figure 4.10.** Computing the similarity of bar charts using Ratio Hashing.

To extract the bar heights from an input image, we process the image as follows:

1. **Conversion to grayscale** because Ratio Hashing does not consider the color of bars, and grayscale images reduce computational effort.
2. **Binarization using global thresholding** to ensure sharp contours.
3. **Padding with white pixels** to ensure bars do not touch a potential border. This property is a prerequisite for the following steps.
4. **Clean artifacts** that the previous steps may introduce. We remove black pixels if they cover a minor share of the overall image area.
5. **Remove a potential image border** by flood filling the image with black pixels, subtracting it from the original image, and checking whether a filled rectangle emerges that covers most of the image area. If such a rectangle exists, the image likely has a border. In that case, we remove the flood-filled area and use the original image otherwise.
6. **Normalizing bar color** by flood filling the image with black pixels and inverting it, thus leaving all bars colored in solid black.
7. **Finding candidates for bars** by determining the lengths of all vertical lines of black pixels.
8. **Determining bars** by clustering all vertical lines. We remove noise from whiskers, labels, and legend entries and assume the average height of the lines in a cluster as the bar height.
9. **Computing the ratio hash** by sorting the bars in decreasing order of their height to speed up the comparison of two ratio hashes. We then calculate the relative bar heights and store the ratio hash, i.e., the sequence of relative bar heights, in the reference database.

The detection process currently compares the ratio hash of an input bar chart to all ratio hashes of bar charts in the database. In the future, one can reduce the computational effort by indexing the descriptors and using filtering steps.

For our initial evaluation, we limited computational effort by requiring bar charts to have the same number of bars. However, the comparison can easily adapt to more exhaustive comparisons that consider the best fit between sets of different sizes if previous filtering steps reduce the analysis scope.

### 4.4.8 Scoring

Each of the four image detection methods described in the previous sections returns the method-specific distances of the input image to all images in the reference collection as an ordered list $D_m$. We make two assumptions to quantify how suspicious, i.e., how indicative of potential image plagiarism these distances are.

First, we assume that the input image can only be suspicious of originating from (an)other image(s) in the collection if it exhibits comparably strong similarities, i.e., small distances, to a small number $c$ of images. Small distances of the input image to other images alone are not necessarily suspicious. The input image could be a logo that the preprocessing step missed to exclude. Such images would exhibit small distances to many images in the collection. Therefore, we additionally require that the input image exhibits small distances to fewer than $c$ images. Our system uses a cutoff parameter $c$ of 10. In essence, $c$ is a filter for false positives that accounts for potential deficits of the detection process or the collection, e.g., standard images or multiple versions of images not eliminated during preprocessing. The parameter should be large enough to rule out any reasonable possibility that strong similarities to more than $c$ images are not false positives. We consider $c = 10$ a conservative estimate to ensure this property even for extensive collections.

Second, we assume that image similarities are comparably strong if a clear separation is observable in the distance scores for the $k < c$ images most similar to the input image and the distance scores of the remaining images in the collection. In other terms, images with strong similarities to the input images must be outliers. If the input image does not fulfill that requirement, the image is either genuine or too dissimilar to a potential source for being detected. Alternatively, the reference collection may not contain the source image, or the analysis method failed to determine a meaningful distance.

Given these assumptions, we identify outlier distances as depicted in **Figure 4.11.**

For each analysis method $m$, the scoring algorithm stores the absolute distances $d_i, 1 \leq i \leq n$ of the input image to all $n$ images in the reference collection as a list $D_m = (d_1, d_2, \ldots d_n)$ in ascending order of $d_i$. The algorithm then transforms each method-specific list of absolute distances $D_m$ into a list $D'_m$ of the relative deltas $d'_i$ between $d_i$ and $d_{i+1}$ as follows

$$d'_i = \frac{(d_{i+1} - d_i)}{d_i} \qquad 0 < i < |D_m|, i \in N$$

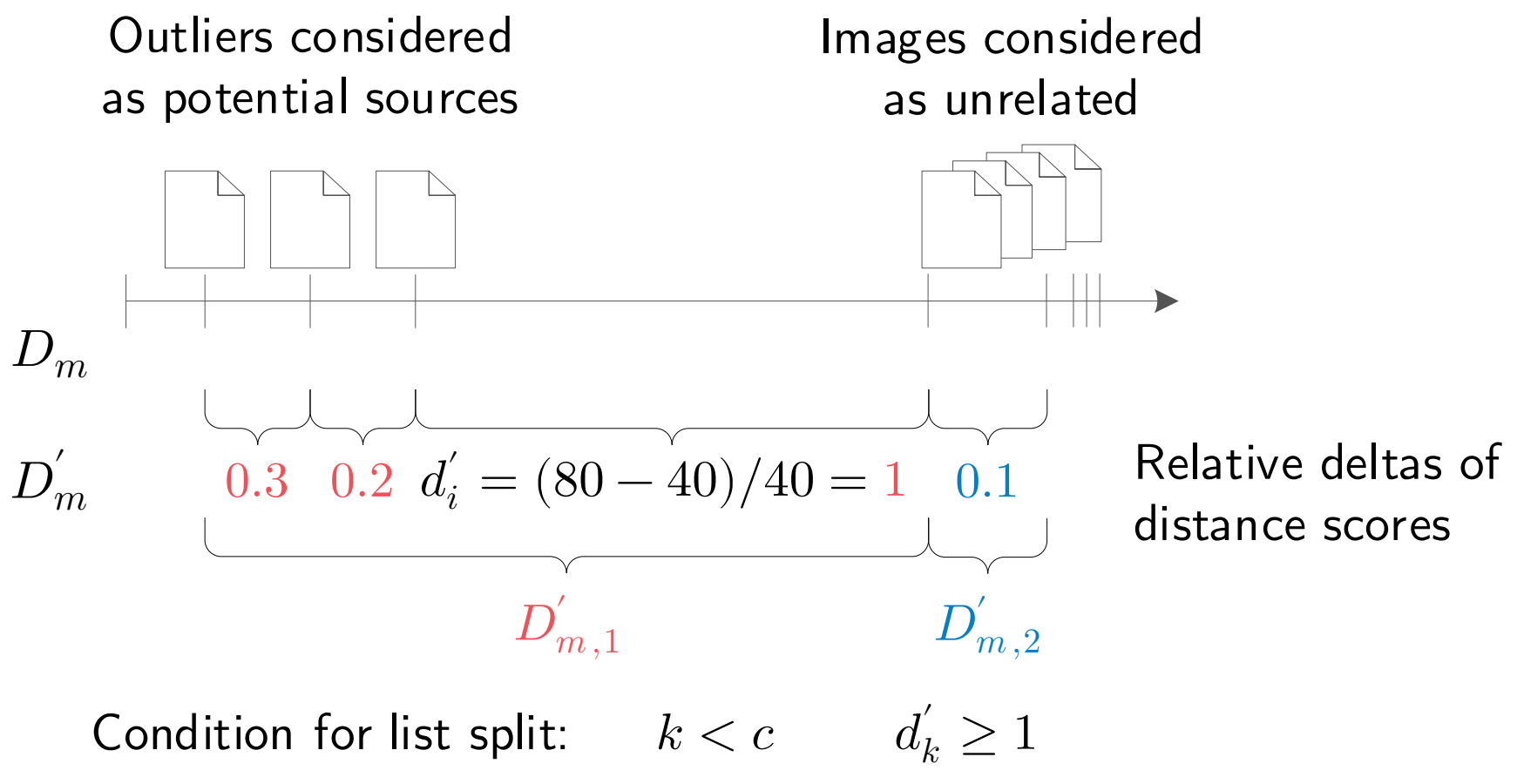

**Figure 4.11.** Illustration of the outlier detection process.

To find outliers, i.e., elements in $D'_m$ clearly separated from succeeding elements, the algorithm sequentially processes $D'_m$ and checks for elements $d'_i$ that exceed a threshold $t$. The threshold is customizable and set to $t = 1$ for all methods except $n$-gram text matching, where we found a threshold of $t = 2$ to yield better results.

In other terms, we require that a pair of distances $(d_i, d_{i+1})$ exists, for which $d_{i+1}$ is at least twice as large as $d_i$ and three times as large as $d_i$ in the case of $n$-gram matching. If an $d'_i$ exceeds these thresholds, $D'_m$ is split into $D'_{m,1}$ and $D'_{m,2}$ at the largest $d'_i$, where $d'_i \in D'_{m,1}$. If $D'_{m,1}$ has less than $c$ elements, the corresponding images are considered potential sources of the input image. The algorithm computes the final similarity scores $s(D_m)$ for each list of distances $D_m$ as

$$s = \frac{\bar{d}}{1 + \bar{d}}$$

where

$$\bar{d} = \frac{\max(d'_i \in D'_{m,1})}{t}.$$

In other terms, the similarity score $s$ considers the relative delta in the distance that separates an identified group of outliers (in our case at most $c - 1 = 9$ images) from the remainder of the collection. The algorithm uses the function

$$y = \frac{x}{x + 1}$$

to normalize the score $s$ to [0,1]. The sublinear normalization function assigns a weight of 0.5 if the image in the outlier group that is least similar to the input image is separated from the remainder of the collection by a margin that is as large as the absolute distance of this least similar outlier to the input image.

For all analysis methods, we set $s = 0.5$ as the reporting threshold to consider an image potentially suspicious and a score $s > 0.75$ as highly suspicious. For the case $s = 0.75$, the least similar outlier has a distance margin to the next similar image three times as large as its absolute distance to the input image.

## 4.5 Evaluation

Due to a lack of evaluation datasets for image plagiarism at the time of our study, we selected 15 image pairs from documents in the VroniPlag collection to evaluate our adaptive image-based detection process. We chose images that reflect the spectrum of image similarities we observed in the collection. We describe some of the test cases in Section 4.2, p. 120. Appendix A, p. 213, lists all test cases.

Most of the selected images are from the life science domain. To create a realistic test collection, we obtained 4,500 random images contained in life science publications from the PubMed Central Open Access Subset [511]. We hid the 15 known source images among the 4,500 obtained images and created the reference database by classifying each image and computing the applicable feature descriptors. After precomputing the reference database, we used each of the 15 reused images individually as the input to the detection process.

**Table 4.1** shows the similarity scores $s(D_m)$ for each input image computed from the distance scores $D_m$ of the four detection methods, perceptual hashing (pHash), character trigram text matching (nTM), Positional Text Matching (posTM), and Ratio Hashing (rHash). The first two columns characterize the input image. Boldface font indicates scores above the reporting threshold $s = 0.5$.

**Table 4.2** complements **Table 4.1** by showing the ranks at which each of the four detection methods retrieved the source image for an input image. Note that the system would not return any results for input images with a score below 0.5, as the images identified as similar would not represent clear outliers in such cases.

To verify the appropriateness of the reporting threshold $s = 0.5$, we selected each input image with scores below 0.5, retrieved the 10 images identified as most similar to that image, and checked whether this set contained the source image. Limiting the set to 10 images is a heuristic that assumes a reviewer might be willing to browse through 10 results, although the system did not identify them as clearly suspicious. For none of those cases did the system retrieve a source image among the top-10 most similar images. This result shows that the scoring algorithm effectively eliminated all false positives, resulting in a precision of one. High precision is essential for plagiarism detection methods to avoid false suspicion.

**Table 4.1.** Similarity scores for input images.

| # | Image Type | Alteration | pHash | nTM | posTM | rHash |
|---|---|---|---|---|---|---|
| 1 | Illustration | Near-duplicate | **0.87** | $< 0.5$ | $< 0.5$ | - |
| 2 | Illustration | Near-duplicate | **1.00** | 0.79 | 0.77 | - |
| 3 | Illustration | Near-duplicate | **0.86** | $< 0.5$ | $< 0.5$ | - |
| 4 | Illustration | Weak | **0.78** | $< 0.5$ | $< 0.5$ | - |
| 5 | Illustration | Weak | **0.57** | $< 0.5$ | $< 0.5$ | - |
| 6 | Illustration | Moderate | $< 0.5$ | **0.87** | $< 0.5$ | - |
| 7 | Illustration | Strong | $< 0.5$ | $< 0.5$ | $< 0.5$ | - |
| 8 | Bar Chart | Near-duplicate | 0.62 | 0.64 | 0.77 | **0.92** |
| 9 | Table | Near-duplicate | $< 0.5$ | $< 0.5$ | $< 0.5$ | - |
| 10 | Table | Near-duplicate | 0.62 | **0.71** | 0.55 | - |
| 11 | Table | Near-duplicate | $< 0.5$ | **0.92** | $< 0.5$ | - |
| 12 | Table | Weak | $< 0.5$ | **0.79** | $< 0.5$ | - |
| 13 | SEM Image | Near-duplicate | $< 0.5$ | $< 0.5$ | $< 0.5$ | - |
| 14 | Line Chart | Weak | $< 0.5$ | $< 0.5$ | $< 0.5$ | - |
| 15 | Line Chart | Strong | $< 0.5$ | **0.70** | $< 0.5$ | - |

**Table 4.2.** Ranks at which the detection process retrieved source images.

| # | Image Type | Alteration | pHash | nTM | posTM | rHash |
|---|---|---|---|---|---|---|
| 1 | Illustration | Near-duplicate | **1** | $> 10$ | $> 10$ | - |
| 2 | Illustration | Near-duplicate | **1** | **1** | **1** | - |
| 3 | Illustration | Near-duplicate | **1** | $> 10$ | $> 10$ | - |
| 4 | Illustration | Weak | **1** | $> 10$ | $> 10$ | - |
| 5 | Illustration | Weak | **1** | $> 10$ | $> 10$ | - |
| 6 | Illustration | Moderate | **1** | **1** | $> 10$ | - |
| 7 | Illustration | Strong | **1** | $> 10$ | $> 10$ | - |
| 8 | Bar Chart | Near-duplicate | **1** | **1** | **1** | **1** |
| 9 | Table | Near-duplicate | $> 10$ | $> 10$ | $> 10$ | - |
| 10 | Table | Near-duplicate | **1** | **1** | **1** | - |
| 11 | Table | Near-duplicate | **1** | **1** | $> 10$ | - |
| 12 | Table | Weak | $> 10$ | **1** | $> 10$ | - |
| 13 | SEM Image | Near-duplicate | **1** | $> 10$ | $> 10$ | - |
| 14 | Line Chart | Weak | $> 10$ | $> 10$ | $> 10$ | - |
| 15 | Line Chart | Strong | $> 10$ | **1** | $> 10$ | - |

As shown in **Table 4.1**, for 11 of the 15 input images, at least one analysis method determined scores above the reporting threshold, thus achieving a recall $R = 0.73$. The cases seven, nine, 13, and 14 are false negatives. However, pHash retrieved the source images for case 13 (an image produced by a Scanning Electron Microscope, SEM) and case seven (a visually sparse sketch exclusively using basic geometric shapes) at the top rank. For these two cases, pHash computed similarities to many unrelated images of the same types. However, the low score assigned to the pHash distances shows that the identified similarities are not clear outliers. For case nine, the low quality of the input image caused poor OCR results, which prevented detecting that the text in the table was copied nearly verbatim. The line chart in case 14 is visually too sparse to be detected by pHash. Little textual content and low image quality also caused the OCR-based methods to fail for this case.

As shown in **Table 4.2**, the detection methods retrieved the correct source images at the top rank for all input images with a score above the reporting threshold.

For near-duplicates and weakly altered images, perceptual hashing combined with sub-image extraction worked well, yielding suspiciously high scores for six of the nine cases falling into these categories.

Text analysis utilizing OCR performed better than perceptual hashing for moderately and strongly altered images if the quality of the image was high enough to perform OCR reliably and if sufficient text content was present. The OCR-based methods identified three of the four cases that involved tables (cases 10, 11, 12), for which they yielded clearly suspicious scores (0.71, 0.92, and 0.79, respectively).

While $n$-gram text matching performed better than the Positional Text Matching for most cases, the Positional Text Matching was more robust to low OCR quality. Therefore, combining both methods allows processing a more significant number of input images in a realistic setting.

The test dataset contained only one case (eight), in which an author reused a bar chart. For this case, Ratio Hashing outperformed all other methods ($s = 0.92$), although the bar chart was rotated and slightly altered. However, reliably determining the performance of Ratio Hashing requires additional evaluations.

Creating the reference database for the 4,515 images took around two hours using single-thread processing on a desktop computer with a 2.70 GHz Intel Core i5-6400 CPU, 8 GB of main memory, and a GeForce GTX960 GPU, which we used to accelerate the CNN classifier. Executing the detection methods took between 1–3 seconds for perceptual hashing and Ratio Hashing and between 2–16 seconds for the OCR-based methods using the same computer. This time includes classifying the input images as well as computing and comparing the feature descriptors.

### 4.5.1 Discussion of Results

Our results demonstrate that an adaptive image-based plagiarism detection process enables identifying a wide range of suspicious image similarities in academic work. While the suitability of detection methods strongly depends on the individual images, the combination of detection methods achieved a good recall of 0.73 in our experiments. The proposed scoring algorithm performed exceptionally well in our experiments. Due to the use of restrictive thresholds, our detection process eliminated false positives and achieved a precision of one.

These results are promising. However, images from life science publications were overrepresented in our small dataset, which limits the generalizability of our results. Future experiments must show whether the properties we assume, i.e., that suspicious image similarities form outliers, are also observable in more extensive collections. The scoring algorithm primarily depends on the distribution of distances between an input image and the images in the collection. In extensive collections, unrelated images may exhibit similarities to the input images that prevent identifying clear outliers. In such cases, reducing the outlier threshold may become necessary and would likely result in the identification of false positives. Because the scoring algorithm operates on a simple list of precomputed distance scores, adjusting the threshold at runtime is feasible. Hence, a frontend application could allow users to interactively adjust the threshold to determine the number of results (including potential false positives) the user is willing to examine.

Our detection process is well-suited for being scaled to allow for an evaluation of more extensive collections. All preprocessing steps allow parallel execution. In the current implementation of the process, only the time required to compare feature descriptors depends on the collection size. We described several easily implementable options to decrease the linear runtime requirement of this step by adding feature indexing and feature selection approaches.

However, as we discuss in Section 2.5, p. 49, and Section 3.4, p. 93, an inherent challenge to conclusive, large-scale evaluations of plagiarism detection methods is the difficulty of compiling test collections. A widely accepted solution is to use collections containing simulated plagiarism instances. For image plagiarism, such collections did not exist at the time of our study. For this reason, we opted to use real cases of image reuse in our experiments.

A technical challenge to the detection effectiveness of the proposed detection process in the realistic detection settings we imposed for our experiments is OCR effectiveness. The OCR-based analysis methods showed the best results for medium to high-level alterations of images. However, poor image quality, especially for older digitized academic publications, reduces OCR performance. Perceptual hashing often performed poorly for visually sparse images. A dilation step might help achieve better results. Although our process for sub-image extraction performed well in most cases, sometimes it failed to extract overlapping sub-images correctly. Specialized post-processing procedures could improve the results.

Aside from improving the analysis methods already included in the process, adding specialized detection methods for image types, such as line graphs, scatter plots, and photographic images, can further augment the detection capabilities.

## 4.6 Conclusion Image-based PD

We introduced an image-based plagiarism detection process that adapts itself to forms of image similarity found in academic work. To derive requirements for our detection process, we examined images in the VroniPlag collection. From these cases, we introduced a classification of the image similarity types that we observed.

Unlike previous work, our detection process can analyze various heterogeneous image types. Primarily to analyze natural images, the detection process integrates perceptual hashing, for which we extended the detection capabilities by including an extraction procedure for sub-images. Because textual labels are common in academic figures, we devised and integrated two methods that use Optical Character Recognition to extract and analyze text from figures, such as graphs, plots, and tables. To address the problem of data reuse, we integrated a method to identify equivalent bar charts. To quantify the suspiciousness of identified similarities, we presented a novel, use-case-specific scoring algorithm. The algorithm searches image similarities that represent outliers within the collection.

The evaluation of our detection process demonstrates reliable performance and extends the detection capabilities of existing image-based detection methods.

We implemented the detection process as a Python application that handles all inputs and outputs via a command-line interface. This setup allows for easy integration into existing information retrieval systems as a loosely coupled module. The code is open source and available together with the data of our experiments at http://thesis.meuschke.org.

5

Chapter 5

# Math-based Plagiarism Detection

**Contents**

This chapter presents Math-based Plagiarism Detection (MathPD)—a new approach we proposed to improve the detection of academic plagiarism, primarily in the science, technology, engineering, and mathematics (STEM) fields.

The structure of this chapter is as follows. Section 5.1 discusses the idea of the new detection approach and how it differs from existing research on Math Information

Retrieval (MathIR). Section 5.2 describes our research approach for devising math-based plagiarism detection methods, which we grounded on the types of similar mathematical content we observed in confirmed cases of plagiarism. Section 5.3 explains the creation of a dataset to evaluate MathPD methods. Section 5.4 summarizes the initial experiments we performed to guide the design of such detection methods. Section 5.5 describes the extension of our initial experiments into a two-stage plagiarism detection process that integrates math-based detection methods with citation-based and text-based detection methods. Sections 5.6 and 5.7 present the methodology and the results of our evaluation of the new detection process. Section 5.8 summarizes the research contributions in this chapter.

## 5.1 Math-based PD Concept

The idea underlying this line of research is that mathematical expressions share many properties of academic citations. Hence they should offer similar benefits regarding the detection of disguised plagiarism in STEM documents. Like academic citations, mathematical expressions are essential components of academic documents in the STEM fields and thus hard to substitute. Furthermore, mathematical expressions are language-independent and semantically rich.

While Citation-based Plagiarism Detection performs well in identifying disguised forms of academic plagiarism for many disciplines, it performs more poorly for physics and mathematics literature. The reason is that publications in these disciplines use academic citations more sparsely. On average, mathematics publications cite only half as many sources as publications in biochemistry [349, p. 178ff.]. A citation-based analysis alone is, therefore, less likely to reveal potentially suspicious content similarity for these disciplines.

Additionally, symbolic language is an integral part of mathematical literature, i.e., authors often substitute parts of speech with mathematical expressions [560, p. 132]. Larson et al. showed that word-based information retrieval methods perform poorly for mathematical document retrieval [296]. Given that many lexical plagiarism detection methods employ word-based similarity assessments, their effectiveness is lower for literature in math-heavy disciplines.

**Figure 5.1** illustrates the challenges that publications in math-heavy disciplines pose for lexical plagiarism detection methods. The figure shows excerpts of two engineering publications. The responsible journal retracted the publication on the left [562] for plagiarizing the source document [154] on the right, including a mathematical model that constitutes a core contribution of the original publication. Blue highlights indicate verbatim text matches with a length of 10 or more characters. Yellow highlights denote semantically identical mathematical expressions.

The example exhibits little textual similarity, which prohibits identifying this plagiarism instance using only lexical detection methods. The significant overlap in mathematics, on the other hand, makes this case discoverable. However, detecting the similarity is not trivial for automated methods because the plagiarized document exhibits differences in mathematical expressions. For example, the plagiarized document uses the identifier $N_U$, whereas the source document uses $I_U$. Equations 38–41 in the plagiarized document represent a split-up of the compound equation 16 in the original document. Equation 42 in the plagiarized document omits intermediate derivational steps in the corresponding equation 17 in the source document.

Examples like the one shown in **Figure 5.1** led us to expect that considering mathematical expressions for document similarity assessment can increase the detection capabilities for literature in math-heavy disciplines.

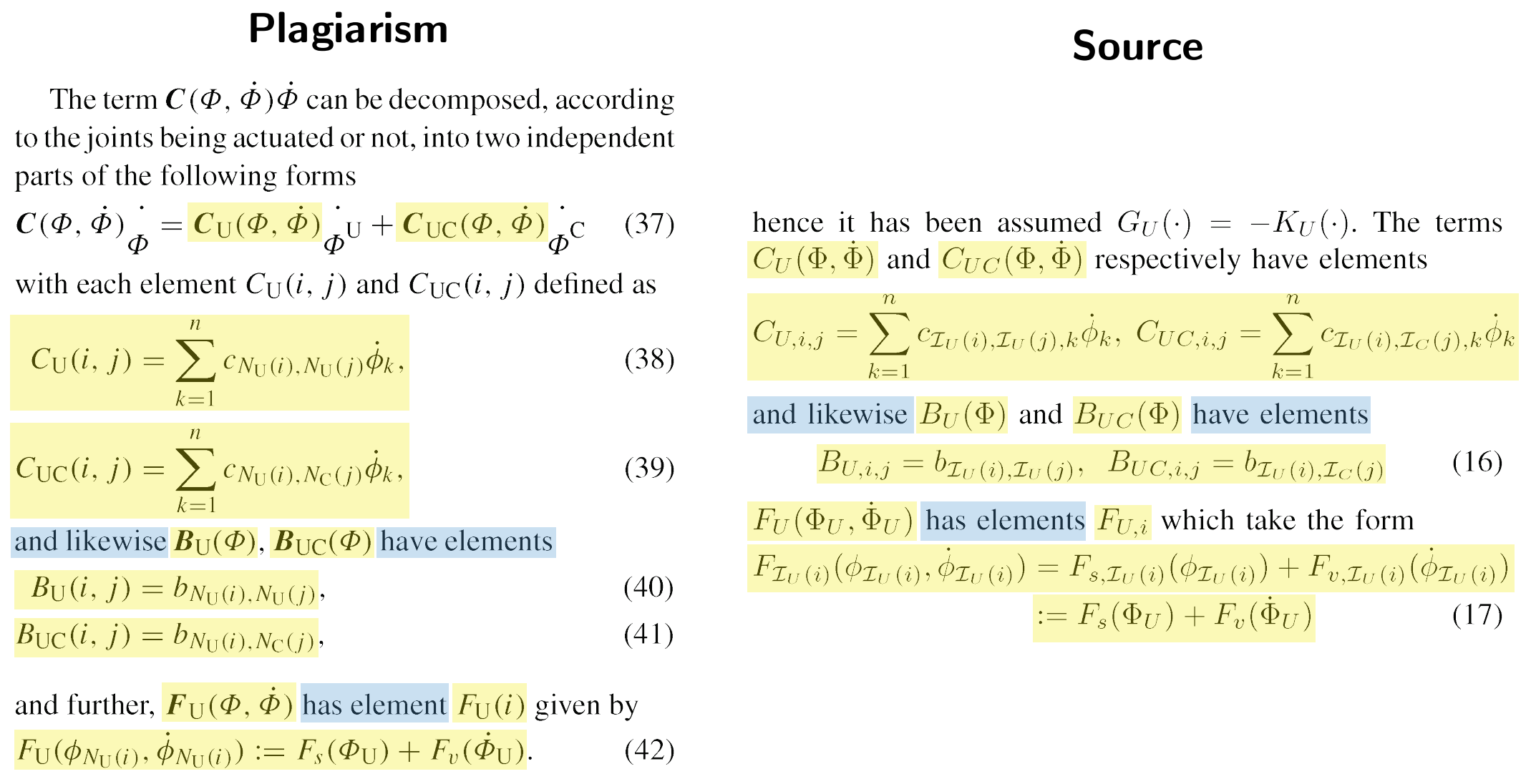
**Plagiarism**

The term $\boldsymbol{C}(\Phi, \dot{\Phi})\dot{\Phi}$ can be decomposed, according to the joints being actuated or not, into two independent parts of the following forms

$$\boldsymbol{C}(\Phi, \dot{\Phi})\dot{\Phi} = \boldsymbol{C}_{\mathrm{U}}(\Phi, \dot{\Phi})\dot{\Phi}_{\mathrm{U}} + \boldsymbol{C}_{\mathrm{UC}}(\Phi, \dot{\Phi})\dot{\Phi}_{\mathrm{C}} \quad (37)$$

with each element $C_{\mathrm{U}}(i, j)$ and $C_{\mathrm{UC}}(i, j)$ defined as

$$C_{\mathrm{U}}(i, j) = \sum_{k=1}^{n} c_{N_{\mathrm{U}}(i), N_{\mathrm{U}}(j)}\dot{\phi}_k, \quad (38)$$

$$C_{\mathrm{UC}}(i, j) = \sum_{k=1}^{n} c_{N_{\mathrm{U}}(i), N_{\mathrm{C}}(j)}\dot{\phi}_k, \quad (39)$$

and likewise $\boldsymbol{B}_{\mathrm{U}}(\Phi)$, $\boldsymbol{B}_{\mathrm{UC}}(\Phi)$ have elements

$$B_{\mathrm{U}}(i, j) = b_{N_{\mathrm{U}}(i), N_{\mathrm{U}}(j)}, \quad (40)$$

$$B_{\mathrm{UC}}(i, j) = b_{N_{\mathrm{U}}(i), N_{\mathrm{C}}(j)}, \quad (41)$$

and further, $\boldsymbol{F}_{\mathrm{U}}(\Phi, \dot{\Phi})$ has element $F_{\mathrm{U}}(i)$ given by

$$F_{\mathrm{U}}(\phi_{N_{\mathrm{U}}(i)}, \dot{\phi}_{N_{\mathrm{U}}(i)}) := F_s(\Phi_{\mathrm{U}}) + F_v(\dot{\Phi}_{\mathrm{U}}). \quad (42)$$

**Source**

hence it has been assumed $G_U(\cdot) = -K_U(\cdot)$. The terms $C_U(\Phi, \dot{\Phi})$ and $C_{UC}(\Phi, \dot{\Phi})$ respectively have elements

$$C_{U,i,j} = \sum_{k=1}^{n} c_{\mathcal{I}_U(i), \mathcal{I}_U(j), k}\dot{\phi}_k, \quad C_{UC,i,j} = \sum_{k=1}^{n} c_{\mathcal{I}_U(i), \mathcal{I}_C(j), k}\dot{\phi}_k$$

and likewise $B_U(\Phi)$ and $B_{UC}(\Phi)$ have elements

$$B_{U,i,j} = b_{\mathcal{I}_U(i), \mathcal{I}_U(j)}, \quad B_{UC,i,j} = b_{\mathcal{I}_U(i), \mathcal{I}_C(j)} \quad (16)$$

$F_U(\Phi_U, \dot{\Phi}_U)$ has elements $F_{U,i}$ which take the form

$$F_{\mathcal{I}_U(i)}(\phi_{\mathcal{I}_U(i)}, \dot{\phi}_{\mathcal{I}_U(i)}) = F_{s,\mathcal{I}_U(i)}(\phi_{\mathcal{I}_U(i)}) + F_{v,\mathcal{I}_U(i)}(\dot{\phi}_{\mathcal{I}_U(i)}) := F_s(\Phi_U) + F_v(\dot{\Phi}_U) \quad (17)$$

**Figure 5.1.** Excerpt from a plagiarized engineering publication and its source.

### 5.1.1 Related Work and Research Gap

Math-based plagiarism detection is related to the research field of Math Information Retrieval yet exhibits characteristic differences. Math Information Retrieval mainly addresses three tasks [190, p. 410ff.]:

1. **Mathematical document retrieval**
2. **Formula retrieval**
3. **Document synthesis**

The objective in mathematical document retrieval and formula retrieval is to process a user query consisting of text, mathematical notation, or both and return a ranked list of documents or formulae that match the query [190, p. 410ff.]. Document synthesis describes the composition of a new document from retrieved fragments, which is mainly relevant for educational purposes. [190, p. 412]

Guidi & Sacerdoti Coen classify mathematical information in ascending order of the difficulty for being accessed by retrieval methods [190, p. 413]:

1. **Presentation**, i.e., information on the layout of mathematical expressions
2. **Content**, i.e., information on the structure of formulae and the ontological relations of symbols and operators in the formulae
3. **Semantics**, i.e., information on the definition of symbols and operators in the specific context

To retrieve mathematical information at the presentation level, researchers typically adapted text retrieval methods, such as specialized keyword indexes. The search engine MIaS (short for Math Indexer and Searcher) exemplifies this approach [473]. The system extends the text-based retrieval system Apache Lucene [31] to enable searching for documents using both text and mathematical queries.

To retrieve mathematical information at the content level, researchers commonly used formula representations that allow comparing the expression trees of the formulae. For example, the MathWebSearch system [214] represents formulae as trie data structures of substitutions, which the system stores in an Elasticsearch index [120]. The hierarchical trie data structures allow retrieving structurally similar formulae independent of the identifiers used in the formula.

To access the semantics of mathematical formulae, this information must be encoded explicitly, e.g., using specialized markup like Content MathML [561], or deduced from the formula's context. To deduce formula semantics from the context, researchers proposed adopting natural language processing methods to analyze the text surrounding mathematical expressions [288], [452].

The objective of Math-based Plagiarism Detection is to compare the mathematical expressions in a query document to the expressions contained in documents within an extensive collection and perform ranked retrieval of all documents with expressions that are similar beyond a chosen threshold.

The main differences between MathPD and mathematical document retrieval are query formulation and query processing. In mathematical document retrieval, the user formulates the query using a combination of search terms, query language operators, and mathematical features [190, p. 410f.]. In MathPD, the query is an entire document. The potential obfuscation of reused content is a threat to retrieval effectiveness specific to the MathPD task. Therefore, feature extraction for MathPD is more challenging than for mathematical document retrieval, as the extracted features should be robust against potential obfuscation.

To the best of our knowledge, the MathPD task has not been investigated before our studies on the subject [340], [344], [456]. Researchers in the Information Retrieval community have discussed the idea for MathPD informally at least since 2015. However, neither our analysis of 376 research papers for the two literature survey articles presented in Chapter 2 nor our contacts into the relevant research communities made us aware of prior research publications addressing this task.

## 5.2 Conceptualization of Detection Methods

To inform our design of math-based detection methods, we investigated confirmed plagiarism cases and performed initial experiments, as we describe hereafter.

### 5.2.1 Investigation of Plagiarism Cases

We collected 44 research publications that had been retracted for plagiarism and involved mathematical content. We found 39 of those publications by searching for documents that contain significant amounts of mathematics among 276 plagiarism cases that Halevi and Bar-Ilan [208] collected. We retrieved three additional cases from the Retraction Watch project [427] and another two cases from the VroniPlag project [536] (see Section 2.5.1, p. 55, for details on the two projects).

Four individuals with degrees in computer science (3), physics (1), and mathematics (1) reviewed the cases. To ensure that the reviewers could judge the appropriateness of similar mathematical content, we limited the collection to publications in computer science (six cases), mathematics (seven cases), and physics (four cases). We also included one case from bioengineering and one case from medical engineering, for which the retraction notices described the plagiarized mathematics.

Our observations from analyzing the 19 cases in our area of expertise were:

1. Most retracted publications contained significant amounts of mathematical expressions that were similar or identical to expressions in the source document and violated scientific practices.
2. Many retracted publications also contained (near-)duplicate text or figures.
3. Most shared mathematics in the retracted publications closely resembled the mathematics in the source and can be categorized as follows.

**Identical**: The mathematical expression in the retracted publication is an exact copy of the expression in the source document.

**Order changes**: The order of expressions in the retracted publication differs.

**Equivalent**: The expression structure in the retracted publication differs due to using the equivalence properties of commutativity, distributivity, and associativity.

**Different presentation**: The mathematical expression in the retracted publication is structurally and semantically identical to an expression in the source document, yet its presentation differs. Typical differences in the presentation we observed were the use of different identifiers, e.g., $v_t$ vs. $\theta_t$, function names, e.g., $\beta(x)$ vs. $f(x)$, or operator symbols, e.g., $\odot$ vs. $\oslash$ for the min-plus deconvolution.

**Splits or merges**: A combination of two or more expressions in the retracted publication is semantically identical to one expression in the source document ("split"), e.g., term substitutions or intermediate steps in a proof. We also found the opposite relation—"merged" expressions.

**Different concepts**: The retracted publication uses different yet semantically (nearly) identical mathematical concepts. Examples include the use of summation over vector components instead of matrix multiplication, discretization of expressions (e.g., transforming integrals into sums), or using multidimensional variables instead of multiple nested single-dimensional variables.

We expect that verbatim and slightly altered copies of mathematical expressions are overrepresented in our sample because they are easier to recognize for humans and likely identified more frequently. In two retracted publications, we encountered similarities of mathematics that are difficult to recognize and for which the legitimacy is hard to assess. In both cases, the authors combined content from two sources. The two retracted publications used their own notation but followed the order of ideas presented in the sources.

### 5.2.2 Devising Mathematical Feature Comparisons

Given the infancy of the research on MathPD, we evaluated options to identify the type of similar mathematics we had observed in the confirmed cases of plagiarism in initial experiments. Because most of the similar mathematics in those cases closely resembled the mathematics in the source, we evaluated methods that compare presentational elements of mathematical notation, such as identifiers, numbers, and operators. We configured our experimental MathPD methods as follows:

**Features**: We select the essential elements of mathematical notation, i.e., identifiers, numbers, and operators.

**Feature descriptors**: Because most mathematics in retracted publications differed slightly, approximate feature comparison methods like vector or set comparisons, histograms, and edit distances seem promising to identify many instances of plagiarized mathematics. Due to their robustness and speed of computation, we evaluated the effectiveness of histograms of the frequency of feature instances within a document or document partition. In other terms, we analyze how often a specific identifier, number, or operator occurs.

**Granularity**: We experimented with two granularities for the feature comparisons:

1. Entire documents
2. Document partitions.

We partitioned documents based on the number of characters into five equally sized partitions. The partitioning roughly reflects the typical research paper structure (introduction, related work, method, evaluation, and conclusion). We added 25% of the length of each partition as overlap to the previous and the following partition.

**Feature comparison**: We used a basic pairwise comparison of all feature descriptors for our initial experiments during the conceptualization phase.

**Similarity measures**: We evaluated two measures to compute the similarity of feature descriptors $K$. First, we computed the difference $d_e$ in the relative occurrence frequencies $f_e$ of features of a specific feature type $e$, i.e., identifiers (ci), numbers (cn), and operators (co)[10], as shown in Equation 5.1. The measure represents the absolute difference of the occurrence frequencies of a feature instance, i.e., the number of times a specific identifier, number, or operator occurs in two documents or partitions normalized by the sum of the larger of the occurrence frequencies of each feature instance in either of the two documents or partitions.

[10] The labels ci, co, and cn reflect the elements that denote identifiers, numbers, and operators in Content MathML markup. Appendix C, p. 231, presents the essentials of the MathML standard.

$$d_e(K, K') = \frac{\sum_{e'} |\, f_{e,K} - f_{e,K'} |}{\sum_e \max(f_{e,K}, f_{e,K'})} \tag{5.1}$$

Second, we computed the aggregated distance $D$ as the sum of the distances for specific feature types, as shown in Equation 5.2.

$$D = \sum_{e \in \{\text{ci},\text{cn},\text{co}\}} d_e \tag{5.2}$$

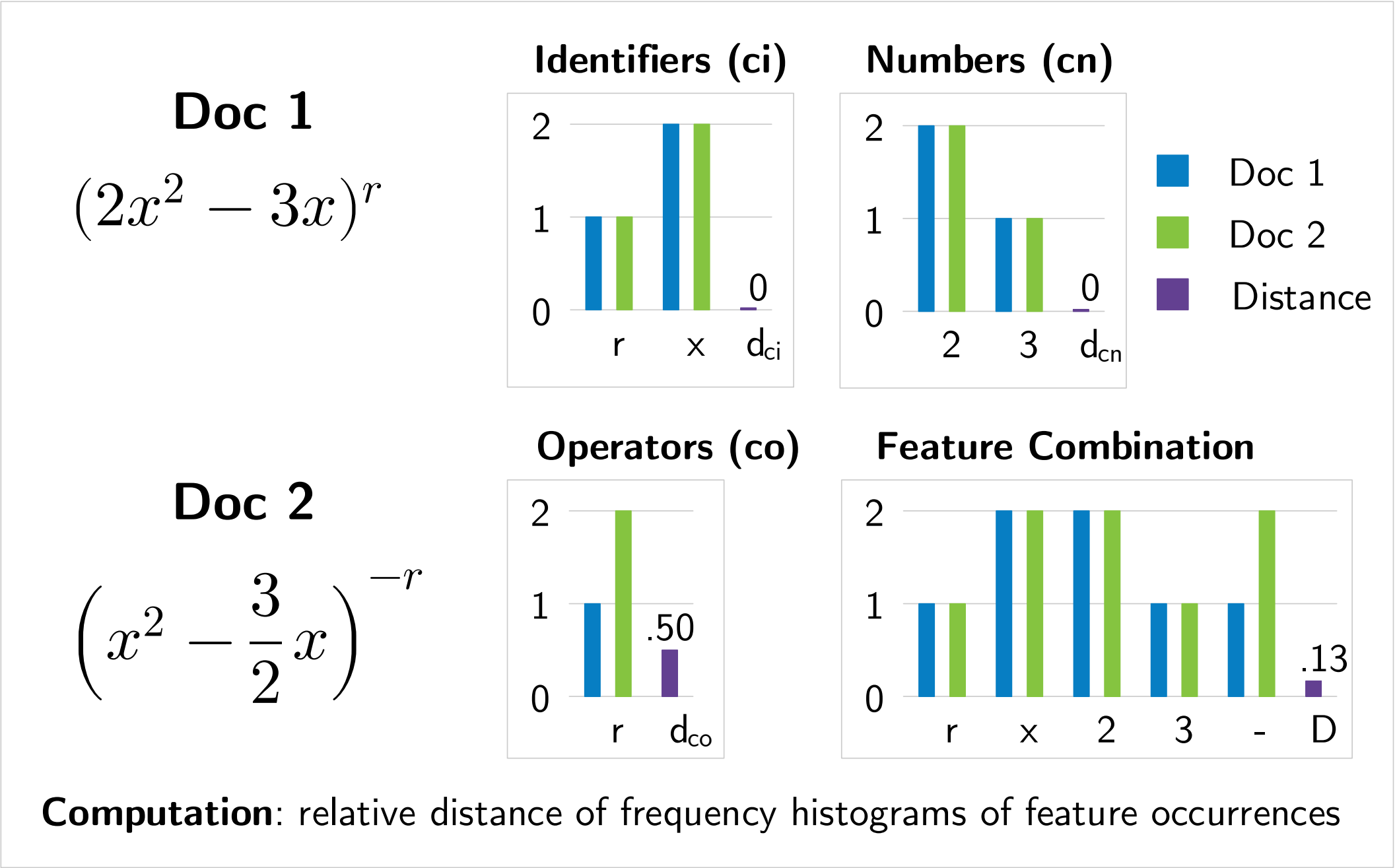

**Figure 5.2.** Illustration of the mathematical feature comparison.

**Figure 5.2** illustrates the feature comparison for two formulae that occur in two documents and consist of two identifiers ($r$ and $x$), two numbers (2 and 3), and one operator ($-$) each. The plots for identifiers, numbers and operators on the right-hand side of the figure show the occurrence frequencies and the relative distance in the occurrence frequencies (cf. Equation 5.1) for each of the three feature types. The plot labeled "Feature Combination" shows the computation of the aggregated distance considering all feature types (cf. Equation 5.2).

## 5.3 Evaluation Dataset

To evaluate math-based detection methods and compare their effectiveness to citation-based and text-based methods, we created a new dataset because no existing dataset offers mathematical content. See Section 2.5.2, p. 56, for summaries of existing datasets. **Figure 5.3** illustrates the process for creating the dataset.

We selected 10 publications we had investigated as test cases. Appendix B, p. 225, describes the test cases to which we refer as C1...C10. Selecting only 10 cases had four reasons. First, we chose cases from research fields within our area of expertise to enable us to assess the relevance of identified similarities. Second, we chose cases that are most representative of the types of mathematical similarity we observed. Third, our preprocessing of documents required manual checks of automatically extracted mathematical content, as we explain in Section 5.3.1. The effort required for this step prevented us from converting more cases. Fourth, we restricted the test cases to disciplines covered by the NTCIR-11 MathIR Task dataset [9].

We used the topically related NTCIR-11 MathIR Task dataset to create a reference collection. The NTCIR dataset includes about 60 million formulae from 105,120 scientific publications in computer science, mathematics, physics, and statistics. The dataset creators retrieved the publications from the arXiv [93] preprint repository in LaTeX format. We embedded the confirmed source documents for each of the test cases into the NTCIR dataset.

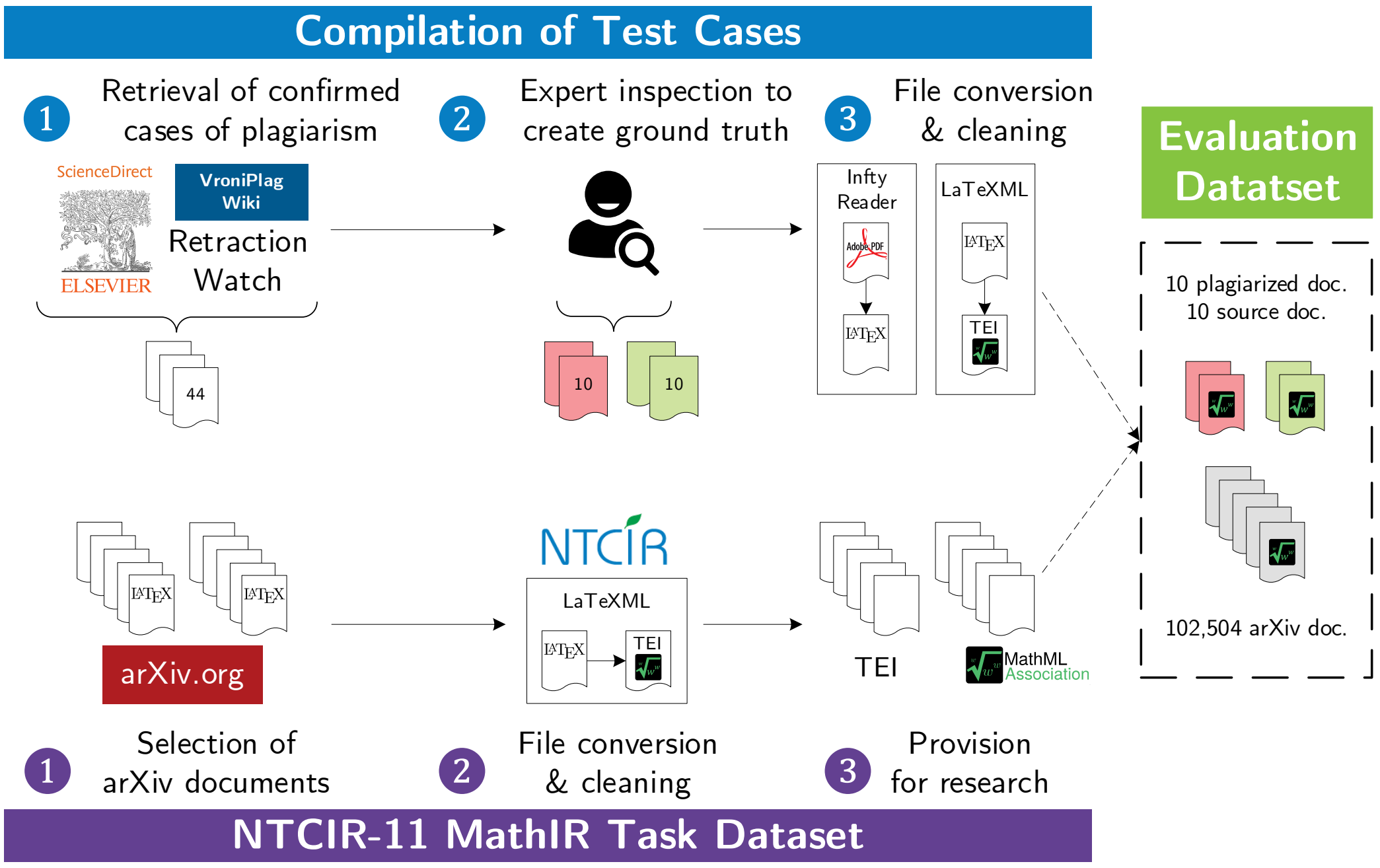

**Figure 5.3.** Creation of the evaluation dataset.

### 5.3.1 Document Preprocessing

We implemented all math-based detection methods in the HyPlag plagiarism detection system prototype, which Chapter 6 presents in detail. HyPlag preprocesses input documents in two steps. In the first step, the system converts the documents to a unified XML format used for the second preprocessing step. Our unified document format uses a subset of the Text Encoding Initiative (TEI) format [502] defined by the information extraction tool GROBID [308] to represent citations and bibliographic references. Additionally, the unified document format employs a subset of the Mathematical Markup Language (MathML) [561] to represent mathematical content. We employ parallel Presentation and Content MathML markup (see Appendix C for an overview of MathML). We refer to our unified document format as **HyPlag TEI (HTEI)**, which Section 6.2, p. 177, presents in detail.

Our evaluation dataset required processing documents in two formats:

1. PDF for the test cases
2. LaTeX source code for the documents of the NTCIR-11 MathIR Task dataset that form the reference collection

We used GROBID to obtain bibliographic references from documents in both formats because the tool achieved excellent results for extracting header metadata, citations, and references [44], [504].

Because GROBID cannot recognize mathematical formulae, we semi-automatically invoked InftyReader [498] to convert the PDFs for our test cases to an intermediate LaTeX format. InftyReader is the most commonly used OCR-based recognition system for mathematical content [241, p. 143]. While the tool typically achieves a recall of at least 0.90, the precision can be as low as 0.13 [241, p. 143]. To prevent bias from recognition errors, we manually checked the LaTeX output of InftyReader. For basic and moderately complex mathematical expressions, the extraction quality was high. For complex expressions involving uncommon notation, some manual cleaning of conversion errors was necessary.

We employed the LaTeXML [363] library to convert LaTeX documents, i.e., the NTCIR-11 dataset and the PDFs for our test cases that InftyReader converted to LaTeX, to the HTEI format. The LaTeXML library offers mathematical content conversion from LaTeX source code to a MathML representation. To enable conversion to our HTEI format, we contributed an XSL style sheet that transforms LaTeXML's native output to TEI. The new conversion option has been included in the LaTexML distribution [287].

To recognize citations, LaTeXML requires the use of LaTeX tags, such as `\cite{}`. Many documents in the dataset did not contain such markup but state citations as plain text. In such cases, our preprocessing pipeline did not recognize the citations. Additionally, many documents do not use citations at all but only reference items in the bibliography. Because of these errors, we obtained fewer unique citations than references for 68,743 documents (67% of the dataset). These **extraction errors** significantly reduced the effectiveness of the citation-based detection methods in our experiments, as we present hereafter.

In the second preprocessing step, HyPlag splits the HTEI document format into separate data structures holding plain text, mathematical formulae, citations, and bibliographic references. The system removes XML structures, images, formulae, and formatting instructions to extract the plain text. Formulae in Content MathML are extracted as they are. Citations are linked to the corresponding reference entries in the bibliography; reference entries are split into author, title, and venue fields.

### 5.3.2 Dataset Statistics

**Table 5.1.** Overview of the document preprocessing results.

| **Criterion** | **No. of documents** |
|---|---|
| NTCIR-11 MathIR Task dataset | 105,020 |
| - Thereof without extractable references (retained) | 6,770 |
| Parsing errors (excluded) | -2,616 |
| Test cases (10 retracted docs., 10 sources) | 20 |
| **Final dataset** | **102,524** |

**Table 5.1** summarizes the results of the preprocessing step. Of the 105,120 documents in the NTCIR-11 MathIR Task dataset, we excluded 2,616 documents, for which LaTeXML or our TEI parser encountered critical processing errors. Approximately one-third of the remaining documents did not contain markup for authors and titles. To achieve the best possible data quality, we used the API of the arXiv preprint repository [92] to obtain author and title information for all documents instead of extracting the information from the LaTeX source files. For 6,770 documents, we were unable to extract bibliographic references due to missing markup. Because the arXiv API does not offer bibliographic reference data, we indexed these documents without references.

**Table 5.2** shows the number of content features we obtained for the final dataset of 102,524 documents. The numbers confirm that this collection of STEM documents contains a significantly higher number of mathematical formulae (52 million) than academic citations (3 million). Therefore, analyzing both mathematical formulae and citations is more promising in these disciplines than analyzing citations alone. The formulae contain more than 156 million identifiers. On average, documents contained 1,513 mathematical identifiers; the average number of different identifiers per document is 70.

**Table 5.2.** Overview of content features in our evaluation dataset.

| Feature | Total | Avg. per doc. |
|---|---:|---:|
| References | 2,201,094 | 21 |
| Citations | 3,068,865 | 30 |
| Text fingerprints | 26,539,276 | 256 |
| Formulae | 52,271,908 | 504 |
| Mathematical identifiers | 156,706,600 | 1,513 |

## 5.4 Preliminary Experiments

In a first step, we evaluated the retrieval effectiveness of the basic comparison of mathematical features presented in Section 5.2.2, p. 149. The goal was to determine which mathematical features are most promising for the MathPD use case.

We computed the distances for all partition-partition and document-document pairs in the collection. We then ranked documents in ascending order of the distance score. For partitions, we only considered the lowest scoring partition pair.

### 5.4.1 Performance Measures

The ground truth for our test cases includes one known item of relevance, i.e., one source document for each retracted document. As is an established practice for known-item retrieval, we report the ranks at which a method retrieved the source documents because ranks are most descriptive of retrieval effectiveness [89].

We also report the Mean Reciprocal Rank, i.e., the average of the reciprocal ranks at which the system retrieves the relevant item for each query $q \in Q$, as shown in Equation 5.3. The 10 retracted documents are the queries. A method achieves the best possible score of 1 if it retrieves the source document at rank one for each test case. Thus, the Mean Reciprocal Rank measure quantifies the average retrieval effectiveness of a method.

$$\mathrm{MRR} = \frac{1}{|Q|}\sum_{i=1}^{|Q|}\frac{1}{\mathrm{rank}_i} \tag{5.3}$$

## 5.4.2 Results

**Table 5.3.** Ranks at which the correct source was retrieved.

| | Partitions | | | | Documents | | | |
|---|---|---|---|---|---|---|---|---|
| **Case** | **ci** | **cn** | **co** | **D** | **ci** | **cn** | **co** | **D** |
| C1 | 1 | 99,201 | 85,418 | 1 | 1 | 30,784 | 27,857 | 3,606 |
| C2 | 1 | 10,277 | 12,266 | 1 | 1 | 90,962 | 88,891 | 1 |
| C3 | 16 | 5,757 | 34,966 | 1 | 2 | 3,144 | 28,415 | 11,628 |
| C4 | 6 | 18,374 | 54,560 | 189 | 1 | 86 | 1,950 | 2,581 |
| C5 | 6 | 16,180 | 92,951 | 1 | 1 | 22,408 | 5,790 | 1 |
| C6 | 3 | 72,687 | 24,405 | 7,976 | 12 | 38,145 | 19,862 | 25,498 |
| C7 | 1 | 14,758 | 67,614 | 19,900 | 1 | 1,627 | 4,690 | 1 |
| C8 | 1 | 9,475 | 21,152 | 1 | 1 | 11,576 | 39,215 | 1 |
| C9 | 1 | 32,687 | 11,519 | 1 | 1 | 35,393 | 13,591 | 1 |
| C10 | 1,223 | 3,280 | 89,703 | 1 | 1 | 30,673 | 76,678 | 1 |
| **MRR** | 0.57 | <0.01 | <0.01 | **0.70** | **0.86** | <0.01 | <0.01 | 0.60 |

**Table 5.3** shows the ranks at which the feature comparison methods for identifiers (ci), numbers (cn), operators (co), and the aggregated features (cf. Equations 5.1 and 5.2, p. 150) retrieved the source partition or source document for each of the 10 test cases (C1...C10). The last row shows the Mean Reciprocal Rank (MRR) as a single-figure summary of a method's average ranking performance.

The two best-performing methods were analyzing the distance for identifiers and analyzing the aggregated distance ($D$). Analyzing the distances for numbers and operators on their own yielded poor results. The frequencies of these features appear to be too unspecific to be helpful for the MathPD use case.

When comparing feature descriptors for entire documents, analyzing the distance measure for identifiers performed best, retrieving eight of the 10 source documents at rank one (MRR=0.86). This result confirms our impression during the manual analysis that many identifiers in the retracted publications matched identifiers in the source documents. Identifier composition seems a valuable indicator of similarity, which in many cases is distinctive enough to retrieve the correct source from the collection of 102,504 documents. The aggregated distance measure failed to rank four of the source documents highly because the distance measures for numbers and operators introduced false positives.

When comparing feature descriptors for document partitions, analyzing the distance for identifiers retrieved five of the 10 source documents at rank one. The combined distance measure retrieved seven of the 10 documents at rank one (MRR=0.70). This result suggests that the pattern of selectively taking over content nearly verbatim in confined parts of a document known as shake and paste [551, p. 8f.] also applies to mathematical content. In such cases, including the distance information on numbers and operators, which are too unspecific at the document level, can improve the similarity assessment for parts of a document.

In summary, our preliminary experiments supported our initial hypothesis that analyzing mathematical similarity holds promise for identifying plagiarism in math-heavy disciplines. An exclusive analysis of basic presentational mathematical features in most cases yielded a similarity score distinctive enough to retrieve the correct source from a collection of more than 100,000 documents at the top rank.

Of the features we analyzed in our preliminary experiments, mathematical identifiers achieved the best overall retrieval effectiveness. Therefore, we focused on analyzing identifiers for designing a complete math-based plagiarism detection process, which we present in the subsequent section.

## 5.5 Math-based Detection Process

Our preliminary experiments used pairwise comparisons of all feature descriptors to determine the maximum retrieval effectiveness achievable by analyzing mathematical features. This approach is computationally too expensive for collections that are significantly larger than our evaluation dataset.

To enable applying Math-based Plagiarism Detection for realistic use cases, we devised a detection process consisting of a candidate retrieval and a detailed analysis stage. We included text-based and citation-based detection methods in the process to enable comparing and combining the different approaches to plagiarism detection. We implemented the detection process in our plagiarism detection system, HyPlag, which we also used to inspect results. **Figure 5.4** illustrates the detection process, whose key components we present in the following subsections.

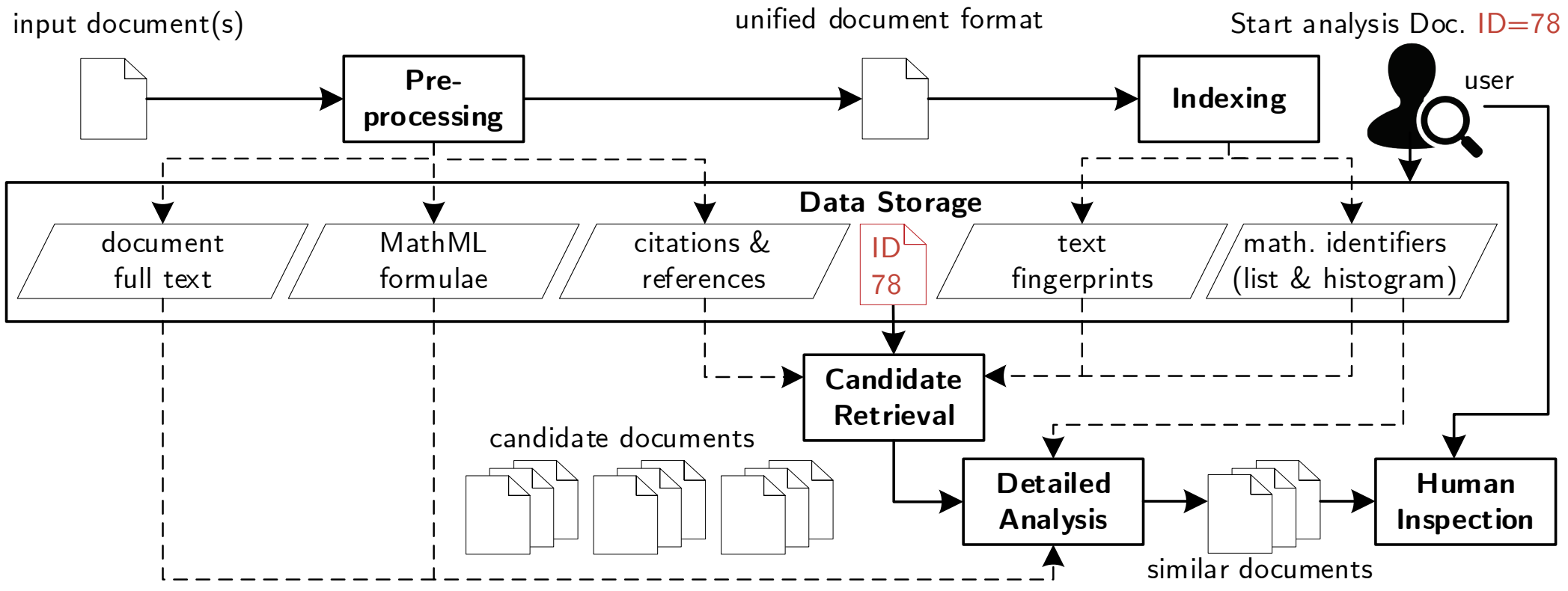

**Figure 5.4.** Overview of the math-based detection process.

## 5.5.1 Indexing

The indexing stage follows the document preprocessing stage presented in Section 5.3.1, p. 152. During the indexing step, HyPlag stores into an Elasticsearch index [120] the following data extracted from the preprocessed documents:

» **Document metadata**: title, authors, publication date, and filename.

» **Mathematical features**: the sequence of all mathematical identifiers in the order of their occurrence in the document and the unordered histogram of the occurrence frequencies of identifiers, i.e., how often an identifier occurs in the document. We used the MathML element `<formula>` to extract formulae, and the `<ci>` elements to extract identifiers in a formula.

» **Citation features**: bibliographic references and citations. The system consolidates the data on referenced documents by comparing the title and author names extracted from reference strings to the data of previously indexed documents while accounting for minor spelling variations utilizing the Levenshtein distance presented in Section 2.4.2, p. 23.

» **Textual features**: the full text and text fingerprints formed by chunking the document into word 3-grams and applying probabilistic chunk selection (average chunk retention rate 1/16). We adapted the Sherlock tool to realize the text fingerprinting method. We also used Sherlock to evaluate citation-based plagiarism detection methods (cf. Section 3.4.20, p. 95).

The indexing process is identical for all documents, i.e., documents that shall be analyzed need to be indexed first.

## 5.5.2 Candidate Retrieval

In the candidate retrieval stage, the system queries the index using mathematical identifiers, citations, and text fingerprints extracted from an input document to retrieve a set of candidate documents for the subsequent detailed analysis.

We employed "Lucene's practical scoring function" implemented in the Elasticsearch server as a computationally efficient, well-performing heuristic to retrieve candidate documents. The scoring function combines a tf-idf-weighted vector space model with a Boolean retrieval approach [119].

We performed three queries, each retrieving the 100 documents with the highest relevance scores. For the citation-based and text-based retrieval of candidate documents, in-text citations, and, respectively, text fingerprints of the input document were the terms of the query. Indexed documents were modeled as sets of citations and text fingerprints. We used the default parameters of Lucene's scoring function.

For the math-based retrieval of candidate documents, the mathematical identifiers in a document were the query. The sequence of mathematical identifiers represented the indexed documents, i.e., identifiers can occur more than once. Using Lucene's default parameters for the relevance scoring yielded unsatisfactory results in the case of mathematical features. This finding is in line with research by Sojka & Líška [473]. Similar to this prior work, we found that query terms, i.e., mathematical identifiers, should receive additional weight for multiple occurrences. Therefore, we set the boost value boost($t$) for the term $t$ in the query, i.e., identifiers, to the number of occurrences of the term (identifier) in the query document.

Because we sought to independently analyze the math-based, citation-based, and text-based detection methods, we did not consolidate the three sets of 100 candidate documents retrieved by the three queries.

## 5.5.3 Detailed Analysis

In the detailed analysis stage, HyPlag compares the input document(s) to each of the three sets of 100 candidate documents. For each document comparison, the system computes the similarity measures we present hereafter.

### Math-based Similarity Measures

HyPlag only computes math-based similarity scores for document pairs that share 20 or more identifiers to prevent high similarity scores resulting from a few shared identifiers, such as the occurrence of $x$ and $y$. For documents that meet this threshold, the system computes the following three similarity measures.

The **Identifier Frequency Histograms (Histo) measure** (see Equation 5.1, p. 150) reflects the global overlap of identifiers in two documents or segments. It achieved the best retrieval effectiveness in our preliminary experiments. The measure is most suitable for documents having comparable numbers of identifiers. Typically, documents that differ significantly in length do not fulfill this criterion.

To evaluate measures that consider the order of identifiers, we adapted the Longest Common Subsequence (cf. Section 3.3.2, p. 86) and the Greedy Tiling measures (cf. Section 3.3.3, p. 87) to the MathPD use case. Both measures achieved good results in our evaluation of citation-based detection methods (cf. Section 3.5.1, p. 103).

The **Longest Common Identifier Sequence (LCIS)**, i.e., the largest number of identifiers that match in two documents $d$ and $d'$ in the same order but not necessarily in a contiguous block, quantifies the global similarity of $d$ and $d'$ as

$$s_{\text{LCIS}}(d, d') = \frac{|L(d, d')|}{I_d}.$$

The score represents the number of identifiers in the query document $I_d$ that are part of the longest common identifier sequence whose length is given by $L$.

The **Greedy Identifier Tiles (GIT)** measure, which considers all individually longest blocks of consecutive shared identifiers in identical order (so-called tiles), quantifies the similarity of two documents as

$$s_{\text{GIT}}(d, d') = \frac{|T_l|}{I_d},$$

where $T_l$ is the set of tiles with a length greater or equal to five matching identifiers and $I_d$ is the number of identifiers in the query document. In other terms, the score reflects the number of identifiers in the query document that are part of identifier tiles with a minimum length of five.

## Citation-based and Text-based Detection Methods

For the citation-based similarity assessment, we used the Bibliographic Coupling **(BC)**, Longest Common Citation Sequence **(LCCS),** and Greedy Citation Tiling **(GCT)** methods (cf. Section 3.3, p. 86). All three methods proved effective in our evaluation of Citation-based Plagiarism Detection.

For the detailed text-based analysis, we used the character 16-gram matching method Encoplot **(Enco)** [187], which we also evaluated in our experiments on Citation-based Plagiarism Detection (cf. Section 3.4.2, p. 95).

## 5.6 Evaluation Methodology

To evaluate the effectiveness of the plagiarism detection methods, we performed two conceptually different investigations. For both investigations, we used the dataset of our preliminary experiments, i.e., 10 confirmed cases of plagiarism embedded into the NTCIR-11 MathIR Task dataset.

The first investigation reflects the typical use case in external plagiarism detection, i.e., checking an input document for similarity to documents in a reference collection. We submitted the retracted publication for each test case to our system. For each query document, the system used the math-based, citation-based, and text-based candidate retrieval algorithms to retrieve three sets of 100 documents. In the subsequent detailed analysis stage, HyPlag compared each query document to all the candidate documents in the three sets without consolidating the sets. Section 5.7 presents the results of this investigation.

The second investigation assesses the effectiveness of combining math-based and citation-based detection methods to discover unknown instances of potentially suspicious document similarity. We submitted each of the $N = 102{,}524$ documents in our dataset to HyPlag. We retrieved the three sets of candidate documents by applying the math-based, citation-based, and text-based candidate retrieval algorithms for all $N$ documents. Opposed to the evaluation of confirmed plagiarism cases, we formed the union of the sets to explore approaches that combine the detection methods. In the detailed analysis stage, we compared each of the $N$ documents in the dataset to its consolidated set of candidate documents $C$. We manually examined the retrieved documents with the highest similarity scores. Section 5.7.2, p. 167, presents the results of this investigation.

## 5.7 Results

We divide the presentation of results according to the two evaluations we performed. First, we describe the results of our small-scale evaluation of confirmed cases of academic plagiarism. Subsequently, we present the findings of our exploratory search for unknown cases of potential academic plagiarism.

### 5.7.1 Confirmed Cases of Plagiarism

**Table 5.4** shows the effectiveness of the **candidate retrieval algorithms**. Plus signs (+) in the table indicate that HyPlag retrieved the source document among the 100 candidate documents when the retracted document for each of the 10 test cases (C1 ...C10) was the query. Minus signs (−) indicate that an algorithm did not retrieve the source document among the candidate documents. The rightmost column in the table shows the recall of the three conceptually different algorithms.

Both the citation-based and the text-based candidate retrieval algorithms achieved a recall of 0.9; the math-based candidate retrieval algorithm achieved a recall of 0.7. Notably, the three conceptually different algorithms failed to retrieve the source document among the candidates for distinct sets of test cases. Combining any two sets of candidate documents would result in a perfect recall.

**Table 5.4.** Recall of candidate retrieval algorithms.

| | C1 | C2 | C3 | C4 | C5 | C6 | C7 | C8 | C9 | C10 | *R* |
|---|---|---|---|---|---|---|---|---|---|---|---|
| Mathematics | + | + | + | − | − | − | + | + | + | + | 0.7 |
| Citations | + | + | − | + | + | + | + | + | + | + | 0.9 |
| Text | + | + | + | + | + | + | − | + | + | + | 0.9 |

**Legend: C1**...**C10** IDs of test cases, ***R*** Recall

To quantify the effectiveness of the detection methods for the **detailed analysis** stage, we performed a score-based and a rank-based assessment.

## Score-based Assessment

This assessment served to determine the significance of similarity scores for our detection methods and dataset, i.e., the thresholds above which we would consider scores potentially suspicious. No prior study had quantified the mathematical similarity that can be expected by chance to derive a significance threshold.

To establish significance thresholds for the scores of all detection methods, we analyzed a random sample of one million document pairs as follows. We randomly picked two documents from the dataset. If the chosen documents had (a) common author(s) or one of the documents cited the other, we discarded the pair. We continued the process until reaching the number of one million document pairs. The selection criteria should eliminate document pairs that exhibit high content similarity for likely legitimate reasons, i.e., authors reusing their own work or referring to the work of others with due attribution.

Our goal was to estimate an upper bound for similarity scores that likely result from random feature matches. To do so, we computed the similarity scores for each of the document pairs using all similarity measures. Then, we manually assessed the topical relatedness of the top-ranked document pairs for each similarity measure. We picked as the significance threshold for a similarity measure the rank of the first document pair for which we could not identify a topical relatedness. **Table 5.5** shows the significance scores we derived using this procedure.

**Table 5.5.** Significance thresholds for method-specific similarity scores.

| | Histo | LCIS | GIT | BC | LCCS | GCT | Enco |
|---|---|---|---|---|---|---|---|
| $s$ | $\geq .56$ | $\geq .76$ | $\geq .15$ | $\geq .13$ | $\geq .22$ | $\geq .10$ | $\geq .06$ |

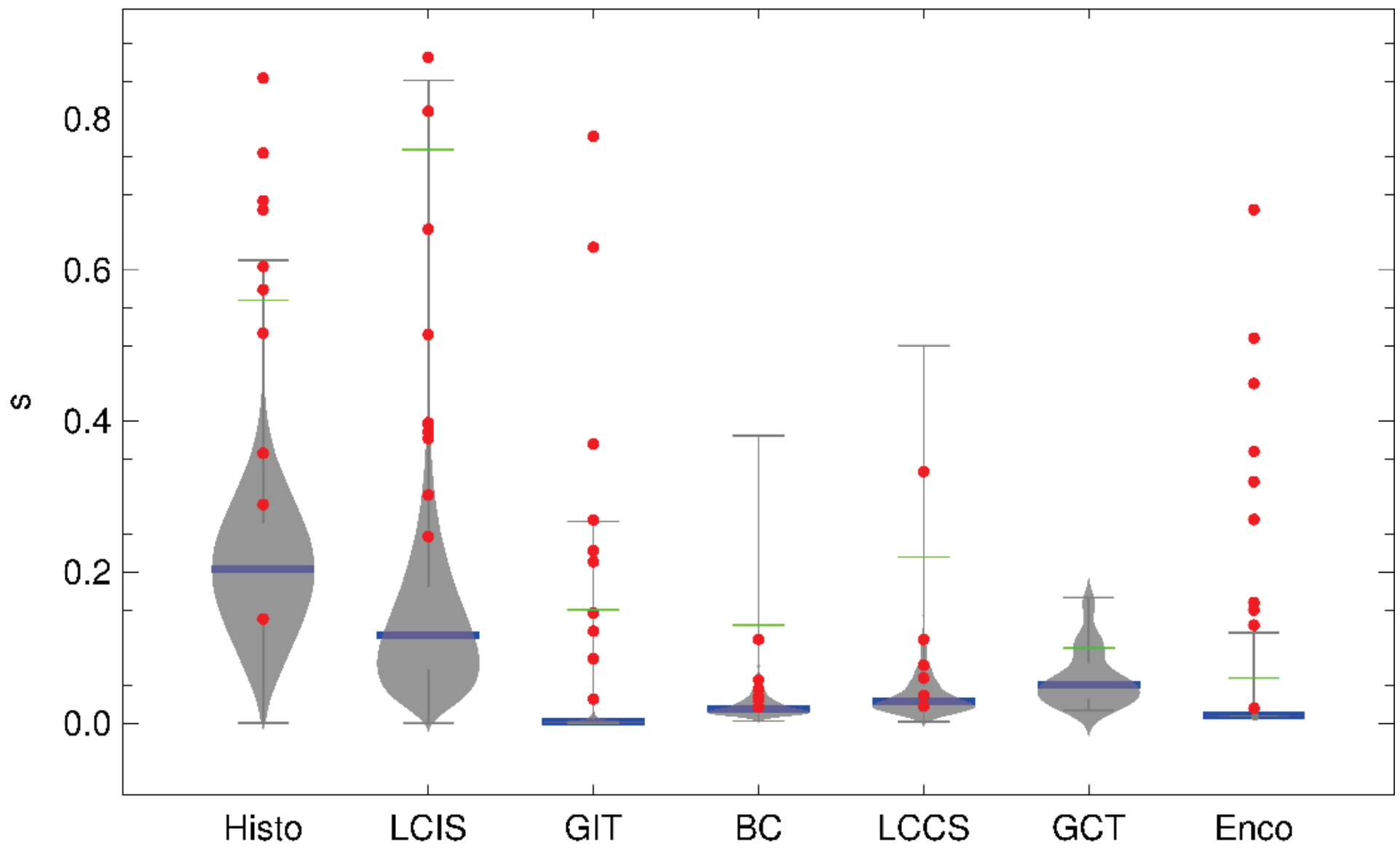

**Figure 5.5.** Distribution of similarity scores in one million document pairs.

**Figure 5.5** shows the distribution of the similarity scores $s$ (vertical axis) computed using each similarity measure for the sample of one million document pairs to investigate the properties of the similarity measures. Horizontal bars indicate the median score (blue shade), minimum and maximum scores (grey shade), and the significance thresholds for each measure (green shade). The grey shapes in the chart show the smoothed **probability density function (PrDF)** of the score frequencies, which we generated by applying a kernel-based density estimation. Red dots in the plot indicate the similarity scores of test cases for which a measure was applicable, i.e., if the document pairs contained enough features to compute a score.

The PrDF of Histo is symmetrical, while the PrDF for all other measures is negatively skewed, i.e., exhibits the highest frequencies at lower scores. The stronger the PrDF of scores is negatively skewed, the more selective the measure is. For the math-based measures (i.e., Histo, LCIS, GIT), considering the order of identifiers increased drastically the selectivity of the measures. The PrDF for the order-agnostic Histo measure is symmetrical. The PrDF of scores for the LCIS measure, which leniently considers the order of identifiers, is slightly skewed towards lower values. In contrast, the PrDF for the GIT measure, which reflects identifier order within confined regions of a document, is strongly skewed towards lower values.

Given the results of our experiments on citation-based similarity (cf. Section 3.5, p. 103), we expected that the citation-based similarity measures would exhibit similar characteristics as the math-based measures. However, as shown in **Figure 5.5**, the order-agnostic BC measure is more selective than the order-considering LCCS measure for our dataset. The reason is the errors in citation extraction (cf. Section 5.3.1, p. 153). The mismatch of references and citations caused the LCCS and GCT measures only to consider a fraction of the citations in the dataset. This fraction is smaller than the fraction of extracted references, which the BC measure uses. Therefore, the BC measure is more selective than the LCCS measure for this dataset because overlaps of the comparably sparse citations increased the LCCS score more than shared references increased the BC score. Unrecognized citations also cause the GCT measure to be overly selective for this dataset. Due to a shortage of data points, the PrDF for scores of the GCT measure shows interpolation artifacts, i.e., the PrDF is not monotonically decreasing for higher scores.

The PrDF of the Encoplot scores shows that the text-based measure is highly selective. Nine of the test cases have scores above the significance threshold, i.e., most of the confirmed cases of academic plagiarism have a significant textual overlap with the respective source. As we explain in Section 2.5.1, p. 49, this characteristic is common for known cases of plagiarism because finding verbatim text overlap is easier for reviewers and better supported by production-grade PDS than identifying concealed content similarity. Therefore, documents with (near-)duplicate text are more likely to be discovered and hence likely overrepresented in our dataset.

### Combined Rank-based and Score-based Assessment

In addition to assessing the significance of the similarity scores, we also examined the ranks $r$ at which the detection methods retrieved the source document for each test case when performing both the candidate retrieval and detailed analysis step. To indicate the average ranking performance of the methods, we computed the Mean Reciprocal Rank (see Equation 5.3, p. 155). In the case of tied ranks, we considered the mean rank, i.e., the pessimistically rounded average of the number of document pairs that share the same rank.

**Table 5.6** shows the results of the rank-based and the score-based assessment. For each test case (C1...C10), the table lists the rank $r$ at which HyPlag retrieved the source document and the score $s$ that the detection method assigned. We mark the mean rank, which we used in the case of tied ranks, with an apostrophe, e.g., 7'. Scores above the significance threshold of a method are underlined. To gauge the performance of the detection methods specifically for the detailed analysis stage, we also state the ranks and similarity scores for the cases not retrieved in the candidate retrieval stage. We mark such entries with parentheses, e.g., (0.15). To compute the ranks and scores for these documents, we compared the query document to all documents in the dataset. Hyphens indicate that HyPlag computed no similarity score due to the exclusion criteria of the detection methods. Because of the incomplete and error-prone extraction of bibliographic data, we state a separate score $s^*$ for the citation-based methods. The score indicates the actual citation-based similarity of the test cases. To compute $s^*$, we manually corrected erroneous data for citations and references before applying the detection methods.

The text-based detection method consisting of word 3-gram fingerprinting (Sherlock) for the candidate retrieval stage and efficient string matching (Encoplot) for the detailed analysis stage achieved the best individual result. The method retrieved nine of the 10 test cases at the top rank. Only test case C7 exhibited a textual similarity that was too low to retrieve the source document in the candidate retrieval stage and mark the document as suspicious in the detailed analysis stage.

The Encoplot scores for six of the 10 test cases exceeded 0.25. Hence, they are clearly suspicious. For C1, C2, and C4, the Encoplot scores exceed our significance threshold of 0.06 yet are lower than 0.20. Reports[11] suggest that a text overlap of 10%–20% is not immediately suspicious but often tolerated by journal reviewers and editors. The practices regarding acceptable text overlap vary between research fields and even between venues. Whether a production-grade plagiarism detection system would flag C1, C2, and C4 as suspicious is thus unclear. The retraction note of C1 names the unattributed reuse of a mathematical model, not the textual overlap with the source, as the reason for the retraction [562]. The scores for Histo (0.68) and GIT (0.21), which both exceed the significance thresholds, reflect this similarity in mathematical content.

---

[11] Higgins et al. used the PDS iThenticate to check submissions to a medical journal for plagiarism. They found that a similarity score of 15% achieved the best tradeoff between sensitivity and specificity for classifying manuscripts as plagiarized or original [227, p. 3]. Other, anecdotal reports support this finding. The question which percentage of similarity in publications is generally treaded as acceptable received 261 replies on the social networking site ResearchGate by June 2020. Most of the replies suggested percentages were in the range of 10%–20% [389].

| | Mathematics | | | | | | Citations | | | | | | | | | | Text | |
|---|---|---|---|---|---|---|---|---|---|---|---|---|---|---|---|---|---|---|
| | Histo | | LCIS | | GIT | | BC | | | LCCS | | | GCT | | | | Enco | |
| Case | ***r*** | ***s*** | ***r*** | ***s*** | ***r*** | ***s*** | ***r*** | ***s*** | ***s****$^*$ | ***r*** | ***s*** | ***s****$^*$ | ***r*** | ***s*** | ***s****$^*$ | | ***r*** | ***s*** |
| C1 | 1 | .68 | 1 | .40 | 1 | .21 | 1 | .06 | .15 | 1 | .06 | .10 | - | - | .04 | | 1 | .13 |
| C2 | 1 | .60 | 1 | .39 | 1 | .12 | 10' | .05 | .28 | 1 | .33 | .42 | - | - | - | | 1 | .16 |
| C3 | 3 | .29 | 1 | .88 | 1 | .78 | - | - | - | - | - | - | - | - | - | | 1 | .36 |
| C4 | (1) | (.36) | (99) | (.37) | (3) | (.03) | - | - | .35 | - | - | .44 | - | - | .25 | | 1 | .15 |
| C5 | (1) | (.57) | (86) | (.30) | (1) | (.23) | 5 | .02 | .18 | 7' | .02 | .23 | - | - | .05 | | 1 | .45 |
| C6 | (19) | (.14) | (98) | (.40) | (1) | (.15) | 2 | .04 | .32 | 1 | .11 | .44 | - | - | .22 | | 1 | .27 |
| C7 | 2 | .52 | 98 | .25 | 1 | .09 | - | - | .04 | - | - | .05 | - | - | - | | (4) | (.02) |
| C8 | 1 | .76 | 1 | .65 | 1 | .37 | 1 | .11 | .37 | - | - | .25 | - | - | - | | 1 | .32 |
| C9 | 1 | .69 | 1 | .51 | 1 | .27 | 1 | .03 | .26 | 1 | .08 | .39 | - | - | - | | 1 | .68 |
| C10 | 1 | .85 | 1 | .81 | 1 | .63 | 1 | .03 | .03 | 1 | .04 | .04 | - | - | - | | 1 | .51 |
| **MRR** | .58 (.79) | | .60 (.60) | | .79 (.93) | | .48 (.48) | | | .60 (.60) | | | .00 (.00) | | | | .90 (.93) | |

**Legend:**
***r*** rank at which the source document was retrieved, ***s*** similarity score, ***s****$^*$ citation-based similarity score without extraction errors, (...) candidate retrieval step did not retrieve the source document, it was added manually to evaluate the detailed analysis step, – no similarity score computed due to method-specific exclusion criteria, 10**'** mean rank considered because the ranks were tied, ### similarity score above the method-specific significance threshold, **MRR** Mean Reciprocal Rank

**Table 5.6.** Retrieval effectiveness for confirmed cases of plagiarism.

The math-based detection methods achieved the second-best result when considering both the candidate retrieval and detailed analysis stages. GIT performed particularly well, retrieving seven cases at the top rank. Considering the detailed analysis stage alone, GIT achieved the same effectiveness as the text-based analysis (nine test cases retrieved at rank one, MRR=0.93). This result could be achieved by simply combining the results of any two candidate retrieval algorithms, i.e., math-based, citation-based, and text-based, as we show in Table 5.4, p. 161.

GIT outperformed the Histo method. In our initial experiments, Histo achieved a slightly higher MRR score of 0.86 compared to 0.79 in the present evaluation. We attribute the difference to using a slightly different preprocessing process. The good performance of GIT further supports our hypothesis that reusing (nearly) identical content in confined parts of a document known as shake and paste [551, p. 8f.] is also observable for mathematical content.

For our test cases, LCIS achieved no significant improvement over the set-based Histo method. Both LCIS and Histo achieved good results for test cases that share a significant fraction of their mathematical content. For such documents, the amount of shared mathematics sufficed to retrieve the documents using the Histo method. The substantial overlap in mathematical content also yielded long identifier subsequences. However, they did not significantly improve the similarity score.

The citation-based methods achieved the lowest overall performance, primarily due to the deficiencies of the extracted data. Nevertheless, the LCCS method retrieved five cases at rank one (MRR=0.60). The similarity scores $s^*$, which assume the bibliographic data would have been extracted and matched correctly, give a better indication of the potential effectiveness of the citation-based methods. Notably, LCCS would yield scores of approximately twice the significance threshold of 0.22 and hence strongly suspicious for C2, C4, C6, and C9. Given that C2 and C4 exhibit a significant but not necessarily suspicious text overlap (0.16 and 0.15), the high LCCS score could indicate suspicious similarity.

For all cases except C7, which none of the measures flagged as suspicious, at least one math-based or citation-based method yielded a similarity score above the respective significance thresholds. For Case C7, the Histo score is closest to the method's significance threshold, making Histo the most likely method to retrieve the case despite the comparably low score.

In summary, the evaluation using confirmed cases of academic plagiarism showed that the combined analysis of math-based and citation-based similarity identified all cases that also a text-based analysis flagged as strongly suspicious. Moreover, the two non-textual detection approaches provide valuable indicators for suspicious document similarity for cases with a comparably low textual similarity.

### 5.7.2 Exploratory Study

This section describes our findings from manually examining the top-ranked documents that HyPlag retrieved when applying math-based and citation-based methods to compare each document of the dataset to its set of candidate documents.

Given the size of the result set (about six million document pairs) and our primary goal of searching for undiscovered cases of academic plagiarism, we employed several filters to focus our investigations on the most critical similarities. To eliminate cases in which authors likely reused their own content, we excluded document pairs that shared at least one author. This exclusion prevents the identification of potential self-plagiarism. Similarly, we pruned document pairs, for which the document published later cites the earlier publication, to reduce results in which authors reproduced previous work with due attribution.

We applied these restrictions for two reasons. First, the definition of self-plagiarism varies significantly in different research fields and even for different venues. The vagueness of the definition prevents a well-founded assessment of the retrieved documents. Second, because we analyze all documents in the dataset, the number of results is much larger than in the typical plagiarism detection use case, i.e., analyzing a single input document.

The primary objective of this evaluation was gauging the benefit a math-based similarity assessment adds to a combined detection process. Therefore, we excluded documents with a Histo score below 0.25, i.e., with little identifier similarity, and sorted the remaining results by their GIT score in descending order. To not exclude cases in which documents contained unequal amounts of identifiers, e.g., because one document was significantly shorter, we did not require a Histo score above the significance threshold of 0.56 but only a score greater or equal to 0.25.

**Table 5.7** shows the 10 top-ranked document pairs and our rating of the similar content we observed in the documents. The case IDs C1...C10 correspond to the confirmed plagiarism cases described in Appendix B, p. 225. We assigned the case IDs C11...C18 to previously unseen document pairs in sequential order of the rank at which HyPlag retrieved the document pairs. Many of these documents represent legitimate content similarity. Therefore, we do not reference the documents directly to prevent academic search engines from associating the documents to a publication on plagiarism detection. Such an association may reflect negatively on the authors. The metadata and full texts of all cases are available at http://thesis.meuschke.org.

**Table 5.7.** Top-ranked documents in an exploratory study.

| Rank | Case ID | Rating |
|---|---|---|
| 1 | C3 | Confirmed plagiarism case |
| 2 | C11 | Suspicious content similarity |
| 3 | C12 | Notable legitimate content reuse |
| 4 | C13 | False-positive detection |
| 5 | C10 | Confirmed plagiarism case |
| 6 | C14 | False-positive detection |
| 7 | C15 | Notable legitimate content reuse |
| 8 | C16 | Notable legitimate content reuse |
| 9 | C17 | Notable legitimate content reuse |
| 10 | C18 | Notable legitimate content reuse |

The highest-ranked document pair is the confirmed case of plagiarism C3. The author of the retracted publication copied three geometric proofs with few changes from a significantly longer publication [1], which caused a high GIT (0.78) but a low Histo score (0.29). HyPlag retrieved another confirmed case of plagiarism (C10) at rank five. The main contribution of the retracted paper in C10, a model in nuclear physics, was taken from the source publication while partially renaming identifiers [109]. Almost all the mathematical content of the retracted publication overlaps with the source document, resulting in the highest Histo score (0.85) in our exploratory study. The differences of identifiers in the source document and the retracted document result in a lower but still suspiciously high GIT score (0.63).

The later publication in C11 (rank two) is a mixture of idea reuse and content reuse. The author of the later publication reused the argumentative structure, sequence of formulae, several cited sources, many descriptions of formulae, and non-trivial remarks about the implications of the research from the earlier paper. By doing so, the author of the later publication derived a minor generalization of an entropy model for a specific type of black holes introduced in the earlier publication. The later publication cites other publications by the author of the earlier publication but not the apparent source publication itself.

We contacted the author of the earlier publication about our findings. In his view, the later publication: "[...] **certainly is a case of plagiarism** [...]." In coordination with the author of the earlier publication, we contacted the journal that published the later publication. The journal's editorial board currently examines the case. Since the journal has not published an official determination about the legitimacy of the publication in question, we classify the document as suspicious. This case exemplifies the benefits of a combined math-based, citation-based, and text-based similarity analysis. Only a combined analysis reveals the full extent of content similarity that encompasses approximately 80% of the document's content.

The five cases of legitimate content reuse (C12, C15, C16, C17, and C18) exhibit similar characteristics. In all five cases, the authors of the later publications reproduce and duly cite extensive mathematical models proposed in the earlier publications. HyPlag failed to recognize the citations and exclude the document pairs due to two challenges. First, the use of severely abridged citation styles, e.g., only stating the author name(s) and the arXiv identifier of a publication. Second, some authors cited the arXiv preprint of a publication, whereas other authors cited the journal version. The journal versions regularly exhibited differences in the order of authors and the title compared to the respective arXiv preprints. Our preprocessing pipeline did not handle either case correctly. HyPlag's procedures for extracting and disambiguating such challenging references require improvement.

However, retrieving these five document pairs at top ranks is justified, given the overlap in mathematical content (typically multiple pages). We expect reviewers would appreciate being made aware of such content overlap, e.g., to verify the correct citation of the previous work.

The two false positive detections, C13 (rank four) and C14 (rank six), reveal potential improvements of the math-based detection methods. C13 comprises two publications in Combinatorics that contain long lists of all possible combinations of the identifiers $a$, $b$, and $c$ according to a set of production rules. The two publications in C14 analyze partition functions and contain long matches entirely made up of the identifiers $p$ and $q$ occurring in large quantities within unrelated formulae.

To prevent such false positives in the future, we plan to devise measures confined to individual formulae. Likewise, we plan to research how an assessment of structural and semantic similarity of formulae can be adapted for the plagiarism detection use case. Research on mathematical information retrieval of the content and semantics of mathematical expressions (cf. Section 5.1.1, p. 146) has provided approaches that could prove valuable for the plagiarism detection use case.

The research on Math-based Plagiarism Detection is infant. Like the early methods for text reuse detection, we explored basic, computationally efficient methods to identify the reuse of identical and slightly altered mathematical content. Our investigations show that a math-based analysis increases the detection capabilities for STEM documents, particularly when combined with other detection methods.

## 5.8 Conclusion Math-based PD

By proposing Math-based Plagiarism Detection, we initiated applied research on analyzing mathematics to identify academic plagiarism in math-heavy STEM publications. By reviewing research on Math Information Retrieval, we showed that MathPD represents a previously unaddressed retrieval task.

By collecting and manually reviewing confirmed cases of academic plagiarism that involve mathematics, we derived insights on common characteristics of plagiarized mathematical content. The findings of this investigation guided the conceptualization process for math-based plagiarism detection methods.

We created a large-scale evaluation dataset by embedding confirmed cases of academic plagiarism involving mathematical content into a topically related collection of 102,504 documents collected from the arXiv preprint repository.

Using the dataset, we evaluated the retrieval effectiveness of order-agnostic comparisons of presentational mathematical features for the detailed analysis stage of the external plagiarism detection process. We found that among the presentational features, identifiers performed best for retrieving typical instances of similar mathematics we observed in confirmed cases of academic plagiarism.

Given the results of our preliminary experiments, we focused on the analysis of mathematical identifiers for designing a math-aware two-stage plagiarism detection process. We devised a computationally efficient candidate retrieval stage that analyzes mathematical features, academic citations, and textual features using production-ready information retrieval technology. Moreover, we created the Greedy Identifier Tiling and Longest Common Identifier Sequence measures, which consider the order of mathematical identifiers, for the detailed analysis stage. We implemented the newly developed math-based detection methods, citation-based, and established text-based detection methods in HyPlag—a working prototype of a hybrid plagiarism detection system.

Using the HyPlag system and our evaluation dataset, we compared the effectiveness of the math-based, citation-based, and text-based detection methods. We showed that a simple unification of the modestly sized sets of candidate documents retrieved by any two candidate retrieval algorithms, e.g., text-based and citation-based or math-based and citation-based, achieved perfect recall for the candidate retrieval stage. For the detailed analysis stage, the GIT method exceeded the effectiveness of the best performing method (Histo) in our preliminary experiments and achieved equal effectiveness as the text-based detection methods.

Errors in the acquisition of citations and bibliographic references decreased the effectiveness of the citation-based detection methods in our experiments. Despite these limitations, citation-based methods contributed significantly to the hybrid detection process, particularly for the candidate retrieval stage. The Longest Common Citation Sequence measure also performed decently for the detailed analysis stage (MRR=0.60 for our test cases). The error-corrected similarity scores showed that the actual effectiveness of the citation-based methods is much higher.

The combined analysis of math-based and citation-based similarity identified all cases that a text-based analysis flagged as strongly suspicious. Moreover, the two non-textual detection methods (mathematics and citations) provided valuable indicators for suspicious document similarity for cases with a comparably low textual similarity. This result indicates that the best detection effectiveness is achievable by combining heterogeneous similarity assessments.

In an exploratory study, we showed the effectiveness of analyzing math-based and citation-based similarity for discovering unknown cases of potential academic plagiarism. We used the GIT and Histo methods in combination with the citation relations of documents to reduce a result set of approximately six million document pairs to 10 document pairs that we investigated manually.

The highest-ranked document pair was a confirmed case of plagiarism. The document retrieved at the second rank was rated as an undiscovered case of academic plagiarism by the author of the apparent source document. The remaining eight cases included one confirmed case of academic plagiarism, five documents with a high but legitimate overlap in mathematical content, and two false positives. The citation-based filter would have eliminated five cases of legitimate content reuse if the bibliographic data had been extracted correctly. These results show the immense potential of analyzing mathematical content and academic citations to complement text-based plagiarism detection methods.

The data and code of our experiments are available at http://thesis.meuschke.org.

# 6

Chapter 6

# Hybrid Plagiarism Detection System

**Contents**

This chapter describes the working prototype of a plagiarism detection system that combines the analysis of citation patterns, images, mathematical content, and text. Our system, HyPlag, implements all detection methods presented in Chapters 3–5 and provides a web-based user interface for examining the detection results.

We structure the presentation of the HyPlag system as follows. Section 6.1 gives a functional and technical overview of the system. Section 6.2 describes the backend that provides all detection functionality and data storage. Section 6.3 presents the frontend that realizes the user interface. Section 6.4 summarizes our contributions presented in this chapter.

## 6.1 System Overview

The main objective of our hybrid plagiarism detection system, HyPlag, is to improve the identification of potentially suspicious content similarity, particularly in research works, such as journal and conference publications, Ph.D. theses, and grant proposals. The system's target audience is reviewers of such works, e.g., journal editors, Ph.D. advisors, and reviewers of grant proposals.

**Figure 6.1** shows a functional overview of the system. HyPlag follows the design principle of a multi-stage detection process consisting of candidate retrieval, detailed comparison, post-processing, and human inspection.

In the candidate retrieval stage, the system employs:

1. The **citation-based detection methods** presented in Section 3.3, p. 86, i.e., Bibliographic Coupling, Longest Common Citation Sequence, Greedy Citation Tiling, and Citation Chunking (both documents chunked, considering consecutive shared citations only, no merging step);
2. **Word $n$-gram fingerprinting** using the Sherlock plagiarism detection tool described in Section 3.4.2, p. 96;
3. The **math-based candidate retrieval** algorithms that employ an Elasticsearch query for mathematical identifiers and an adaption of the Apache Lucene scoring function as presented in Section 5.5.2, p. 158;
4. The **image-based detection process** described in Section 4.4, p. 127.

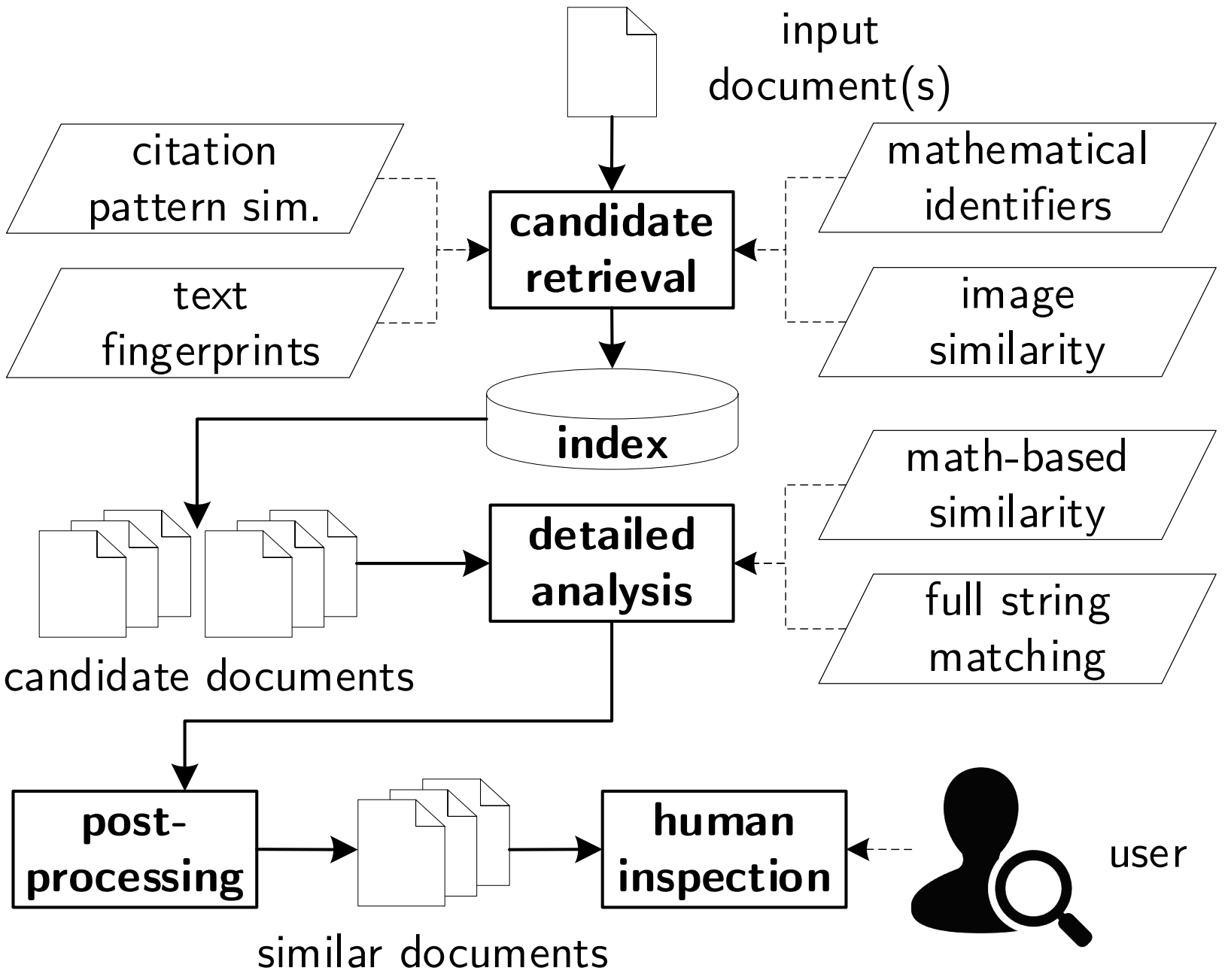

**Figure 6.1.** Functional overview of the hybrid plagiarism detection system.

Users can select which detection methods the system applies. For each of the four citation-based methods, the text-based method, and the math-based method, HyPlag retrieves up to 100 documents in the candidate retrieval stage.

For the image-based candidate retrieval method, the number of retrieved candidates depends on the image type. As described in Section 4.4.2, p. 129, HyPlag selectively applies the image-based detection methods depending on their suitability for the specific image type. The system retrieves up to nine candidate documents for photographs and bar charts as it exclusively analyzes such images using perceptual hashing or Ratio Hashing, respectively. For other image types, the system retrieves up to 27 candidate documents (up to nine candidates each for perceptual hashing and both OCR text matching methods). The lower numbers of image-based candidate documents are due to the scoring function included in the image-based detection process (see Section 4.4.8, p. 135). The function pre-filters candidate documents according to their likelihood of being suspicious. The other methods do not include a comparable assessment, which is why we retrieve more candidates for these methods. The system forms the union of all method-specific sets of candidate documents and passes them to the detailed analysis stage.

In the detailed analysis stage, HyPlag performs pairwise comparisons of the input document(s) to each candidate document using:

1. The **math-based similarity measures** Identifier Frequency Histograms (Histo), Longest Common Identifier Sequence, and Greedy Identifier Tiles presented in Section 5.5.3, p. 158.
2. The character 16-gram text-matching algorithm of **Encoplot**, which ignores repeated matches as described in Section 3.4.2, p. 95. To enable identifying also repeated matches, we included an adaption of the Boyer-Moore string matching algorithm [60] that finds all matching substrings of six or more tokens if they include at least 12 characters.

In the postprocessing stage, HyPlag applies several heuristics to prevent frequently occurring issues, such as excluding text-based matches in formulae.

HyPlag stores the detection results to enable asynchronous communication with the frontend and the caching of results. Communicating asynchronously with the frontend allows performing long-running jobs typical for the plagiarism detection use case. Long runtimes may result from performing computationally expensive analyses or processing many input documents as a batch job, e.g., checking all submissions for an assignment. Results caching reduces the computational effort for repeated analysis requests for the same document. Users can choose whether to store input documents or results and, if so, for how long.

**Figure 6.2** shows the main components and data flow in our system from a technical perspective. The system consists of a Java-based backend and a web-based frontend implemented in Ruby on Rails. The backend and frontend are loosely coupled via a REST API through which the backend exposes all detection functionality and offers access to the stored data.

The frontend allows users to explore the detection results through interactive visualizations, which Section 6.3, p. 179, presents in detail.

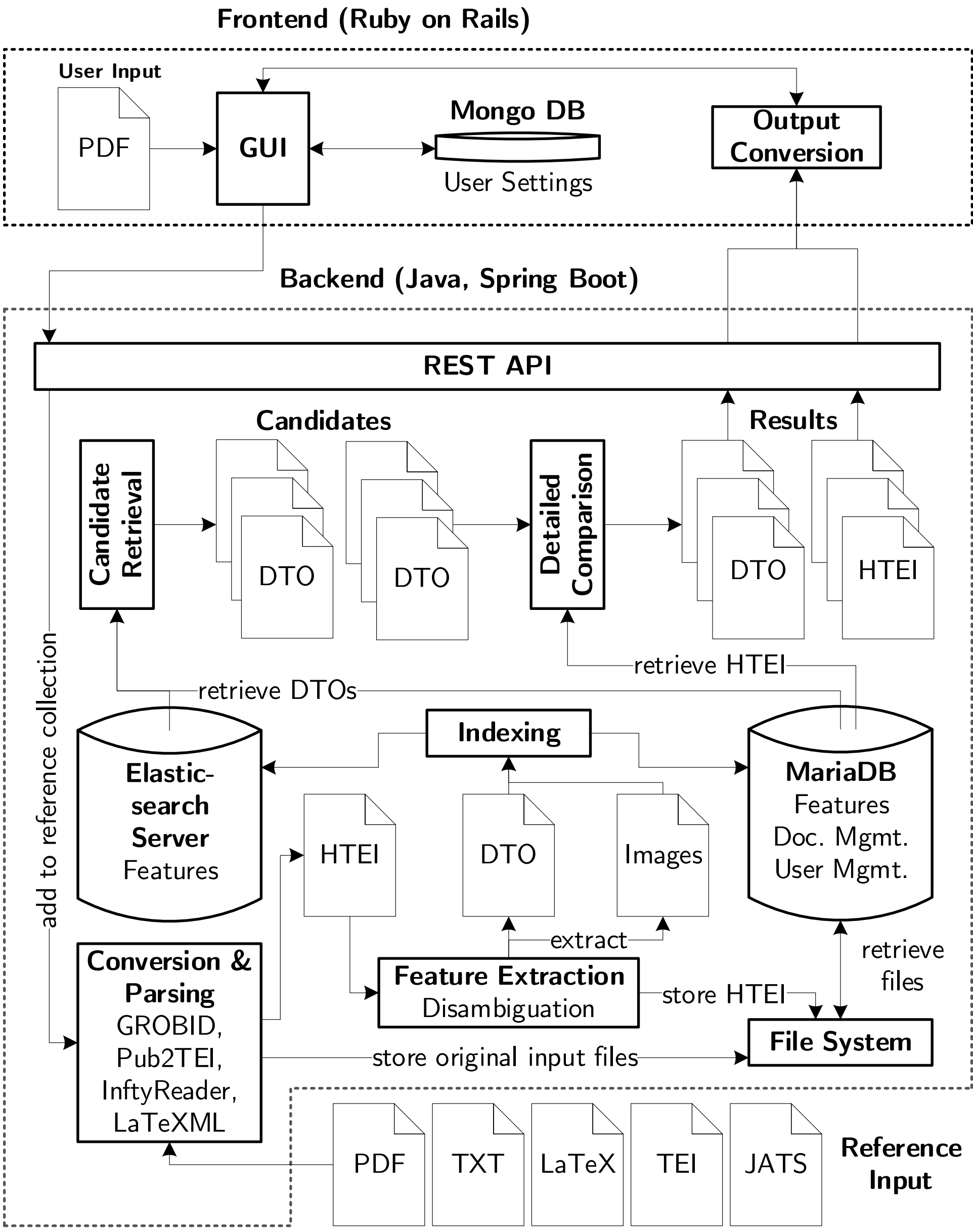

**Figure 6.2.** Technical system architecture.

## 6.2 Backend

The backend's fundamental software architecture derives from the Spring Boot [478] application framework for the Java [376] programming language. The backend includes a MariaDB [324] relational database as its primary data storage and an Elasticsearch server [120] as a search engine.

### 6.2.1 Supported Document Formats

The backend can currently process documents in five formats: PDF, plain text, LaTeX, Text Encoding Initiative Guidelines for Electronic Text Encoding and Interchange (TEI), and Journal Article Tag Suite (JATS).

JATS [360] is an XML format that the National Center for Biotechnology Information (a subunit of the US National Library of Medicine) [362] developed to represent documents in their open access scientific document repository PubMed Central [361]. In 2012, the American National Standards Institute (ANSI) and the US National Information Standards Organization (NISO) adopted JATS as a joint standard [28]. JATS focuses on describing typical content elements of academic journal publications for presentation, archiving, and interchange purposes.

TEI is an XML format developed by the Text Encoding Initiative consortium [501]. The format has a broader scope and offers more customization options than JATS [502]. While JATS focuses on scientific journal publications, TEI originates from efforts to digitally encode texts in literary and linguistic disciplines [502, p. xxiii]. For example, TEI includes XML tags for verses and transcribed speech. Today, TEI allows representing virtually any type of document and content. The format also allows customizing the extensive default tag collection for specific use cases.

HyPlag uses a customized TEI subset [311] defined by the information extraction tool GROBID [308] to represent documents for storage and display. If applicable, we include in this format Parallel MathML markup (see Appendix C, p. 231) for mathematical expressions. We refer to this format as HyPlag TEI (HTEI).

### 6.2.2 Conversion and Parsing

HyPlag converts documents in any supported format to the HTEI format, extracts the content features necessary for analysis, and adds the documents to its index. This process also applies to documents a user submits for checking since the system adds these documents to the reference collection before triggering the analysis.

To convert PDF and plain text documents to the HTEI format, HyPlag uses the GROBID[12] information extraction library [308]. GROBID repeatedly achieved excellent results for extracting and parsing metadata and content, such as authors, affiliations, citations, and bibliographic references, from unstructured document formats, such as PDF and plain text [44], [504]. The tool employs a machine learning approach based on Conditional Random Fields to label content in unstructured document formats with the tags defined by the tool's custom TEI subset. Numerous well-known applications and services, such as ResearchGate, Mendeley, HAL Research Archive, the European Patent Office, and the Internet Archive, use the GROBID library as part of their production settings for document conversion and information extraction [310]. Extracting mathematical expressions from PDF documents currently requires a semi-automated conversion of the PDF to LaTeX using InftyReader [498] (as we describe in Section 5.3.1, p. 152).

To process documents in JATS format, HyPlag employs the Pub2TEI library [309]. The library provides Extensible Stylesheet Language Transformations (XSLT) schemas and stylesheets for converting several other XML formats for encoding academic documents into the GROBID TEI subset.

To convert LaTeX documents to the HTEI format, we employ the LaTeXML library [363], which we extended with a use-case-specific XSLT stylesheet [287].

### 6.2.3 Feature Extraction & Indexing

After converting input documents to the HTEI format, HyPlag extracts the main text, header metadata (e.g., authors, title, and venue), citations, references, images, and mathematical content using the respective HTEI tags. The features are stored as internal **data transfer objects (DTO)**.

HyPlag disambiguates documents and bibliographic references, which it also stores as document entities, during the extraction step to prevent storing duplicates. For this purpose, we check whether the newly added item, i.e., document or reference, includes identifiers, such as a Digital Object Identifier (DOI), an International Standard Book Number, or a PubMed ID that matches items in our reference collection. Additionally, we perform approximate string matching using the Levenshtein distance on authors and titles to identify likely duplicates.

The indexing step propagates the extracted content features to the Elasticsearch server and the relational database. The primary task of the Elasticsearch server is performing the candidate retrieval step for all but the image-based detection process. Consequently, HyPlag adds the information on citations, bibliographic references, mathematical identifiers, and text fingerprints (hashed word 3-grams) to the Elasticsearch index. We chose Elasticsearch because it is an open-source, production-grade search engine that is scalable to virtually any collection size.

[12] GROBID is an abbreviation for GeneRation Of BIbliographic Data

The relational database serves as the primary storage for all document data, cached results, and user account information. The indexing step transfers all data, e.g., images, formulae in MathML, document metadata, and the main text of documents from the feature-specific DTOs to the database. The database records also link to the original files and the converted HTEI files in the file system.

### 6.2.4 Detection Process

The detection process realizes the method-specific candidate retrieval, detailed analysis, and postprocessing steps for each detection method. The backend allows limiting the analysis to specific documents. The frontend uses this option to enable a collusion check, i.e., letting the user pick the documents the backend then compares pairwise instead of querying the input document to the entire collection.

The system sends the results of the detection process as HTTP responses via the REST API to the frontend. The results include the ranked list of documents with similar content and the identified similarities in JSON format. Additionally, the backend provides the HTEI files of all documents involved in the analysis. Most browsers can display the XML-based HTEI files natively.

## 6.3 Frontend

The HyPlag frontend is a Ruby on Rails web application that uses JavaScript for interactive content, Cascading Stylesheets for styling, and a MongoDB no-SQL database for managing front-end user accounts and their settings.

The user interface consists of three functional views:

1. The **dashboard view** for managing documents, analyses, and results
2. The **results overview** for browsing retrieved candidate documents
3. The **detailed analysis view** for performing a side-by-side comparison of the input document and one of the retrieved candidate documents

### 6.3.1 Dashboard View

**Figure 6.3** shows the initial dashboard view presented to users after they log into the frontend. The asymmetric two-column layout of the view consists of a navigation panel ① and an organization panel ② on the left, a central main column ③, and a minimalistic top menu ④. The top menu contains links to the dashboard, the about page, and the user profile settings.

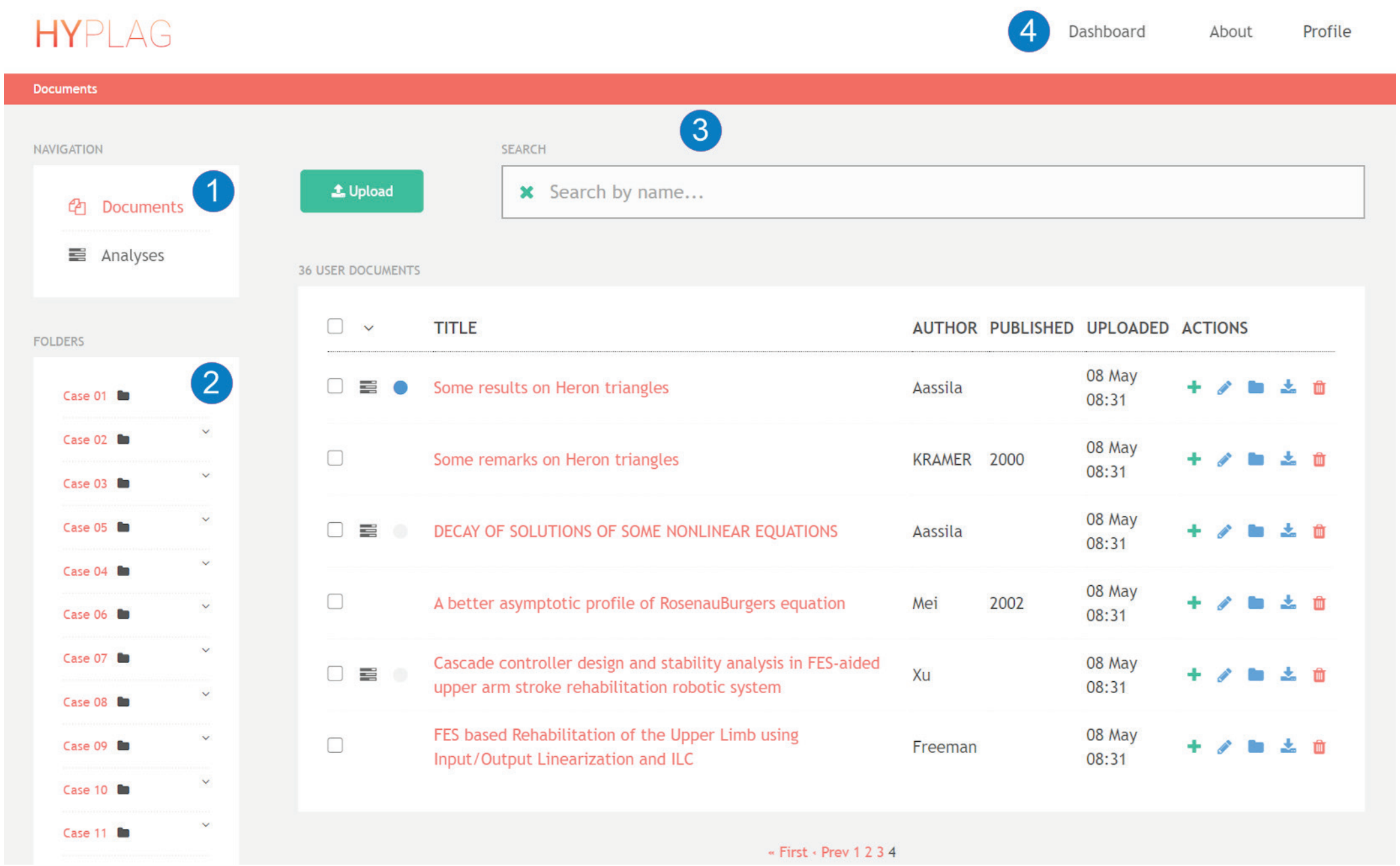

**Figure 6.3.** Dashboard view in the HyPlag system.

The main component of the dashboard is a paginated table showing the documents submitted for analysis (default view) or the analyses triggered for documents (available via the navigation panel ①). The upload button allows submitting documents (PDF or plaintext) to the system. The search box allows finding documents within the table by searching for authors, title, or year of publication.

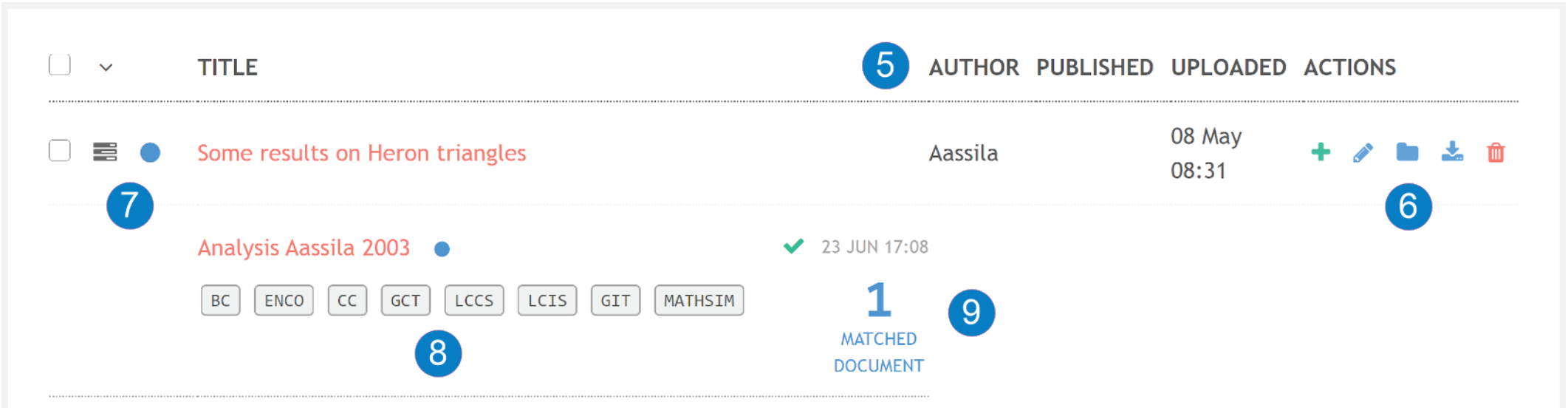

**Figure 6.4.** Document entry in the dashboard view.

**Figure 6.4** shows a close-up view of an expanded document entry in the central table. For each document, the table lists metadata, such as authors, title, and year of publication 5. The last column of the table contains action buttons 6 that allow starting a news analysis for the selected document (+), editing the document's metadata ( ), placing the document into a folder ( ), downloading the original file ( ), and deleting the document ( ). The status icons 7 indicate whether one or more analyses are available for the document ( ) and whether the results are unseen ( ). Clicking on the analysis icon ( ) expands the document entry and shows a list of all analyses the system performed for the document. For each analysis, the system shows which detection methods were applied 8 and summarizes the results, i.e., whether the analysis was successful (✓) and how many documents with similar content were retrieved 9. Clicking on the analysis title opens the results overview for the document (cf. Section 6.3.2, p. 182).

Clicking the icon + in the list of action buttons for a document 6 opens a dialog window for starting a new analysis. In the first step, the user can select the detection methods the system shall apply (see **Figure 6.5**)

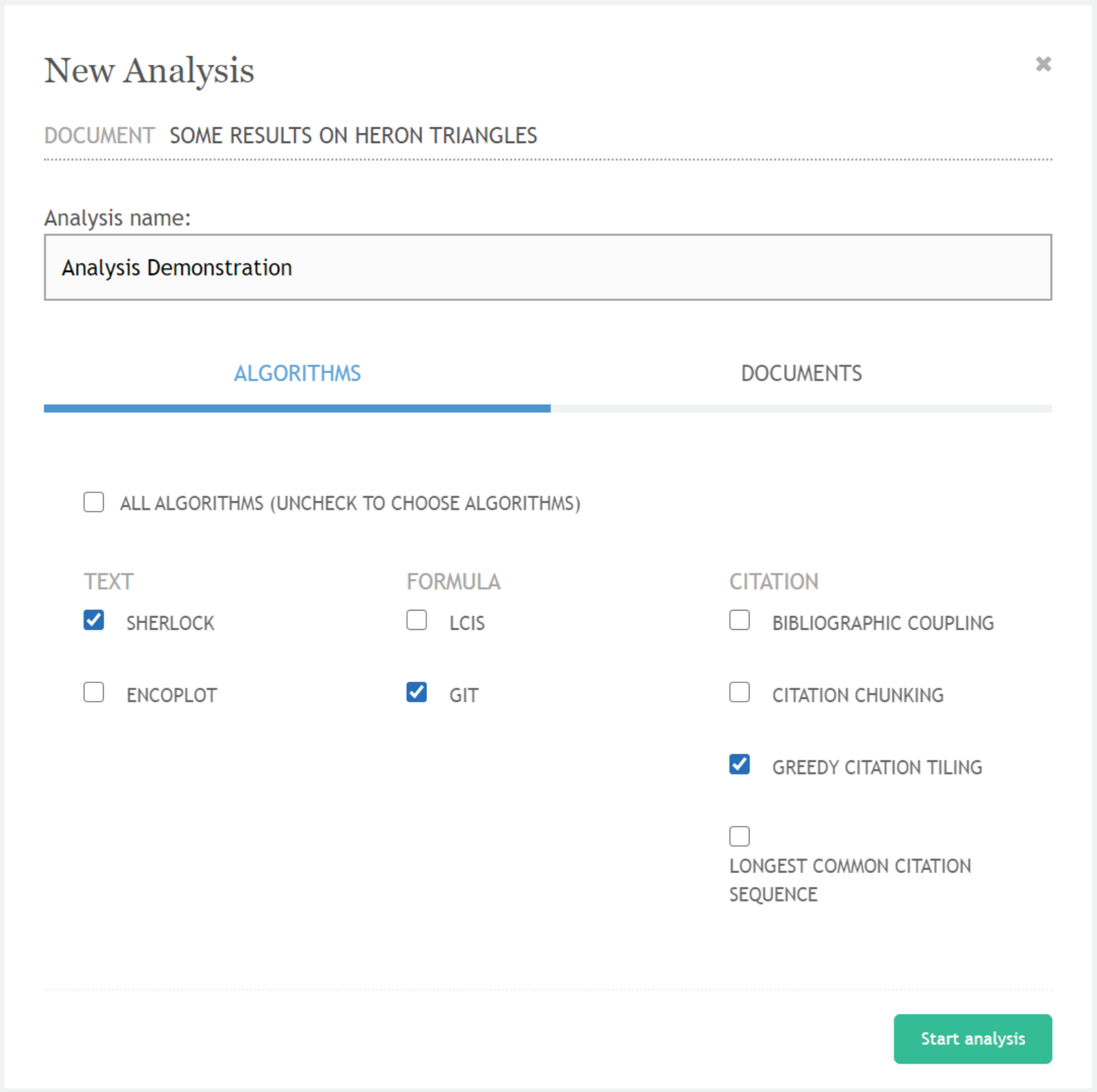

**Figure 6.5.** Starting a new analysis: Selection of detection methods.

In a second step, the user can choose whether the system shall compare the selected document to the entire collection, which is the default setting, or perform pairwise comparisons to one or more selected documents (see **Figure 6.6**). The latter option allows performing collusion checks.

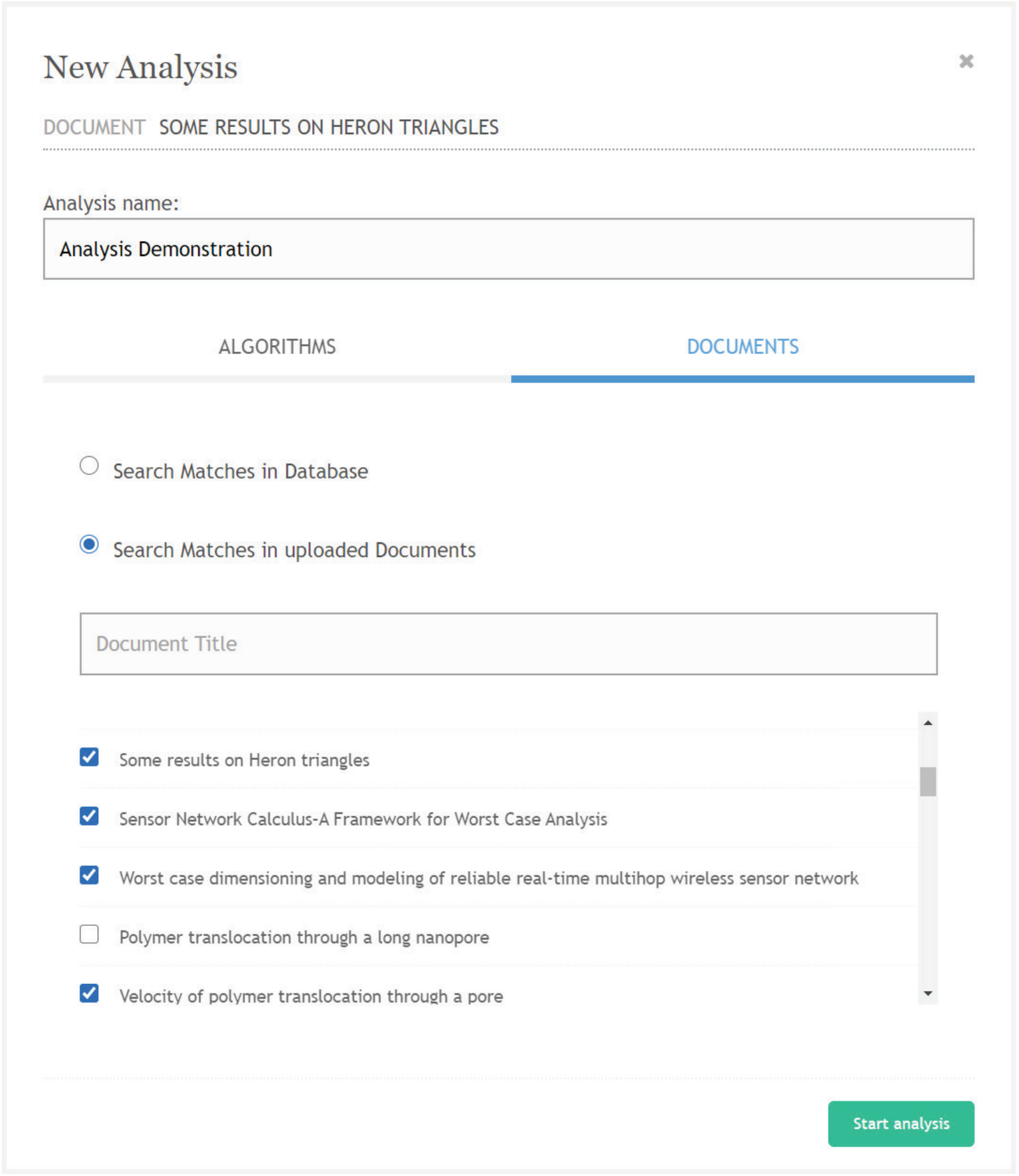

**Figure 6.6.** Starting a new analysis: Selection of comparison documents.

## 6.3.2 Results Overview

**Figure 6.7** shows the results overview, which is the first screen a user sees after selecting to examine the results of an analysis. The left part of the view shows the full text of the input document 1. The right part shows a list of abstract representations (so-called result summaries 2) for all documents the system retrieved.

Each result summary includes one or more match views 3. Each match view has two panels and represents the similarities that a detection method identified, e.g., matching citations or similar formulae. The left panel 4a represents the input document, and the right panel 4b the comparison document. Matching features appear in the match views connected by lines. The positions of features in the match views reflect their relative positions in the documents. This visualization displays similar features in the same order as parallel lines. Such patterns are a strong indicator of potentially undue content similarity. The system displays features in each match view using a unique color.

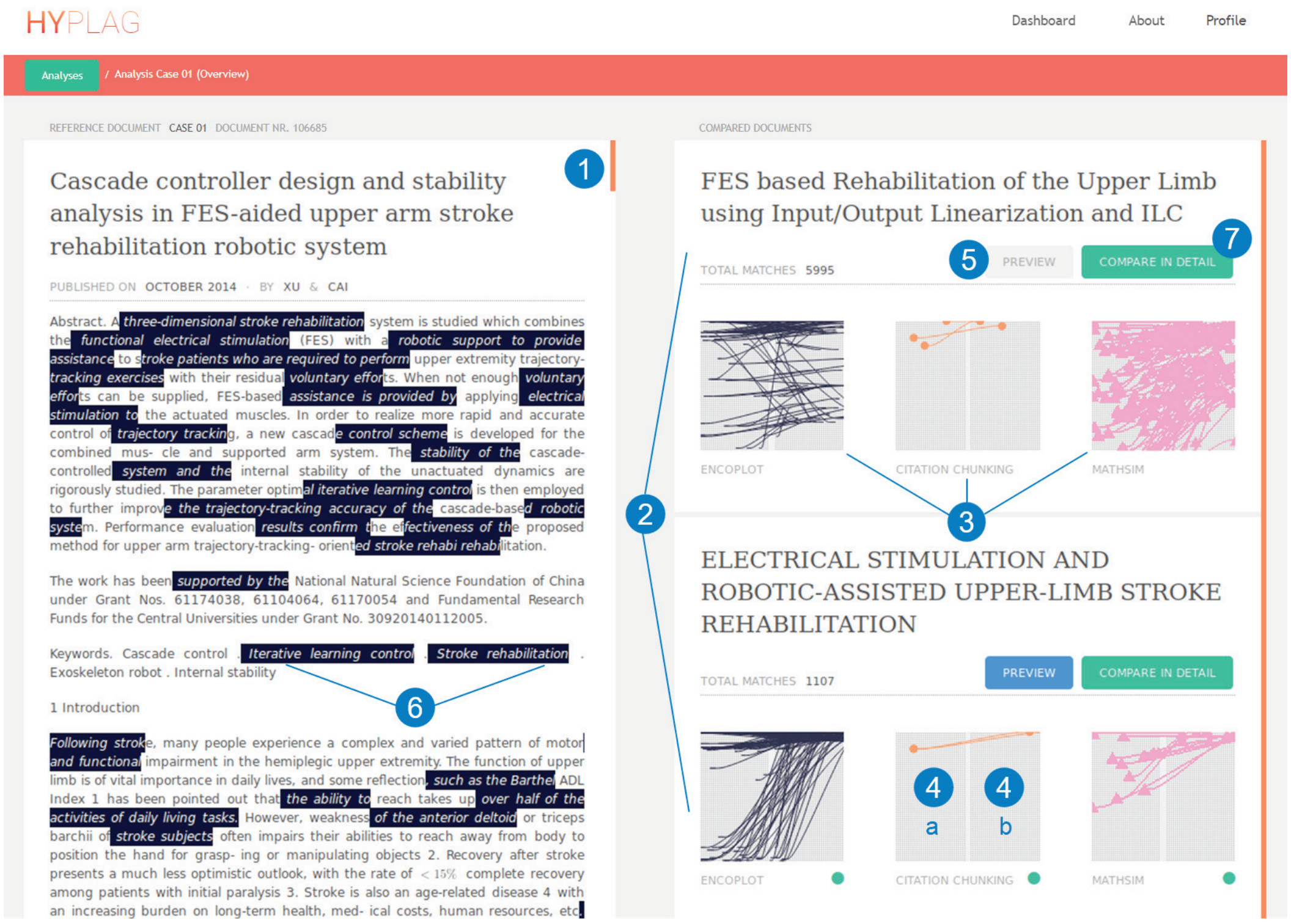

**Figure 6.7.** Results overview in the HyPlag system.

The user can activate the preview 5 of matches for one comparison document at a time. In that case, the system highlights all features in the input document that match features in the currently active comparison document within the full text of the input document using the unique color of the feature 6.

The results overview enables users to quickly browse all identified similarities and check which parts of the input document they affect. By clicking the button 7 available in each result summary, a user can switch to the detailed analysis view that exclusively displays the selected document and the input document.

We explain the functionality of the results overview and the detailed analysis view using a confirmed case of plagiarism in a bioengineering journal article [562]. We also used the case with the identifier C1 to evaluate the math-based detection methods (see Appendix B, p. 225, for a summary of the case). The retraction note explains that the journal retracted the article because it reused a three-page mathematical analysis without attribution from an earlier publication. We used HyPlag to compare the retracted article with the source indicated in the retraction note and other publications by the first author of the source.

For our example case, the match views in **Figure 6.7** show the similarity of text (left), citations (middle), and mathematical content (right) in the retracted article and two publications with overlapping author sets. The top-most result summary represents the source publication named in the retraction note. The match views for text indicate moderate similarity of the retracted article, particularly in the introduction, to both comparison documents. This similarity is primarily due to overlapping keywords and general scientific phrases that likely would not have caused suspicion for either of the two comparison documents. However, the match view for mathematical content (right) in the top-most result summary shows a suspicious similarity that should prompt a user to review the documents in detail.

### 6.3.3 Detailed Analysis View

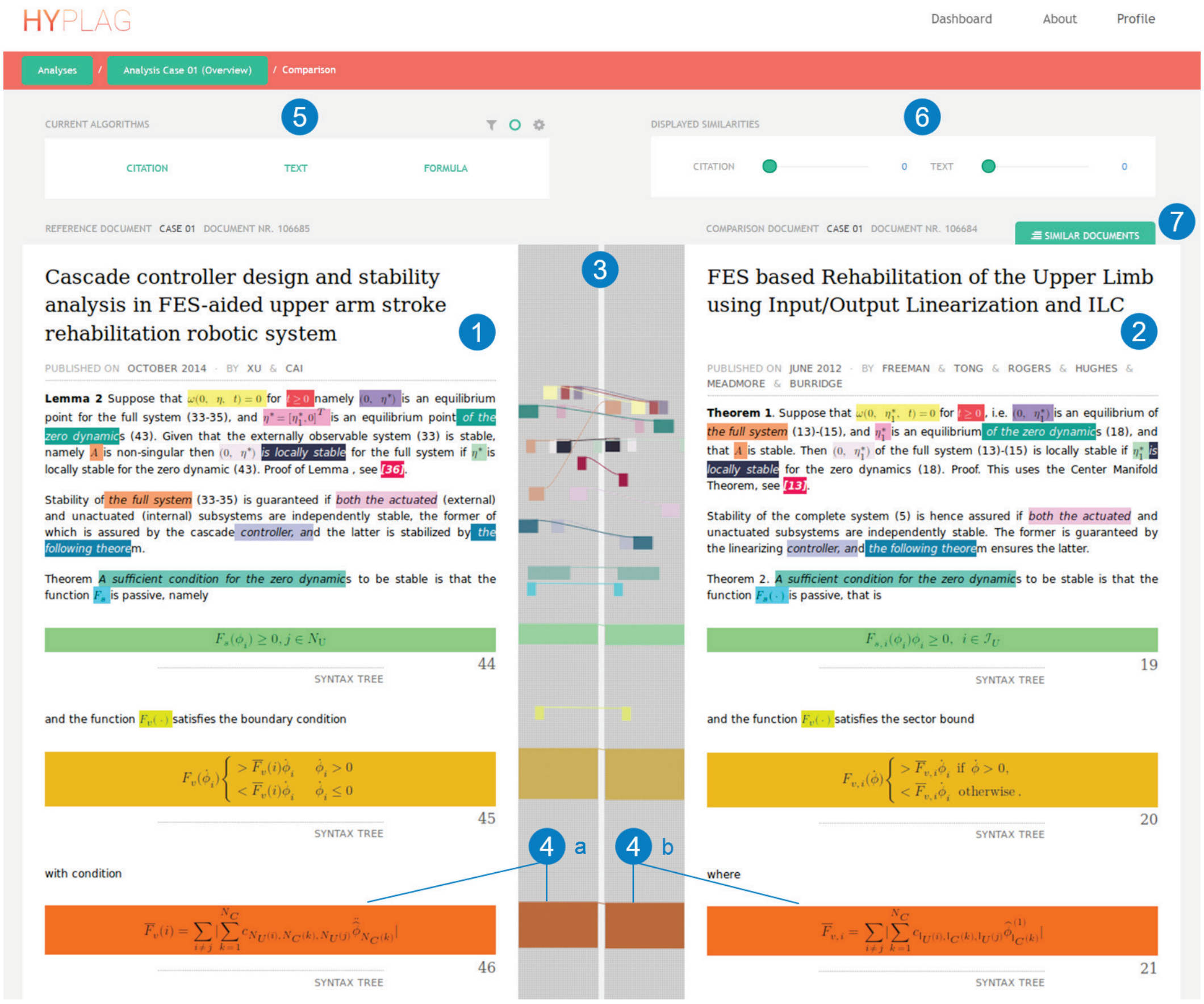

Figure 6.8. Detailed analysis view in the HyPlag system.

**Figure 6.8** shows the detailed analysis view, whose main component is a side-by-side comparison of the input document 1 and one candidate document 2 the user selected in the results overview. Between the full texts, a match view 3, which resembles the match views in the results overview, highlights all matching features in both documents. However, in the detailed analysis view, the system assigns a different color for each feature match 4a,b instead of using one color per feature type as in the results overview. Clicking on any highlight in the full-text panels or the central match view aligns the respective matches. Since the central match view represents the entire document, the current viewport, i.e., the text segment visible in the adjacent full-text panel and its position within the document, is indicated using a darker shade. To improve the legibility of the screen capture, we manually selected a passage from our example case that does not exceed the screen.

For our example case C1, the combined visualization of similar content features in **Figure 6.8** shows that in addition to dispersed keyword matches, particularly the mathematical formulae in both documents exhibit a high similarity and occur in nearly identical order. Also, the only source cited in the shown segments (reference 36 on the left and 13 on the right) is identical.

The panels above the full texts control the type 5 and length 6 of content matches that the system displays. Using the left quick filter panel 5, the user can enable or disable the display of matches by feature type (default view) or detection method (available via the filter button). Using the settings button, the user can select to show the results of up to five detection methods.

The user can control the visibility of additional methods via the detailed filter controls shown in **Figure 6.9**. Furthermore, the button "show inactive matches" allows displaying matches by any detection method, even if the user previously deselected the detection method in the quick filter panel. The sliders in the right panel 6 allow choosing a minimum length for matches to be displayed.

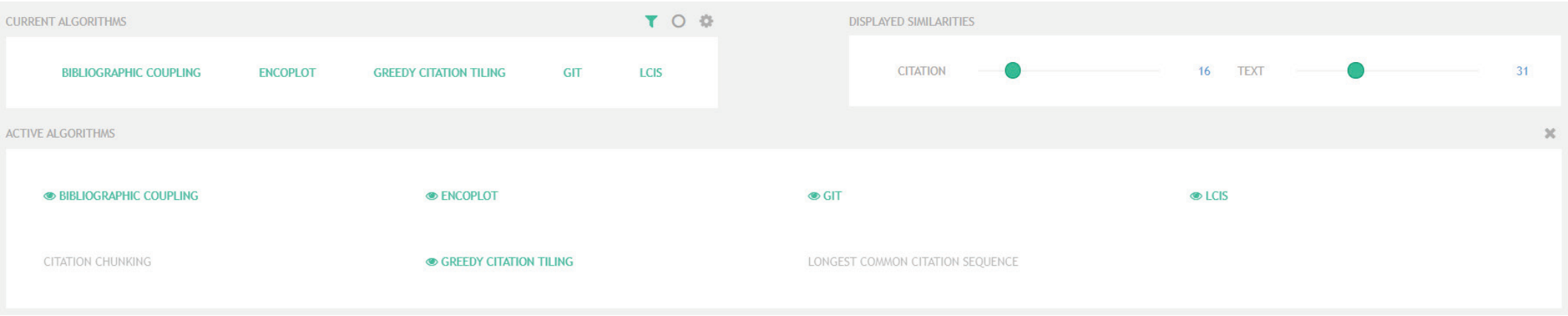

**Figure 6.9.** Detailed display filter controls.

The button "similar documents" (7 in **Figure 6.8**) opens a list of all candidate documents and the number of matches identified in those documents, as shown in

**Figure 6.10**. Using the list, the user can select another candidate document to compare in the detailed analysis view without going back to the results overview.

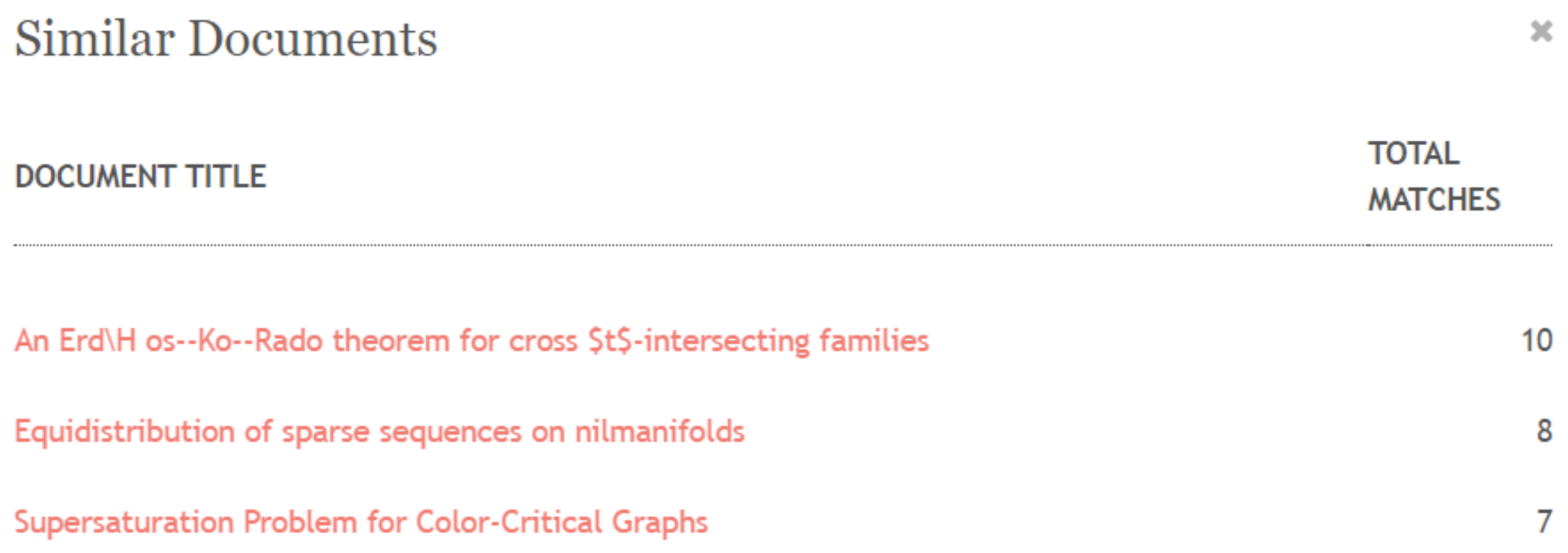

**Figure 6.10.** List of similar documents in the detailed analysis view.

We first devised and implemented the side-by-side layout and interaction concept of the detailed analysis view in the predecessor system CitePlag [172], [337]. This system exclusively visualized matching text and citation patterns in a one-to-one comparison, i.e., did not include a view supporting the candidate retrieval phase. We employed CitePlag to collect relevance judgments during our evaluation of the citation-based detection methods (cf. Section 3.4.6, p. 101). The participants appreciated the visualization concept. They rated the exclusive visualization of similar citation patterns or a combined visualization of similar citation patterns and matching text as most effective for all forms of user-perceived plagiarism except for copy and paste. The combined visualization of textual and non-textual similarity reduced the time required for examination for all but copy-and-paste-type cases of user-perceived plagiarism (cf. Section 3.5.3, p. 110).

### Access Management

The backend allows serving multiple frontends via its REST API while keeping the data of each frontend separate. For example, we employ HyPlag's back-end functionality also to recommend academic literature. Reference [63] describes a frontend tailored to this use case. The system divides the access management between the backend and frontend(s).

To manage the accessibility and retention of data items on the back-end side, the data models of the relational database and the Elasticsearch server maintain the attributes **technical account** and **scope** for each data item.

Technical accounts identify frontends and are the only attribute that the backend uses for checking access permissions. In other terms, each frontend, by default, has full access to all items assigned to the technical account associated with the frontend. This design simplifies the user and access management in the backend by leaving the implementation of more complex permission schemes to the frontend.

**Scopes** are the attributes that frontends can use to realize permission schemes. The idea of scopes is comparable to the file permissions in UNIX-like file systems. The frontend can define the permissions of users and user groups by assigning the items to corresponding scopes. Other than UNIX-like files systems, the HyPlag backend application allows assigning arbitrarily many scopes to items.

The scope system allows creating complex, front-end-specific permission schemes. For example, the frontend supporting the plagiarism detection use case employs scopes to identify the owner of items and create institutional collections.

To identify the owner of a data item, i.e., a document or an analysis, the plagiarism detection frontend creates and associates a unique scope for each front-end user account. The owner has full control over the item, i.e., can edit the item's metadata, manage the item's accessibility, and delete the item either immediately or after the user-defined retention period has expired. By default, the frontend makes items accessible exclusively to the owner. Owners can release items into any other scope, including a unique public scope. Items in the public scope are available for checks to anyone. Only system administrators can delete items in the public scope.

Items in institutional scopes are exclusively available to users who are members of the scope. The frontend allows elevating user accounts to scope administrators who can manage all items and users belonging to a scope.

HyPlag's data and access management addresses concerns regarding the confidentiality of data and transparency of data management frequently voiced for commercial plagiarism detection providers (see Section 2.4.8, p. 47).

## 6.4 Conclusion Hybrid PD System

With HyPlag, we presented the first plagiarism detection system that jointly analyzes the similarity of citations, images, mathematical content, and text. The system consists of a server-side backend and a web-based frontend.

The backend relies on production-grade, open-source software, e.g., the Spring Boot application framework and an Elasticsearch server, to ensure the adaptability and scalability of the system. We modularized all vital back-end components using software containers and programming interfaces to facilitate changing or adding detection methods and processing steps, such as information extraction or postprocessing components. With GROBID, the backend uses the current best-of-breed software library for converting and extracting information from scholarly documents.

By providing access to all core functionalities via a REST interface, the backend can serve multiple frontends. Aside from its primary purpose of supporting plagiarism detection, the backend also serves as the data processing backbone of a novel literature recommendation service [63].

The frontend consists of a dashboard area and two analysis views tailored to the candidate retrieval and detailed analysis stages of the plagiarism detection process. The dashboard area allows users to upload and organize documents, as well as create and manage custom plagiarism checks. The customization options include the selection of detection methods and the definition of the document set to analyze.

The results overview allows selecting one or more documents to compare in detail by browsing summaries of the feature matches in all candidate documents. The detailed analysis view provides an interactive side-by-side comparison of the input document and one selected candidate document. The view visualizes similar citation patterns, images, text, and mathematical content. The user can customize the display of similar features by filtering according to feature type, detection method, and match length. The functionality of the detailed analysis view reflects the preferences of domain experts, which criticized that few available systems offer a side-by-side comparison [145, p. 22] (see also Section 2.5.3, p. 72). In our evaluation of citation-based detection methods, the interactive visualizations of similar features significantly reduced the effort for assessing the severity of similarities.

A demonstration system and our code are available at http://thesis.meuschke.org.

# 7

Chapter 7

# Conclusion and Future Work

**Contents**

This chapter concludes the thesis by summarizing our research in Section 7.1, presenting an overview of our research contributions in Section 7.2, and discussing ideas for future research in Section 7.3.

## 7.1 Summary

This thesis presented novel approaches to address an open research problem in Information Retrieval—identifying disguised academic plagiarism, such as strong paraphrases, sense-for-sense translations, and the appropriation of ideas.

A comprehensive review of the literature showed that plagiarism detection methods proposed so far perform unsatisfactorily for identifying strongly disguised forms of academic plagiarism. Methods presented in the literature primarily analyze lexical, syntactic, and semantic text similarity. In ideal laboratory settings (i.e., all sources are accessible and no limits on computing time exist), such methods achieved $F_1$ scores of 88%–96% for identifying plagiarism forms with little or no disguise but $F_1$ scores of less than 60% for disguised plagiarism forms. In evaluations under realistic conditions, the detection rate of production-grade plagiarism detection systems was even lower. The effectiveness of detection methods that exclusively analyze text has reached a plateau in recent years.

To address the shortcomings of current detection methods, we proposed analyzing non-textual content elements of academic documents in addition to text to enable a better assessment of the semantic content of documents. We introduced three approaches that implement this idea by analyzing citations, images, and mathematical content in academic documents. These three content types have in common that they occur in many, if not all, academic documents, contain much semantic information, and are language-independent. Also, changing or replacing these types of content requires more expertise and effort than paraphrasing text, which makes disguising plagiarism harder. To devise methods that analyze non-textual content elements to identify plagiarism, we initially examined confirmed cases of academic plagiarism that entail the reuse of citations, images, or mathematics. Then, we devised methods that search for the characteristic properties of reused citations, images, and mathematical content, we observed for common forms of plagiarism.

Citation-based plagiarism detection analyzes the placement of citations within academic documents for distinctive patterns of shared citations that indicate a high semantic similarity of the documents. The number, order, and proximity of shared citations determine the distinctiveness of a pattern. To identify the citation patterns unique to common forms of academic plagiarism, we adapted established similarity measures, e.g., Longest Common Subsequence and Greedy Tiles, and devised Citation Chunking as a new, use-case-specific class of detection methods.

The adaptive image-based plagiarism detection process we presented addresses the disadvantages of prior works that focused on specific image types. Our process integrates well-performing content-based image retrieval methods, such as perceptual hashing, with newly proposed detection methods that target specific image types, e.g., Ratio Hashing for bar charts. We introduced a novel scoring function to quantify the likelihood that identified image similarities represent plagiarism.

Math-based plagiarism detection is the first approach to identify the plagiarism of concepts and ideas expressed as mathematical expressions. The approach primarily targets plagiarism in math-heavy STEM documents for which existing detection methods often perform unsatisfactorily. In preliminary experiments, we found that identifiers achieved the best detection effectiveness of all presentational mathematical features. Consequently, we devised a two-stage plagiarism detection process that analyzes mathematical identifiers. The process employs a retrieval model that combines tf-idf-weighted vector space and Boolean retrieval for identifiers in the candidate retrieval stage. In the detailed analysis stage, the process employs set-based and sequence-based similarity measures for identifiers.

We evaluated our citation-based detection methods in two distinct experiments. First, we analyzed the confirmed instance of translated plagiarism in the doctoral thesis of K. T. zu Guttenberg. Of the 16 translated passages in the thesis, the citation-based detection methods identified 13, while text-based methods detected none. Second, we analyzed 185,170 bioscience publications from the PubMed Central Open Access Subset using citation-based and text-based detection methods. We performed a user study to assess the retrieved documents. The results show that citation-based detection methods significantly outperformed text-based methods in ranking highly user-perceived instances of disguised plagiarism. Also, the citation-based methods revealed nine previously undiscovered cases of plagiarism.

We demonstrated the effectiveness of our image-based detection process by showing its ranking performance for retrieving the sources of 15 plagiarized images from a collection of 4,500 topically related images. Our process identified 11 of the 15 source images as suspiciously similar and retrieved them at rank one.

We showed the retrieval performance of our math-based detection methods in two conceptually different experiments. First, we compared the effectiveness of math-based, citation-based, and text-based methods for identifying 10 confirmed cases of plagiarism in a collection of 102,524 topically related publications. The combination of any two retrieval approaches, e.g., math-based and text-based or citation-based and text-based, achieved optimal recall for the candidate retrieval stage, performing better than text-based methods alone. For the detailed analysis stage, the math-based methods performed equally well as the text-based methods. However, they offered advantages for cases the text-based methods could not identify.

In a second experiment, we analyzed the documents that a combination of math-based and citation-based detection methods ranked most highly when analyzing each document in the collection. The top-10 results included two confirmed cases of plagiarism, a previously unknown case that the author of the earlier document considers plagiarism, three documents with notable yet legitimate content similarity, and two false positives. These results show that our non-textual content analysis methods can identify previously undiscovered cases of academic plagiarism and similarities in the content of which examiners should be made aware.

In summary, our evaluations showed that the detection approaches we proposed effectively identify confirmed cases of academic plagiarism, including cases that were previously non-machine detectable. Our detection methods also revealed previously undiscovered cases of academic plagiarism.

To demonstrate and evaluate the newly proposed detection approaches in practice, we implemented them together with well-performing text-based detection methods in a prototype of a hybrid plagiarism detection system called HyPlag. The system consists of a backend that provides all functionality for preprocessing, storing, and analyzing documents and a web-based frontend that provides the user interface.

**Figure 7.1** shows the main view for analyzing documents in the HyPlag frontend. The view presents the input document and a potential source side-by-side. Between the documents, an abstract match view highlights the identified similarities in any part of the documents. The user can change the displayed matches by filtering for specific detection methods or the length of the matches. Clicking on any highlight aligns the matching features in both documents.

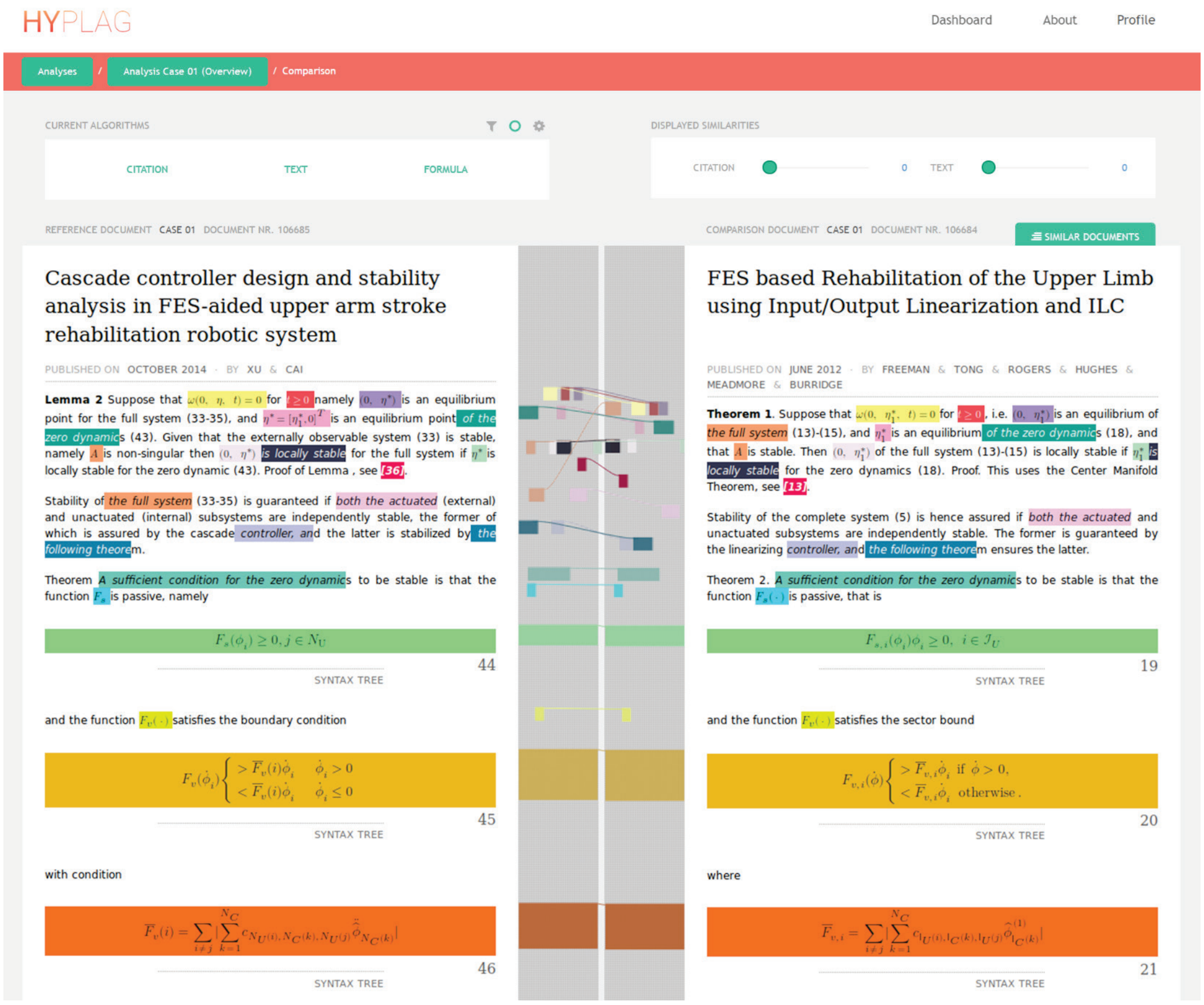

**Figure 7.1.** HyPlag detailed analysis view.

We evaluated the visualization concept of the detailed analysis view as part of the user study we performed to evaluate Citation-based Plagiarism Detection. The combined visualization of textual and non-textual similarity reduced the time users needed to spot the first two instances of perceived plagiarism for all forms of plagiarism, except copy and paste. The most significant time savings from visualizing non-textual similarity were observable for structural and idea plagiarism.

We conclude from our investigations that analyzing non-textual content elements in academic documents increases the capabilities to detect academic plagiarism, particularly for disguised plagiarism forms. The citation-based, image-based, and math-based detection methods we introduced complement current text-based detection methods, which excel in identifying plagiarism instances with no or moderate disguise. Non-textual content analysis proved beneficial for:

1. Identifying plagiarism forms with low textual similarities, such as strong paraphrases, translation, and idea plagiarism;
2. Increasing recall in the candidate retrieval stage of the plagiarism detection process, which determines the achievable detection effectiveness overall;
3. Detecting plagiarism in STEM documents that interweave natural language with mathematics and technical notation. The intermittent elements reduce the effectiveness of text-based detection methods, which often prevented adequate plagiarism checks for these disciplines so far.

**Figure 7.2** visualizes our research contributions by depicting the applicability of detection methods for different forms of academic plagiarism.

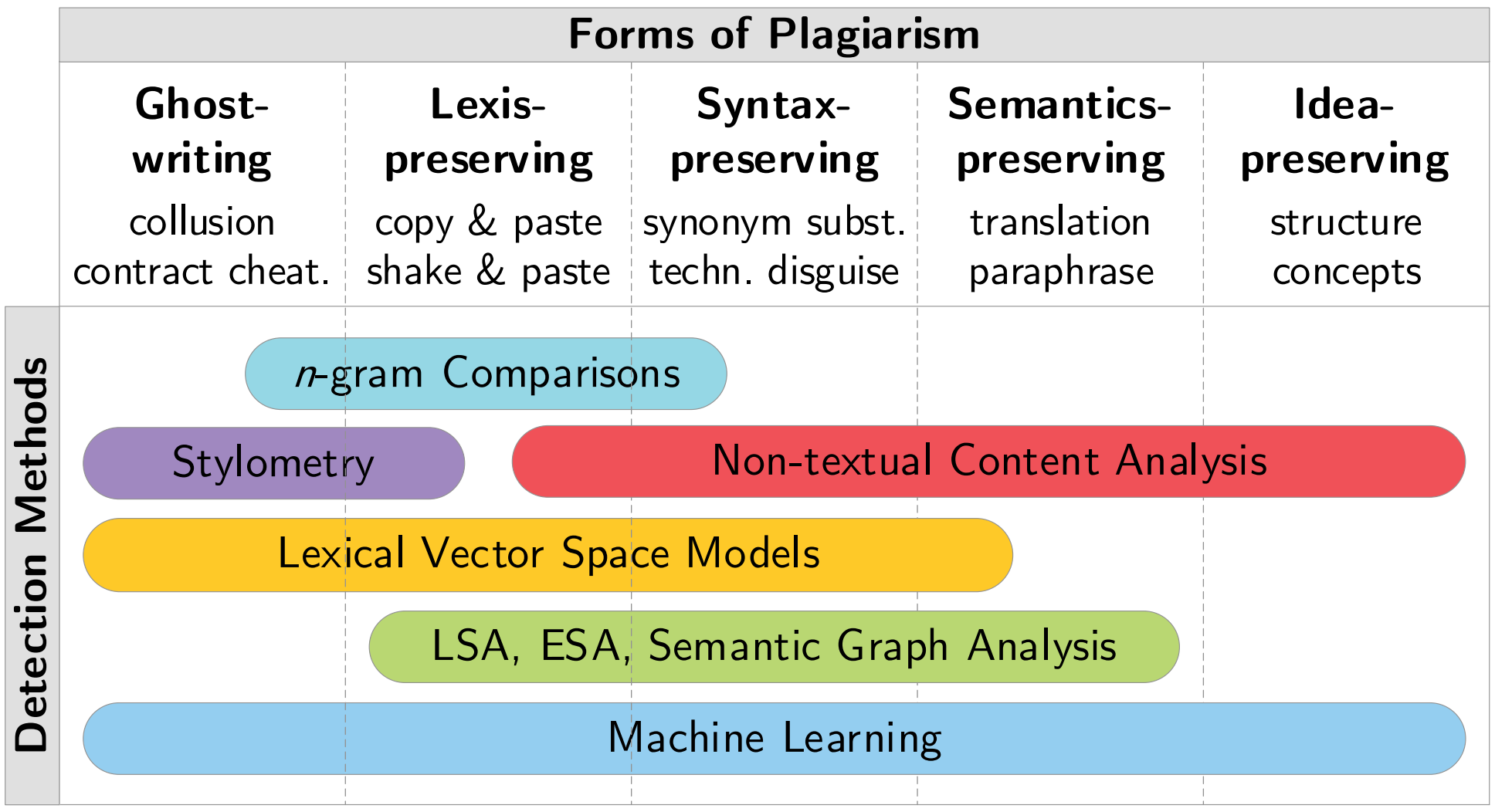

**Figure 7.2.** Applicability of plagiarism detection methods.

$n$-gram comparisons are highly effective and efficient for detecting lexis-preserving plagiarism and collusion. Intrinsic detection methods apply stylometry, which can hint at the presence of plagiarism with no or little disguise. Stylometric comparisons at the document level can reveal outliers, which may represent documents written by ghostwriters [255]. Lexical vector space models have many applications but are not particularly beneficial for detecting idea plagiarism. Semantics-based detection methods, such as ESA, LSA, and Semantic Graph Analysis, are tailored to detecting semantics-preserving plagiarism, yet also perform well for lexis-preserving and syntax-preserving plagiarism. Machine learning is a universal approach whose applicability depends on the chosen features. Non-textual content analysis is particularly beneficial for detecting strongly disguised idea-preserving plagiarism.

Our research on non-textual content analysis increases the likelihood that disguised academic plagiarism is detected and the effort required of plagiarists to avoid detection. The widespread use of non-textual content analysis in plagiarism detection software would force plagiarists not only to rephrase text but also to change citations, images, and mathematics to minimize the chance that plagiarism is discovered. Altering these types of non-textual content without raising the suspicion of domain experts like peer reviewers requires more subject expertise and effort than paraphrasing text. Consequently, we hope that non-textual content analysis will contribute to preventing academic plagiarism by convincing potential plagiarists that the effort for masking plagiarism exceeds the potential rewards.

However, past advancements in plagiarism detection technology did not exclusively result in positive behavioral changes. For example, increased use of plagiarism detection software led to increased contract cheating, i.e., using professional ghostwriting services [366]. We expect that the widespread adoption of non-textual content analysis will have similarly mixed effects. While the new detection methods will incentivize some authors to abstain from plagiarizing, others will search for possibilities to defeat the technology.

Consequently, continued research on plagiarism detection technology and plagiarism policy is necessary to keep pace with unwanted behavior changes of plagiarists. Plagiarism detection methods and systems must be integrated with policy frameworks that promote and enforce academic integrity. Examples of effective policy actions include teaching research skills to prevent plagiarism and defining guidelines for investigating and prosecuting plagiarism. In Section 7.3, p. 197, we present ideas for future research that can help to improve plagiarism detection technology and related applications further. Before that, the following section presents a concise summary of the research contributions we made in this thesis.

## 7.2 Contributions of the Thesis

This section summarizes the contributions of the thesis for each of the four research tasks defined in the Introduction (cf. Section 1.3, p. 4)

| | |
|---|---|
| **Research Task 1** | Identify the strengths and weaknesses of state-of-the-art methods and systems to detect academic plagiarism. |

To accomplish Research Task 1, we performed the most comprehensive literature review on plagiarism detection technology to date. The review covers 25 years (1994–2019) and includes 376 research publications. We proposed a new classification of the literature addressing academic plagiarism and a novel technically oriented typology of plagiarism forms.

We found that detection methods proposed so far almost exclusively analyze text similarity. These methods are highly effective and efficient in identifying copied and moderately altered text within web-scale document collections. However, they perform unsatisfactorily for finding strong paraphrases, sense-to-sense translations, and the plagiarism of ideas and non-textual content.

| | |
|---|---|
| **Research Task 2** | Devise detection approaches that address the identified weaknesses. |

To address the deficiency of existing plagiarism detection approaches, we initiated the research on analyzing non-textual content in addition to text. The idea is to model and compare the semantics of documents by analyzing semantically rich elements that are language-independent and difficult to obfuscate or replace.

With Citation-based Plagiarism Detection and Math-based Plagiarism Detection, we introduced two novel approaches that implement this idea. Our research on Image-based Plagiarism Detection extended prior work by enabling the analysis of multiple image types, devising new detection methods, and improving the scoring of results. Initial investigations of confirmed cases of academic plagiarism guided our research on all three detection approaches. We devised mutually complementary detection methods for each approach to identify the typical properties of reused citations, images, and mathematics we observed for common plagiarism forms.

| Research Task 3 | Evaluate the effectiveness of the proposed detection approaches. |
|---|---|

To validate the effectiveness of our detection approaches, we performed five evaluations using real cases of academic plagiarism embedded into realistic collections. We chose this evaluation approach for two reasons. First, we consider the ability to identify real cases of plagiarism committed by expert researchers with strong incentives to disguise their actions as essential for assessing the benefit of any new plagiarism detection approach. Second, existing evaluation datasets for plagiarism detection technology do not include the content features, i.e., citations, images, and mathematical content, our detection approaches analyze.

We showed that Citation-based Plagiarism Detection considerably outperformed text-based detection methods in identifying translated, paraphrased, and idea plagiarism instances. Moreover, our citation-based detection methods found nine previously undiscovered cases of academic plagiarism.

We demonstrated that our image-based plagiarism detection process effectively finds frequently observed forms of image plagiarism for a large variety of image types that authors of academic documents typically use.

We verified that Math-based Plagiarism Detection reliably retrieves confirmed cases of academic plagiarism involving reused mathematical content and can identify previously undiscovered cases. Particularly in combination with citation-based detection methods, math-based methods offered advantages for identifying plagiarism cases that text-based methods could not detect.

Our experiments required creating three large-scale datasets suitable for evaluating citation-based, image-based, and math-based detection methods. We make all datasets and code for our experiments available at http://thesis.meuschke.org.

| Research Task 4 | Implement the proposed detection approaches in a plagiarism detection system capable of supporting realistic detection use cases. |
|---|---|

With HyPlag, we provided the first plagiarism detection system that integrates the analysis of citations, images, mathematical content, and text. Our system consists of a server-side backend and a web-based frontend.

The backend can process PDF, LaTeX, plaintext, and two XML formats commonly used for representing academic documents (JATS and TEI). The index component, data storage, and application logic of the backend rely on production-grade software frameworks that allow scaling the system to massive collection sizes.

The frontend allows users to upload and manage documents in PDF or plaintext format, perform customized plagiarism checks, and inspect the retrieved results using two interactive views.

## 7.3 Future Work

The research presented in this thesis yielded numerous ideas to improve plagiarism detection technology and other information retrieval applications. We briefly discuss these ideas hereafter. Section 7.3.1 describes options to increase the effectiveness of the non-textual detection approaches we presented. Section 7.3.2 presents ideas to increase the effectiveness and usability of plagiarism detection systems. Section 7.3.3 motivates how the research we presented could benefit applications beyond plagiarism detection. Section 7.3.4 briefly discusses the implications that non-textual content analysis could have on plagiarism policy.

### 7.3.1 Increase Detection Effectiveness

Our evaluations of the citation-based, image-based, and math-based detection methods identified promising ideas to increase the effectiveness of the approaches.

#### Math-based Plagiarism Detection

Of the three non-textual detection approaches we presented, Math-based Plagiarism Detection exhibits the largest need and most extensive opportunities for future research. Compared to text-based information retrieval and citation analysis, math information retrieval is a nascent research field. Therefore, math information retrieval lacks many technologies to access, extract, process, and retrieve information that are well-established standards in the other two fields.

To address this lack in retrieval technology for mathematics, we extend the research presented in this thesis as part of **two DFG-funded research projects**. The first project[13] (DFG1) researches fundamental methods and tools for making mathematical knowledge accessible to information retrieval systems. The objective of DFG1 is to enable the extraction of mathematical concepts, i.e., allowing automated access to the semantics of mathematical expressions. To achieve this objective, we develop approaches to identify and reliably differentiate mathematical expressions from similar or neighboring content elements as well as perform type detection and tokenization of mathematical expressions. We will make the developed technologies publicly available to facilitate future research on math retrieval problems.

[13] DFG1: Methods and Tools to Advance the Retrieval of Mathematical Knowledge from Digital Libraries for Search-, Recommendation- and Assistance-Systems (https://gepris.dfg.de/gepris/projekt/350192710)

The second DFG-funded project[14] (DFG2) aims to improve the math-based plagiarism detection approach presented in this thesis by making two contributions. First, we research methods to extract mathematical expressions from PDF documents. For this task, the project teams of DFG1 and DFG2 collaborate closely. DFG1 researches methods to extract and semantically augment mathematical content from structured document formats. DFG2 seeks to adapt these methods to the challenging PDF format. Reliably extracting the presentation, structure, and semantics of mathematical content from PDF documents is a crucial prerequisite for making Math-based Plagiarism Detection effective in practice.

The second contribution we work on as part of DFG2 is improving the math-based detection methods presented in this thesis. We expect that we can increase the effectiveness of the candidate retrieval algorithms by devising better retrieval models. Currently, we exclusively use mathematical identifiers to build a retrieval model. As part of DFG2, we investigate including more information in the model, such as function type (e.g., sinus, logarithm, factorial), expression type (e.g., if features represent a term, equation, or definition), as well as positional information of mathematical features. Moreover, we investigate adapting successful text retrieval models, such as feature embeddings, to mathematics.

To improve the math-based similarity measures employed for the detailed analysis stage, we investigate analyzing the structural and semantic similarity of mathematical expressions in addition to presentational similarity. The combined analysis is necessary to identify cases in which authors obfuscated reused mathematical content. For example, we plan to consider expression trees for quantifying structural similarity and knowledge graph representations of mathematical concepts to determine semantic similarity. Additionally, we plan to devise improved pattern detection methods, e.g., using the clustering of mathematical features. To improve the scoring of mathematical expressions retrieved as similar, we plan to devise outlier detection algorithms similar to those we employ for our image-based plagiarism detection process (cf. Section 4.4.8, p. 135).

---

[14] DFG2: Analyzing Mathematics to Detect Disguised Academic Plagiarism (https://gepris.dfg.de/gepris/projekt/437179652)

## Citation-based Plagiarism Detection

To increase the effectiveness of citation-based plagiarism detection methods, we see experimenting with new pattern identification algorithms and scoring functions as promising directions for future research.

To identify citation patterns, we currently use well-established sequence-based pattern detection methods, such as the Longest Common Subsequence and Greedy Tiling algorithms. The patterns identified by these algorithms reflected the patterns we observed in confirmed cases of academic plagiarism. Conceptually, we employed a supervised approach to defining pattern identification methods.

Future research could investigate unsupervised approaches to identifying citation patterns. Applying Sequential Pattern Mining (SPM) to the problem could prove beneficial. SPM seeks to determine the relationships between occurrences of sequential elements or events [574, p. 7]. SPM algorithms identify interesting subsequences in a set of sequences. The interestingness of subsequences can derive from their occurrence frequency, length, or other criteria, e.g., profit [146, p. 55]. Sequential pattern mining is a prolific research field (the References [146], [353], [574] present surveys) with numerous applications, e.g., in computational biology to analyze DNA sequences or business intelligence to investigate customer behavior. El-Matarawy et al. applied SPM to text-based plagiarism detection [124].

Sequential pattern mining could improve our handcrafted Citation Chunking heuristics by searching for all subsequences that fulfill specific criteria. Conceptually, sequential patterns combine the desirable properties of the longest common subsequences and greedy tiles. Like the longest common subsequences, sequential patterns allow intermittent non-matching elements. As in the case of greedy tiles, multiple sequential patterns can exist, and a criterion for their detection could be that the algorithm exclusively considers the individually longest patterns. Furthermore, the textual distance of citations could be a criterion for pattern detection.

A central open question is whether SPM is computationally viable for the plagiarism detection use case. An essential assumption of SPM is that frequent subsequences are interesting. The opposite is the case for plagiarism detection. Citation sequences that occur in many documents typically indicate that the authors cite seminal publications representing the state of the art in a research field. In plagiarism detection, infrequent yet otherwise distinctive citation patterns are interesting. The minimum occurrence frequency of sequential patterns is an essential user-defined parameter that determines the computational effort required for SPM. The opposing assumptions of SPM and plagiarism detection raise two questions. First, can tuning, e.g., by requiring minimum lengths of patterns, make SPM algorithms computationally feasible? Second, will the tuned algorithms produce more informative patterns than the patterns that our computational modest Citation Chunking methods identify? If the answer to these research questions would be positive, SPM might also be an interesting direction for future research on Math-based Plagiarism Detection. Since mathematical features, like identifiers, operators, and functions, are more heterogeneous and occur in larger quantities than citations, the viability of SPM for the math-based approach to plagiarism detection is more questionable than for the citation-based approach.

To improve the scoring functions for citation patterns, we consider analyzing the frequency of a pattern and the citations it contains as promising. The idea for all options we describe hereafter is that frequent citation patterns likely represent publications that authors cite routinely to describe the state of the art. A basic approach could be to exclude citation patterns or reduce their score if a document shares the pattern with many other documents. However, differences in presenting seminal publications likely lead to transpositions, scaling, and incompletely matching citation patterns. Therefore, excluding specific patterns reliably appears futile.

Co-Citation Proximity Analysis (CPA) could serve as a proxy to realize the score reduction for frequent citation patterns despite the challenges to pattern identification. Documents have a high CPA score if many other documents cite them together in close textual proximity. Therefore, citation patterns containing citations with high mutual CPA scores indicate that the documents are often co-cited in proximity to each other and thus are likely unsuspicious. Unlike the exclusion of specific patterns, this approach could handle incomplete pattern matches.

Moreover, one could investigate reducing the score of all citation pattern matches between documents that have a high CPA score [173, p. 214]. The rationale is that many of the citing authors likely read both documents carefully and considered their contributions valid. Consequently, we expect that documents with high CPA scores likely do not contain illegitimately shared content because the citing authors would probably have discovered such content overlaps.

Likewise, employing CPA as part of the scoring functions for citation patterns could help recognize attempts to obfuscate plagiarism by substituting citations [173, p. 214]. If plagiarism checks routinely include citation pattern analysis in the future, plagiarists could try to replace the citations taken from a source with citations to related documents. Future citation-based detection methods could consider citations to related documents occurring in a citation pattern as “soft” matches to deter such behavior. This scoring approach requires determining possibly interchangeable references. Text-based retrieval models like vector representations, citation-based models, or a combination can serve to find the set of possibly interchangeable references in the database. Using the citation-based CPA approach for this task has the advantage that CPA leverages the expertise of researchers on the subject matter of documents to determine the semantic relation of the documents. CPA has proven its ability to identify semantic relations that text-based models cannot reflect [168]. The major challenges for realizing the soft-matching approach would be managing the increase in computational effort and false positives.

### Image-based Plagiarism Detection

The evaluation of our image-based plagiarism detection process revealed several technical challenges in the realistic scenario we imposed for our experiments. Perceptual hashing often performed poorly for visually sparse images. Including a dilation step before performing perceptual hashing could help to achieve better results. Low OCR quality reduced the effectiveness of the detection methods that analyze labels and other textual content in figures. This problem was particularly severe for older digitized academic publications. We expect that advancements in deep-learning-based OCR technology will help to alleviate these weaknesses in the future. The approach to sub-image extraction we proposed sometimes failed to extract overlapping sub-images correctly. Specialized post-processing procedures could improve the results of our method. Alternatively, employing specific image recognition and segmentation software could prove beneficial.

Aside from improving the effectiveness of detection methods included in the adaptive image-based process, adding specialized detection methods for more image types can augment the detection capabilities. We expect that analyzing image types, such as line graphs, scatter plots, and microscopic images, for potential plagiarism of ideas or data will require detection methods tailored to the specific properties of the image types. The design of our detection process allows for the inclusion of additional detection methods with comparably little effort. Adding classifiers for image types that new detection methods analyze would contribute to spending computational resources most effectively. Our flexible scoring function could account for the similarity scores of additional detection methods without changes. However, future research could also investigate novel scoring functions.

## Hybrid Plagiarism Detection Approaches

Given our observations during the development and evaluation of the non-textual detection approaches, we expect that combining textual and non-textual content analysis promises the most significant improvement in detection effectiveness.

Realizing this potential requires, as a first step, additional large-scale evaluations of non-textual plagiarism detection approaches. Compared to the maturity of text-based detection methods, the non-textual detection approaches are in an infant state. Additional investigations of confirmed plagiarism cases and exploratory searches for unknown cases are necessary to:

- Better understand the typical properties of plagiarized non-textual content;
- Establish baselines on the levels of legitimate content similarity (textual and non-textual) expectable by chance for different research fields;
- Create large-scale datasets that include non-textual content for the development and test of plagiarism detection approaches.

In a second step, future research should investigate combinations of the similarity assessments for text and non-textual content. If enough real cases can be collected or realistic artificial cases generated, machine learning approaches could improve the scoring of identified similarities. Machine learning methods could find the combinations of similarity scores returned by textual and non-textual detection methods that maximize the probability that a document is a case of plagiarism.

An important factor that future scoring functions should consider is the position of identified similarities in the document. For example, mathematics and theoretical physics publications often reuse descriptions of research problems consisting of standardized natural language and mathematical content as part of their introductory or related work sections. Depending on the complexity of the problem, such descriptions can span multiple pages in length and do not always cite a source. This type of content reuse is an accepted practice in these fields but poses a challenge for plagiarism detection methods. Couzin-Frankel & Grom reported on comparable practices regarding the reuse of text in biomedical publications [94]. Many researchers described the reuse of text that they or collaborators published previously as an accepted practice if the reused text occurs in review articles, the introduction, related work section, or the description of experimental settings [94]. These observations suggest that reducing the score of content similarities in the introduction and related work sections could help distinguish documents that are suspicious of plagiarism from documents containing likely legitimate content reuse.

## 7.3.2 Improve Plagiarism Detection Systems

In the following, we present our plans for improving our prototype of a hybrid plagiarism detection system, HyPlag, and ideas to increase the usability, availability, and security of plagiarism detection systems in general.

### Improvements to the HyPlag System

Using HyPlag for the evaluations presented in Chapter 6 revealed technical issues we plan to address in our future work. First, we will fully integrate the extraction of mathematical content from PDF documents into the preprocessing phase of HyPlag. So far, we employed the standalone tool InftyReader [498] to perform the extraction semi-automatically. In the DFG2 project (cf. Section 7.3.1, p. 197), we will develop new extraction methods and fully integrate them into HyPlag.

Second, we will improve the extraction and disambiguation of references for publications in math-heavy STEM disciplines like mathematics and theoretical physics. HyPlag's preprocessing steps, and in consequence, the citation-based detection methods yielded unexpectedly poor results (cf. Section 5.7, p. 160) for such documents. The reasons were that many documents used severely abridged citation styles, which, e.g., omit the title in favor of stating an arXiv identifier. Another problem was that some authors cite the arXiv preprint, whereas other authors cite the published journal article. The two versions of a publication regularly exhibited differences in the order of authors and the title. The differences prevented our disambiguation methods from recognizing the different versions as representing the same document. We plan to increase the integration of public APIs, like Crossref [95], arXiv [92], and ORCID [378], to obtain additional information on publications and authors that can improve the disambiguation of references.

Aside from fixing technical weaknesses, we plan to extend HyPlag by adding detection methods that analyze syntactic and semantic text similarity and increasing thy system's reference collection. Currently, HyPlag includes lexical detection methods, such as character $n$-gram fingerprinting and string matching. These methods are representative of the methods production-grade plagiarism detection systems employ. To extend HyPlag's text-based detection capabilities, we plan to integrate detection methods that analyze syntactic and semantic text similarity. We consider dense vector representations and neural language models as promising semantic text analysis methods. Such methods require a candidate retrieval step to be computationally feasible. Our evaluations of citation-based and math-based detection methods showed that the methods are computationally modest but can retrieve documents with high semantic and low lexical similarity. These properties make them ideal for complementing text-based candidate retrieval methods. To expand HyPlag's reference collection, we plan to integrate arXiv [93], CORE [91], and possibly other open collections of research publications.

## Novel Visualization Concepts

The new detection approaches we presented in this thesis improve the capabilities of plagiarism detection systems. However, these new capabilities also increase the complexity of the systems and the cognitive load they impose on the user. Highlighting identical text is the only similarity visualization that today's production-grade plagiarism detection systems offer. Most users do not have difficulties with inspecting and judging such verbatim text matches. However, examining text matches typically does not suffice to identify disguised plagiarism. Deciding whether similar content beyond verbatim text matches is problematic requires inspecting various additional information, such as paraphrased text, similar mathematical formulae, suspicious citation patterns, and similar figures.

The visualization concepts implemented in the frontend of our HyPlag prototype are a first step towards enabling users to examine this broad spectrum of possible content similarity in academic documents. To manage the cognitive load and avoid visual clutter, we use two visualizations—to provide an overview and allow a detailed inspection. The detailed analysis view offers filter functions for the type and length of similar content to avoid information overload.

While we consider our visualizations an improvement over existing systems, their current limitation is explaining non-binary content similarities, such as similar mathematical formulae or images. Unlike characters or citations, which can either be identical or different, the semantic and syntactic similarity of text and mathematical formulae, as well as the visual similarity of images, are continuous. So far, our visualization treats these types of similarity as binary by highlighting a formula or image as a match if the similarity score exceeds certain thresholds.

In terms of Shneiderman's well-known Visual Information Seeking Mantra: "Overview first, zoom and filter, then details-on-demand" [464, p. 336], HyPlag is currently missing visualizations that present details on demand. Such visualizations need to explain to the user, e.g., the elements of a mathematical formula that caused the system to consider the formula as similar to another formula and why. Similarly, visualizing why the system determined the semantics of two text passages to be suspiciously similar is a non-trivial task.

We proposed an interactive visualization of pairwise formula similarity [454], which we also tested as a details-on-demand visualization in HyPlag. The visualization shown in **Figure 7.3** displays the MathML expression tree for the formula in the input document in light blue shading and the formula in the comparison document in light green shading. Hovering over nodes in the tree visualization highlights the corresponding elements in the inline formula at the top. The system highlights identical (e.g., 1) and similar leaf nodes (e.g., 2). A layout algorithm that minimizes edge crossings aligns the formulae to emphasize structural similarity. To facilitate the structure analysis, the user can collapse and freely arrange nodes.

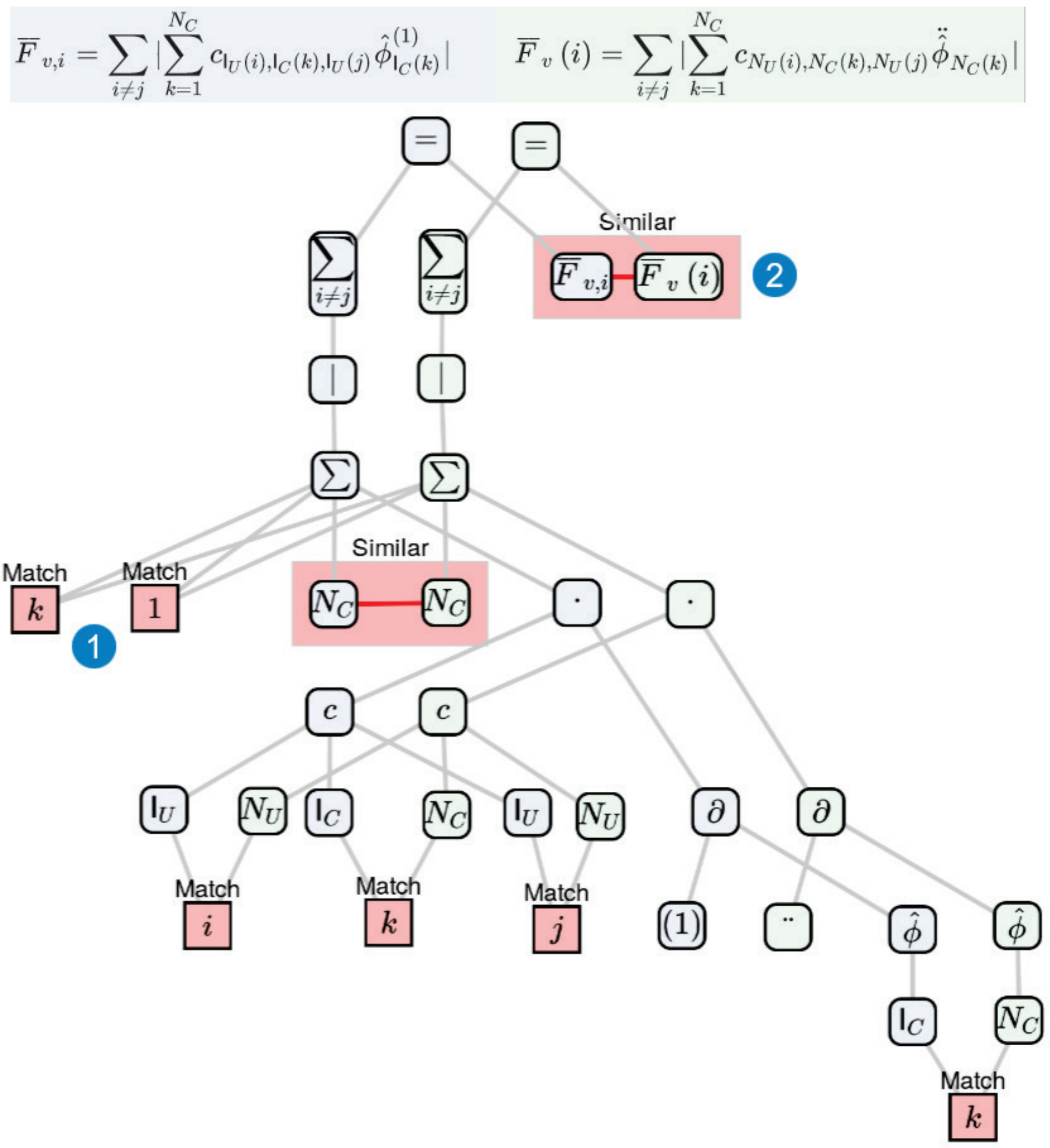

**Figure 7.3.** Visualization of pairwise formula similarity.

Mathematical practitioners appreciated the visualization for tasks like editing MathML as part of the digital publishing process. However, many users of HyPlag found the visualization too complicated and unintuitive for the plagiarism detection use case. Therefore, we deactivated the functionality after a while. The anecdote exemplifies the difficulty of visualizing the similarity of complex structures like mathematical formulae and meeting the task's and users' requirements.

The heterogeneity of users is a challenge that future visualization concepts for plagiarism detection systems should address. For example, the information need of a schoolteacher will most likely differ significantly from that of a journal reviewer. While schoolteachers typically want to find out quickly whether student assignments contain copy-and-paste plagiarism, the journal reviewer is likely willing to invest more time to check for disguised plagiarism carefully. Consequently, selecting the most informative features and similarities for individual users is an open research problem. Moreover, the experience of users in operating the system could be a factor that future visualizations consider. For example, to avoid information overload for occasional users, a system could offer different modes, e.g., for first-time users, typical users, and power users like academic integrity experts.

In summary, plagiarism detection systems that adapt (ideally automatically) the type and amount of information they present to the requirements and expertise of the user are an exciting and challenging area for future research.

### Decentralized Confidentiality-Preserving PDS

Currently, any production-grade plagiarism detection software requires full access to a document's content to run checks for plagiarism. This requirement introduces concerns about the privacy of users' data and the risk of confidential intellectual property being leaked. Moreover, a few private corporations, primarily outside of Europe, control today's plagiarism detection services. This consolidation of services further aggravates legal concerns, facilitates potential data misuse, and can lead to oligopoly effects, such as prohibitively high licensing fees (cf. Section 2.4.8, p. 47).

To address these issues, we will research the use of textual and non-textual content analysis, secure multi-party computation, and blockchain technology. Our goal is to devise the first fully confidential, decentralized plagiarism detection system. The system will identify similar content without users having to share their intellectual property or data in plaintext.

To achieve this objective, we will develop methods capable of creating confidential "semantic fingerprints" representative of the content to be checked for plagiarism. The system will extract various textual and non-textual content features, obscure them, and secure them using a hash function. All these steps are performed exclusively on a user's computer. Only after securing the features, the system transmits them to a distributed plagiarism detection system running on a blockchain.

The blockchain-backed system architecture allows for the distributed operation of computing nodes, e.g., at universities or research institutions. The redundant, cryptographically secured storage of the data on numerous computers prevents data leaks and subsequent data manipulation. Also, the distributed architecture we propose makes it impossible for individual entities to control the system. Our research lays the foundation for a non-commercial "Academic Blockchain" that allows:

1. Performing plagiarism checks without transmitting documents in plaintext;
2. Proving that intellectual property or data was not retrospectively copied or falsified without requiring public disclosure. Intellectual property, e.g., grant applications or research data, can be assigned a tamper-proof timestamp stored on the blockchain, which anyone can verify without requiring access to the data.

A **DFG grant proposal** for the project is currently in review. We published initial results on methods to securely obscure and subsequently analyze secured citation patterns as part of a check for plagiarism [236]. Analyzing obfuscated citation patterns achieved the same effectiveness as analyzing plaintext citations. With Decentralized Trusted Timestamping, we devised a key technology of the future system [177]. The approach embeds hash values for arbitrary digital data into the blockchain data structure underlying distributed ledgers, such as Bitcoin. The user in possession of the data associated with the hash can use the tamperproof timestamp of the blockchain transaction to prove that the data existed at the transaction time. The characteristics of the blockchain data structure allow that the user can remain anonymous and does not need to disclose the data to create the timestamps. Furthermore, no central authority can censor or manipulate the process.

### 7.3.3 Other Applications

The research on non-textual content analysis and pattern identification we presented could improve several other information retrieval applications.

Literature search and recommender (LSR) systems, particularly for scientific literature, are a natural application of the similarity assessments we presented. Finding and reviewing research publications is a necessity for researchers to stay informed about the progress in their field. However, the large volume and rapid growth of publications[15] make discovering relevant literature tedious and time-consuming. LSR systems facilitate the retrieval process. Current LSR systems predominantly rely on text and basic citation analysis [48]. The systems do not yet use the semantic information contained in images, mathematical content, and the combination of textual and non-textual analysis approaches.

We expect that extending literature search and recommender systems with adaptions of the non-textual content analysis approaches we presented could significantly increase their retrieval effectiveness. The systems could, for example, allow users to upload their publications and search for related work that cites similar sources in similar sections or discuss related mathematical problems. Specialized image analysis methods could enable the retrieval of documents whose figures show similar results. This functionality could, e.g., facilitate retrieving medical studies supporting a hypothesis during the compilation of a systematic literature review.

Extending the content analysis methods of LSR systems would call for research on novel user interfaces to effectively use the novel capabilities. We recently presented the first prototype of a possible interface for an LSR system that uses the content analysis capabilities of the HyPlag backend [63]. Instead of using ranked lists, which are the de-facto standard visualization of LSR systems, we employed an interactive force-directed graph layout. The users can manipulate the layout by choosing the types of content similarity (text, citation, mathematics, or combinations thereof) that are most relevant to their information needs.

Aside from LRS systems, extending the assessments of mathematical content we presented could improve a range of specialized retrieval systems for domain experts. For example, the ability to extract mathematical concepts from their mathematical representations occurring in publications, i.e., accessing the semantics of mathematical content, could improve mathematical expert search systems. As part of the DFG1 project, we research methods to achieve this extraction by simultaneously analyzing mathematical expressions and the text surrounding the expression. The goal is to automatically label expressions like $\zeta(s) = 0 \Rightarrow \Re\, s = \frac{1}{2} \vee \Im s = 0$ with the corresponding concepts, in this case, <Riemann hypothesis>. Methods to assess

15 Jinha estimated that the number of journal articles surpassed 50 million in 2009 [252]. Other studies consistently found that the annual growth rate of journal articles is about 3% [58], [237, p. 5]. By extrapolation, we estimate that 67 million journal articles existed at the end of 2019 and journals currently publish about 2 million new articles per year. This number underestimates the true number of publications as it does not consider, e.g., conferences papers, books, or preprints.

mathematical content's lexical and structural similarity could improve systems supporting theorem search or definition lookup. Likewise, tutoring assistance tools for students and patent search systems would benefit from such analysis methods.

The core idea underlying our research, i.e., a combined analysis of patterns in the textual and non-textual content, is undoubtedly relevant for other retrieval tasks. An example of a more distant application, for which we consider applying the principles investigated in this thesis, is the automated identification of media bias, i.e., slanted news coverage. Media bias is a ubiquitous phenomenon in news coverage that can have severely detrimental effects on individuals and society [219]. Bias by omission or commission of information, i.e., by leaving out or adding information, is a common form of media bias. Identifying entities, such as persons, places, actions, and events, in a text and comparing their occurrence patterns in numerous news articles could help identify media bias. Mapping the text to semantic concepts, e.g., using Explicit Semantic Analysis, and analyzing patterns in the occurrence of concepts, could allow this type of analysis across languages. We demonstrated the approach for the plagiarism detection use case [341].

### 7.3.4 Plagiarism Policy Implications

Increasing the detection capabilities for disguised forms of academic plagiarism through non-textual content analysis further aggravates a pressing policy issue. Currently, no widely accepted definition of what constitutes plagiarism exists.

The content similarity that is considered plagiarism differs significantly between academic disciplines, research institutions, journals, and individual academics [138, p. 1f.], [551, p. 3ff.]. Debnath & Cariappa found that 55% of the 320 PubMed journals they analyzed did not have any plagiarism policy or statement by early 2017. Another 9% of the journals made "negligible mentions" to plagiarism [102, p. 146]. Numerous researchers and journal editors reported that journals typically accept a literal text overlap of 10–20% [389].

From our experience, we know that other journals have a strict zero-tolerance policy to improperly attributed text overlap. When submitting a 35-page-long literature review on plagiarism detection technology [140], we received a desk-reject notification. The reason was that we incorrectly quoted the bulleted list of questions shown in **Figure 7.4**. We stated the source but did not enclose the list in quotation marks or set it apart visually from the rest of the text. The figure shows the problematic text passage in the similarity report of the journal's text-matching system. Thankfully, the journal allowed us to rectify our mistake and published the article.

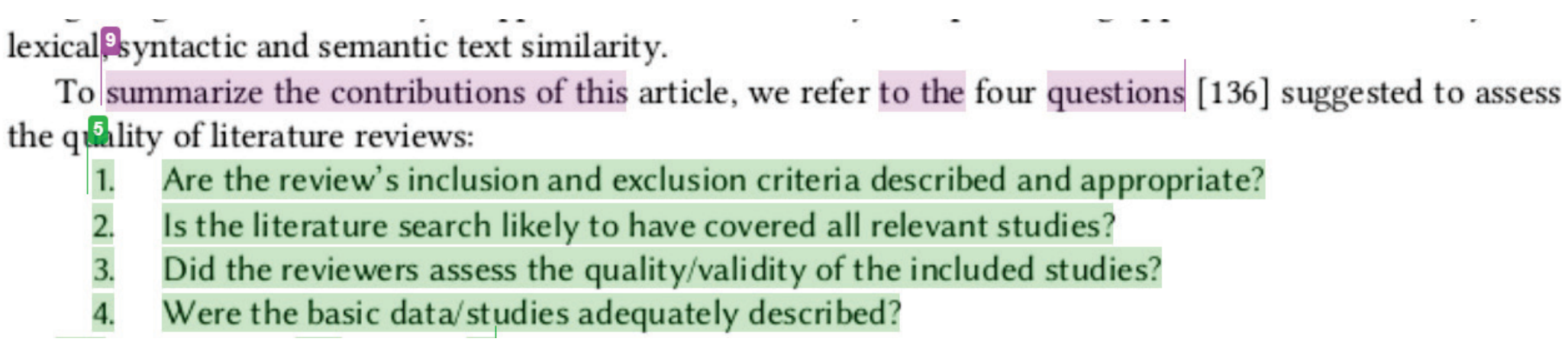
lexical, syntactic and semantic text similarity.
To summarize the contributions of this article, we refer to the four questions [136] suggested to assess the quality of literature reviews:
1. Are the review's inclusion and exclusion criteria described and appropriate?
2. Is the literature search likely to have covered all relevant studies?
3. Did the reviewers assess the quality/validity of the included studies?
4. Were the basic data/studies adequately described?

**Figure 7.4.** Incorrectly marked literal quote causing a desk-reject of an article.

The vast differences in handling text overlap in academic documents indicate the difficulties that will arise when non-textual content analysis finds widespread use. Future research on plagiarism policy should find definitions of academic plagiarism that specifically include non-textual content similarity and answer questions like:

» Does the selection and composition of citations constitute an intellectual contribution that authors must attribute if they use the same citations?

» What constitutes image similarity that authors must cite?

» Which amount of similar content, i.e., text, citations, images, mathematics, and combinations thereof, should a) warrant citing a source and b) represent the threshold above which the similarity is unacceptable?

# Appendices

# Appendix A

# Test Cases for Image-based Plagiarism Detection

This section presents the 15 cases of image reuse we obtained from the VroniPlag collection and used to evaluate our adaptive image-based plagiarism detection process, as described in Section 4.5, p. 137.

## Case 1: Illustration (Near-Duplicate)

<table>
<tr>
<td>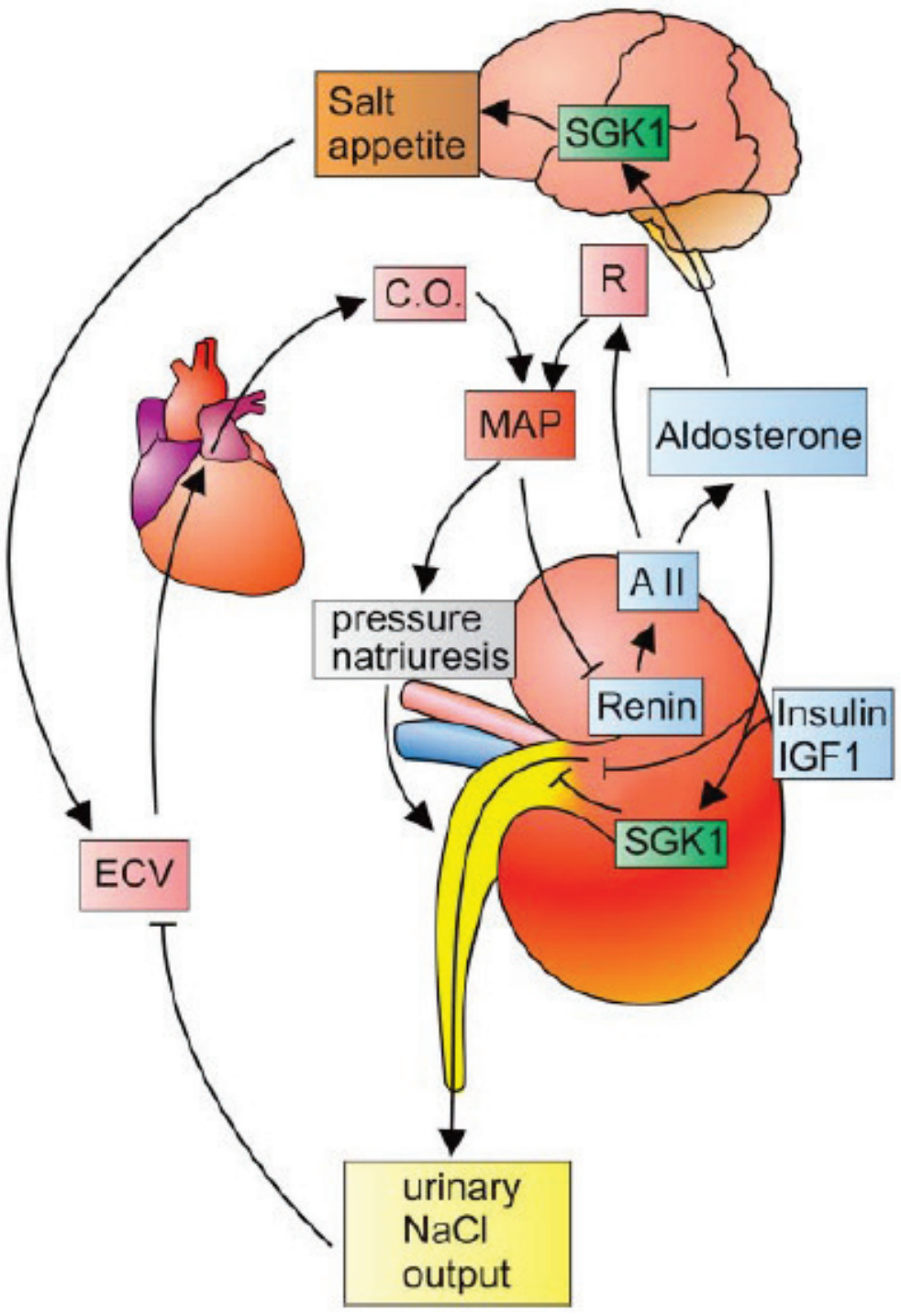
</td>
<td>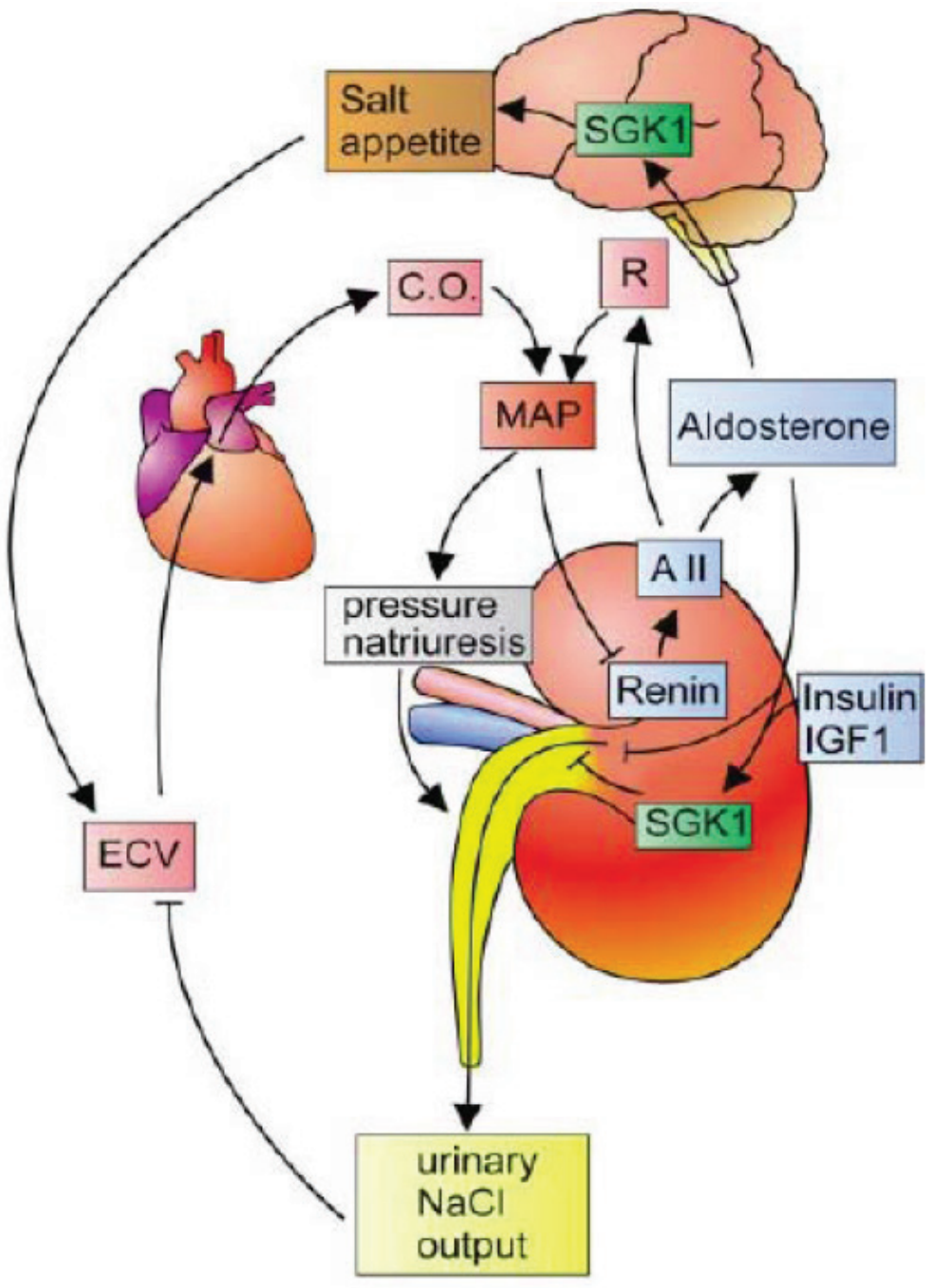
</td>
</tr>
<tr>
<td>Original image: [294, p. 1163]</td>
<td>Reused image: [440, p. 25]</td>
</tr>
<tr>
<td colspan="2">VroniPlag report on the finding: [530]</td>
</tr>
</table>

## Case 2: Illustration (Near-Duplicate)

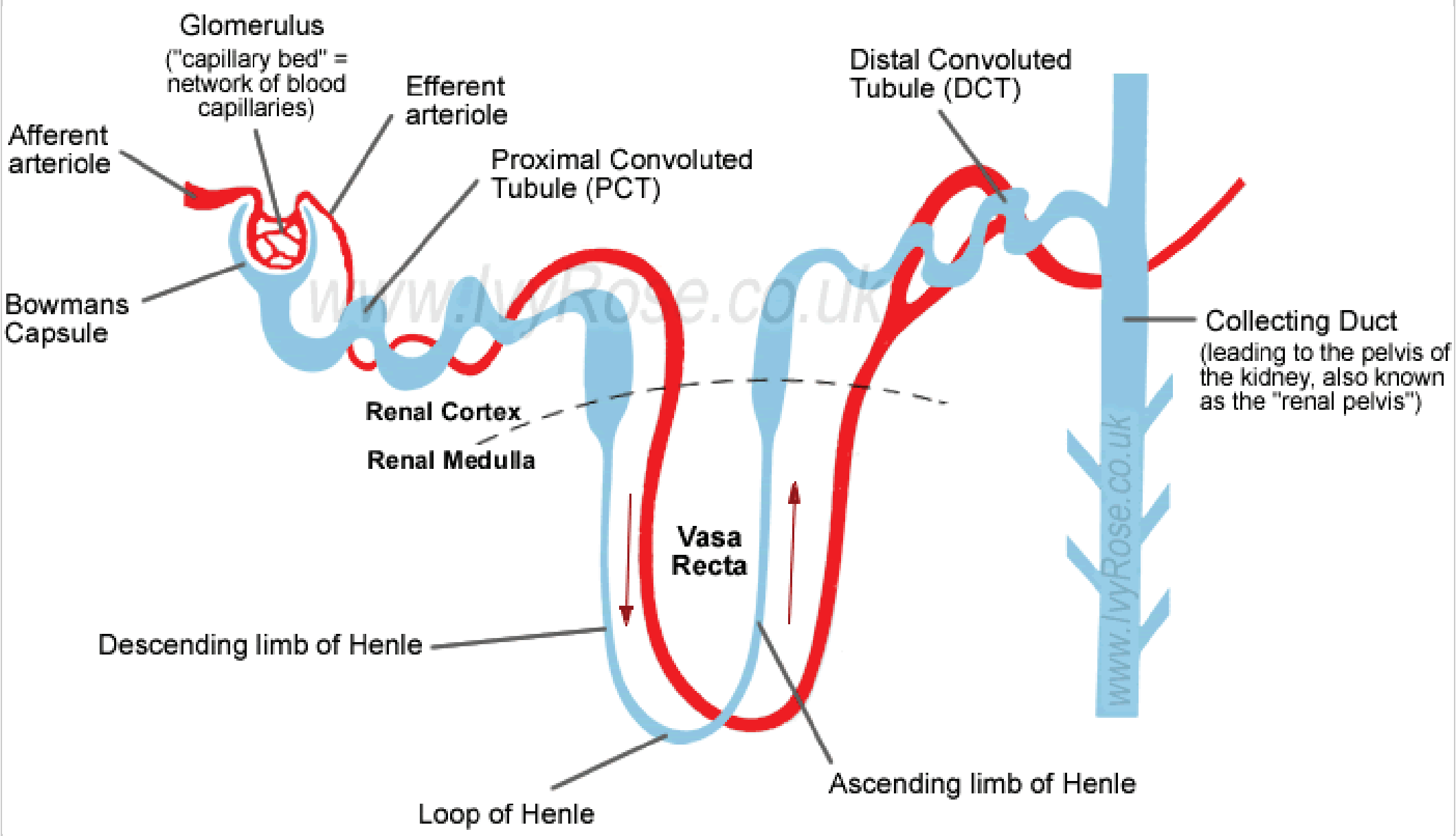

Original image: [239]

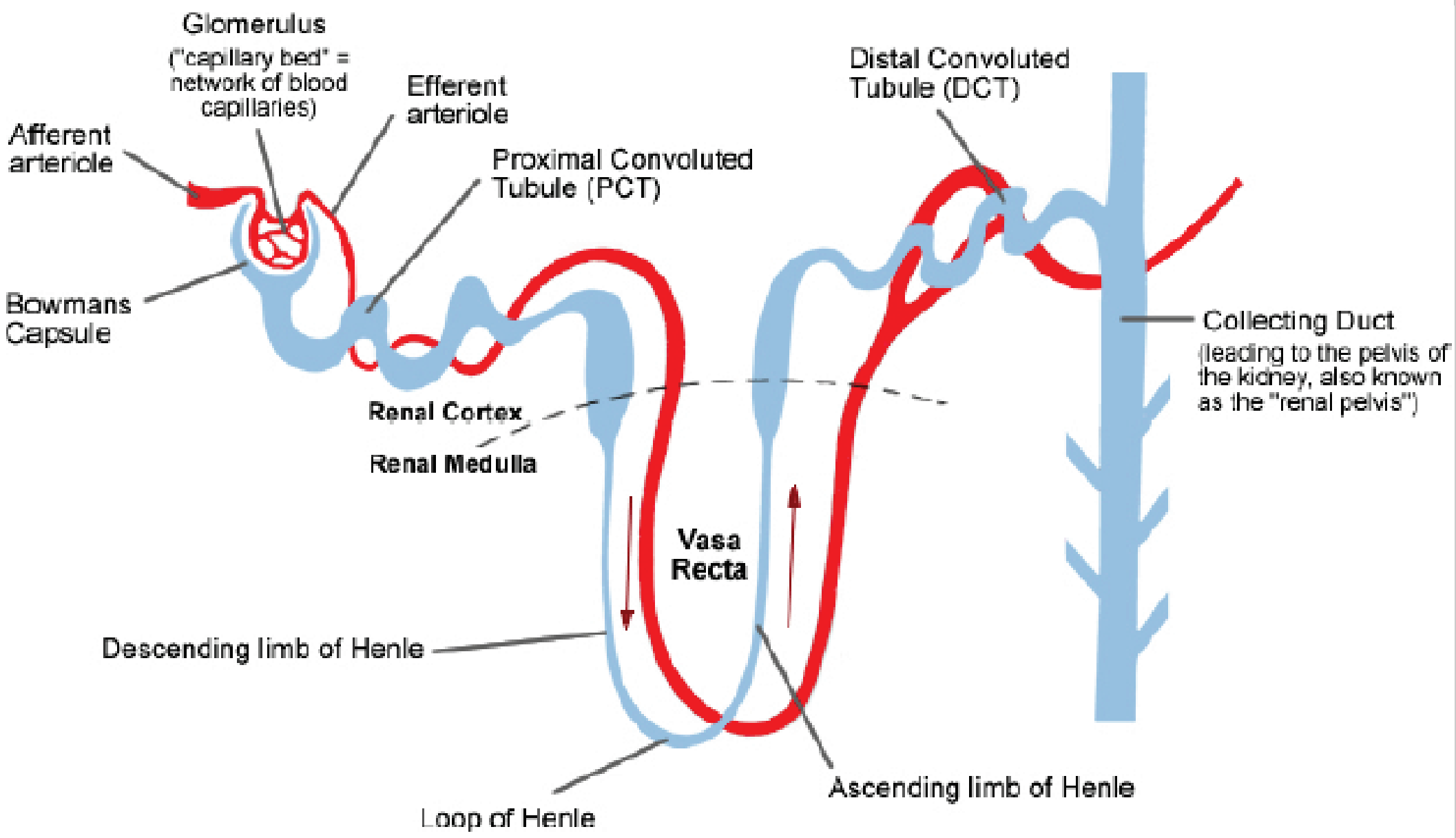

Reused image: [440, p. 15]

VroniPlag report on the finding: [529]

## Case 3: Illustration (Near-Duplicate)

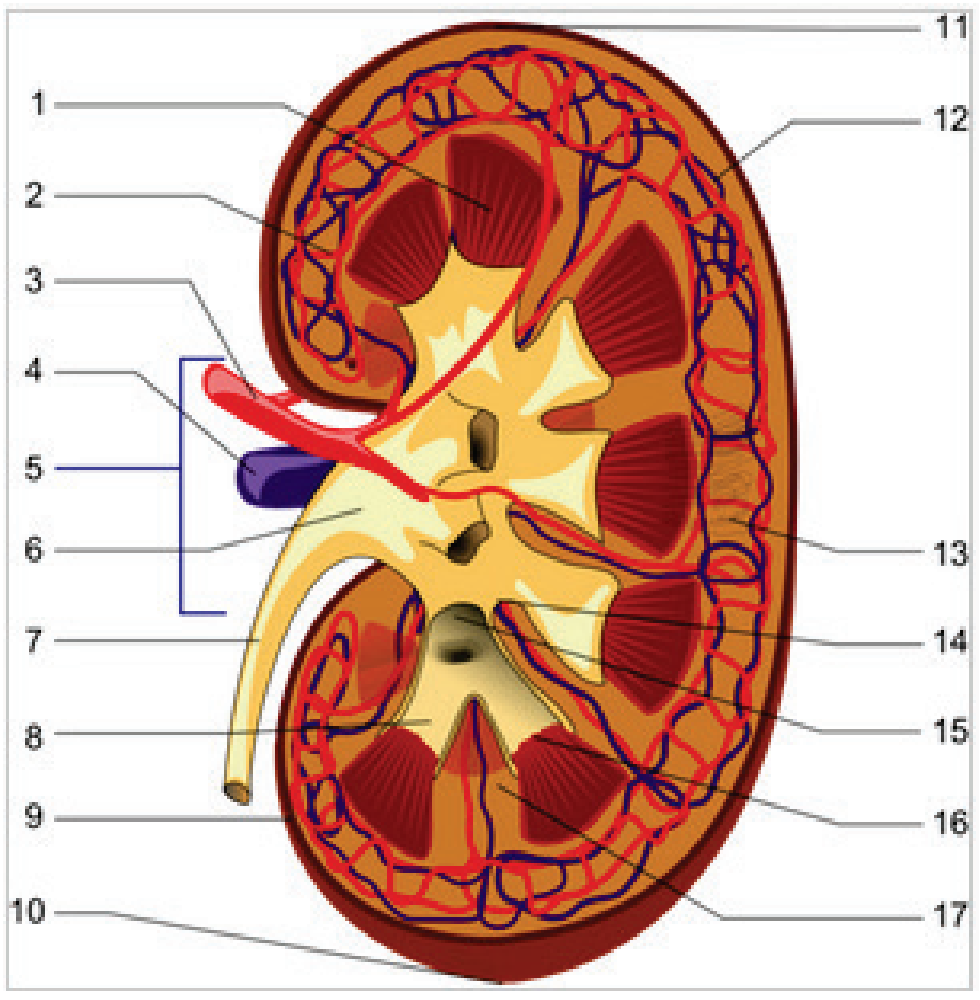

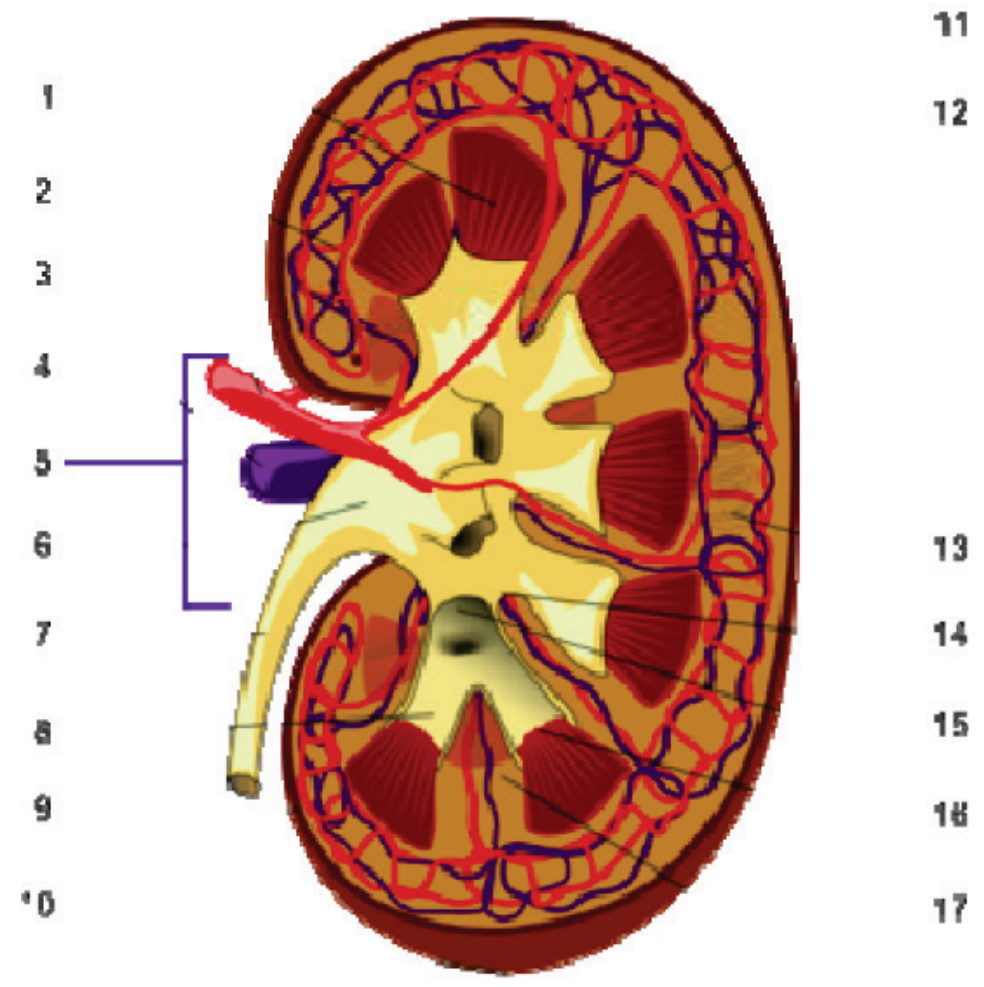

| | |
|---|---|
| Original image: [246] | Reused image: [440, p. 14] |
| VroniPlag report on the finding: [531] | |

## Case 4: Illustration (Weak Alteration)

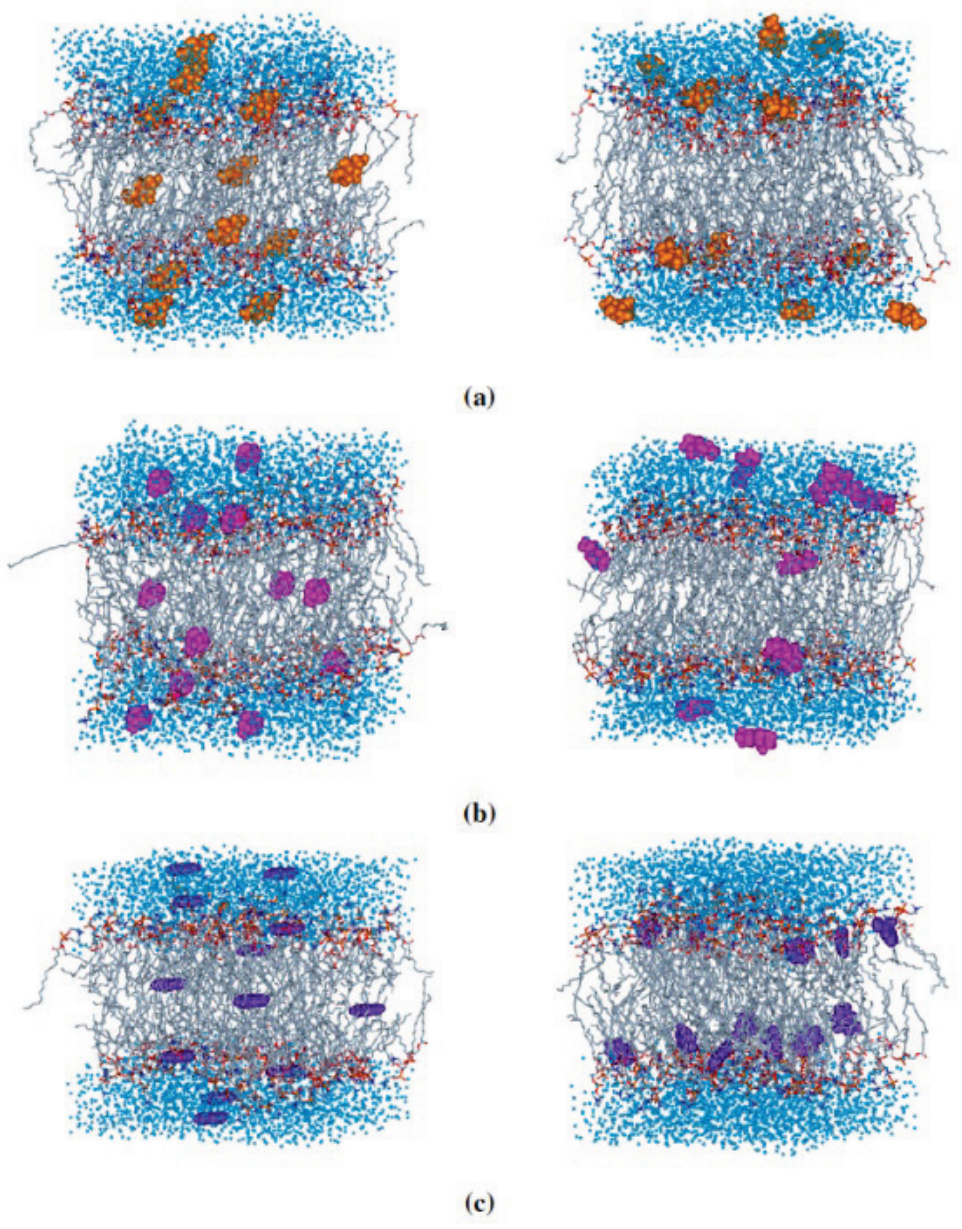

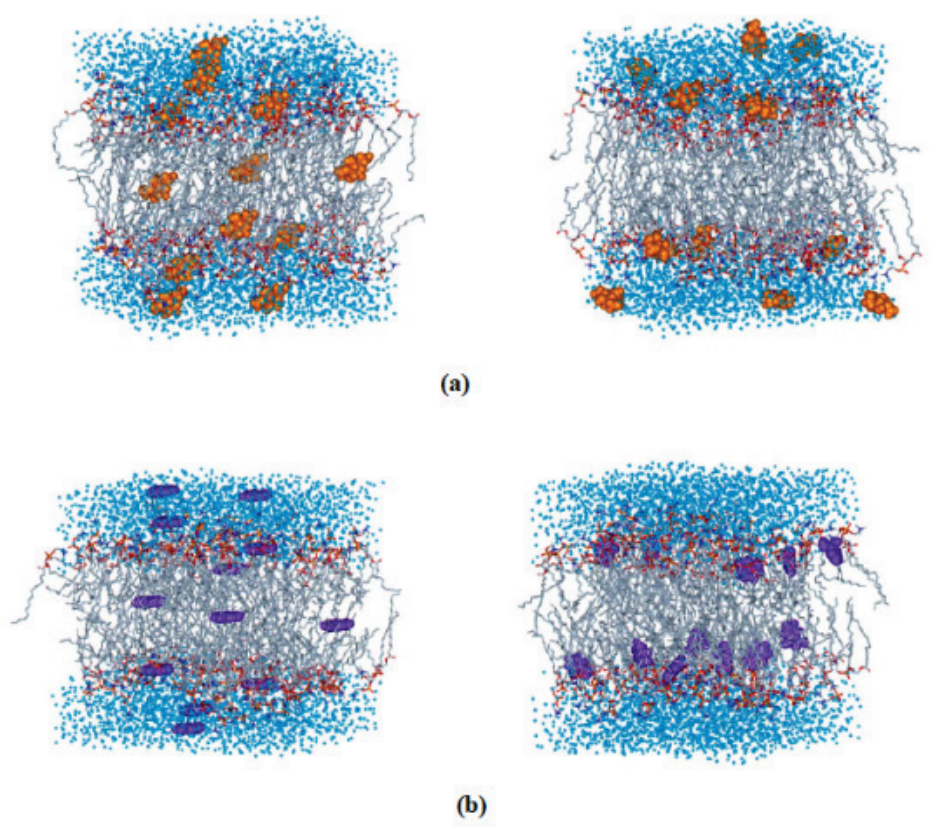

| | |
|---|---|
| Original image: [30, p. 160] | Reused image: [567, p. 73] |
| VroniPlag report on the finding: [537] | |

## Case 5: Illustration (Weak Alteration)

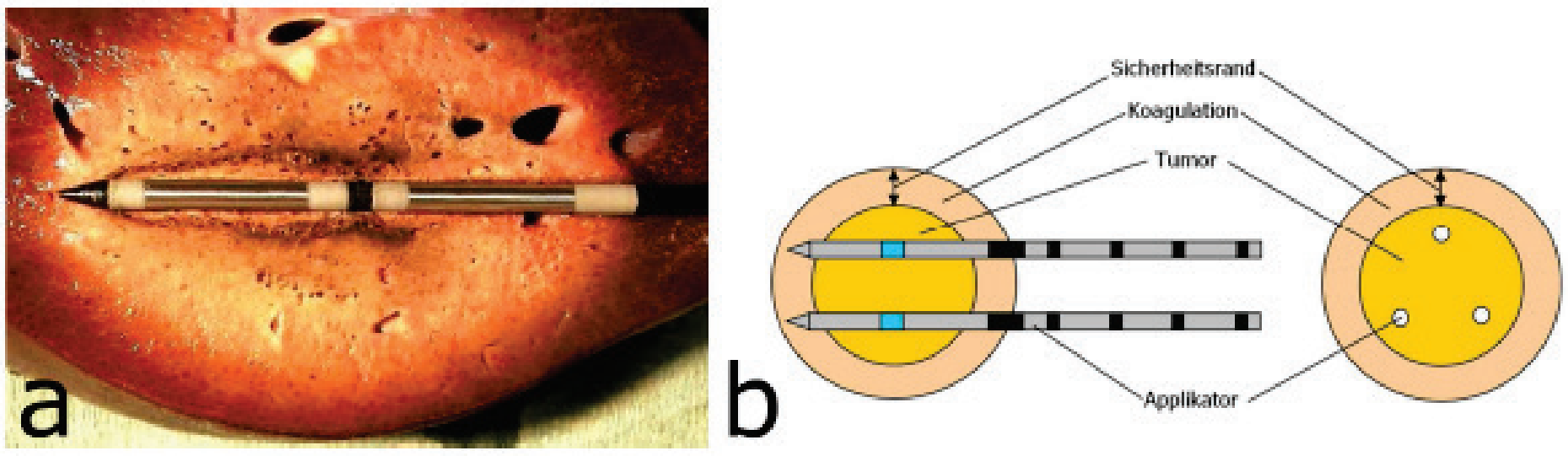

Original image: [302, p. 9]

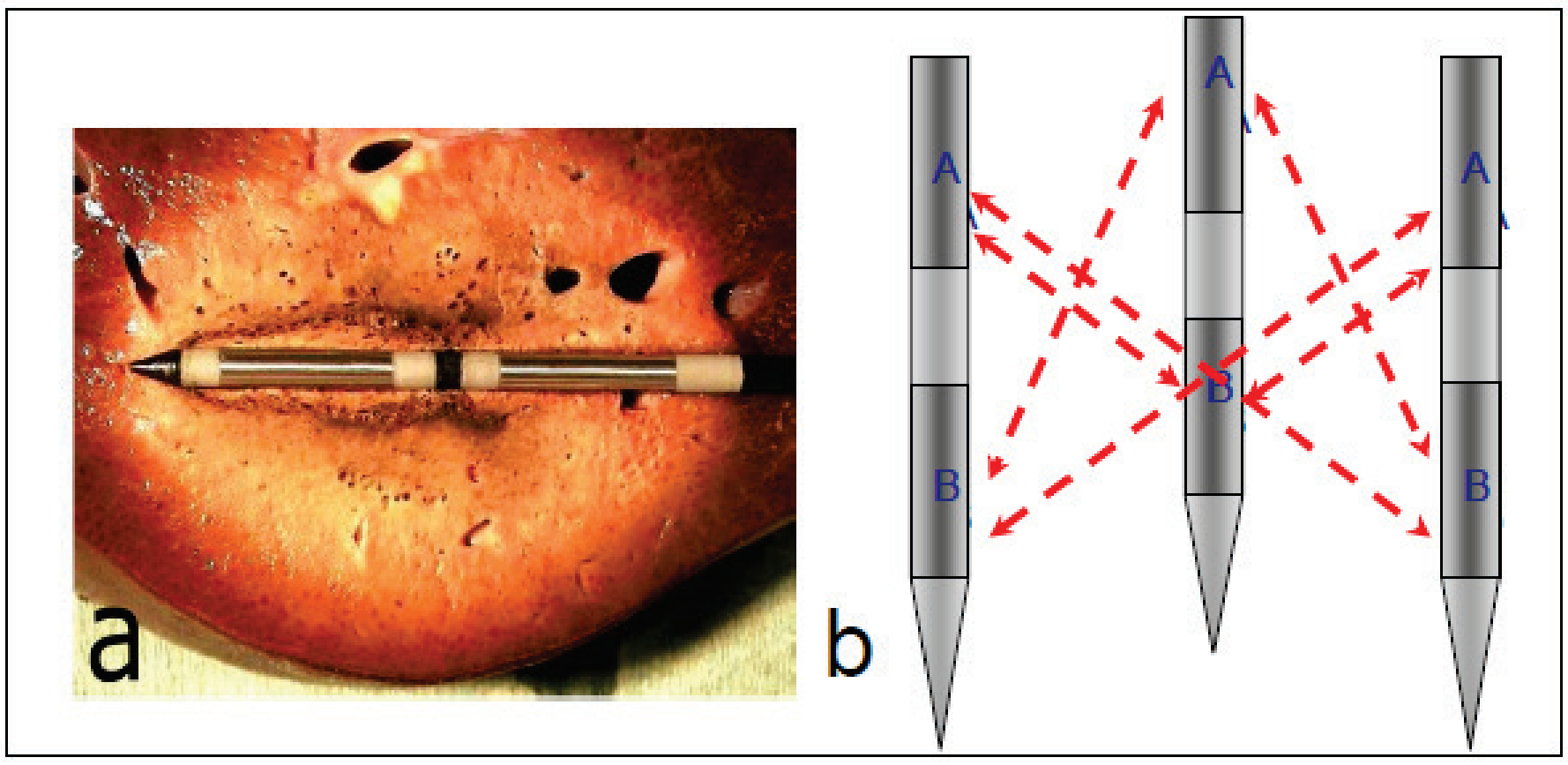

Reused image: [229, p. 5]

VroniPlag report on the finding: [523]

## Case 6: Illustration (Moderate Alteration)

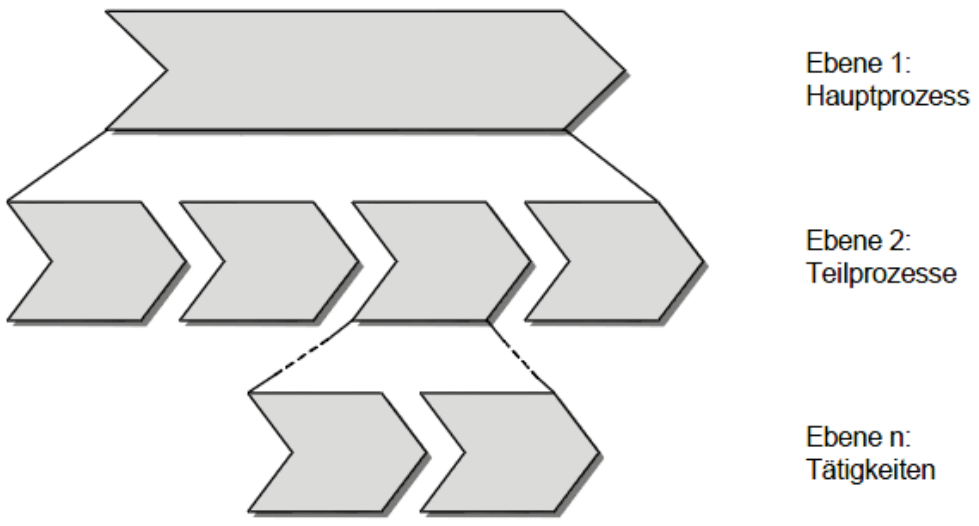

Original image: [163, p. 23]

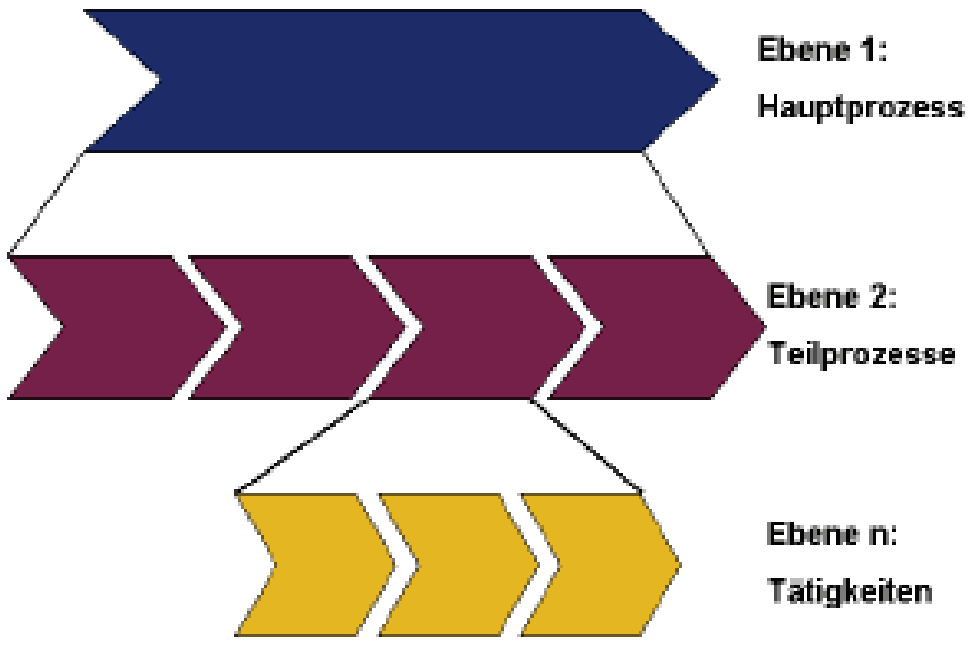

Reused image: [45, p. 14]

VroniPlag report on the finding: [520]

## Case 7: Illustration (Strong Alteration)

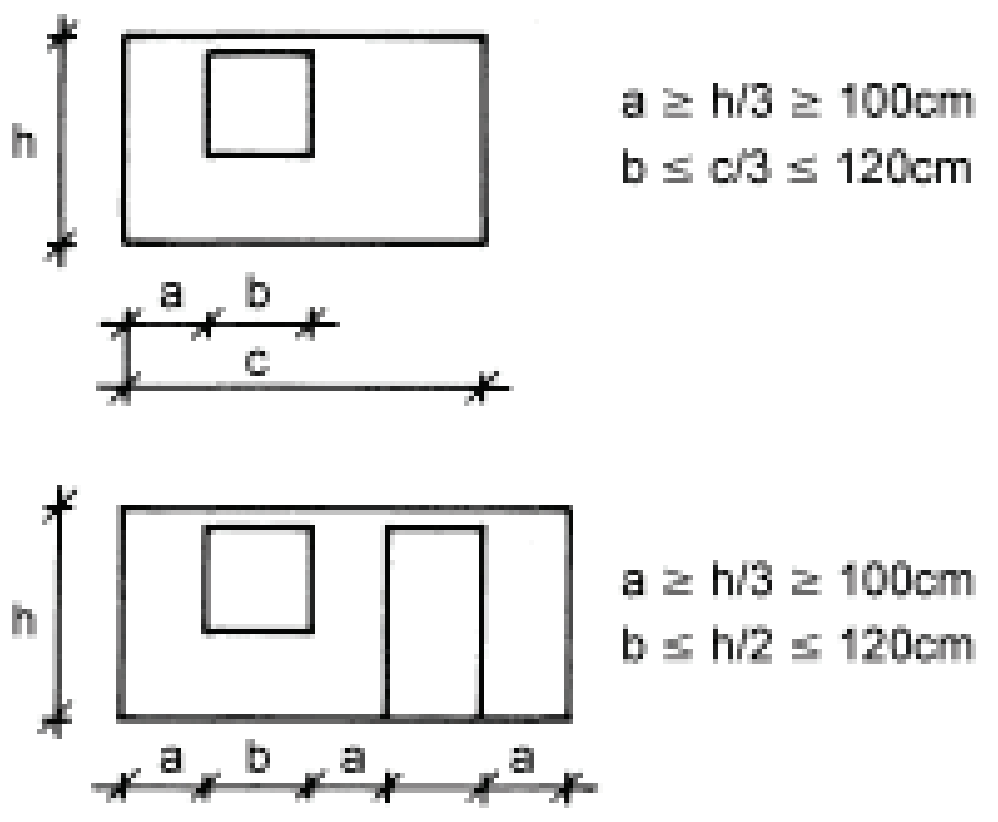

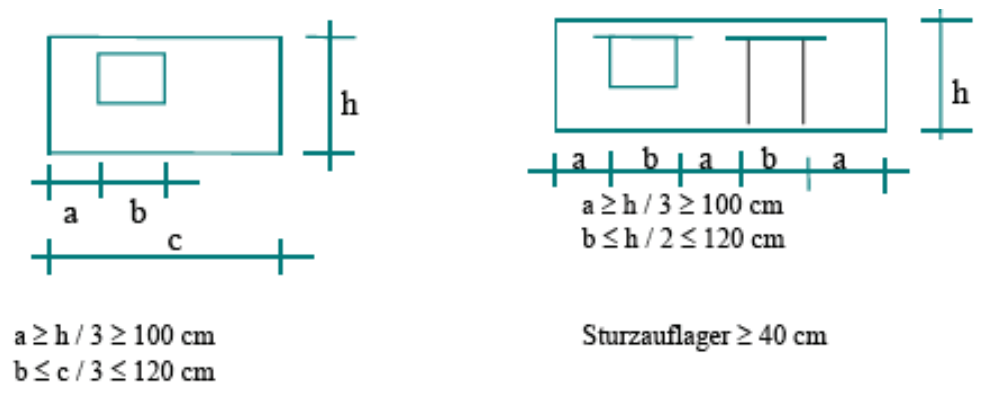

| Original image: [347, p. 3] | Reused image: [435, p. 193] |
|---|---|

VroniPlag report on the finding: [522]

## Case 8: Bar Chart (Near-Duplicate)

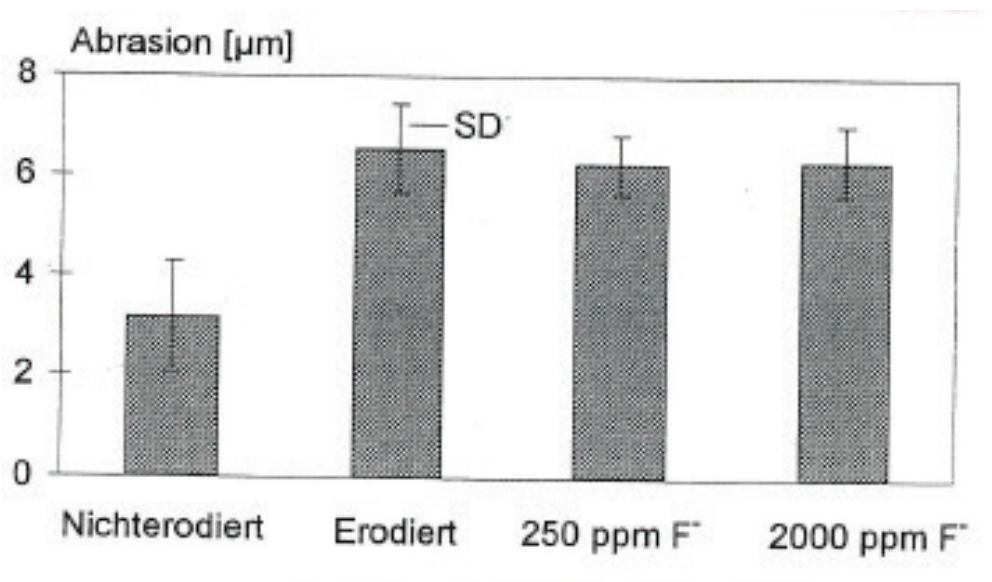

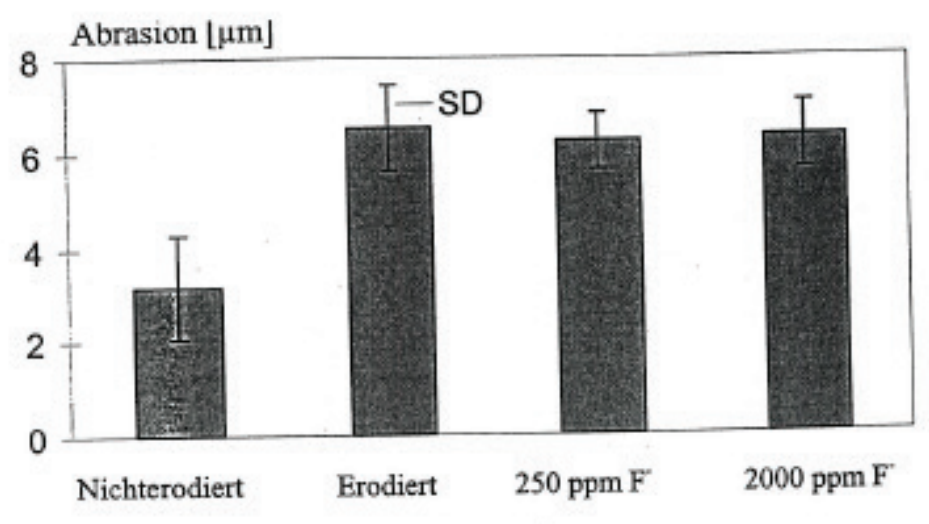

| Original image: [36, p. 117] | Reused image: [575, p. 47] |
|---|---|

VroniPlag report on the finding: [527]

## Case 9: Table (Near-Duplicate)

| | |
|---|---|
| Vertikale Auflösung | 5 Å |
| Horizontale Auflösung | 500 Å |
| Digitale Rauschunterdrückung | möglich |
| Meßbereich vertikaler Strukturen | 100 Å - 655.000 Å |
| Vertikale Auflösung | 5 Å |
| Horizontale Auflösung | 500 Å |
| Digitale Rauschunterdrückung | möglich |
| Meßbereich vertikaler Strukturen | 100 Å - 655.000 Å |

| Original image: [36, p. 81] | Reused image: [575, p. 39] |
|---|---|

VroniPlag report on the finding: [525]

## Case 10: Table (Near-Duplicate)

**Tab.4.1.1.**
Konzentrationsreihen von DaunoXome®
bei Langzeitversuchen

| µg/ml | µM |
|---|---|
| 0,000 | 0,000 |
| 0,010 | 0,018 |
| 0,025 | 0,044 |
| 0,050 | 0,089 |
| 0,100 | 0,177 |
| 0,250 | 0,443 |
| 0,500 | 0,887 |
| 1,000 | 1,773 |
| 2,500 | 4,433 |
| 5,000 | 8,865 |
| 10,000 | 17,730 |
| 25,000 | 44,326 |

**Abb.4.1.1.** Schematische Darstellung
des Versuchsaufbaus (entspricht Abb.3.4.1.)

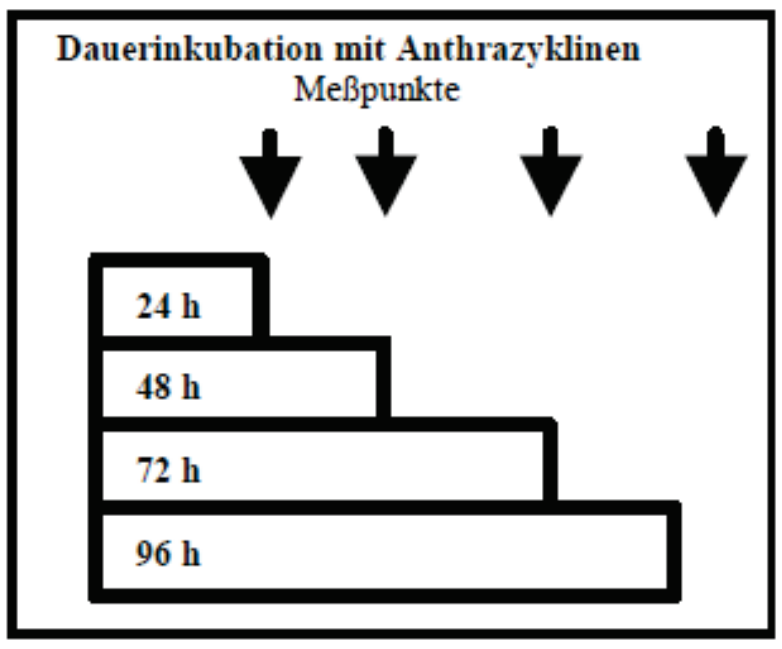

Original image: [450, p. 51]

**Tab. 4.1.1.**

Konzentrationsreihen von Daunorubicin bei Langzeitversuchen

| [µg/ml] | [µM] |
|---|---|
| 0,000 | 0,000 |
| 0,010 | 0,018 |
| 0,025 | 0,044 |
| 0,050 | 0,089 |
| 0,100 | 0,177 |
| 0,250 | 0,443 |
| 0,500 | 0,887 |
| 1,000 | 1,773 |
| 2,500 | 4,433 |
| 5,000 | 8,865 |
| 10,000 | 17,730 |
| 25,000 | 44,326 |

**Abb. 4.1.1.** Schematische Darstellung
des Versuchsaufbaus (entspricht Abb.
3.4.1.)

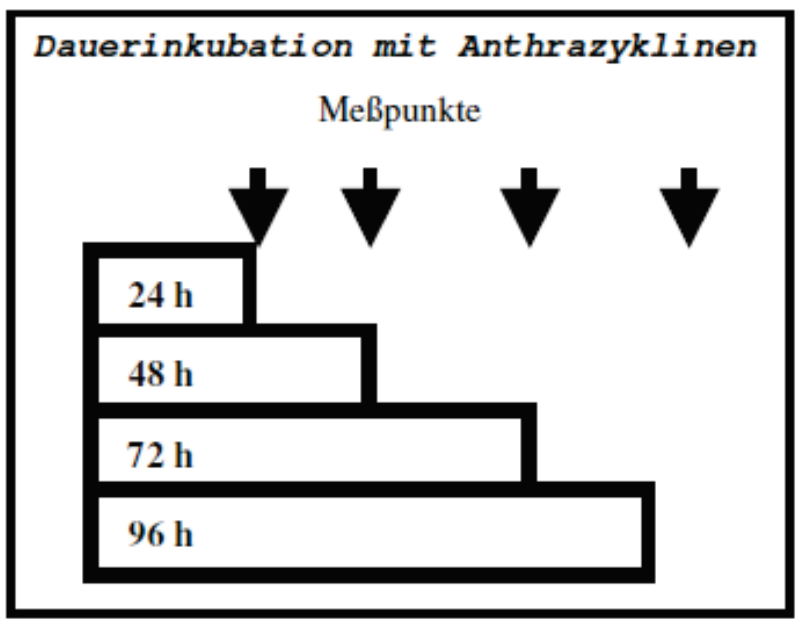

Reused image: [439, p. 29]

VroniPlag report on the finding: [535]

## Case 11: Table (Near-Duplicate)

| Zelllinie | Verdopplungszeit in der exponentiellen Wachstumsphase |
|---|---|
| Cado ES-1 | ca. 32 Stunden |
| STA-ET-1 | ca. 48 Stunden |
| STA-ET-2.1 | ca. 76 Stunden |
| VH-64 | ca. 20 Stunden |

Original image: [450, p. 43]

| Zelllinie | Verdopplungszeit in der exponentiellen Wachstumsphase |
|---|---|
| Cado ES-1 | ca. 32 Stunden |
| STA-ET-1 | ca. 48 Stunden |
| STA-ET-2.1 | ca. 76 Stunden |
| VH-64 | ca. 20 Stunden |

Reused image: [439, p. 22]

VroniPlag report on the finding: [534]

## Case 12: Table (Weak Alteration)

| Inhaltsstoffe | Abrasionseigenschaften |
|---|---|
| Wasser, Sorbitol, Hydroxy-ethylcellulose, Geschmacks-stoffe, Titandioxid, Saccharin<br>Silica-Putzkörper:<br>mittlere Größe 8,1 µm; 90 % ≤ 2 µm; 3 % ca. 30 µm (= maximale Größe) | REA-Wert 4,3 ± 0,3<br>RDA-Wert 77,0 ± 2,0 |

Tab. 4-3: Zusammensetzung (Herstellerangaben) und Abrasionseigenschaften (BARBAKOW *et al.*, 1989) der Zahnpasta Elmex®.

REA-Wert = Radioactive Enamel Abrasion
RDA-Wert = Radioactive Dentin Abrasion

Original image: [36, p. 80]

| Inhaltsstoffe | Abrasionseigenschaften |
|---|---|
| Wasser, Sorbitol, Hydroxy-ethylcellulose, Geschmacks-stoffe, Titandioxid, Saccharin<br>Silica-Putzkörper:<br>mittlere Größe 8,1 µm; 90 % ≤ 2 µm; 3 % ca. 30 µm (= maximale Größe) | REA-Wert 4,3 ± 0,3<br>RDA-Wert 77,0 ± 2,0 |

REA-Wert = Radioactive Enamel Abrasion
RDA-Wert = Radioactive Dentin Abrasion

Tab.4-2: Zusammensetzung (Herstellerangaben) und Abrasionseigenschaften (BARBAKOW et al., 1989) der Zahnpasta Elmex®

Reused image: [575, p. 35]

VroniPlag report on the finding: [524]

## Case 13: SEM Image (Near-Duplicate)

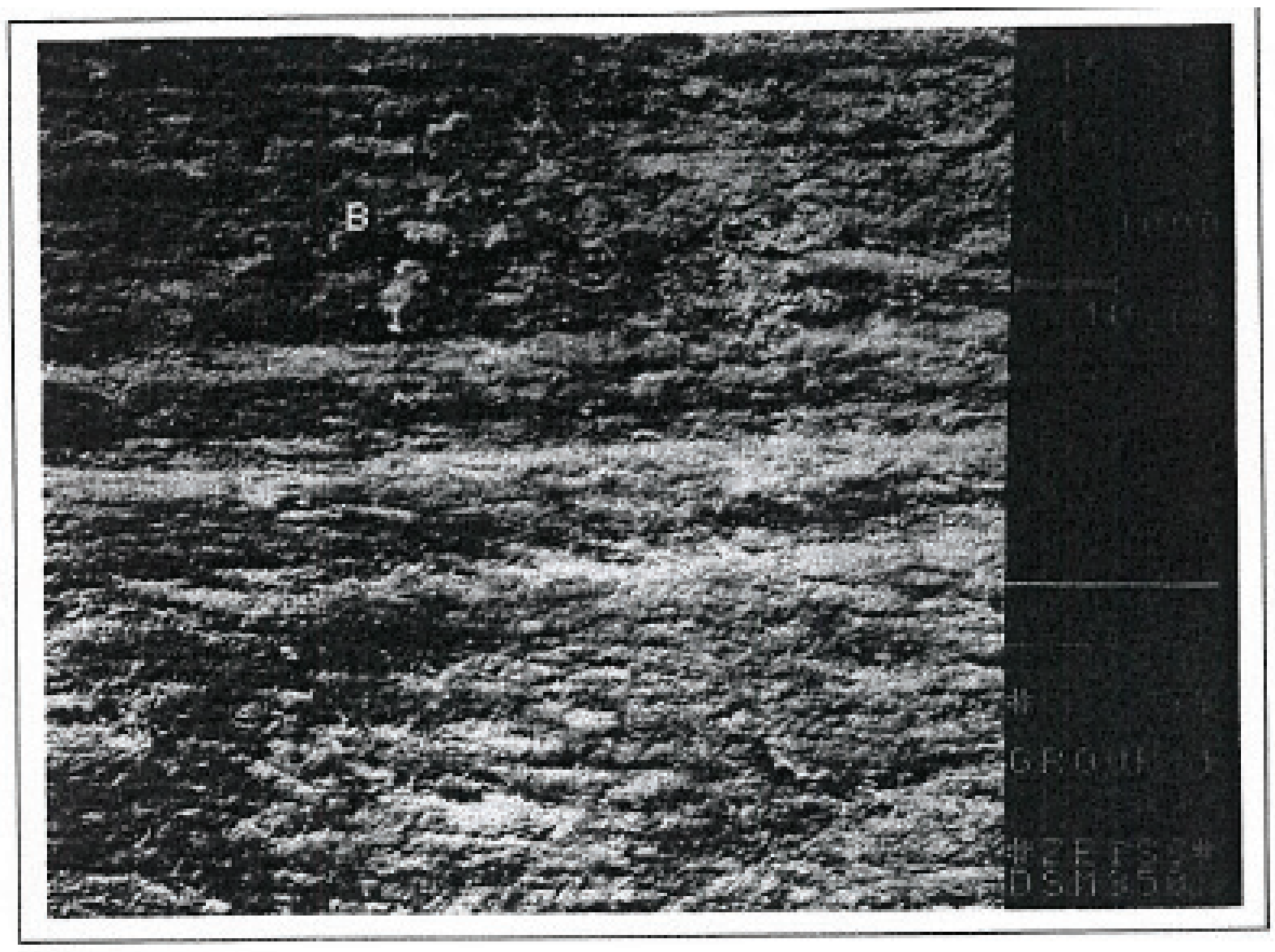

Original image: [36, p. 119]

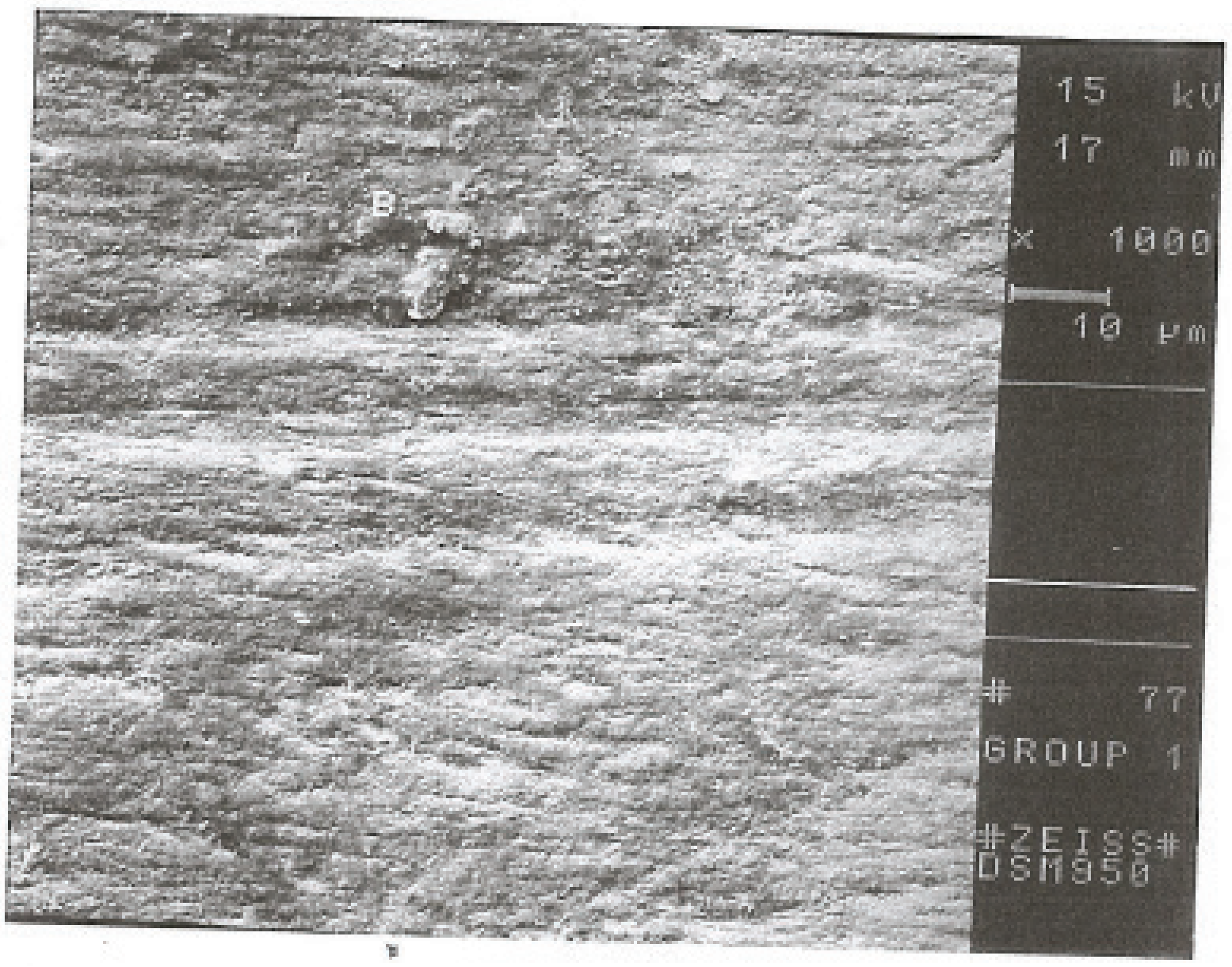

Reused image: [575, p. 48]

VroniPlag report on the finding: [528]

# Case 14: Line Chart (Weak Alteration)

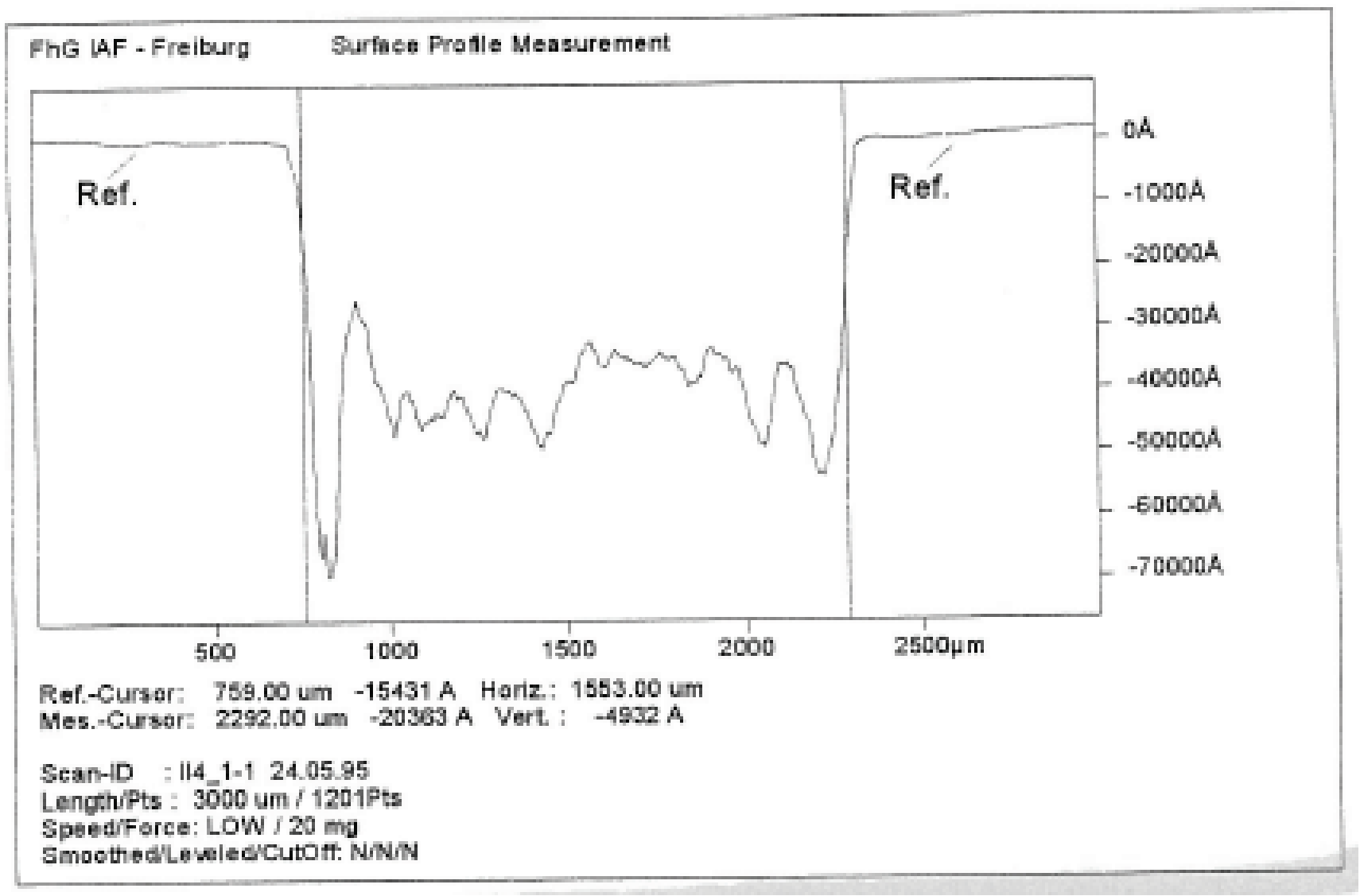

Original image: [36, p. 116]

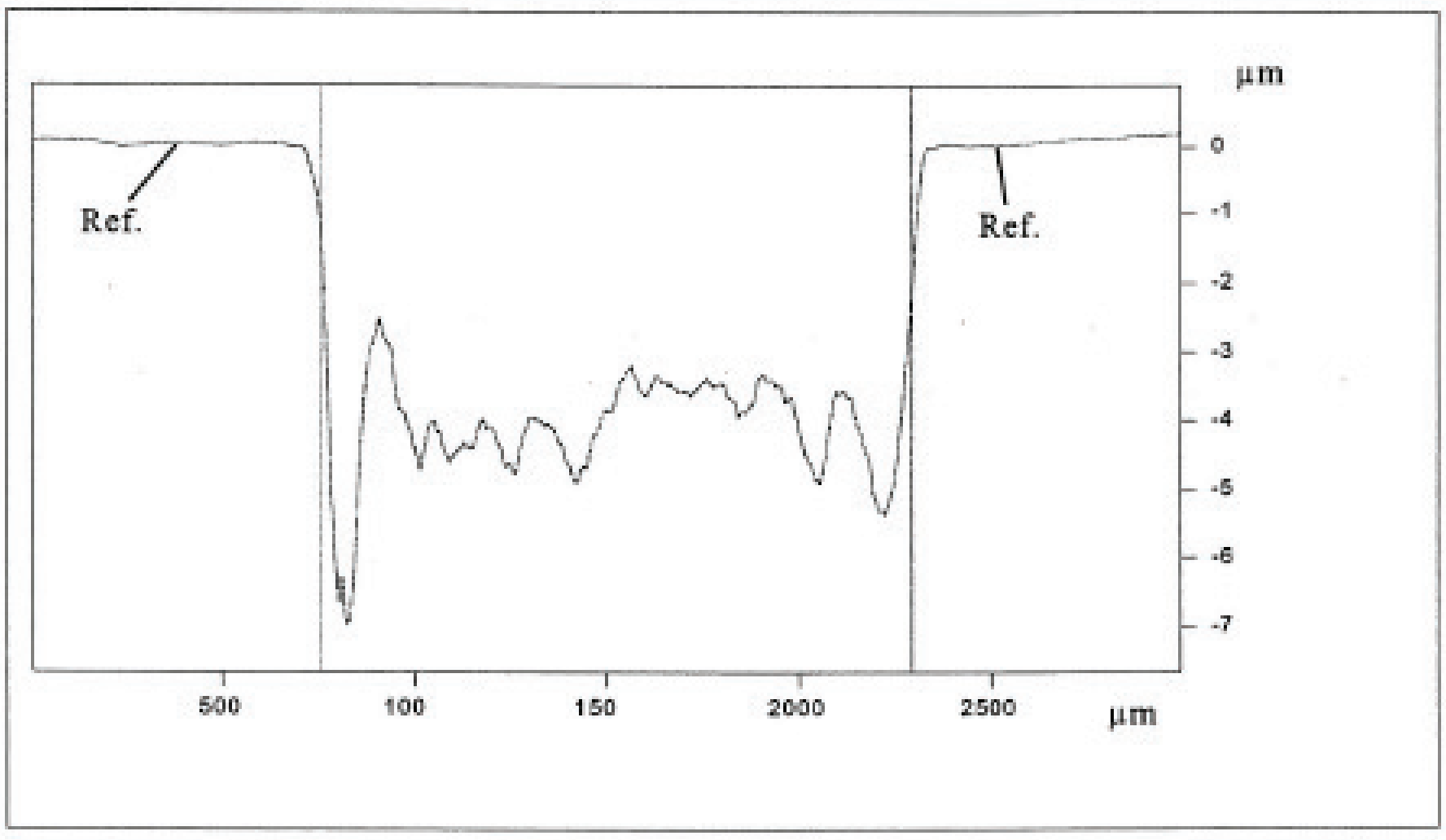

Reused image: [575, p. 44]

VroniPlag report on the finding: [526]

## Case 15: Line Chart (Strong Alteration)

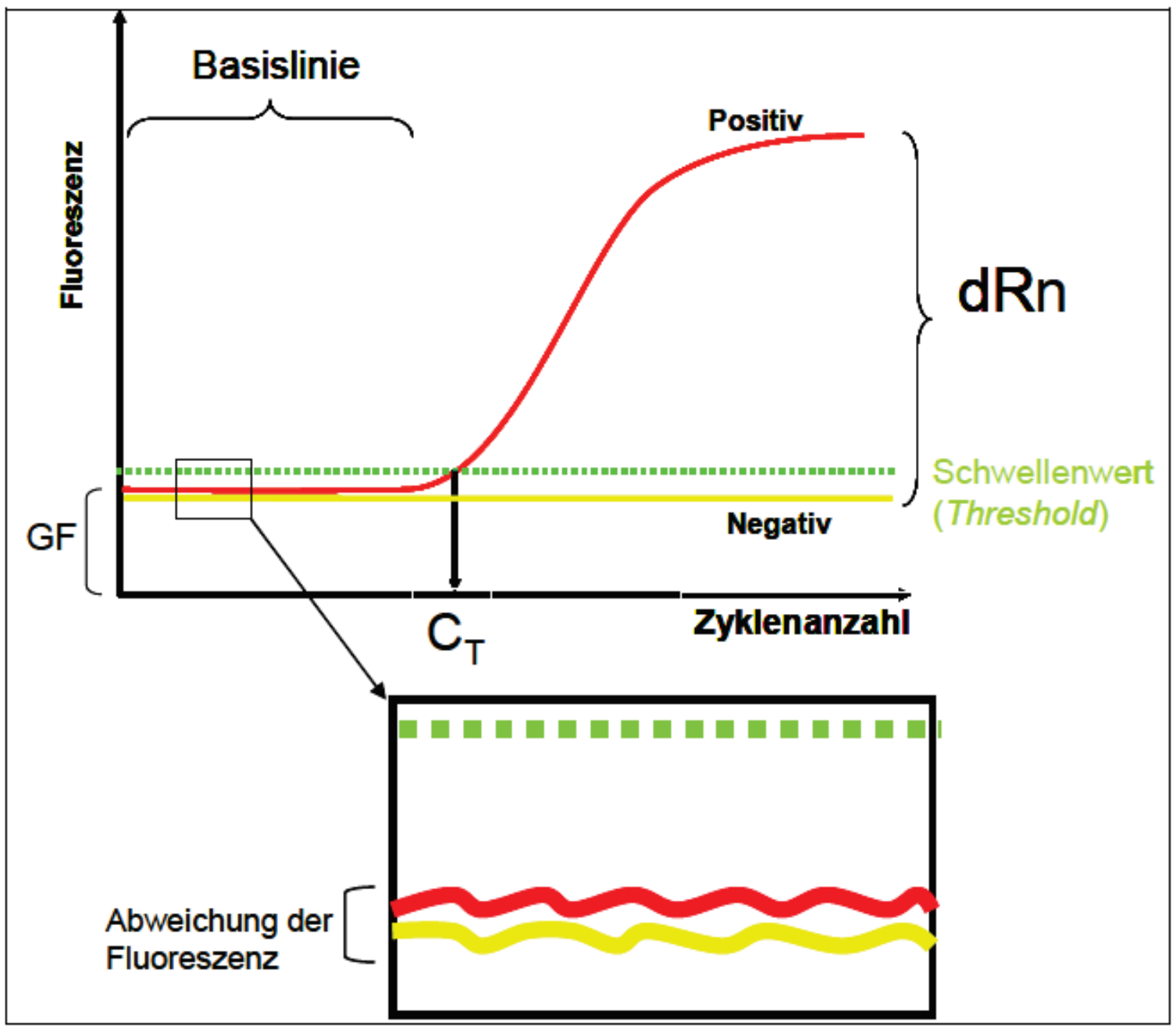

Original image: [52, p. 54]

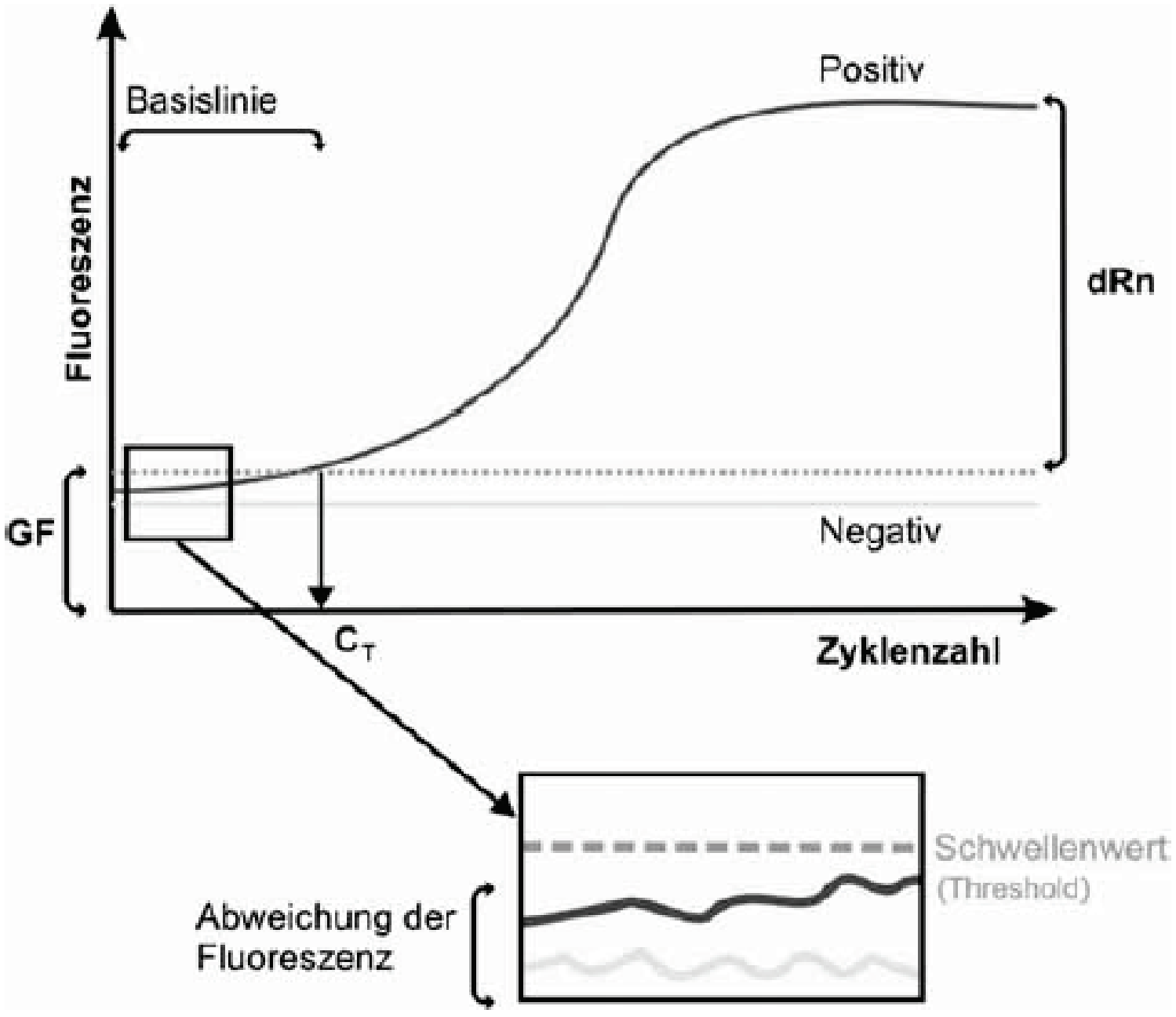

Reused image: [108, p. 68]

VroniPlag report on the finding: [521]

B

Appendix B

# Test Cases for Math-based Plagiarism Detection

**Table B.1** summarizes the 10 confirmed cases of academic plagiarism that we used to evaluate our math-based plagiarism detection methods. The case IDs in the first column correspond to the IDs used in Chapter 5. The second column indicates the research field of the retracted publication. The third column briefly characterizes the type of similar mathematics we observed in the retracted publications. In the fourth column, we cite the retraction note for each case.

**Table B.1.** Overview of test cases for MathPD evaluation.

| Case | Research Field | Description of Similar Mathematics | Ref. |
|---|---|---|---|
| C1 | Medical Engineering | **Summary:** Highly similar mathematics throughout an entire section (four pages).<br><br>**The retracted publication contains:**<br>- Identical formulae<br>- Near identical formulae with renamed identifiers<br>- Use of equivalent notation, partially rather complex and not easy to determine<br>- Split-up formulae, partially using equivalent notation | [562] |

| Case | Research Field | Description of Similar Mathematics | Ref. |
|---|---|---|---|
| C2 | Mathematics | **Summary:** Highly similar mathematics throughout the retracted publication.<br><br>**The retracted publication contains:**<br>- Identical formulae<br>- Near identical formulae with renamed identifiers<br>- Near identical formulae with inserted variables, e.g., $a+b=c$ vs. $b = x+y$ and $a+x+y=c$<br>- Use of equivalent notation, partially rather complex and not easy to determine<br>- Switch from inline formulae in the source to block formulae in the retracted publication<br>- Use of different brackets, e.g., round brackets vs. square brackets, small curly braces vs. large curly braces<br>- $(\ldots)^{-4}$ vs. $1/(\ldots)^4$<br>- Some intermediate formulae from the source are missing in the retracted publication (interesting pattern: often, every second formula is missing.) | [404] |
| C3 | Mathematics | **Summary:** Highly similar mathematics throughout the retracted publication.<br><br>**The retracted publication contains:**<br>- Identical formulae<br>- Near identical formulae with renamed identifiers<br>- Switch from inline formulae in the source to block formulae in the retracted publication | [1] |

| Case | Research Field | Description of Similar Mathematics | Ref. |
|---|---|---|---|
| C4 | Bioengineering | **Summary**: Reuse of an entire model, which is at the core of the publication.<br><br>**Observations:**<br>- Identifiers partially differ in the source and the retracted publication.<br>- Sometimes, multiple formulae in the source are merged into one formula in the retracted publication, but the sequence of formulae is nearly identical. | [268] |
| C5 | Computer Science (Sensor Networks) | **Summary:** Highly similar mathematics but also significant new content.<br><br>**The retracted publication contains:**<br>- Identical formulae<br>- Identical formulae with added parameter to the declared function, which also caused the addition of higher-level parentheses<br>- Changes in the order of formulae<br>- Use of different character to denote product operation | [396] |
| C6 | Computer Science (Image Analysis) | **Summary:** Clear reuse of mathematics and text.<br><br>**The retracted publication contains:**<br>- Identical formulae<br>- Near identical formulae with renamed identifiers<br>- Split-up formulae<br>- Integral range defined as new formula, while in the source, the range is inline | [257] |

| Case | Research Field | Description of Similar Mathematics | Ref. |
|---|---|---|---|
| C7 | Computer Science (Image Analysis) | **Summary:** The retracted publication references two sources for mathematics. The content apart from the formulae is clearly reused inappropriately; however, the retracted publication makes somewhat of a contribution regarding the mathematics. The formulae are substantially altered.<br><br>**The retracted publication contains:**<br>- Identical formulae<br>- Similar formula structure as in the sources, but some components are different beyond simple formula editing. | [267] |
| C8 | Applied Mathematics | **Summary:** Some equations correspond to equations in the source.<br><br>**Observations:**<br>- Similar equations are nearly identical to the source; we found only one instance of differing function names and some formatting differences, mostly inline fractions vs. display fractions<br>- Identical order of formulae as the source, but many equations are also not from the source. Only certain chunks of equations match those in the source. | [238] |

| Case | Research Field | Description of Similar Mathematics | Ref. |
|---|---|---|---|
| C9 | Mathematics | **Summary:** At several positions, mathematics is highly similar to mathematics in the source.<br><br>**Observations:**<br>- Some equations contain additional variables, e.g., a 2D vector in the source corresponds to a 3D vector in the retracted publication by splitting a variable C into two variables C1 and C2<br>- Some equations do not occur in the source, e.g., p4, or they are heavily modified. Some of the variables in those equations are defined in the same order, but the equations are mostly different.<br>- Some equations are expanded compared to the source, e.g., $a(b+c)$ vs. $ab+ac$<br>- The numerical example in the retracted publication uses the same equations and values as the source | [96] |
| C10 | Mathematics | **Summary:** Almost all equations in the retracted publication correspond to equations in the source.<br><br>**Observations:**<br>- Similar equations in the retracted publication are in the same order, and derivational steps are identical.<br>- Sometimes, the retracted publication uses slightly different notation than the source, e.g., $j\ (!=\ i)$ becomes $j\ !=\ i$ .<br>- The formatting of equations sometimes differs slightly, e.g., different line breaks.<br>- Two tables are exactly alike (contain the same values).<br>- One table exhibits a different order of columns and rows than the source. | [109] |

C

Appendix C

# Overview of the MathML Standard

Mathematical Markup Language (MathML) is a W3C and ISO standard (ISO/IEC DIS 40314) [561] for representing mathematical content using XML syntax. MathML is part of HTML5 and enables serving, receiving, and processing mathematical content on the World Wide Web. MathML allows users to describe the notation and the meaning of mathematical content using two vocabularies: Presentation MathML and Content MathML.

Presentation MathML describes the visual layout of mathematical content. The vocabulary contains elements for basic mathematical symbols and structures. Each element specifies the role of the presentation element, e.g., the element `<mi>` represents identifiers, and the element `<mo>` represents operators. The structure of Presentation MathML markup reflects the two-dimensional layout of the mathematical expression. Elements that form semantic units are encapsulated in `<mrow>` elements, which are comparable to `<div>` elements in HTML. **Figure C.1** exemplifies Presentation MathML markup for the expression $f(a+b)$.

```
<math xmlns="http://www.w3.org/1998/Math/MathML">
  <semantics>
    <mrow id="r1">
      <mi id="i1">f</mi>
      <mo id="o1">(</mo>
      <mrow id="r2">
        <mi id="i2">a</mi>
        <mo id="o2">+</mo>
        <mi id="i3">b</mi>
      </mrow>
      <mo id="o3">)</mo>
    </mrow>
```

**Figure C.1.** Presentation MathML encoding of the expression *f(a+b)*.

Content MathML explicitly encodes the semantic structure and the meaning of mathematical content using expression trees. In other terms, the Content MathML vocabulary specifies the frequently ambiguous mapping from the presentation of mathematical content to its meaning. For example, the presentation MathML markup of the expression $f(a+b)$ represents two possible syntactic structures because the symbol $f$ could represent either an identifier or a function. Content MathML uses `<apply>` elements to make explicit which elements represent functions. Subordinate elements represent the arguments of the functions. **Figure C.2** illustrates Content MathML markup for the expression $f(a+b)$.

```
1 <annotation-xml encoding="MathML-Content">
2       <apply xref="r1">
3         <ci xref="b">f</ci>
4         <apply xref="r2">
5           <plus xref="o2"/><!-- <csymbol cd="arith1">plus</csymbol> in strict encoding -->
6           <ci xref="i2">a</ci>
7           <ci xref="i3">b</ci>
8         </apply>
9       </apply>
```

**Figure C.2.** Content MathML encoding of an expression *f(a+b).*

Content MathML offers two subsets of elements to specify function types: Pragmatic Content MathML and Strict Content MathML. Pragmatic Content MathML uses a large set of predefined functions encoded as empty elements, e.g., `<plus/>`, as used in Line 5 in **Figure C.2**, or `<log/>` for the logarithm. Strict Content MathML uses a minimal set of elements, which are further specified by referencing extensible content dictionaries. For example, the plus operator (+) is defined in the content dictionary arith1. In Strict Content MathML, the operator is encoded using the element for symbols `<csymbol>` and declaring that the specification of the symbol is available under the term `plus` in the content dictionary `arith1`. Line 5 in **Figure C.2** shows this option as a comment (grey font color).

The Presentation MathML and Content MathML vocabularies can be used individually and independently, or in conjunction. For example, Presentation MathML is frequently used without content markup to display mathematical content on websites. Content MathML without presentation markup can, for instance, be used to exchange data between computer algebra systems. However, Presentation MathML and Content MathML markup can also be used in conjunction to simul-taneously describe the presentation, structure, and semantics of mathematical expressions. The combined use of Presentation MathML and Content MathML is commonly referred to as parallel MathML.

In parallel MathML markup, presentation and content elements are mutually interlinked by including `xref` arguments that point to the corresponding element in the other vocabulary. The Presentation and Content MathML markup in **Figure C.1** and **Figure C.2** contain `xref`-links to create parallel MathML markup.

# Glossary

**Apache Lucene**:
A free and open-source search engine software library (https://lucene.apache.org)

**API — Application Programming Interface**:
A computing interface for software-to-software communication; specifies the possible interactions, workflow, and data exchanged

**arXiv**:
An open-access repository of research publication preprints, primarily covering STEM disciplines (https://arxiv.org)

**BC — Bibliographic Coupling**:
A similarity measure representing the number of identical bibliographic references in two academic documents

**CbPD — Citation-based Plagiarism Detection**:
An approach to identify academic plagiarism by analyzing citations in academic documents for similar patterns (see Chapter 3)

**CC — Citation Chunking**:
A collection of citation-based algorithms to identify citation patterns regardless of whether the order of citations differs

**CL-ASA — Cross-language Alignment-based Similarity Analysis**:
A method that uses a bilingual word-unigram dictionary derived from a parallel corpus to quantify the likelihood that a text is a translation of another text

**CLEF — Conference and Labs of the Evaluation Forum**:
A series of scientific events addressing information access and information retrieval technology (http://www.clef-initiative.eu)

**CoCit — Co-Citation**:
A similarity measure representing the number of times documents are cited together in later documents

**CPA — Co-Citation Proximity Analysis**:
A measure representing how often and with which textual proximity later documents cite two earlier documents together

**DOI — Digital Object Identifier**:
A persistent identifier for digital data maintained and resolved by a registrar; frequently assigned to research publications

**DTO — Data Transfer Object**:
An object defined in an object-oriented programming language that carries data between processes to reduce method calls

**Elasticsearch**:
A free and open-source full-text search engine based on the Apache Lucene software library (https://www.elastic.co/elastic-stack)

**ENAI — European Network for Academic Integrity**:
An association of 30 universities and research institutions from Europe and Asia engaging in actions to improve academic and educational integrity (http://www.academicintegrity.eu)

**Encoplot**:
A text-matching tool using character 16-grams and pairwise document comparisons; ignores repeated matches to achieve O(n) complexity

**ESA — Explicit Semantic Analysis**:
A retrieval model that represents documents as vectors of semantic concepts derived from an external knowledgebase

**External plagiarism detection**:
A paradigm specifying that approaches following the paradigm compare documents to an extensive collection to identify plagiarism

**$F_1$-measure**:
The harmonic mean of Precision and Recall

**False positive**:
A non-relevant item a method retrieved

**Feature point methods**:
A class of content-based image retrieval methods that identify and match visually interesting areas of a scene

**Fingerprinting**:
Algorithms that map arbitrary input data to much shorter bit strings that uniquely identify the original data typically using one-way hash functions

**Greedy Tiling**:
A hypernym for methods that identify individually longest blocks of consecutive identical items occurring in the same order within two sequences;

**Greedy Citation Tiling (GCT)**, **Greedy String Tiling (GST)**, and **Greedy Identifier Tiling (GIT)** are implementations for citations, text strings, and mathematical identifiers, respectively.

**GROBID — Generation of Bibliographic Data**:
A software library for extracting, parsing, and re-structuring unstructured documents, such as PDF, into structured formats, such as XML or TEI (https://github.com/kermitt2/grobid)

**GuttenPlag**:
A crowdsourced project of volunteers that investigated the doctoral thesis of Karl-Theodor zu Guttenberg (former German Minister of Defense) for plagiarism and documented the results in a wiki (https://guttenplag.wikia.org/de)

**Hashing/hash function**:
Applying any function that maps arbitrarily sized input data to fixed-sized values; We refer to the values as hashes or hash values.

**Histo — Identifier Frequency Histograms**:
A math-based similarity measure; analyzes the difference in the relative frequencies of mathematical identifiers in the analyzed document

**HTEI — HyPlag TEI**:
A structured document format used in our hybrid plagiarism detection system HyPlag; uses a **TEI** subset and Parallel **MathML** markup

**HTTP — Hypertext Transfer Protocol**:
A request-response protocol defining the exchange of data between servers and clients via the transmission of hypertext messages

**HyPlag — Hybrid Plagiarism Detection System**:
An open-source prototype of a plagiarism detection system that combines the analysis of citations, images, mathematics, and text to identify academic plagiarism (see Chapter 6)

**InftyReader**:
Commercial software for recognizing and extracting mathematical content from unstructured documents, such as PDF, and images to structured formats, such as LaTeX and MathML (http://www.inftyreader.org)

**Intrinsic plagiarism detection**:
A paradigm specifying that approaches following the paradigm only analyze the input document for stylistic differences to identify plagiarism

**IR — Information Retrieval**:

A research field in computer science studying methods to find unstructured information relevant to an information need in extensive collections

**JATS — Journal Article Tag Suite**:

An XML format, primarily intended for encoding academic journal articles

**JSON — JavaScript Object Notation**:

A standardized, human-readable plaintext format to store and transmit data objects as attribute-value pairs and array data types

**KGA — Knowledge Graph Analysis**:

A retrieval model that represents documents as a weighted directed graph of semantic concepts typically derived from an external knowledgebase

**LaTeX**:

A markup language for typesetting documents in the homonymous document preparation system

**LaTeXML**:

A public-domain software to convert LaTeX documents to structured formats like XML, HTML, JATS, and TEI (https://dlmf.nist.gov/LaTeXML)

**LCS — Longest Common Subsequence**:

The largest number of elements occurring in a set of sequences in the same order but not necessarily at consecutive positions; often used to quantify the similarity of sequences; **Longest Common Citation Sequence (LCCS)** and **Longest Common Identifier Sequence (LCIS)** consider academic citations and mathematical identifiers as items.

**LSA — Latent Semantic Analysis**:

A method to derive descriptive semantic concepts for a collection of documents by reducing the dimensionality of the word-document occurrence matrix

**MAP — Mean Average Precision**:

A measure representing the mean of the average Precision scores a method achieves for a set of queries

**MathML — Mathematical Markup Language**:

A technical standard for representing mathematical content using XML syntax

**MathPD — Math-based Plagiarism Detection**:
An approach to identify academic plagiarism particularly in STEM disciplines by analyzing the similarity of mathematical content; see Chapter 5

**MRR — Mean Reciprocal Rank**:
A measure typically employed to evaluate methods for known-item retrieval; represents the average of the reciprocal ranks at which the method retrieves the relevant item for each query

***n*-gram**:
A contiguous sequence of $n$ items of the same type, e.g., characters, words, citations, mathematical identifiers

**NLP — Natural Language Processing**:
An interdisciplinary research field investigating automated methods to process and analyze human language

**NTCIR — NII Testbeds and Community for Information Access Research**:
A series of evaluation workshops for information access and information retrieval technology (http://ntcir.nii.ac.jp)

**OCR — Optical Character Recognition**:
A hypernym for methods to convert images of typed, handwritten or printed text into machine-readable text

**PAN — Plagiarism Analysis, Authorship Identification, and Near-Duplicate Detection**:
A series of scientific events on plagiarism detection, digital text forensics, and stylometry (https://pan.webis.de)

**Perceptual Hashing**:
A class of methods that map the perceived content of images, videos, or audio files to a fixed-size value

**PlagDet — Plagiarism Detection Score**:
A measure to evaluate the effectiveness of plagiarism detection methods; represents the harmonic mean of Precision and Recall normalized by a granularity score quantifying whether a method identified coherent plagiarism instances as multiple instances

**PMC OAS — PubMed Central Open Access Subset**:
A collection of publicly accessible biomedical research publications provided by the US National Library of Medicine
(https://www.ncbi.nlm.nih.gov/pmc/tools/openftlist)

**PoS — Part of speech**:
A category of words with similar grammatical properties, e.g., noun, pronoun, adjective, determiner, verb, adverb, preposition, conjunction

**Positional Text Matching**:
A detection method tailored to analyzing text in images by comparing both the value and position of text $n$-grams

**PrDF — Probability density function**:
A function whose value represents the likelihood of a series of outcomes for a discrete random variable

**Precision**:
A performance measure representing the fraction of retrieved items that are relevant to a query

**Ratio Hashing**:
An image-based method to identify similar bar charts by computing a descriptor containing the relative bar heights sorted in decreasing order

**Recall**:
A performance measure representing the fraction of all relevant items in a collection that were retrieved

**Reference**:
An entry in the list of cited works that is part of academic documents

**REST — Representational State Transfer**:
A software architecture style requiring to provide the functionality of web services as predefined, uniform, and stateless operations to facilitate interoperability of services on the Internet

**Retraction Watch**:
A non-profit project reporting on retractions in scientific publications in the form of a blog and a publicly accessible database (https://retractionwatch.com)

**SCA — Semantic Concept Analysis**:
A hypernym describing methods that analyze the meaning of documents by mapping the documents into a space of semantic concepts, e.g., derived from an external knowledge base

**SemEval — Semantic Evaluation**:
A series of conferences and workshops evaluating methods for computational semantic analysis (https://aclweb.org/anthology/venues/semeval)

**Sherlock**:

A text-matching tool using word $n$-gram fingerprinting with semi-random fingerprint selection

**SPM — Sequential Pattern Mining**:

A research field investigating methods to identify interesting subsequences in sequences for a large variety of applications

**SRL — Semantic Role Labeling**:

An approach to determine the semantic function of words in a sentence, e.g., actor, action, or goal, by querying linguistic databases

**STEM — Science, Technology, Engineering, and Mathematics**:

A collective term for referring to these academic disciplines

**SVD — Singular Value Decomposition**:

A matrix factorization method typically applied to reduce the dimensionality of large matrices while retaining the dominant relations in the matrix

**Synset**:

A set of synonyms that are semantically equivalent and interchangeable in many contexts

**TEI — Text Encoding Initiative**:

A professional association that maintains a homonymous XML format for document encoding

**Tf-idf — Term frequency-inverse document frequency**:

A numerical score typically computed to reflect how representative a term is for the content of a document in a collection

**TREC — Text Retrieval Conference**:

A series of evaluation workshops for information retrieval technology (https://trec.nist.gov)

**True positive**:

A relevant item a method retrieved

**VroniPlag**:

A crowdsourced project of volunteers that investigate doctoral and habilitation theses submitted to German universities for plagiarism and document the results in a wiki (https://vroniplag.wikia.org)

**VSM — Vector space model**:

A representation of documents as numeric vectors, e.g., using raw or weighted term counts as the vector elements

**WordNet**:
A lexical database of semantic relations, such as synonymy, meronymy, and hypernymy, between words in more than 200 languages (http://globalwordnet.org)

**XML — Extensible Markup Language**:
A standard for encoding documents in a format that is readable for machines and humans published by the World Wide Web Consortium

**XSLT — Extensible Stylesheet Language Transformations**:
A programming language to transforming XML documents into other XML documents or other structured document formats